 \documentclass[final,3p,sort,compress,times]{elsarticle}

\usepackage{multirow,setspace,times,amssymb,amsmath,graphicx,color,rotating,subfigure,url}
\usepackage{lineno}
\usepackage{natbib}
\usepackage{etoolbox}
\apptocmd{\sloppy}{\hbadness 10000\relax}{}{}
\usepackage[CJKbookmarks=true,colorlinks,linkcolor=red,anchorcolor=blue,citecolor=green,unicode]{hyperref} %
\usepackage{bookmark}

\journal{Physics Reports}

\begin{document}

\begin{frontmatter}



\title{Multifractal analysis of financial markets}
\author[RCE,BS]{Zhi-Qiang Jiang\fnref{cofirstauthor}}
\author[RCE,BS]{Wen-Jie Xie\fnref{cofirstauthor}}
\author[RCE,BS,SS]{Wei-Xing Zhou \corref{cor}}
\ead{wxzhou@ecust.edu.cn} %
\cortext[cor]{Corresponding author.}

\author[ETH,SFI]{Didier Sornette}
\fntext[cofirstauthor]{These authors contribute equally.}
\ead{dsornette@ethz.ch} %

\address[RCE]{Research Center for Econophysics, East China University of Science and Technology, Shanghai 200237, China}
\address[BS]{Department of Finance, School of Business, East China University of Science and Technology, Shanghai 200237, China}
\address[SS]{Department of Mathematics, School of Science, East China University of Science and Technology, Shanghai 200237, China}
\address[ETH]{Department of Management, Technology and Economics, ETH Zurich, Zurich, Switzerland}
\address[SFI]{Swiss Finance Institute, c/o University of Geneva, 40 blvd. Du Pont d'Arve, CH 1211 Geneva 4, Switzerland}

\begin{abstract}
  Multifractality is ubiquitously observed in complex natural and socioeconomic systems. Multifractal analysis provides powerful tools to understand the complex nonlinear nature of time series in diverse fields. Inspired by its striking analogy with hydrodynamic turbulence, from which the idea of multifractality originated, multifractal analysis of financial markets has bloomed, forming one of the main directions of econophysics. We review the multifractal analysis methods and multifractal models adopted in or invented for financial time series and their subtle properties, which are applicable to time series in other disciplines. We survey the cumulating evidence for the presence of multifractality in financial time series in different markets and at different time periods and discuss the sources of multifractality. The usefulness of multifractal analysis in quantifying market inefficiency, in supporting risk management and in developing other applications is presented. We finally discuss open problems and further directions of multifractal analysis.
\end{abstract}


\end{frontmatter}

\tableofcontents



\section{Introduction}
\label{S1:Intro}

\subsection{A very brief history of econophysics}

Physicists have long being interested in financial markets both practically and academically. On the investment side, Sir Isaac Newton, one of the greatest scientists in history, is reported to have invested in the South Sea Company stock in the 1720s and to have lost 20000 pounds (about 3 million US dollars in today's money)
as a result of the burst of the infamous South Sea bubble. His speculative endeavors made for the famous quote that ``I can calculate the motions of heavenly bodies, but not the madness of people.''
In 1991, Doyne Farmer co-founded the Prediction Company together with Norman Packard and James McGill to design automatic statistical arbitrage strategies. Prediction Company was quite successful and was sold to UBS in 2006 and was then re-sold to Millenium Management in 2013.
In 1994, Jean-Philippe Bouchaud and Didier Sornette founded together a research company called Science \& Finance, which was merged with Capital Fund Management in 2000. In the mean time, Sornette left Science \& Finance and Bouchaud became Chairman and Chief Scientist of Capital Fund Management. Many other examples abound of physicists flirting with Wall Street, Emmanuel Derman being a role model \cite{Derman-2004}.

Intellectually, there are deep relationships (as well as crucial differences) between physics and finance \cite{Sornette-2014-RPP,Huber-Sornette-2016-EPJst}
that have inspired generations of physicists as well as  economists. In general,
physicists apprehend financial markets as complex systems and, as such, they conducted numerous scientific investigations. In the 1930s, Ettore Majorana wrote a paper entitled ``The value of statistical laws in physics and social sciences'', which was published by Giovanni Gentile Jr in 1942 after his disappearance \cite{Majorana-1942-Scientia}, and its English translation by Mantegna was presented in the journal of Quantitative Finance in 2005 \cite{Majorana-2005-QF}.
In the 1960s, Benoit B. Mandelbrot pioneered an empirical approach, which would be later viewed as a precursor of
econophysics, in a series of seminal works of income distributions \cite{Mandelbrot-1960-IER,Mandelbrot-1961-Em,Mandelbrot-1962-QJE,Mandelbrot-1963-JPE,Mandelbrot-1963-IER}, price variation distributions of speculative assets \cite{Mandelbrot-1963-JB,Mandelbrot-1967-JB}, and long-term correlations in financial and economic time series using the rescaled range (R/S) analysis \cite{Mandelbrot-1970-Em,Mandelbrot-1971-Em}.

The current flourishing econophysics has multiple seeds that date from the 1990s.
One can identify four main directions of research that are among the most actives in the field of econophysics in the past two decades.
Following Mandelbrot's ideas,
the first stream of studies focused on the distribution of financial returns, as well as conceptual formulations.
In 1991, Rosario Mantegna studied the L{\'e}vy-walk-like superdiffusive behavior of indices of the Milan stock exchange \cite{Mantegna-1991-PA}, and with Gene Stanley the scaling behavior of the S\&P 500 index in 1995 \cite{Mantegna-Stanley-1995-Nature,Mantegna-Stanley-1996-Nature}.
In 1996, Ghashghaie et al. studied the evolving return distributions at different timescales and the multifractal behavior in the structure function of the US dollar - German mark exchange rates \cite{Ghashghaie-Breymann-Peinke-Talkner-Dodge-1996-Nature}. In 1998, Laherr{\`e}re and Sornette proposed stretched exponential distributions for time series in nature and economy as a complement to the often used power-law distributions \cite{Laherrere-Sornette-1998-EPJB}, while Gopikrishnan et al. observed the inverse cubic law \cite{Gopikrishnan-Meyer-Amaral-Stanley-1998-EPJB,Gopikrishnan-Plerou-Amaral-Meyer-Stanley-1999-PRE,Plerou-Gopikrishnan-Amaral-Meyer-Stanley-1999-PRE}. We note that, compared with power laws, stretched exponentials \cite{Malevergne-Pisarenko-Sornette-2005-QF,Malevergne-Pisarenko-Sornette-2006-AFE,Malevergne-Sornette-2006} and lognormals \cite{Malevergne-Pisarenko-Sornette-2011-PRE} are competing alternatives for return distributions that share hard-to-distinguish tail properties.

A second stream was devoted to the macroscopic modelling of financial markets. In 1994, Bouchaud and Sornette generalized the Black-Scholes option pricing problem for a large class of stochastic processes \cite{Bouchaud-Sornette-1994-JPIF} using the functional integration and derivative approach well-known in field theory.
In 1996, Sornette et al. proposed the log-periodic power-law singularity (LPPLS) model to study financial crashes modelled as critical events \cite{Sornette-Johansen-Bouchaud-1996-JPIF},. This approach was soon exploited further by Sornette and Johansen \cite{Sornette-Johansen-1997-PA,Johansen-Sornette-1999-IJMPC}, Feigenbaum and Freund \cite{Feigenbaum-Freund-1996-IJMPB,Feigenbaum-Freund-1998-MPLB}, Vandewalle et al. \cite{Vandewalle-Ausloos-Boveroux-Minguet-1998a-EPJB,Vandewalle-Boveroux-Minguet-Ausloos-1998-PA}, Dro{\.z}d{\.z} et al. \cite{Drozdz-Ruf-Speth-Wojcik-1999-EPJB}, Sornette and Zhou \cite{Sornette-Zhou-2002-QF,Zhou-Sornette-2003b-PA,Sornette-Zhou-2006-IJF,Zhou-Sornette-2006b-PA} and many others. Identification of financial bubbles and forecasting of market turning points advanced successfully in diverse markets \cite{Sornette-2003,Sornette-2003-PR,Jiang-Zhou-Sornette-Woodard-Bastiaensen-Cauwels-2010-JEBO}.

The third stream is concerned with computational econophysics. In 1992, Takayasu et al. \cite{Takayasu-Miura-Hirabayashi-Hamada-1992-PA} developed the first agent-based model in econophysics with the goal to provide an alternative to dynamical stochastic general equilibrium (DSGE) models by incorporating agents' heterogeneous characteristics and the role of extended networks. In 1994, Palmer et al. designed an artificial stock market model composed of adaptive agents with bounded rationality \cite{Palmer-Arthur-Holland-LeBaron-Tayler-1994-PD}.
In 1996, Bak et al. introduced a multi-agent model based on diffusion and annihilation \cite{Bak-Paczuski-Shubik-1997-PA}.
In 1997, Challet and Zhang introduced the minority game \cite{Challet-Zhang-1997-PA}.
In 1999, Lux and Marchesi proposed an agent-based model with heterogenous chartists and fundamentalists \cite{Lux-Marchesi-1999-Nature}.
In general, there are three basic ingredients for the price formation process in financial markets, including global news that influence
more or less all market participants, mutual imitations among traders, and idiosyncratic preferences of traders \cite{Sornette-Zhou-2006-PA,Zhou-Sornette-2007-EPJB}. Overconfidence of traders in interpreting the predictive power of global news, which mis-attributes the success of news to predict returns to traders' own stock picking ability and leads to positive feedbacks and herding behavior, plays a crucial role in generating the main stylized facts \cite{Sornette-Zhou-2006-PA,Zhou-Sornette-2007-EPJB,Sornette-2014-RPP}, such as the fat-tailed distribution of returns, the absence of linear correlations in returns, the long memory in volatilities, and the multifractal nature of financial time series.

A four stream of investigations has been involved in the more recent use of the emerging field of network theory applied to finance.
In 1999, Mantegna studied the hierarchical structure of the US stock market from the angle of minimal spanning trees \cite{Mantegna-1999-EPJB}.
Laloux et al. and Plerou et al. applied random matrix theory to stock return correlation matrices to identify the information contents hidden in eigenvalues and their corresponding eigenvectors \cite{Laloux-Cizean-Bouchaud-Potters-1999-PRL,Plerou-Gopikrishnan-Rosenow-Amaral-Stanley-1999-PRL,Plerou-Gopikrishnan-Rosenow-Amaral-Guhr-Stanley-2002-PRE}.

\subsection{Stylized facts}

\begin{figure}[tb]
\centering
  \includegraphics[width=0.33\linewidth]{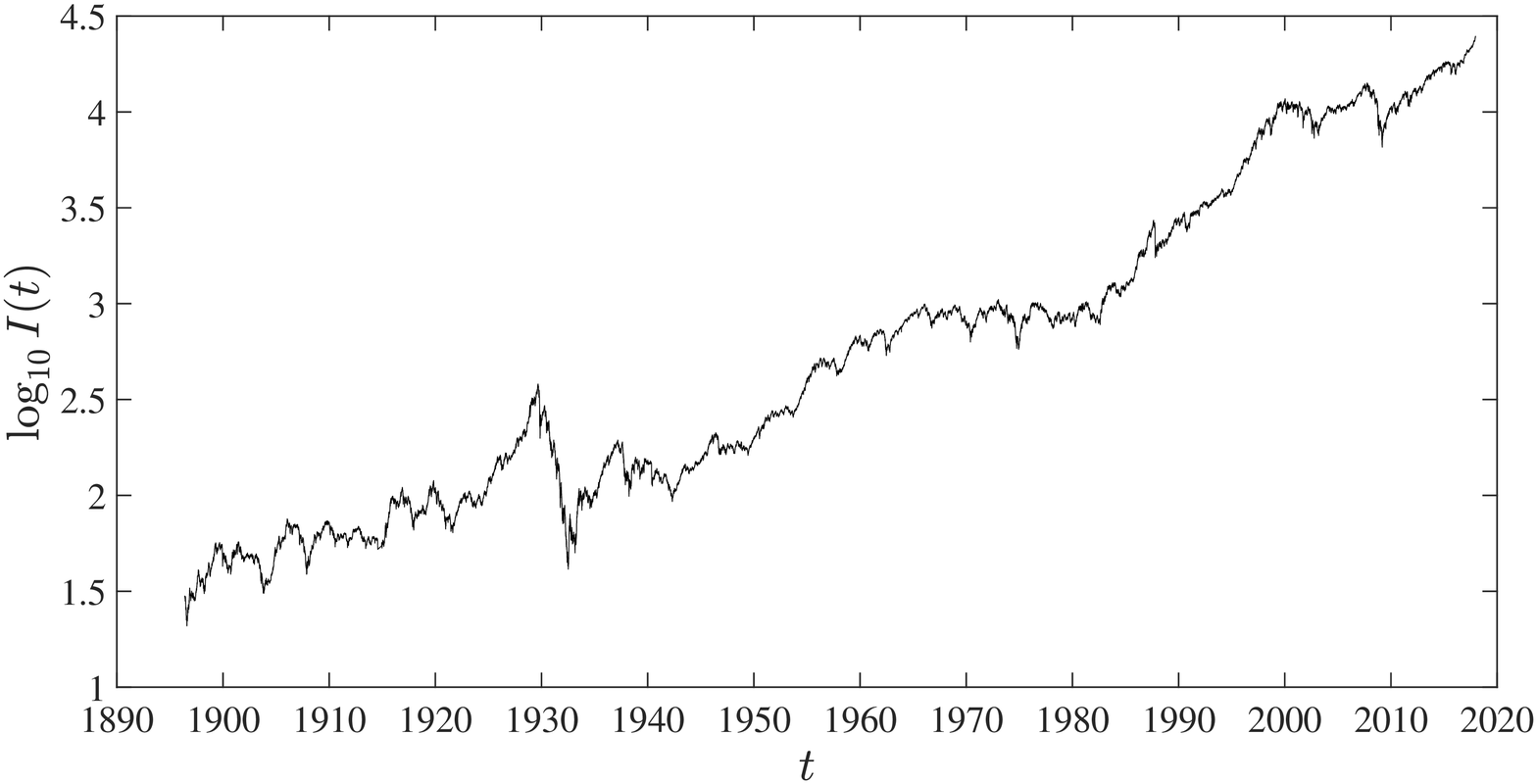}
  \includegraphics[width=0.33\linewidth]{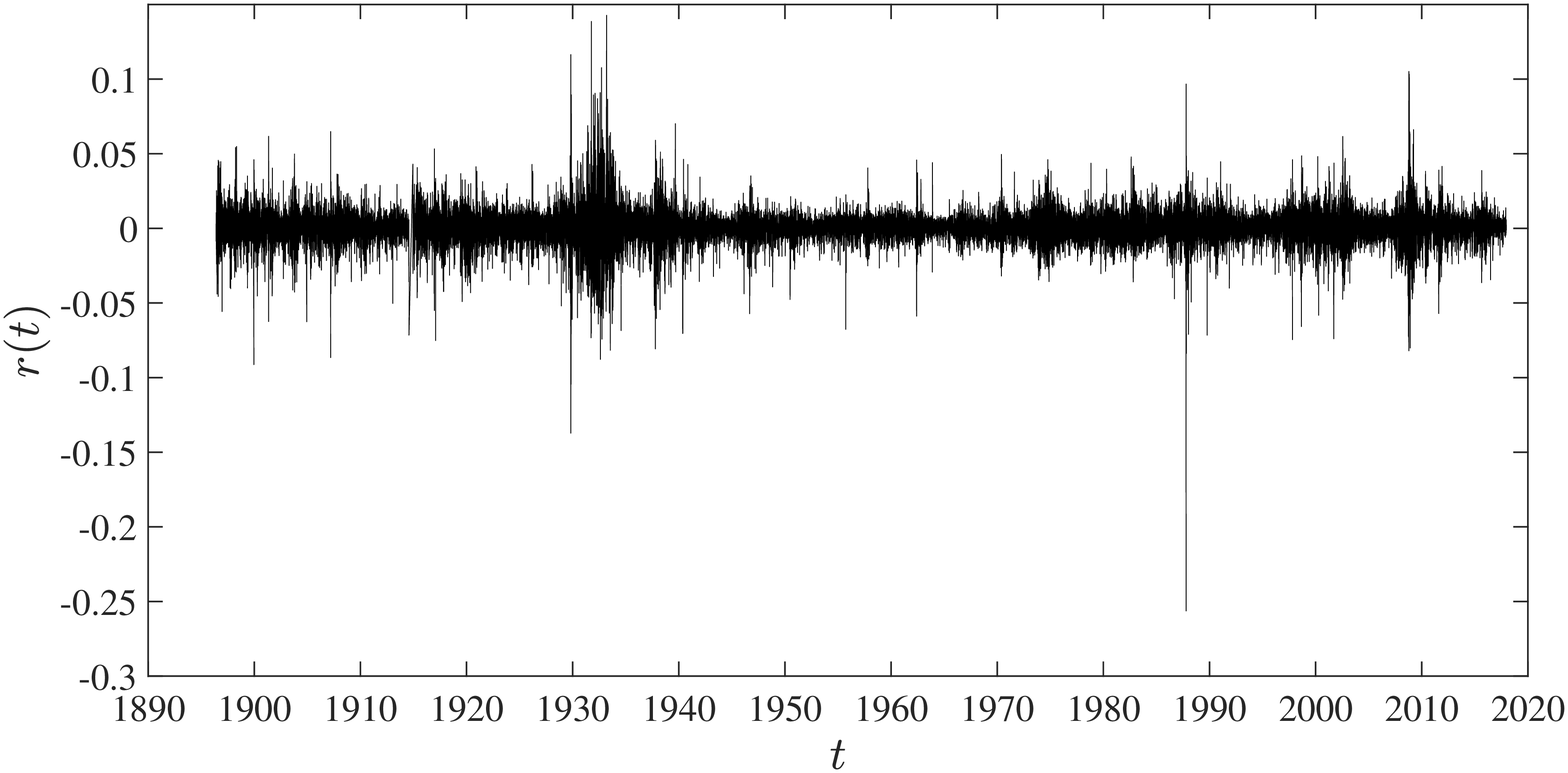}
  \includegraphics[width=0.33\linewidth]{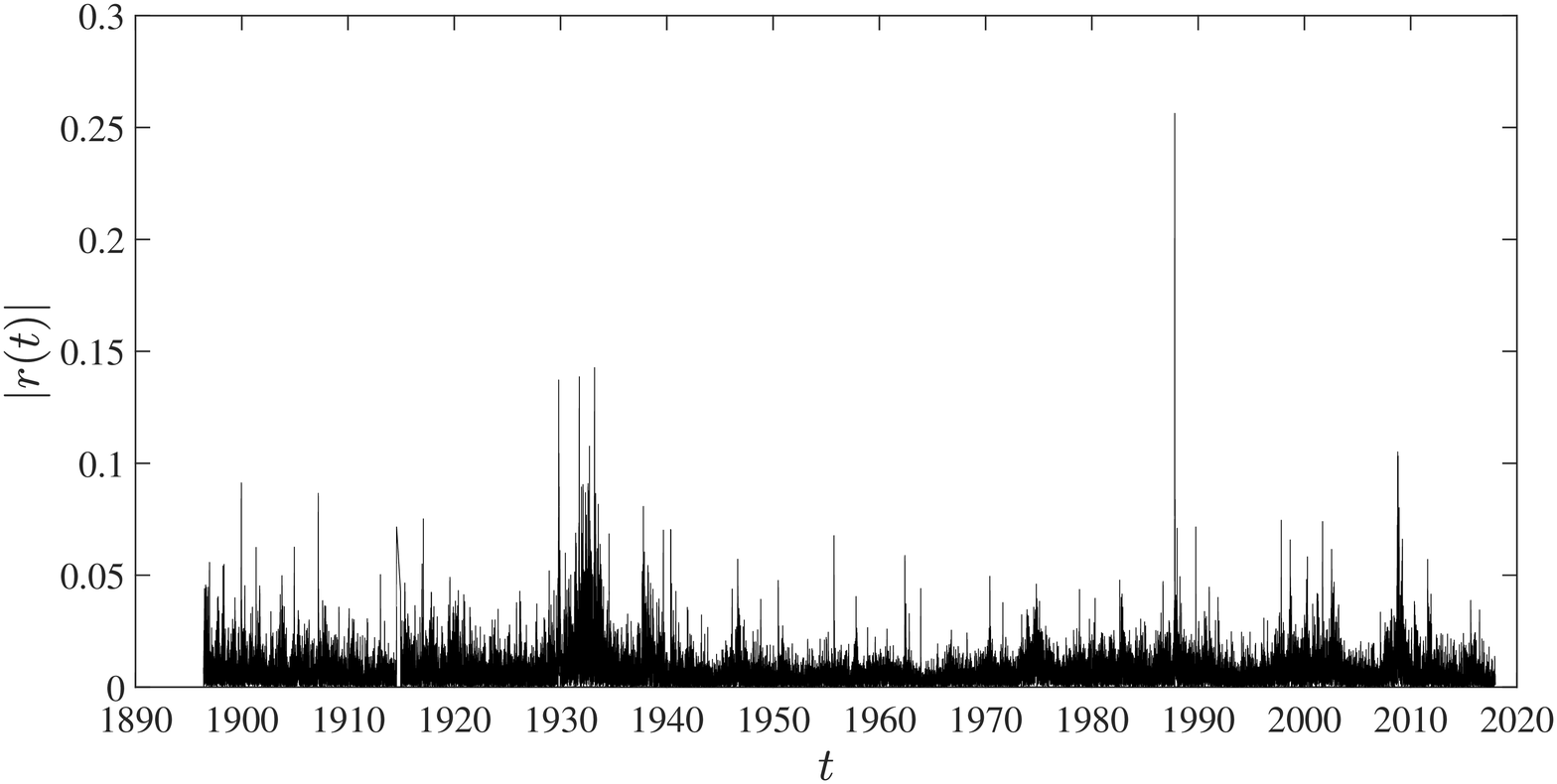}
  \includegraphics[width=0.33\linewidth]{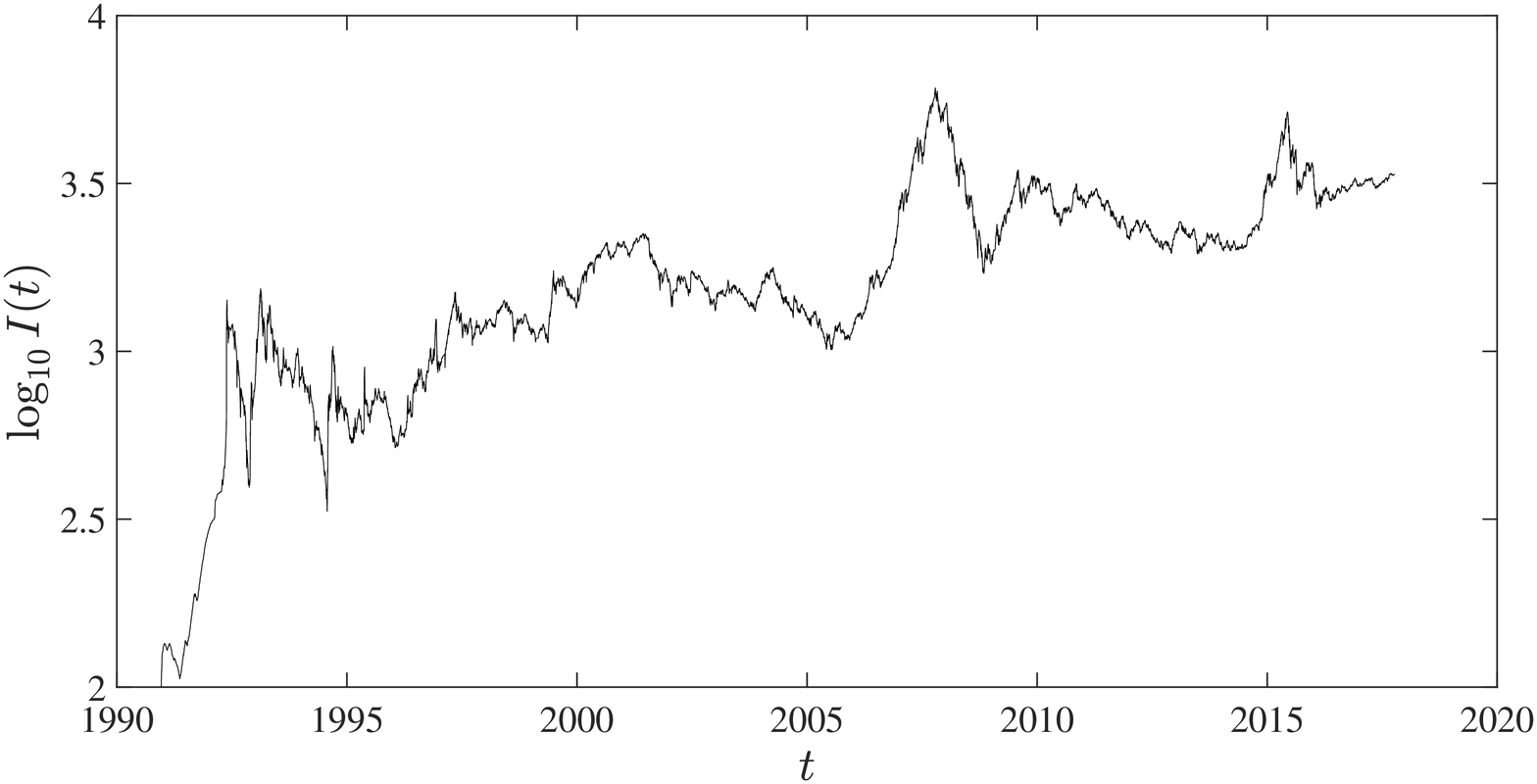}
  \includegraphics[width=0.33\linewidth]{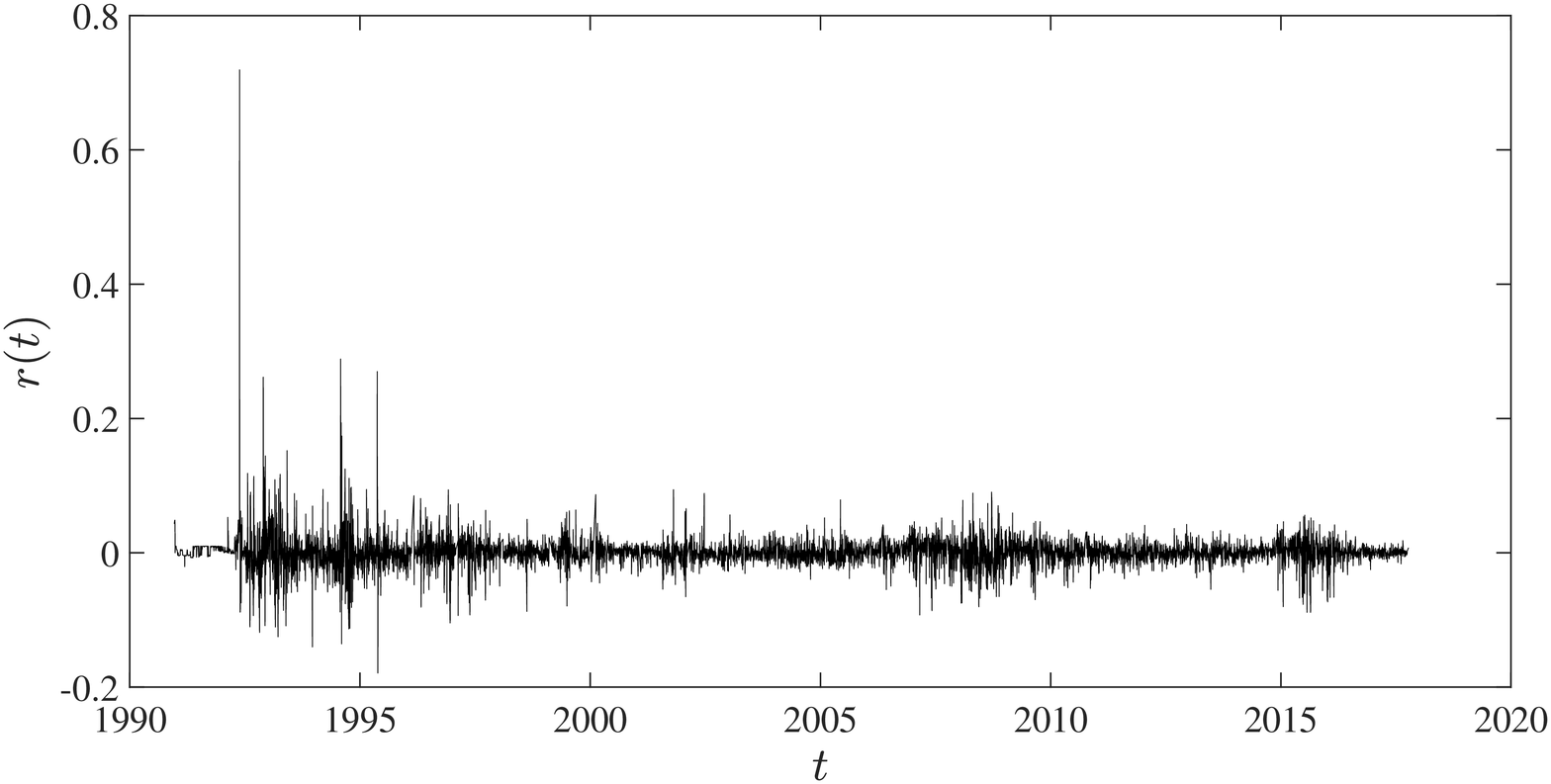}
  \includegraphics[width=0.33\linewidth]{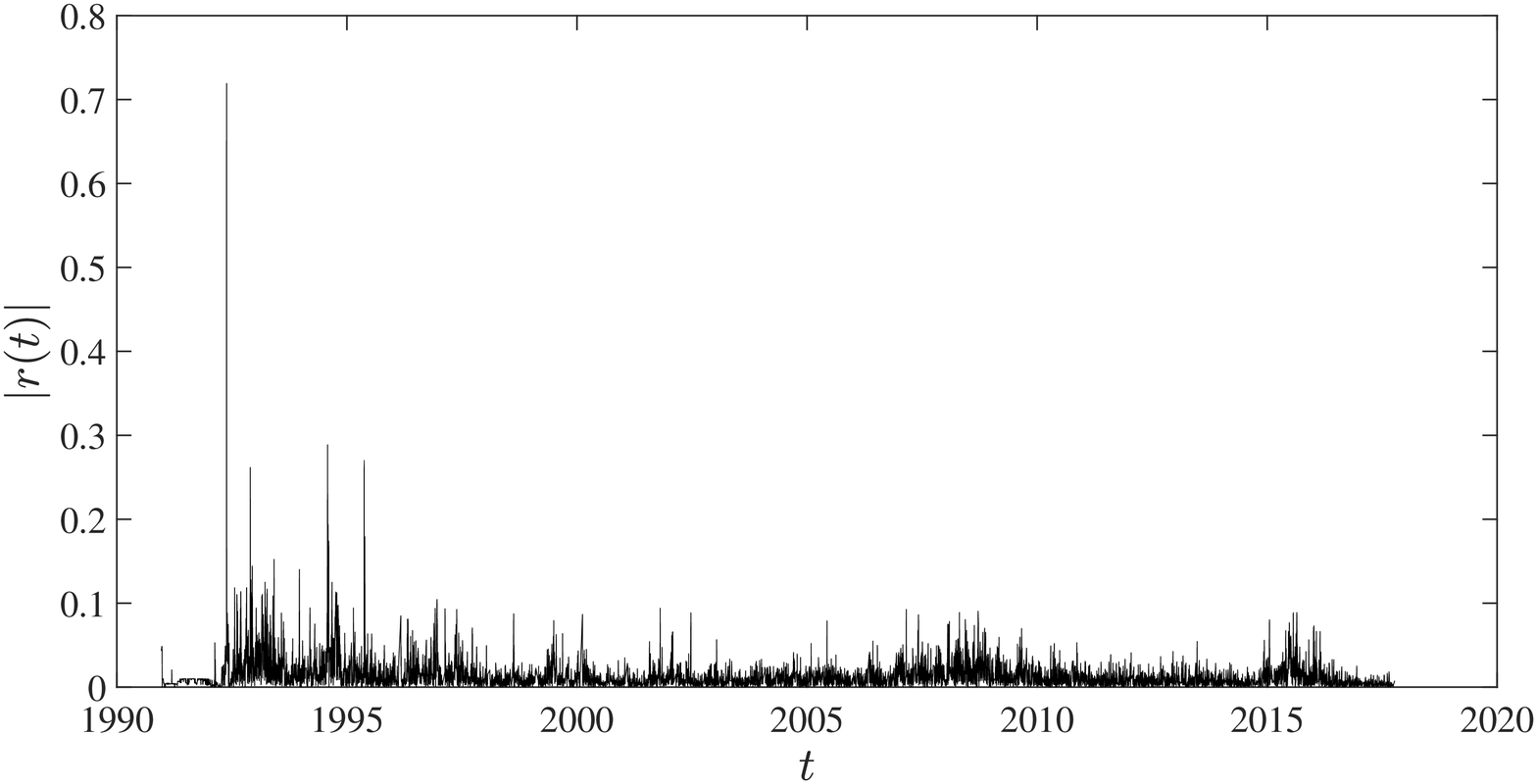}
  \caption{Time series of daily closing prices (left), daily returns (middle) and daily volatilities (right) of two stock market indices. The upper panel is for the Dow Jones Industrial Average index and the lower panel is for the Shanghai Stock Exchange Composite index.}
  \label{Fig:StylizedFacts:TimeSeries}
\end{figure}

Financial markets are complex adaptive systems, in which universal and non-universal statistical laws emerge at the macroscopic level from the microscopic behavior of heterogeneous traders through reactions to external stimuli in a self-organized manner \cite{Mantegna-Stanley-2000,Bouchaud-Potters-2000,Sornette-2003,Sornette-2003-PR,Challet-Marsili-Zhang-2005,Zhou-2007,Zhang-Zhang-Xiong-2010,Huang-2015-PR}.
These laws can be explored from the rich dataset of financial stock market price time series.
As an illustration, the left pannel of Fig.~\ref{Fig:StylizedFacts:TimeSeries} shows the daily price trajectories $I(t)$ of two stock market indices, the Dow Jones Industrial Average (DJIA) index from 26 May 1896 to 29 December 2017 and the Shanghai Stock Exchange Composite (SSEC) index from 19 December 1990 to 29 December 2017. These two stock markets are representative of developed and emerging markets that are of great interest.
The logarithmic return of $I(t)$ over a time interval $\Delta{t}$ is defined by
\begin{equation}
 r_{\Delta{t}}(t)=\ln I(t)- \ln I(t-\Delta{t}).
 \label{Eq:Return}
\end{equation}
The middle panel of Fig.~\ref{Fig:StylizedFacts:TimeSeries} illustrates the daily return time series $r(t)$ with $\Delta{t}=1$ day and the right panel shows the corresponding volatility time series. Here, the volatility for a given day is defined as the square root of the sum of
the squares of intraday returns of that day, where
the intraday time scale needs to be sufficiently short (say 1 minute) to provide a good  convergence of the volatility estimator.
Clear bursty behaviors can be observed in the return time series, which are associated with volatility clustering.

Numerous statistical regularities or stylized facts have been reported in different financial markets \cite{Cont-2001-QF,Chakraborti-Toke-Patriarca-Abergel-2011a-QF,Gould-Porter-Williams-McDonald-Fenn-Howison-2013-QF,Liang-Yang-Huang-2013-FoP}. We first recall three main stylized facts, which are universal properties of financial markets. The first stylized fact is the fat-tailed distributions of financial returns.
The form of the distribution of asset price fluctuations plays crucial roles in asset pricing and risk management \cite{Mantegna-Stanley-2000,Bouchaud-Potters-2000,Malevergne-Sornette-2006}. Early empirical and theoretical works argued that asset prices follow geometric Brownian motions \cite{Bachelier-1900,Osborne-1959a-OR}, which is the main assumption of the Black-Scholes option pricing model \cite{Black-Scholes-1973-JPE}. In the 1960s, Mandelbrot suggested that incomes and speculative price returns follow the Pareto-L{\'{e}}vy distribution, which has power-law tails
\begin{equation}
 \Pr(|r|) \sim |r|^{-\gamma-1},
 \label{Eq:PL}
\end{equation}
whose exponents $1<\gamma<2$ claimed to be in the domain of attraction of stable L{\'{e}}vy laws
 \cite{Mandelbrot-1960-IER,Mandelbrot-1961-Em,Mandelbrot-1963-IER,Mandelbrot-1963-JPE,Mandelbrot-1963-JB}.
 Recall that L{\'{e}}vy laws generalise the Gaussian distribution, and their family exhausts all possible distributions that are stable
 under convolution, with no or only weak dependence between the random variables.

The possibility that financial returns might be distributed according to L{\'{e}}vy laws rather than being Gaussian
immediately attracted the interest of notable financial economists, such as Fama \cite{Fama-1965-JB}, 
Fama and Roll \cite{Fama-Roll-1968-JASA},
Samuelson \cite{Samuelson-1967-JFQA}, Sargent \cite{Blattberg-Sargent-1971-Em},
and others. But there was also immediate resistance to abandon
the statistical theory that exists for the normal case, which was non-existent for the other members
of the L\'evy laws, with the like of Cootner \cite{Cootner-1964} and Granger and Orr \cite{Granger-Orr-1972-JASA}
opposing strong arguments. Further in-depth studies
concluded to the error of Mandelbrot, in the sense that, while the distributions of returns are fat-tailed, they
are not so heavy as to be described by L\'evy laws: more and more statistical tests
brought increasing and unavoidable evidence that the variance of the return is not infinite (and thus
the exponent $\gamma$ is larger than $2$), thus
irremediably excluding the heavy tail regime of L\'evy laws with tail exponent less than $2$ \cite{Officer-1972-JASA,Hsu-Miller-Wichern-1974-JASA,Fama-1976,MacKenzie-2008}.
The tools based on variance and covariance could still be used after all.
This led to a lull in the interest of fat-tailed distributions, which was interrupted by the
re-discovery by econophysicists in the early 1990s \cite{Mantegna-1991-PA,Mantegna-Stanley-1995-Nature,Mantegna-Stanley-1996-Nature}.

More recent empirical investigations with more data converge to  $\gamma\approx3$ (or more prudently that $2 < \gamma <4$)
\cite{Gopikrishnan-Meyer-Amaral-Stanley-1998-EPJB,Gopikrishnan-Plerou-Amaral-Meyer-Stanley-1999-PRE,Plerou-Gopikrishnan-Amaral-Meyer-Stanley-1999-PRE}.
Further empirical investigations have been conducted on financial returns at different time scales $\Delta{t}$ and reported that distributions vary from exponential to stretched exponential to power law
\cite{Laherrere-Sornette-1998-EPJB,Makowiec-Gnacinski-2001-APPB,Bertram-2004-PA,Matia-Pal-Salunkay-Stanley-2004-EPL,Yan-Zhang-Zhang-Tang-2005-PA,Coronel-Hernandez-2005-RMF,Qiu-Zheng-Ren-Trimper-2007-PA,Drozdz-Forczek-Kwapien-Oswicimka-Rak-2007-PA,Pan-Sinha-2007-EPL,Pan-Sinha-2008-PA,Tabak-Takami-Cajueiro-Petitiniga-2009-PA,Eryigit-Cukur-Eryigit-2009-PA,Jiang-Li-Cai-Wang-2009-PA,Queiros-2005-QF,Queiros-Moyano-deSouza-Tsallis-2007-EPJB,Zhang-Zhang-Kleinert-2007-PA,Gu-Chen-Zhou-2008a-PA,Fuentes-Gerig-Vicente-2009-PLoS1,Gerig-Vicente-Fuentes-2009-PRE}.
These observations are in agreement with the fact that the fat-tailed nature of the distributions of returns at small scale with $\gamma>2$ implies
that the domain of attraction at large scales is the Gaussian distribution. One should thus observe and we do observe a
progressive transition from a power law like regime at small time scales to a Gaussian regime at large scales \cite{Ghashghaie-Breymann-Peinke-Talkner-Dodge-1996-Nature}. Note that the stretched exponential distribution serves as a convenient bridge between exponential and power-law distributions \cite{Laherrere-Sornette-1998-EPJB,Sornette-2004}. In fact, the power law family of distributions can be shown to be
asymptotically nested in the stretched exponential family, the later converging to the former in the limit of vanishing stretched
exponential exponent \cite{Malevergne-Pisarenko-Sornette-2005-QF,Malevergne-Pisarenko-Sornette-2006-AFE,Malevergne-Sornette-2006}.
This apparently esoteric property is very useful practically as it provides the robust Wilks test to compare power laws and
stretched exponentials for any data set of interest.

\begin{figure}[tb]
\centering
  \includegraphics[width=0.33\linewidth]{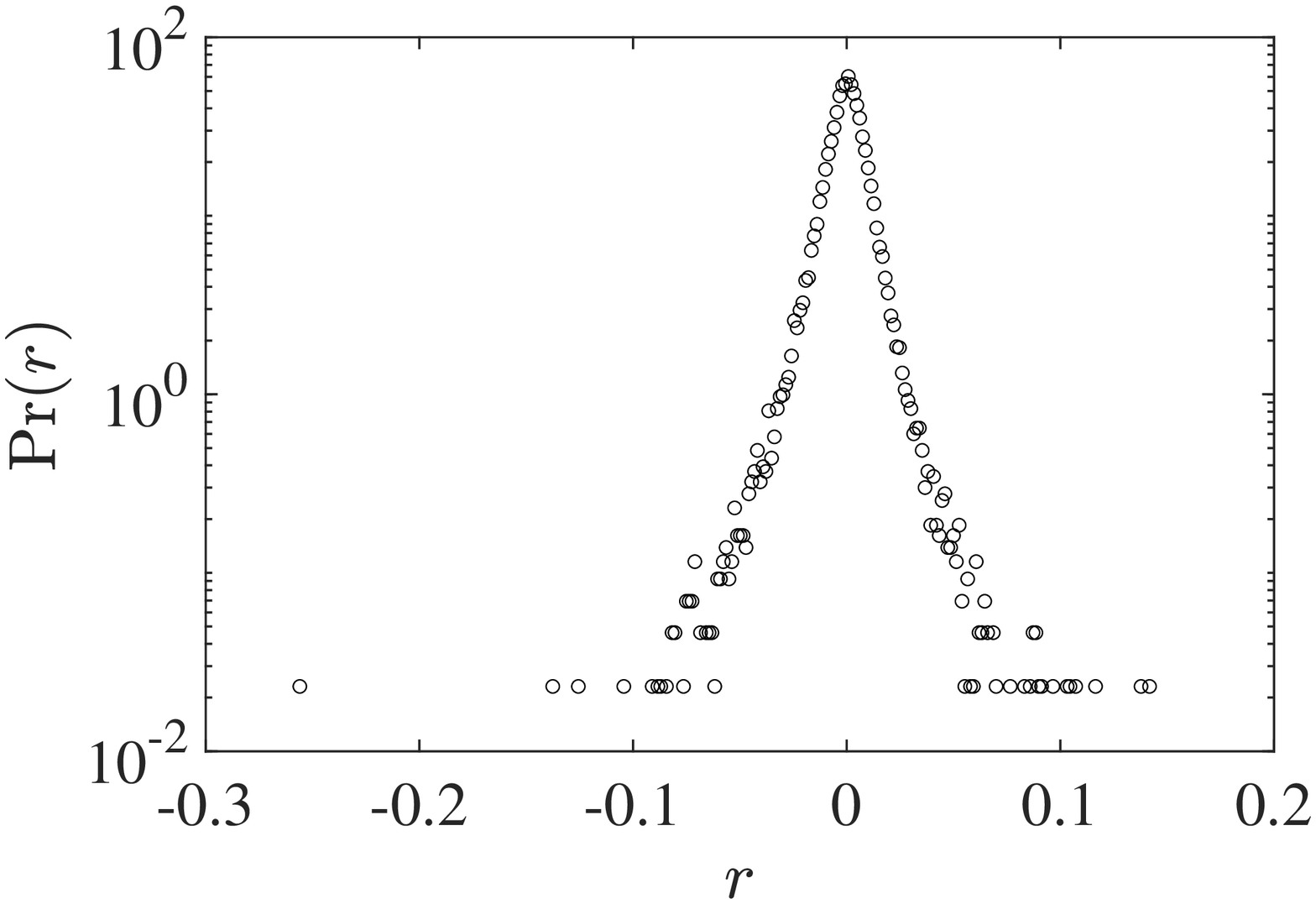}
  \includegraphics[width=0.33\linewidth]{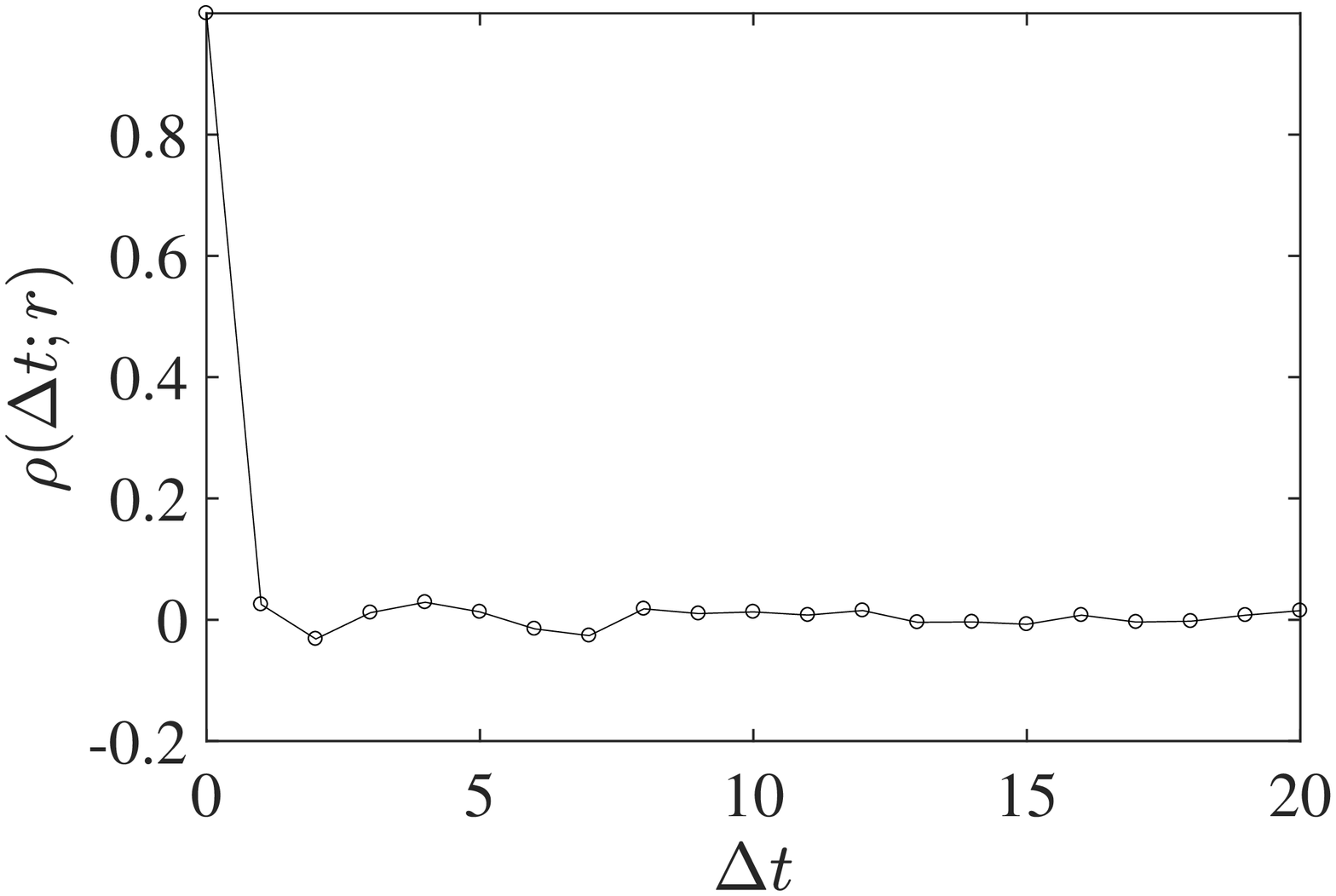}
  \includegraphics[width=0.33\linewidth]{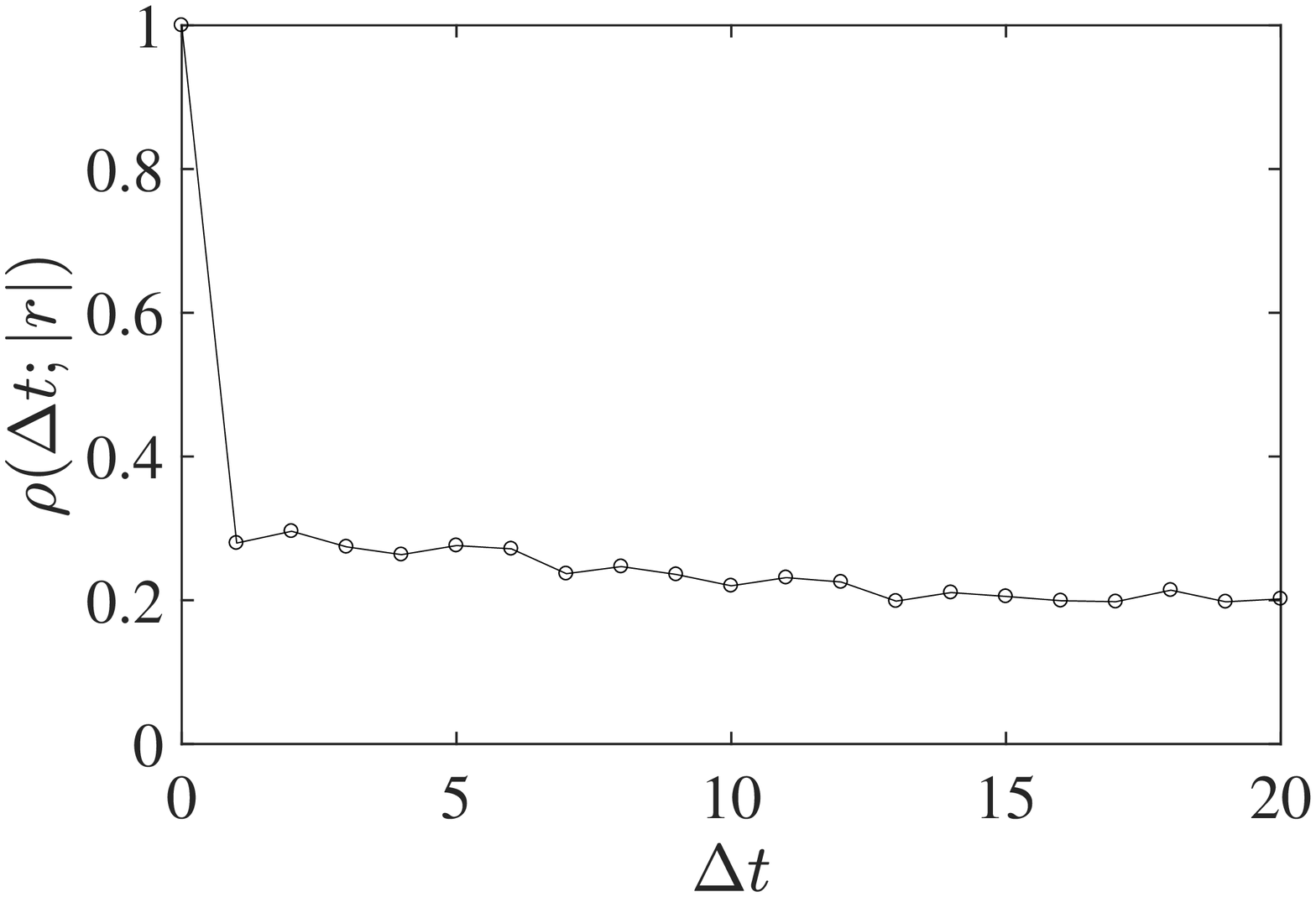}
  \includegraphics[width=0.33\linewidth]{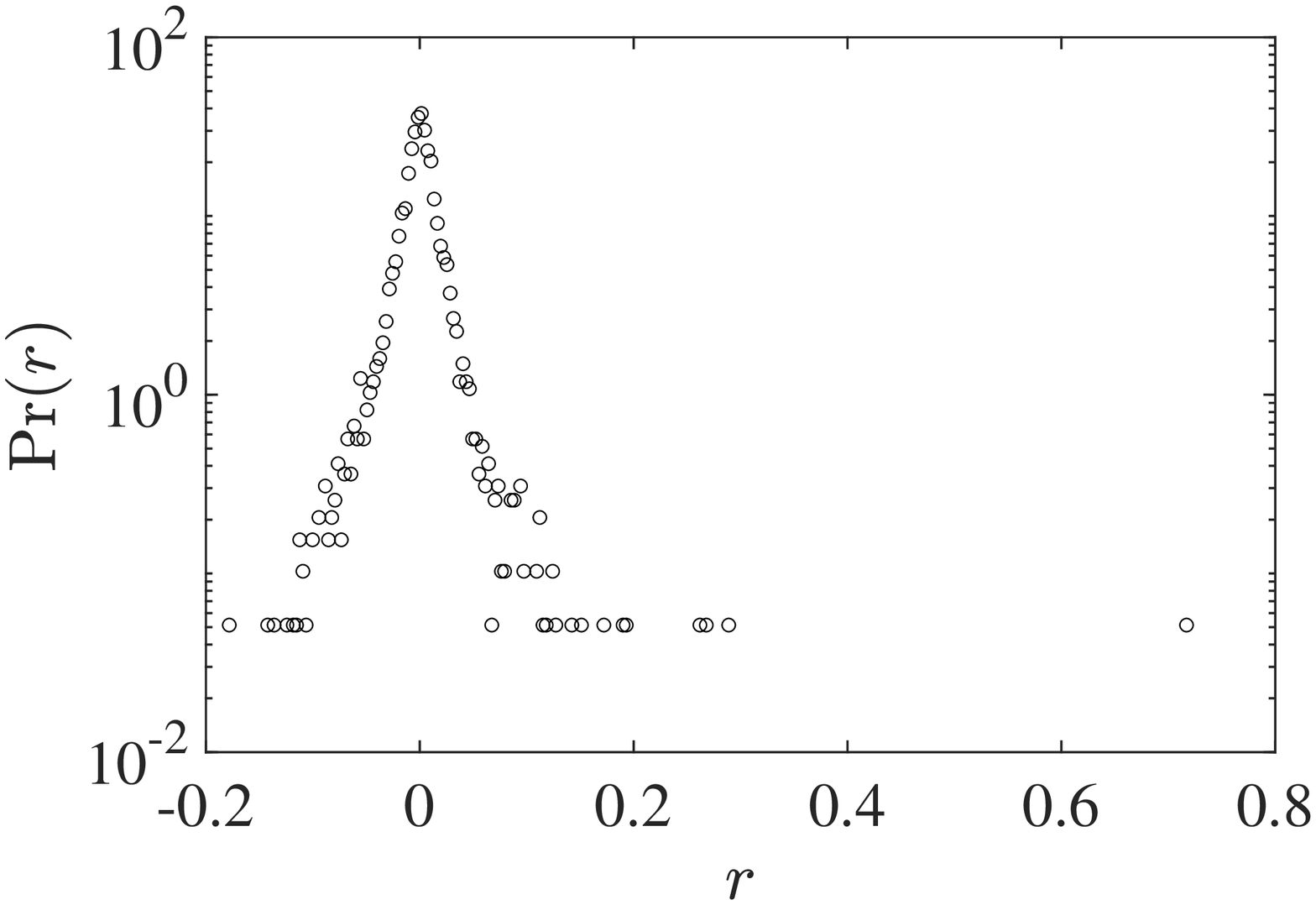}
  \includegraphics[width=0.33\linewidth]{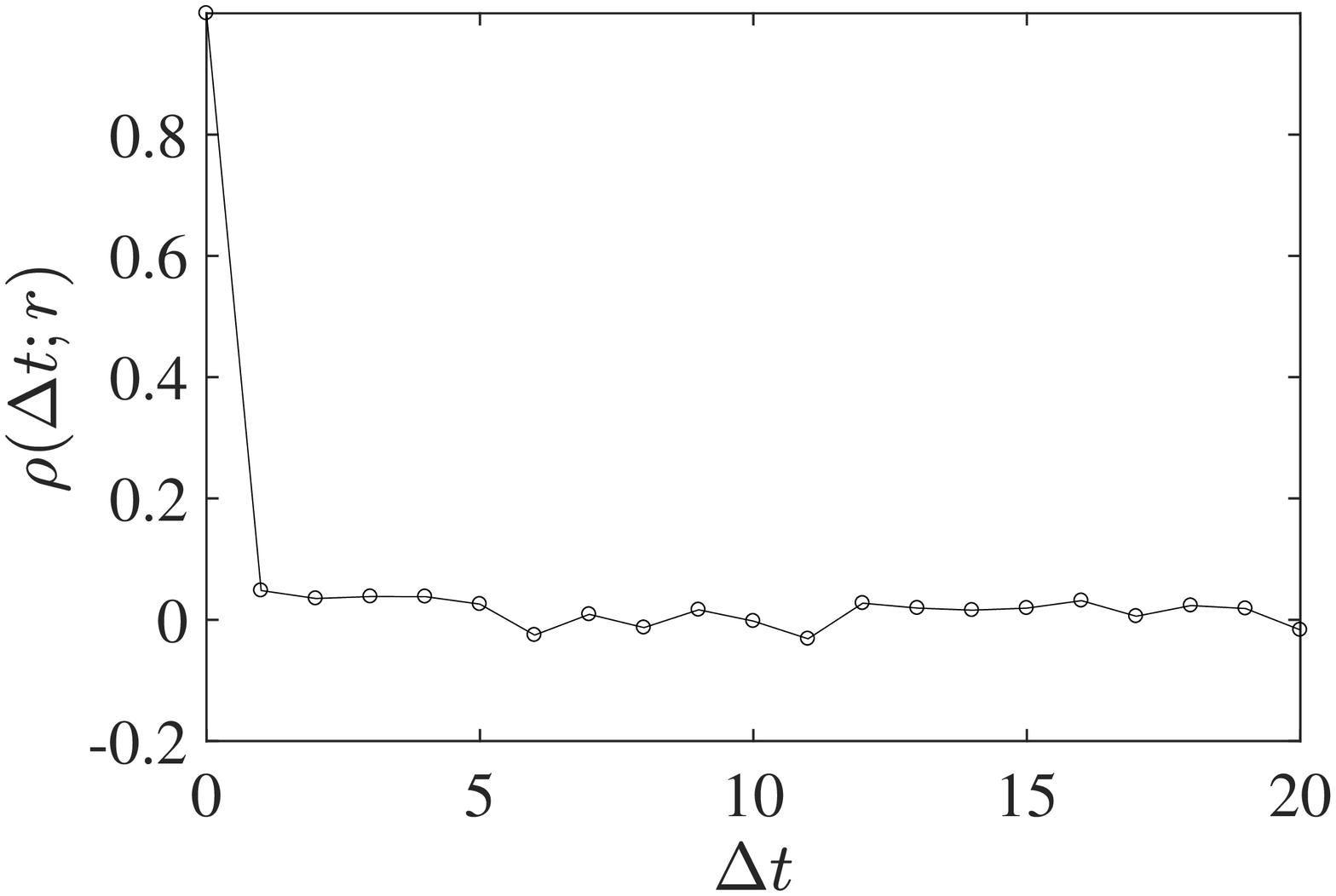}
  \includegraphics[width=0.33\linewidth]{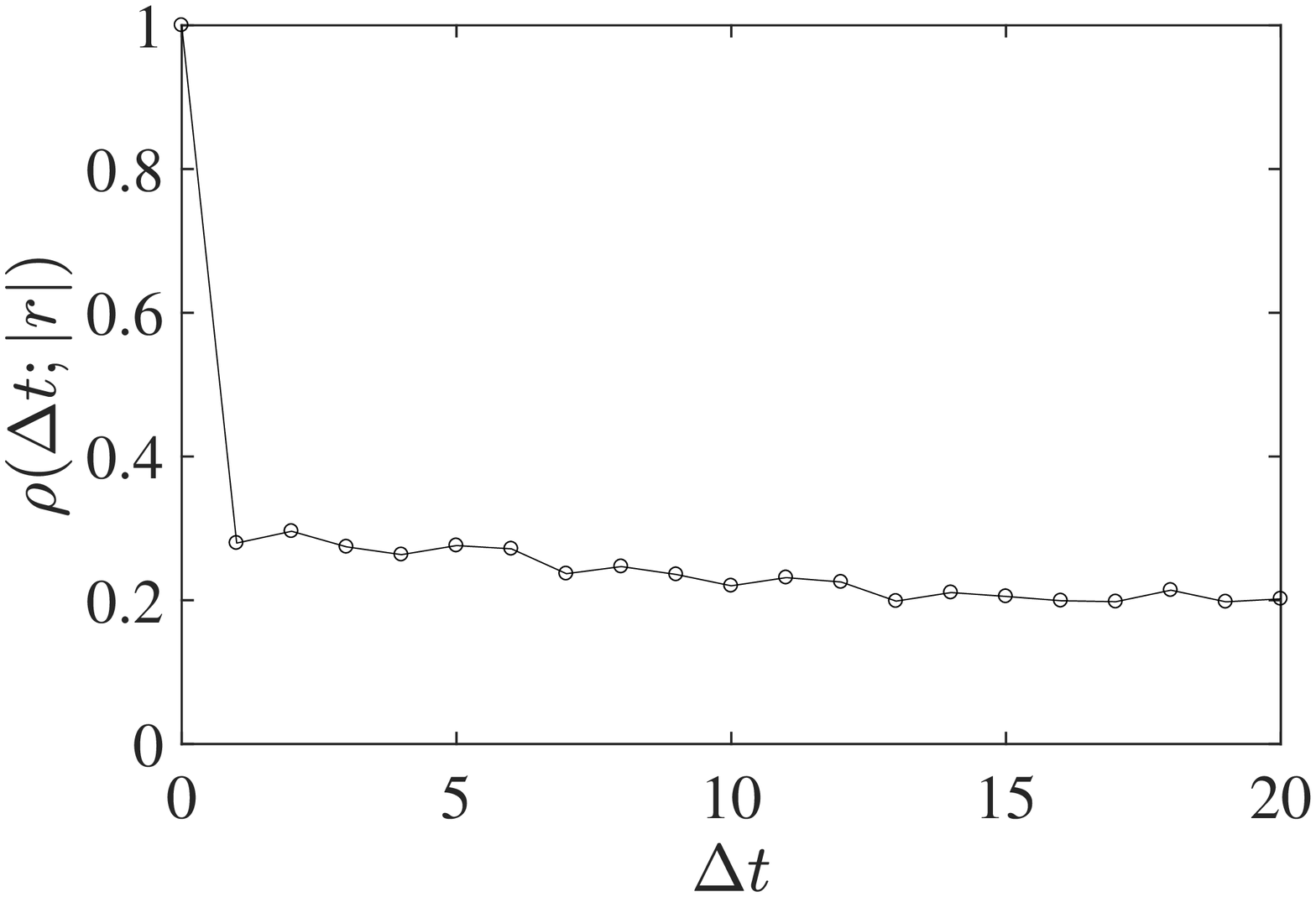}
  \caption{Three main stylized facts: Fat-tailed distribution of returns (left), absence of autocorrelation of returns (middle), and long-range correlations of volatilities (right). The upper panel is for DJIA and the lower panel is for SSEC.}
  \label{Fig:StylizedFacts:TimeSeries:3}
\end{figure}

As illustrations of the above, the left column of Fig.~\ref{Fig:StylizedFacts:TimeSeries:3} shows the empirical distributions of daily returns of the DJIA and SSEC indices. There are evident fat tails and outliers, compared with the inverted parabola shape that would be expected
under the Gaussian hypothesis.

Two other main stylized facts are the absence of long memory in returns and the presence of long memory in volatility, as shown respectively in the middle and right panels of Fig.~\ref{Fig:StylizedFacts:TimeSeries:3}. Quantitatively, performing detrended fluctuation analysis or detrending moving average analysis of the return and volatility time series allows one to obtain the corresponding Hurst exponents.

\subsection{Benchmark econometric models}
\label{S2:Models:Econometric}

In econometric finance, there are several cornerstone models, such as the autoregressive conditional heteroskedasticity (ARCH) model \cite{Engle-1982-Em}, the generalized autorregressive conditional heteroskedasticity (GARCH) model \cite{Bollerslev-1986-JEm}, the exponential GARCH (EGARCH) model \cite{Nelson-1991-Em}, the integrated GARCH (IGARCH) model \cite{Engle-Bollerslev-1986-EmR}, the fractionally integrated generalized autoregressive conditional heteroskedasticity (FIGARCH) model \cite{Baillie-Bollerslev-Mikkelsen-1996-JEm}, and the autoregressive fractionally integrated moving average (ARFIMA) model \cite{Granger-Joyeux-1980-JTSA,Hosking-1981-Bm}.

An autoregressive model of order $p$, AR($p$), is defined as
\begin{equation}
  r(t) = a_0 + \sum_{k=1}^p a_kr(t-k) + \epsilon_t,
  \label{Eq:AR}
\end{equation}
where $a_1, \cdots, a_p$ are the parameters of the model, $a_0$ is a constant, and $\epsilon_t$ is white noise.
To model a financial time series using an ARCH process \cite{Engle-1982-Em}, the error term $\epsilon _{t}$ is split into a stochastic piece $z_{t}$ and a time-dependent standard deviation $\sigma_t$ such that
\begin{equation}
  \epsilon _{t}=\sigma_tz_{t}.
  \label{Eg:ARCH:eps}
\end{equation}
The random variable $z_{t}$ is a strong white noise process. The series $\sigma_t^{2}$ is modelled by
\begin{equation}
  \sigma_{t}^{2}
  =\alpha _{0}+\alpha _{1}\epsilon _{t-1}^{2}+\cdots +\alpha _{q}\epsilon _{t-q}^{2}
  =\alpha _{0}+\sum _{i=1}^{q}\alpha _{i}\epsilon _{t-i}^{2},
  \label{Eg:ARCH:sigma}
\end{equation}
where $\alpha_0>0$ and $\alpha _{i}\geq 0$ for $i>0$.

Considering a regression with coefficient $b$ between a variable $x_t$ and $y_t$,
the GARCH$(p, q)$ model of the residuals is given by \cite{Bollerslev-1986-JEm}
$$y_{t}=b x_{t}+\epsilon_t$$
$$\epsilon _{t}|\psi _{t}\sim {\mathcal {N}}(0,\sigma_t^{2})$$
\begin{equation}
\sigma_t^{2}=\omega +\alpha _{1}\epsilon _{t-1}^{2}+\cdots +\alpha _{q}\epsilon _{t-q}^{2}+\beta _{1}\sigma _{t-1}^{2}+\cdots +\beta _{p}\sigma _{t-p}^{2}=\omega +\sum _{i=1}^{q}\alpha _{i}\epsilon _{t-i}^{2}+\sum _{i=1}^{p}\beta _{i}\sigma _{t-i}^{2}~,
\label{rthj3ythgwq}
\end{equation}
where $p$ is the order of the GARCH terms $\sigma^2$ and $q$ is the order of the ARCH terms $\epsilon ^{2}$.
In finance, one usually applied the GARCH model to the return time series formulated as (\ref{Eg:ARCH:eps}) with
$\sigma$ given by (\ref{rthj3ythgwq}).

In the ARCH, GARCH and EGARCH models, the autocorrelation of volatility decays at exponential rates. In contrast, the volatility of the integrated models has long memory. Empirical analyses show that GARCH-type models without long memory in volatility can hardly capture the multifractal nature in financial time series, while other models with long memory component can capture partly the apparent multifractality, which is however spurious by definition of the integrated models \cite{Schmitt-Schertzer-Lovejoy-1999-ASMDA,Calvet-Fisher-2002-RES,Jiang-Zhou-2011-PRE,Lv-Shan-2013-PA,Wang-Wu-2013-CE, Zhao-Shang-Shi-2014-PA,Oswiecimka-Drozdz-Forczek-Jadach-Kwapien-2014-PRE,Wang-Yang-Wang-2016-PA, Thompson-Wilson-2016-MCS,He-Wang-2017-PA}.
Therefore, benchmark economic models fail to capture an important dimension of market complexity, namely the multifractal nature of financial market  \cite{Mandelbrot-1999-SA,Buonocore-Musmeci-Aste-DiMatteo-2016-EPJst}.

\subsection{Introduction to multifractality}


Multifractality has been observed in many diverse complex systems \cite{Paladin-Vulpiani-1987-PR,McCauley-1990-PR,Coleman-Pietronero-1992-PR,Janssen-1998-PR,Olemskoi-Klepikov-2000-PR,Touchette-2009-PR,Kwapien-Drozdz-2012-PR}, including financial markets \cite{Mandelbrot-1999-SA}. The term ``multifractal'' was coined by Frisch and Parisi in 1983 \cite{Frisch-Parisi-1985}, as confirmed by Benoit B. Mandelbrot \cite{Mandelbrot-1989-PAG}, who is the well respected Father of Fractals \cite{Mandelbrot-1983}. The concept of multifractality can be traced back to Novikov \cite{Novikov-1971-JAMM} and Mandelbrot \cite{Mandelbrot-1972-LNP,Mandelbrot-1974-JFM} in their study of turbulence in fluid mechanics. On the other hand, two groups led by Grassberger and Procaccia generalized the fractal dimension, the information dimension and the correlation dimension into a unified expression of generalized dimensions through the R{\'e}nyi entropy \cite{Grassberger-1983-PLA,Hentschel-Procaccia-1983-PD,Grassberger-1985-PLA}, with the initial aim of describing strange attractors in nonlinear dynamics \cite{Grassberger-Procaccia-1984-PD}. Halsey et al. also made seminal contributions in the early development of multifractal theories, with applications to fractal growth processes  \cite{Halsey-Jensen-Kadanoff-Procaccia-Shraiman-1986-PRA}.


%
%

There are two equivalent descriptions of multifractals, {\textit{i.e.}}, via the function $\tau(q)$ or the function $f(\alpha)$, where $q$ is the order of certain moments, $\tau(q)$ is the mass exponent function, $\alpha$ is the singularity strength, and $f(\alpha)$ is the singularity spectrum. These two
representations are related through the Legendre transform \cite{Frisch-Parisi-1985,Halsey-Jensen-Kadanoff-Procaccia-Shraiman-1986-PRA} such that
\begin{subequations}
\begin{equation}
  \alpha = {{\mbox{d}}\tau(q)}/{{\mbox{d}}q}
  \label{Eq:MF:Intro:alpha:dtau:dq}
\end{equation}
and
\begin{equation}
  f(\alpha) =  q\alpha -\tau(q).
  \label{Eq:MF:Intro:f:alpha}
\end{equation}
\label{Eq:MF:LegendreTrandform}
\end{subequations}
Concerning the $\tau(q)$ representation, there are two additional equivalent functions $D_q$ and $H(q)$. The $D_q$ function presents the generalized dimensions that are determined by
\begin{equation}
    D_q = \lim_{q'\to{q}}\frac{\tau(q')}{(q'-1)},
    \label{Eq:MF:Intro:tau:Dq}
\end{equation}
while $H(q)$ is defined by
\begin{equation}
    H(q) = \lim_{q'\to{q}}\frac{\tau(q')+1}{q'},
    \label{Eq:MF:Intro:tau:Hq}
\end{equation}
and corresponds to the generalized Hurst exponents.
The detailed meaning of these quantities and there relations will be made clear later.

In the study of turbulence, there are two classical methods for the analysis of multifractals, {\it{i.e.}}, the structure function approach and the partition function approach \cite{Kolmogorov-1962-JFM,VanAtta-Chen-1970-JFM,Novikov-1971-JAMM,Mandelbrot-1972-LNP,Mandelbrot-1974-JFM,Anselmet-Gagne-Hopfinger-Antonia-1984-JFM,Frisch-Parisi-1985,Benzi-Paladin-Parisi-Vulpiani-1984-JFM}. Due to the striking similarities between turbulence and financial markets, though the analogy has its limitations \cite{Arneodo-Bouchaud-Cont-Muzy-Potters-Sornette-1996-XXX}, the multifractal nature of financial time series has attracted much interest \cite{Ghashghaie-Breymann-Peinke-Talkner-Dodge-1996-Nature,Mantegna-Stanley-1996-Nature}. The structure function approach dominated in the first wave of multifractal analysis in econophysics \cite{DiMatteo-2007-QF}. Since the seminal work of Kantelhardt et al. in 2002 \cite{Kantelhardt-Zschiegner-KoscielnyBunde-Havlin-Bunde-Stanley-2002-PA}, the multifractal detrended fluctuation analysis soon became the dominant method not only for financial time series but also for other time series. In 2008, Podobnik and Stanley introduced the detrended cross-correlation analysis for non-stationary time series \cite{Podobnik-Stanley-2008-PRL}, which was extended by one of us (Wei-Xing Zhou) to analyze multifractal time series \cite{Zhou-2008-PRE}. These milestone works triggered the third wave of simultaneous multifractal analysis of two time series and the invention of many variant types of multifractal analyses.

In financial markets, risk is a central concept in all financial activities. Large financial fluctuations, especially finance ``tsunamis'' such as the one
that broke out in 2008 \cite{Jiang-Zhou-Sornette-Woodard-Bastiaensen-Cauwels-2010-JEBO}, usually cause tremendous economic distress and
are followed by vigorous changes in risk perception and regulations \cite{Sornette-2003,Sornette-2003-PR}. Identifying, measuring and forecasting financial risks are of great theoretical and practical significance in risk management. Econophysicists have shown that multifractal analysis provides an alternative way in studying market risks. Indeed, large singularity strengths qualify the behavior of small fluctuations, while small singularity strengths characterize large fluctuations. Therefore, quite a few efforts have been made to apply multifractal analysis to quantifying market inefficiency, measuring financial volatility, and so on.

This review provides an extensive survey of the multifractal analysis of financial time series. We start with different methods of multifractal analysis for univariate and multivariate time series in Section \ref{S1:MF} and Section \ref{S1:MF-X}. We review also the methods that were invented in other fields,
because they have potential application value in econophysics. We discuss important mathematical and econophysical models in Section \ref{S1:Models} that can deepen our understanding of the multifractal nature of financial markets. Important properties and subtle issues
associated with the algorithms used in empirical multifractal studies are discussed in Section \ref{S1:MF:Properties}, many of which are often overlooked by researchers leading to unreliable results. In Section \ref{S1:EmpAnal}, we give an extensive survey of the literature on empirical multifractal analysis of financial time series, which overall confirms the presence of multifractality in financial markets. We raise in Section \ref{S1:MF:Sources} the important issue of apparent and effective multifractality and discuss different sources causing apparent multifractality. Different applications of the multifractal nature of financial time series are reviewed in Section \ref{S1:Applications}. We discuss open problems and provide perspectives for future research in Section \ref{S1:Perspectives}.

Although this review focuses on the multifractal analysis of financial markets, most of its contents are suitable for multifractal time series analysis in all other fields. Moreover, a large part of this review can be usefully read to inspire multifractal analyses of measures on surfaces and in higher dimensional spaces.

\section{Multifractal analysis}
\label{S1:MF}

%
%
%
%

\subsection{Partition function approach (MF-PF)}
\label{S2:MF-PF}

\subsubsection{Generalized dimensions and mass exponents}

Consider a measure $m$ embedded in a geometric support ${\cal{F}}$, whose density is $\dot{m}(t')$ at position $t'$. By definition, $\int_{t'\in{\cal{F}}} \dot{m}(t')dt'=1$. The measure in the neighbourhood of $t'$ is $\dot{m}(t')dt'$. Using the idea of the classic box-counting method \cite{Mandelbrot-1983}, we cover the geometric support ${\cal{F}}$ using boxes of size $s$. The integrated measure $m(s,t)$ in the $t$th box ${\cal{B}}(s,t)$ is
\begin{equation}
  m(s,t) = \int_{t'\in{\cal{B}}(s,t)} \dot{m}(t')dt',
\end{equation}
where $\sum_t \dot{m}(s,t)=1$ for any $s$. The fractal dimension of ${\cal{F}}$ can be determined as follow
\begin{equation}
  D_f := D_0= \lim_{s\to0}\frac{\ln\sum_t[m(s,t)]^0}{\ln{1/s}},
\end{equation}
where the term $\sum_t[m(s,t)]^0$ gives the number of non-empty boxes needed to cover the support ${\cal{F}}$. Note that fractal dimension is also called similarity dimension or capacity dimension. It is obvious that, if there is no measure distributed in a box, the box should not be counted. In most cases in handling financial time series, we have $D_0=1$. In the empirical determination of $D_0$, we plot $\sum_t[m(s,t)]^0$ against $s$ in log-log scales and the slope of the linear part in a proper scaling range is regarded as its estimate.

The information entropy or Shannon entropy of the measure can be defined as \cite{Shannon-1948a-BSTJ,Shannon-1948b-BSTJ}
\begin{equation}
  I(s) = -\sum_t m(s,t)\ln[m(s,t)],
\end{equation}
and one can define the information dimension $D_1$ as follows
\begin{equation}
  \sigma = \lim_{s\to0}\frac{I(s)}{\ln{(1/s)}},
\end{equation}
which was introduced by Balatoni and R{\'e}nyi in 1956 \cite{Balatoni-Renyi-1956-PMIHAS,Renyi-1959-AMASH} and used to study strange attractors \cite{Farmer-1982-PD,Farmer-Ott-Yorke-1983-PD}.

A third index used to characterize the measure is the correlation dimension \cite{Grassberger-Procaccia-1983-PRL,Grassberger-Procaccia-1983-PD}. The correlation dimension was originally introduced to study the scaling behavior of the correlation integral of time series  of length $N$:
\begin{equation}
  C(s) = \lim_{N\to\infty}\frac{1}{N^2}\sum_{i,j=1}^N{\cal{H}}(s-|\vec{x}_i-\vec{x}_j|) \approx \sum_t[m(s,t)]^2,
\end{equation}
where ${\cal{H}}(x)$ is the Heaviside function. The correlation dimension is defined as follows
\begin{equation}
  \nu  = \lim_{s\to0}\frac{\ln\sum_t[m(s,t)]^2}{\ln{1/s}}.
\end{equation}

These three dimensions $D_f$, $\sigma$ and $\nu$ can be unified into one framework of generalized dimensions \cite{Hentschel-Procaccia-1983-PD,Grassberger-1983-PLA,Grassberger-1985-PLA}. We define the $q$-order R{\'{e}}nyi entropy as \cite{Renyi-1961-Proc,Renyi-1970}
\begin{equation}
    I_q(s) = \lim_{p\to{q}}\frac{1}{1-p}\ln\sum_t[m(s,t]^p,
    \label{Eq:RenyiEn}
\end{equation}
and the generalized dimension as \cite{Csiszar-1962-AMASH}
\begin{equation}
    D_q = \lim_{s\to0}\frac{I_q(s)}{\ln(1/s)} = \lim_{s\to0} \lim_{p\to{q}}\frac{1}{p-1}\frac{\ln{\chi(p,s)}}{\ln{s}}
    =\left\{
    \begin{aligned}
       &\lim_{s\to0} \frac{1}{p-1}\frac{\ln{\chi(p,s)}}{\ln{s}},   & q\neq1\\
       &\lim_{s\to0} \frac{\sum_t m(s,t)\ln[m(s,t)]}{\ln{s}},      & q=1
    \end{aligned}
    \right.
    \label{Eq:MF-PF:Dq:Iq:s}
\end{equation}
where
\begin{equation}
  \chi(q,s)= \sum_t\left[m(s,t)\right]^q
  \label{Eq:MF-PF:chi:s}
\end{equation}
is the $q$-order partition function of the measure. It is easy to verify that $D_0=D_f$, $D_1=\sigma$ and $D_2=\nu$. In practice, we can compute $D_q$ according to Eq.~(\ref{Eq:MF-PF:Dq:Iq:s}). For a given $q$, we plot $I_q(s)$ against $\ln{s}$ for various box sizes $s$ and perform linear regression in a proper scaling range to obtain the slope $D_q$.

We can rewrite Eq.~(\ref{Eq:MF-PF:Dq:Iq:s}) as follows
\begin{equation}
    (q-1)D_q = \lim_{s\to0} \frac{\ln{\chi(q,s)}}{\ln{s}}
    \label{Eq:MF-PF:Dq:chi}
\end{equation}
or
\begin{equation}
    \chi(q,s) \sim s^{\tau(q)},
    \label{Eq:MF-PF:tau:chi:s}
\end{equation}
where
\begin{equation}
    \tau(q) = (q-1)D_q.
    \label{Eq:MF-PF:tau:Dq}
\end{equation}
In practice, we can compute $\tau(q)$ according to Eq.~(\ref{Eq:MF-PF:tau:chi:s}). For a given $q$, we can compute $\chi(q,s)$ for various box sizes $s$ and perform a linear regression of $\ln\chi(q,s)$ against $\ln{s}$ in a proper scaling range to obtain $\tau(q)$. Alternatively, $\tau(q)$ can be transformed from $D_q$ using Eq.~(\ref{Eq:MF-PF:tau:Dq}).

Recently, Xiong and Shang proposed a variance-weighted partition function approach and established the relationship between the corresponding multifractal functions \cite{Xiong-Shang-2017-ND}. Numerical simulations show that the variance-weighted partition function approach performs comparatively as the standard partition function approach.

\subsubsection{Scaling behavior of partition functions}

Based on the box-counting idea, the geometric support is partitioned into non-overlapping boxes of size $s$. The local singularity strength $\alpha$ in the $t$th box ${\cal{B}}(s,t)$ is defined according to the following relationships \cite{Halsey-Jensen-Kadanoff-Procaccia-Shraiman-1986-PRA}:
\begin{equation}
  m(s,t) \sim s^{\alpha},
  \label{Eq:MF-PF:m:s:alpha}
\end{equation}
which may differ for different boxes. Let $N_s(\alpha)$ denote the number of boxes in which the singularity strengths are within $[\alpha,\alpha+d\alpha]$. Hence, the fractal dimension of the set is determined according to
\begin{equation}
  N_s(\alpha) = \rho(\alpha')s^{-f(\alpha')}{\rm{d}}\alpha',
  \label{Eq:MF-PF:Ns:s:f:alpha}
\end{equation}
where $\rho(\alpha)$ is the density of the singularity strength and $f(\alpha)$ is the fractal dimension of the boxes under consideration \cite{Halsey-Jensen-Kadanoff-Procaccia-Shraiman-1986-PRA}, which is usually called multifractal spectrum or singularity spectrum.
One can directly obtain $\alpha$ from Eq.~(\ref{Eq:MF-PF:m:s:alpha}) and $f(\alpha)$ from Eq.~(\ref{Eq:MF-PF:Ns:s:f:alpha}) using local and pointwise singularity methods \cite{Meneveau-Sreenivasan-1989-PLA,Pont-Turiel-PerezVicente-2006-PRE,Salat-Murcio-Arcaute-2017-PA}.

Inserting Eq.~(\ref{Eq:MF-PF:m:s:alpha}) into the partition function (\ref{Eq:MF-PF:chi:s}) and rewriting the sum into an integral over $\alpha$, we have
\begin{equation}
  \chi(q,s) = \int{s}^{q\alpha'}\rho(\alpha')s^{-f(\alpha')}{\mbox{d}}\alpha' = \int \rho(\alpha')s^{q\alpha'-f(\alpha')}{\mbox{d}}\alpha'.
  \label{Eq:MF-PF:chi:integral}
\end{equation}
Assume that $\rho(\alpha)$ is nonzero and non-singular. It therefore contributes a constant proportional factor to the
leading behavior of the integral.
Because $s$ is small, according to the method of steepest descent, the integral will be dominated by the value of $\alpha(q)$ that makes the power $q\alpha'-f(\alpha')$ smallest. We thus replace $\alpha'$ by $\alpha(q)$, which is defined by the extremal condition:
\begin{subequations}
\begin{equation}
    {{\mbox{d}}\left[q\alpha'-f(\alpha')\right]}/{{\mbox{d}}\alpha'}\big|_{\alpha'=\alpha(q)}=0
    \label{Eq:MF-PF:MinCond1}
\end{equation}
and
\begin{equation}
    {{\mbox{d}}^2\left[q\alpha'-f(\alpha')\right]}/{{\mbox{d}}\alpha'^2}\big|_{\alpha'=\alpha(q)}>0.
    \label{Eq:MF-PF:MinCond2}
\end{equation}
\label{Eq:MF-PF:MinCond}
\end{subequations}
Then, Eq.~(\ref{Eq:MF-PF:chi:integral}) becomes
\begin{equation}
  \chi(q,s) \sim s^{q\alpha-f(\alpha)}.
  \label{Eq:MF-PF:chi:s:qalpha:f}
\end{equation}
It follows from Eq.~(\ref{Eq:MF-PF:MinCond}) that
\begin{subequations}
\begin{equation}
  f'(\alpha) =  {{\mbox{d}}f(\alpha)}/{{\mbox{d}}\alpha} = q
  \label{Eq:MF-PF:df:dalpha:q}
\end{equation}
and
\begin{equation}
  f''(\alpha) < 0,
  \label{Eq:MF-PF:d2f:dalpha2}
\end{equation}
\end{subequations}
which means that the multifractal spectrum $f(\alpha)$ is a concave function and its slope at point $(\alpha(q),f(\alpha(q))$ is $q$.

Comparing Eq.~(\ref{Eq:MF-PF:tau:chi:s}) and Eq.~(\ref{Eq:MF-PF:chi:s:qalpha:f}), we have
\begin{equation}
  \tau(q) = q\alpha - f(\alpha).
  \label{Eq:MF-PF:tau:alpha:f}
\end{equation}
Taking the derivative of Eq.~(\ref{Eq:MF-PF:tau:alpha:f}) with respect to $q$, we have
\begin{equation}
  \frac{{\mbox{d}}\tau(q)}{{\mbox{d}}q} = \alpha +q\frac{{\mbox{d}}\alpha}{{\mbox{d}}q}-\frac{{\mbox{d}}f(\alpha)}{{\mbox{d}}\alpha}\frac{{\mbox{d}}\alpha}{{\mbox{d}}q} = \alpha.
  \label{Eq:MF-X-PF:dtau:dq:alpha}
\end{equation}
\begin{subequations}
Rewriting Eq.~(\ref{Eq:MF-PF:tau:alpha:f}) and Eq.~(\ref{Eq:MF-X-PF:dtau:dq:alpha}), we have
\begin{equation}
  \alpha = {{\mbox{d}}\tau(q)}/{{\mbox{d}}q}
  \label{Eq:MF-X-PF:alpha:dtau:dq}
\end{equation}
and
\begin{equation}
  f(\alpha) =  q\alpha -\tau(q)~.
  \label{Eq:MF-PF:f:alpha}
\end{equation}
\end{subequations}
This shows that $ f(\alpha)$ is the Legendre transform of  $\tau(q)$ \cite{Frisch-Parisi-1985,Halsey-Jensen-Kadanoff-Procaccia-Shraiman-1986-PRA}. Therefore, after obtaining $\tau(q)$, we can determine numerically $\alpha(q)$ using Eq.~(\ref{Eq:MF-X-PF:alpha:dtau:dq}) and $f(\alpha)$ using Eq.~(\ref{Eq:MF-PF:f:alpha}). To qualify a multifractal measure, we can use either $(q,\tau)$ or $(\alpha,f)$, which are equivalent.

\subsubsection{Direct determination of a multifractal spectrum}
\label{S3:MF-PF:DirectDetermine}

From the canonical perspective, we can obtain the $f(\alpha)$ function directly \cite{Chhabra-Jensen-1989-PRL,Chhabra-Meneveau-Jensen-Sreenivasan-1989-PRA}. We can define the canonical measures
\begin{equation}
  \mu(q,s,t) = \frac{[m(s,t)]^q}{\sum_t [m(s,t)]^q}. 
  \label{Eq:MF-PF:mu:q}
\end{equation}
The singularity strength $\alpha(q)$ and its spectrum $f(q)$ can be computed by linear regressions in log-log scales using the following equations:
\begin{subequations}
\begin{equation}
  \alpha(q) = \lim_{s\to0} \frac{\sum_t \mu(q,s,t) \ln[m(s,t)]}{\ln{s}},
  \label{Eq:MF-PF:alpha:mu:m}
\end{equation}
and
\begin{equation}
  f(q)\triangleq f(\alpha(q))  = \lim_{s\to0}\frac{\sum_t \mu(q,s,t) \ln\left[\mu(q,s,t)\right]}{\ln{s}}.
  \label{Eq:MF-PF:fq:mu}
\end{equation}
\label{Eq:MF-PF:alpha:fq:mu}
\end{subequations}
To compute $\alpha(q)$, we can plot $\sum_t \mu(q,s,t) \ln{[m(s,t)]}$ against $\ln{s}$, determine the scaling range and obtain the slope by linear regression. The $f(q)$ function can be determined similarly based on Eq.~(\ref{Eq:MF-PF:fq:mu}). Combining Eqs.~(\ref{Eq:MF-PF:tau:chi:s}) and (\ref{Eq:MF-PF:alpha:fq:mu}), we can verify Eq.~(\ref{Eq:MF-PF:tau:alpha:f}) easily. The mass exponent function $\tau{(q)}$ can be obtained by using Eq.~(\ref{Eq:MF-PF:tau:alpha:f}).

\subsubsection{Inverse partition function and inversion formula}
\label{S3:MF-PF:inverse}

For financial volatilities, we can further define the inverse partition function \cite{Jiang-Zhou-2009-CPL}. We start with an illustration of the partition function approach using a high-frequency volatility time series.

\begin{figure}[htb]
  \centering
  \includegraphics[width=0.32\linewidth]{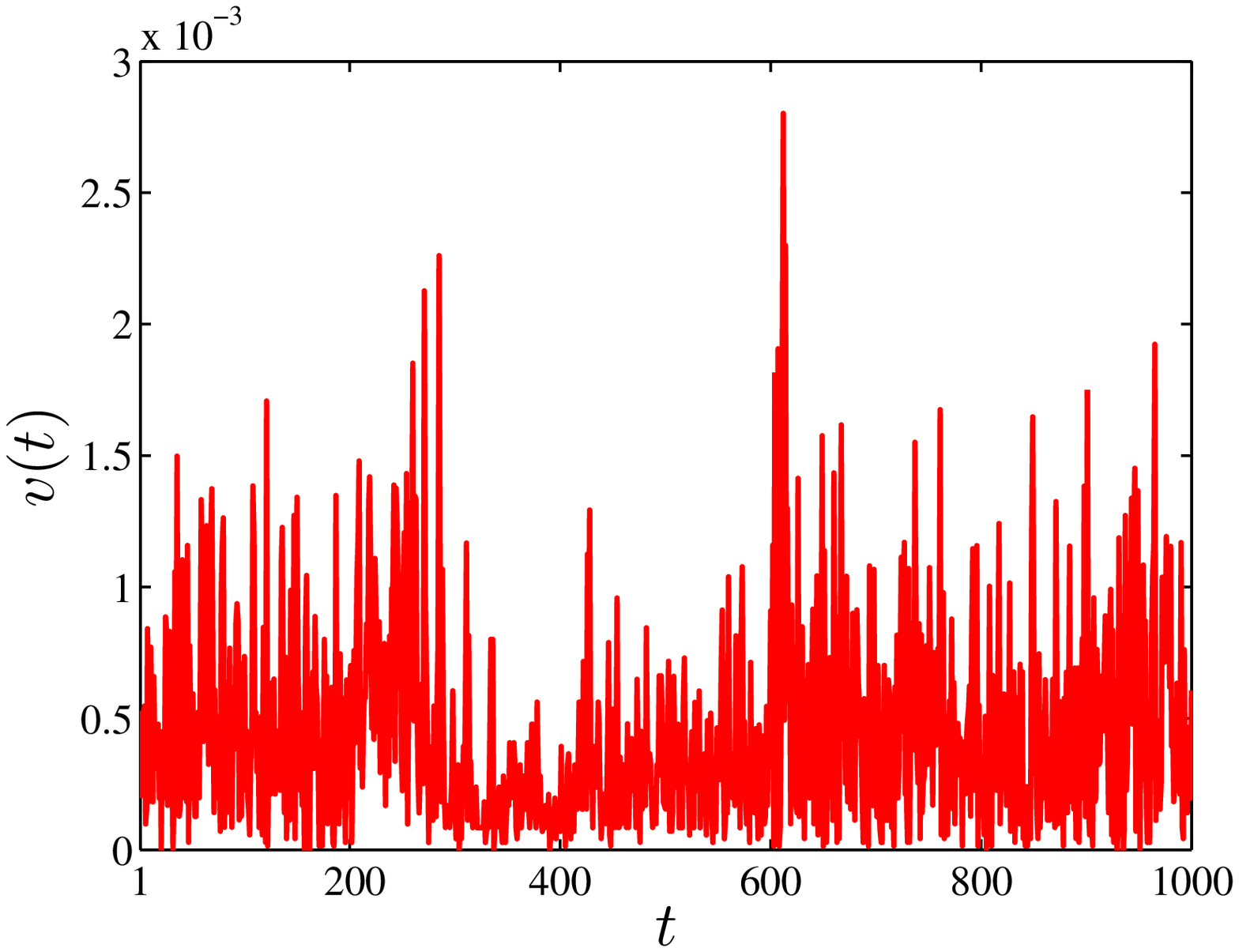}
  \includegraphics[width=0.32\linewidth]{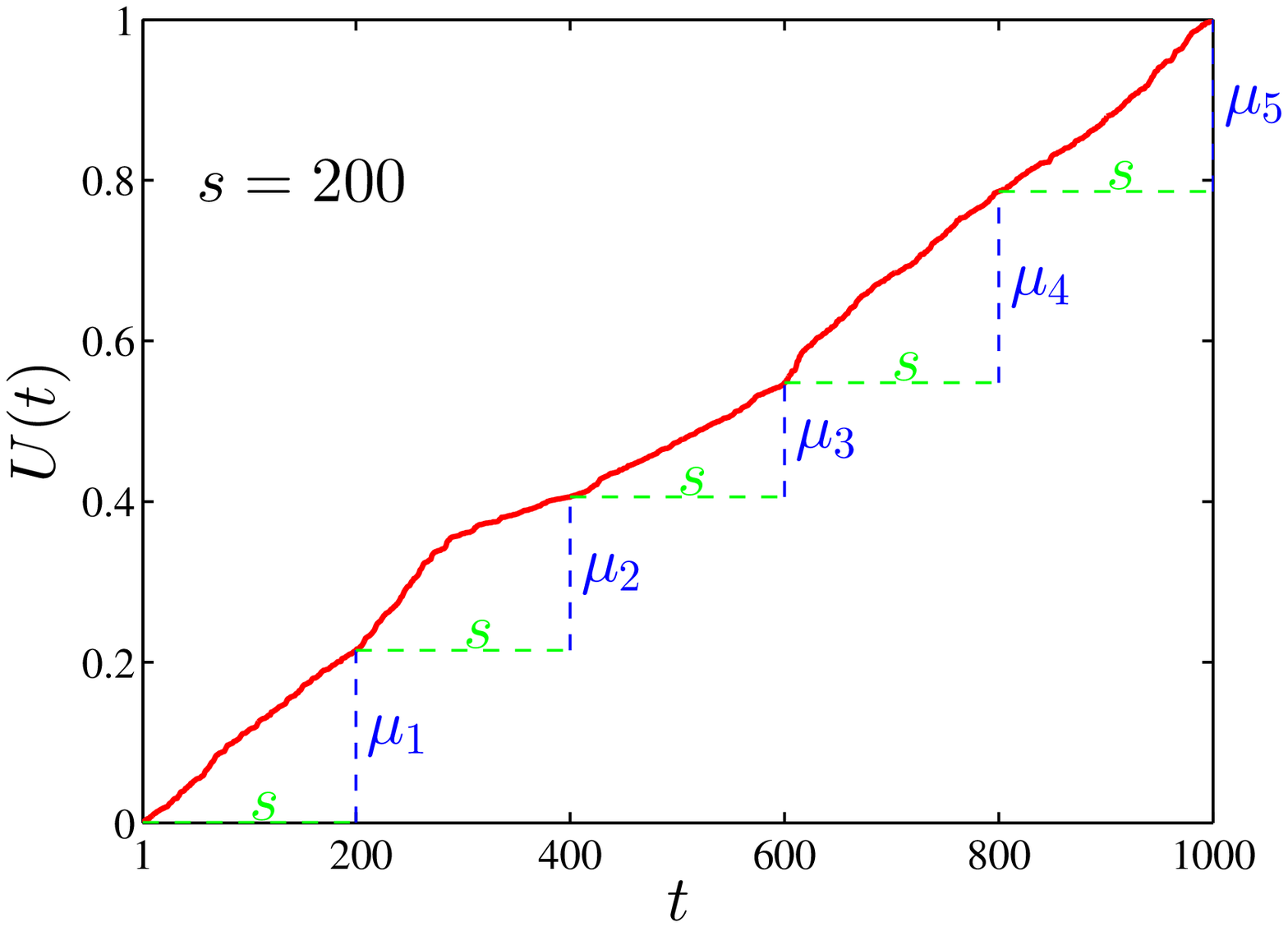}
  \includegraphics[width=0.32\linewidth]{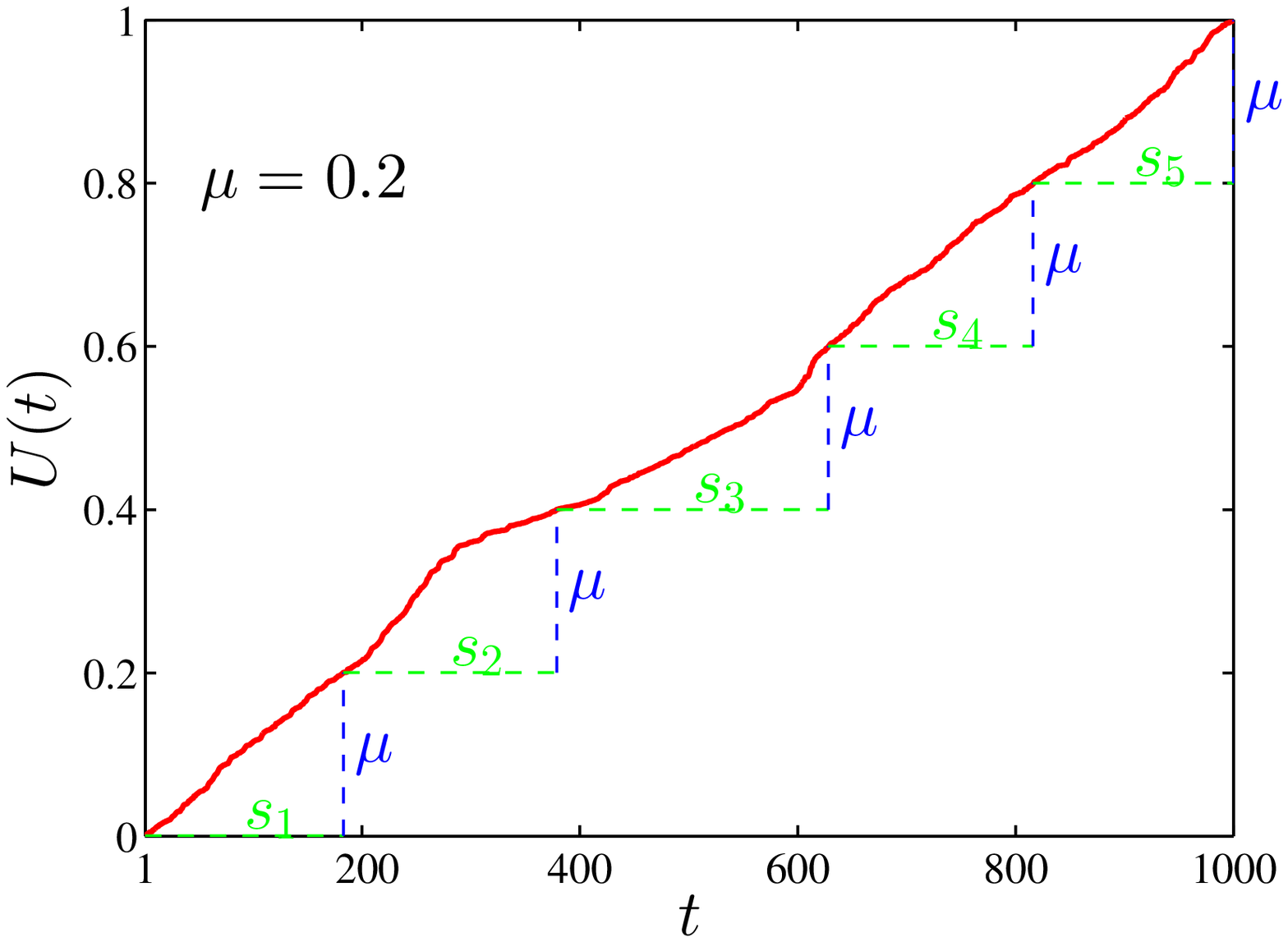}\\
  \includegraphics[width=0.32\linewidth]{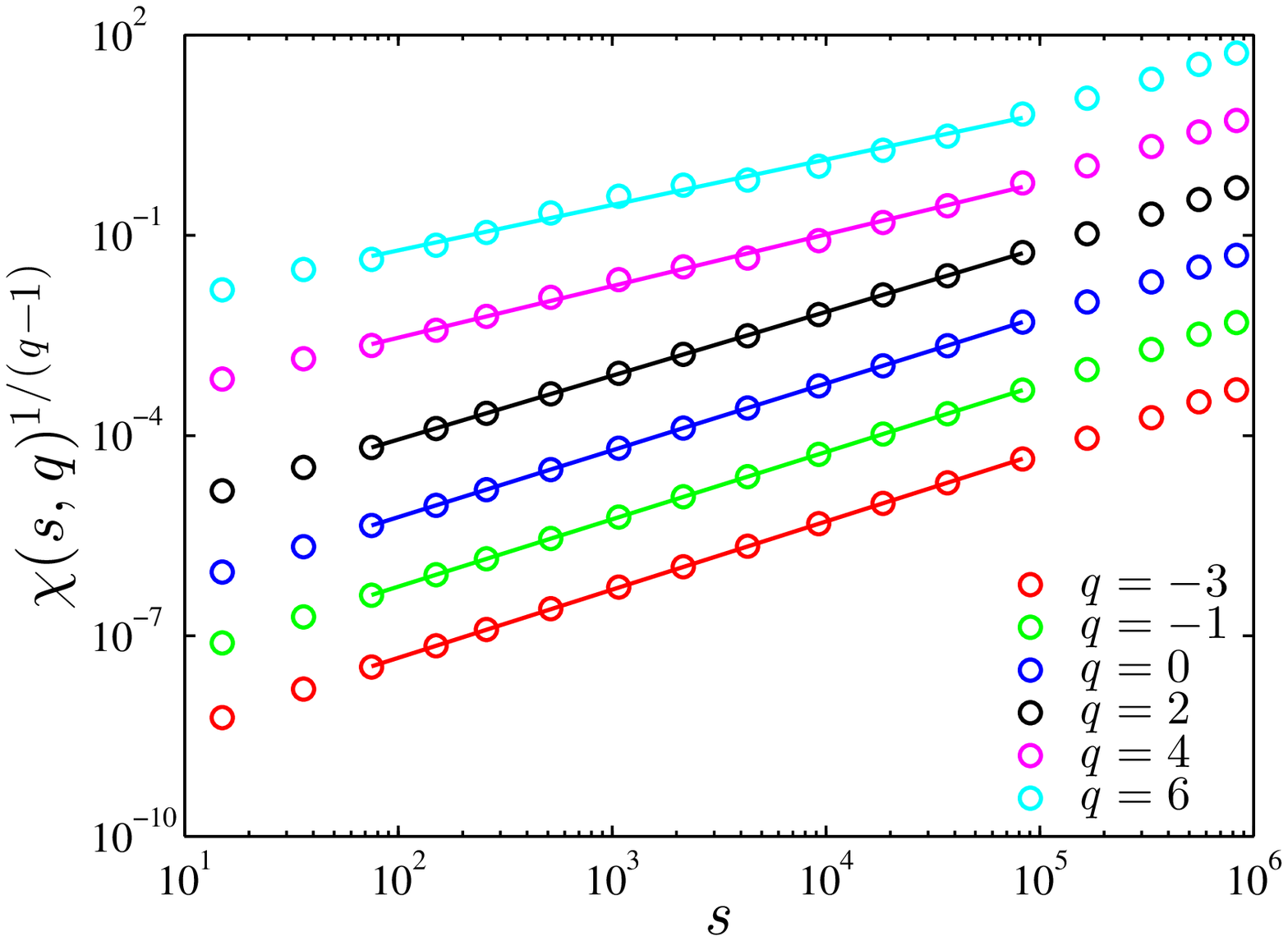}
  \includegraphics[width=0.32\linewidth]{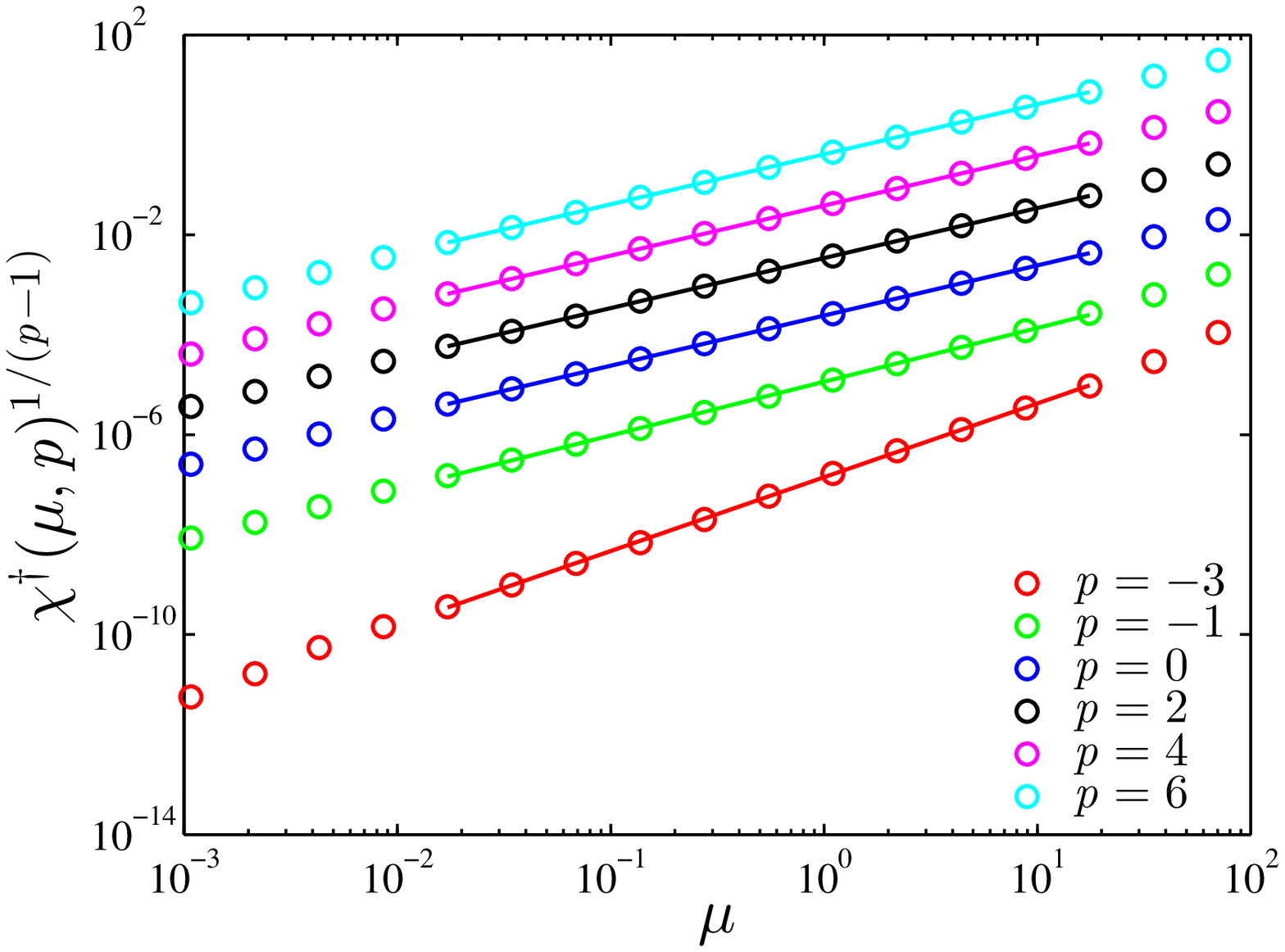}
  \includegraphics[width=0.32\linewidth]{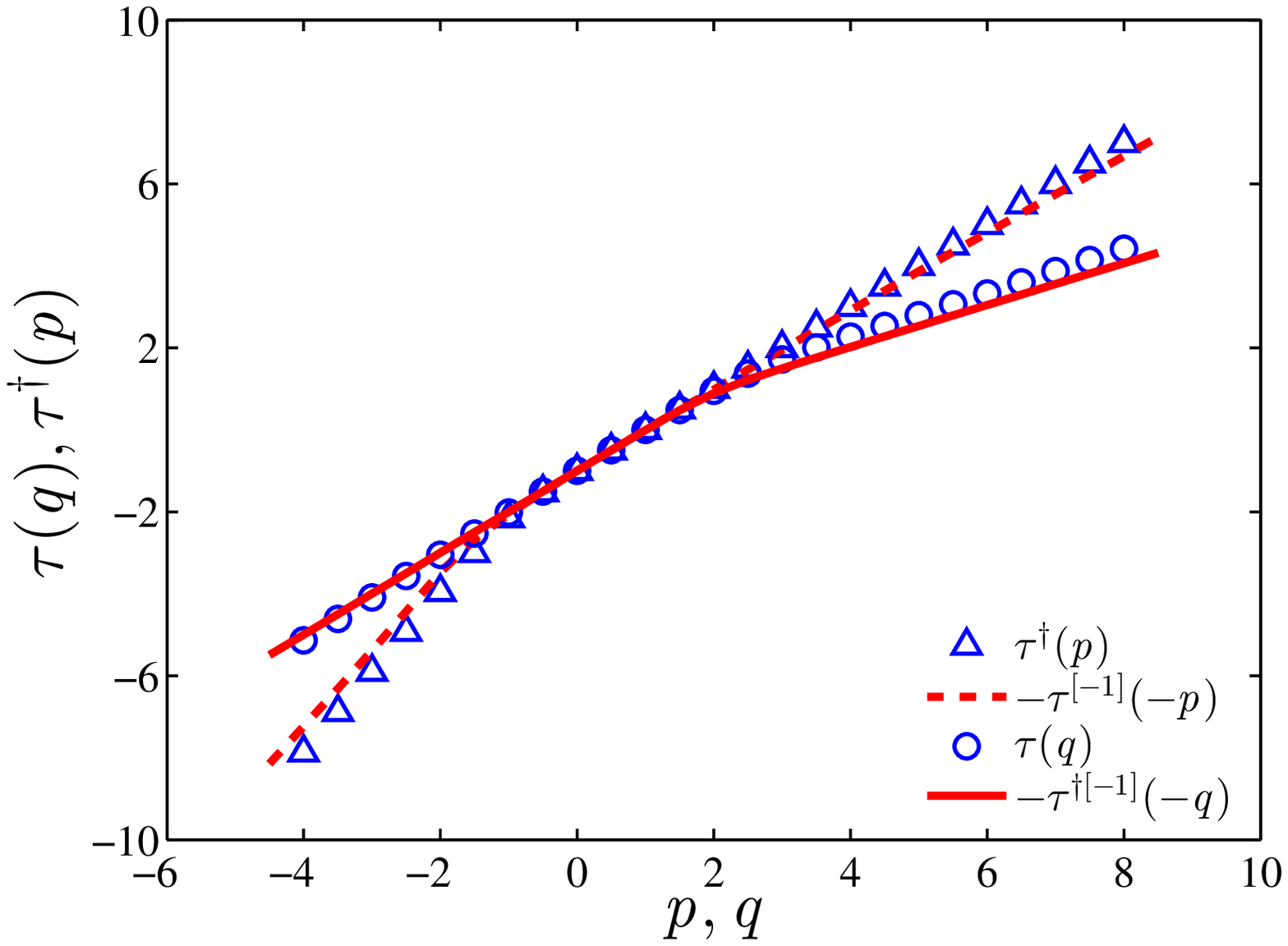}
  \caption{(color online) Multifractal analysis of high-frequency financial volatility based on the partition function approach and the inverse partition function approach. The time series contains about 1.7 million data points of the S\&P 500's 1-min volatility from 1 January 1982 to 31 December 1999. (a) A segment of volatility time series $\{v(t): t= 1:1000\}$. (b) Cumulative volatility measure $U(t)$ showing the determination of $\mu$ for a fixed scale $s$. (c) Cumulative volatility $U(t)$ showing the determination of $s_k$ for a fixed measure $\Delta{v}$. (d) Dependence of $[\chi(s,q)]^{1/q-1}$ as a function of box size $s$ for $q = -3$, $q = -1$, $q = 0$, $q =4$, and $q = 6$. The curves have been translated vertically by a factor of 0.001, 0.01, 0.1, 10, and 100 in turn for better visibility. The solid lines are power-law fits in the scaling range. (e) Dependence of $[\chi^\dag(\mu,p)]^{1/p-1}$ on the thresholds $\mu$. The curves have been translated vertically for clarity. The solid lines are the best fits to the data. (f) Testing the inversion formula in financial volatility.}
  \label{Fig:MF-PF:iPF}
\end{figure}

Denote $\{I(t): t=1,\cdots, T\}$ the price time series of a financial asset. The logarithmic return $r(t)$ at the finest time resolution is calculated by
\begin{equation}
  r(t) = \ln{I(t)} - \ln{I(t-1)},
  \label{Eq:Def:return}
\end{equation}
where $t=2,\cdots,T$. The absolute return is utilized as a proxy for volatility according to
\begin{equation}
  v(t) = |r(t)|.
  \label{Eq:Def:volatility}
\end{equation}
Figure \ref{Fig:MF-PF:iPF}(a) illustrates a segment of the 1-min volatility time series of the S\&P 500 index, in which, according to the partition function method, the original volatility time series is divided into $N$ boxes with identical size $s=T/N$. The box sizes are chosen such that the number of boxes of each size is an integer to cover the whole time series.

We define the continuous volatility measure $u(t')$ for any $t'\in (t-1, t]$ by
\begin{equation}
  u(t') = \frac{v(t)}{\sum_{t=1}^Tv(t)},
  \label{Eq:continue:volatility}
\end{equation}
where $t=1,\cdots,T$. The cumulative function of the volatility measure is obtained as
\begin{equation}
  U(t) = \int_0^t u(t')dt'.
  \label{Eq:CumSum:measure}
\end{equation}
The  function $U(t)$ of $v(t)$ in Fig.~\ref{Fig:MF-PF:iPF}(a) is presented in Fig.~\ref{Fig:MF-PF:iPF}(b).
The measure in each box of size $s$ is determined by
\begin{equation}
  \mu_n(s) = U(ns)-U(ns-s),
  \label{Eq:direct:measure}
\end{equation}
which is also illustrated in Fig.~\ref{Fig:MF-PF:iPF}(b). Usually, one computes the measure in a discrete way,
\begin{equation}
  \mu_n(s) = \sum_{t=(n-1)s+1}^{ns} u(t),
  \label{Eq:disctrete:measure}
\end{equation}
in which $T/s$ must be an integer to avoid edge effects (see Section~\ref{S2:MF:N/s} for more information). When the continuous measure $u(t)$ is used, more choices for $s$ values are possible as $N$ can be any integer.

For a given order $q$, the direct partition function $\chi_q(s)$ can be estimated by using
\begin{equation}
\chi_q(s) =  \sum_{n=1}^{N} [\mu_n(s)]^q.
 \label{Eq:dpf}
\end{equation}
In Fig.~\ref{Fig:MF-PF:iPF}(d), we plot $[\chi_q(s)]^{1/q-1}$ as a function of $s$ for different $q$'s. The scaling exponent function $\tau(q)$ can be obtained through power-law regressions to the data points in the scaling range, which spans about three orders of magnitude. We find from Fig.~\ref{Fig:MF-PF:iPF}(f) that $\tau(q)$ is a nonlinear function of $q$, confirming the presence of multifractality in the volatility measure.

We now turn to investigate the inverse partition function. For each threshold $\mu$, a sequence of exit times $\{s_j(\mu): j=1,\cdots,J\}$ can be determined successively by
\begin{equation}
  \sum_{j=1}^J s_j = \inf \left\{t : U(t) \geqslant j \mu \right\},
  \label{Eq:exittime}
\end{equation}
where $J = 1/\mu$ is an integer. Graphically, Fig.~\ref{Fig:MF-PF:iPF}(c) shows how the exit times $s_j$ can be determined. The inverse measure is defined as the normalized exit time
\begin{equation}
  \mu^\dag_j(\mu) = s_j/T,
  \label{Eq:imeasure}
\end{equation}
and the inverse partition function can be determined as follows:
\begin{equation}
  \chi^\dag(\mu,p) =  \sum_{j=1}^J \left[\mu^\dag_j(\mu)\right]^p,
 \label{Eq:MF-PF:iPF}
\end{equation}
Figure~\ref{Fig:MF-PF:iPF}(e) shows the dependence of $[\chi^\dag_p(\Delta v)]^{1/(p-1)}$ as a function of the thresholds $\Delta{v}$ for different $p$ values. Power-law scaling can be observed over about three orders of magnitude:
\begin{equation}
  \chi_p^\dag(\mu) \sim \mu^{\tau^\dag(p)}.
  \label{Eq:MF-PF-chip2}
\end{equation}
The scaling exponent function $\tau^\dag(p)$ can be obtained through power-law regressions to the data points in the scaling range, as is shown in Fig.~\ref{Fig:MF-PF:iPF}(f). The nonlinearity of the $\tau^\dag(p)$ function confirms the presence of multifractality in the exit time measure.

An important issue arises concerning the relationship between $\tau(q)$ and $\tau^\dag(p)$. Mandelbrot and Riedi proved an elegant inversion formula mathematically for both discontinuous and continuous multifractal measures \cite{Mandelbrot-Riedi-1997-AAM,Riedi-Mandelbrot-1997-AAM}. Roux and Jensen independently proved the exact relation between the direct and inverse scaling exponents for the classical binomial measures \cite{Roux-Jensen-2004-PRE}. Xu et al. provided a simple ``proof'' of the inversion formula for multinomial measures \cite{Xu-Zhou-Liu-Gong-Wang-Yu-2006-PRE}, which is presented below.

Let $\mu$ be a probability measure on $[0,1]$ with its integral function $M(t) = \mu([0,t])$. Then its inverse measure can be defined by
\begin{equation}
  \mu^\dag = M^\dag(s) = \left\{
  \begin{array}{lll}
    \inf\{t:M(t)>s\}, && {\mbox{if }} s<1\\
    1, && {\mbox{if }} s=1
  \end{array}
  \right.,
  \label{Eq:MF-PF:invMu}
\end{equation}
where $M^\dag(s)$ is the inverse function of $M(t)$. If $\mu$ is self-similar, then the relation $\mu = \sum_{i=1}^n p_i\mu(m_i^{-1}(\cdot))$ holds, where the $m_i$'s are similarity maps with scale contraction ratios $r_i\in(0,1)$ and $\sum_{i=1}^np_i=1$ with $p_i>0$. The multifractal spectrum of measure $\mu$ is the Legendre transform $f(\alpha)$ of $\tau$, which is defined by the generating function
\begin{equation}
  \sum_{i=1}^n p_i^qr_i^{-\tau}=1.
  \label{Eq:MF-PF:iPF:tau}
\end{equation}
It can be shown \cite{Mandelbrot-Riedi-1997-AAM,Riedi-Mandelbrot-1997-AAM} that the inverse measure $\mu^\dag$ is also self-similar with ratio $r_i^\dag=p_i$ and $p_i^\dag=r_i$, whose multifractal spectrum $f^\dag(\alpha^\dag)$ is the Legendre transform of $\tau^\dag$, which is defined implicitly by
\begin{equation}
  \sum_{i=1}^n (p_i^\dag)^p (r_i^\dag)^{-\tau^\dag}=1.
  \label{Eq:MF-PF:iPF:theta}
\end{equation}
It is easy to verify that the following inversion formula holds
\begin{equation}
  \left\{
  \begin{array}{lll}
    \tau(q) &=& -p\\
    \tau^\dag(p) &=& -q
  \end{array}
  \right..
  \label{Eq:MF-PF:Tau:Theta}
\end{equation}
Two equivalent testable formulae follow immediately
\begin{equation}
  \left\{
  \begin{array}{lll}
    \tau(q) = -\tau^{\dag[-1]}(-q)\\
    \tau^\dag(p) = -\tau^{[-1]}(-p)
  \end{array}
  \right.,
  \label{Eq:MF-PF:InvForm}
\end{equation}
where the superscript ${[-1]}$ denotes the inverse function operator. Figure~\ref{Fig:MF-PF:iPF}(f) verifies that the inversion formulae (\ref{Eq:MF-PF:InvForm}) hold for the high-frequency volatility time series of the S\&P 500 index \cite{Jiang-Zhou-2009-CPL}.

We can also relate the direct and inverse singularity strengths and the direct and inverse singularity spectra. From Eq.~(\ref{Eq:MF-PF:Tau:Theta}), we have
\begin{equation}
  \alpha(q)=\frac{{\mbox{d}}\tau(q)}{{\mbox{d}}q}=\frac{-{\mbox{d}}p}{-{\mbox{d}}\tau^\dag(p)}=\frac{1}{\alpha^\dag(p)}.
  \label{Eq:MF-PF:InvForm:alpha}
\end{equation}
Combining Eq.~(\ref{Eq:MF-PF:Tau:Theta}), Eq.~(\ref{Eq:MF-PF:InvForm:alpha}) and the Legendre transform, we have
\begin{equation}
  f(\alpha) = q\alpha-\tau = -\tau^\dag(p)\alpha+p = \alpha[-\tau^\dag(p)+p\alpha^\dag] = f^\dag(\alpha^\dag)/\alpha^\dag = \alpha f^\dag(1/\alpha).
  \label{Eq:MF-PF:InvForm:falpha}
\end{equation}
Rigorous proofs of these results can also be found in Refs.~\cite{Mandelbrot-Riedi-1997-AAM,Riedi-Mandelbrot-1997-AAM}. It is easy to obtain the relation between direct and inverse generalized dimensions \cite{PastorSatorras-1997-PRE}:
\begin{equation}
  \left\{
  \begin{array}{lll}
    p &=& -(q-1)D(q)\\
    D^\dag(p) &=& q/[1+(q-1)D(q)]
  \end{array}
  \right.,
  \label{Eq:MF-PF:Dq:DP}
\end{equation}
where $D^\dag(p)$ is the inverse generalized dimension function.

\subsubsection{Ensemble averaging}
\label{S3:MF-PF:Ensemble}

In the multifractal analysis of financial time series, almost all studies with very few exceptions focused on individual time series. One can further investigate an ensemble of financial time series of individual assets \cite{Jiang-Zhou-2008b-PA}. Ensemble averaging was developed to determine the multifractal dimensions of
Diffusion-Limited Aggregations \cite{Cates-Witten-1987-PRA,Halsey-Leibig-1992-PRA,Halsey-Honda-1994-IFIPTA,Hentschel-1994-PRE,Halsey-Honda-Duplantier-1996-JSP}, and unearthed a subtle discrete hierarchy quantified by complex fractal dimensions \cite{Sornette-Johansen-Arneodo-Muzy-Saleur-1996-PRL}.

One defines the quenched and annealed mass exponents $\tau_\mathrm{quen} (q)$ and $\tau_\mathrm{ann} (q)$ as follows,
\begin{align}
\left\langle \ln \chi_{q}(s) \right\rangle &\sim -\tau_\mathrm{quen}(q) \ln s, \label{Eq:quenched}\\
\ln \left\langle \chi_{q}(s) \right\rangle &\sim -\tau_\mathrm{ann}(q) \ln  s, \label{Eq:Annealed}
\end{align}
where the angular brackets $\langle \cdot \rangle$ signify the ensemble average over all time series. The annealed exponents are more sensitive to rare samples of the ensemble with unusual values of $\chi_q(s)$, while the quenched exponents are more characteristic of typical members of the ensemble \cite{Halsey-Duplantier-Honda-1997-PRL}.

The reasoning under the ensemble averaging lies in the consideration that one performs the measurement many times of the dynamics of an ``ensemble'' of financial instruments. The ensemble averaging method provides an alternative way to measure market risks from individual equities other than the market index. It is also different from directly averaging the risk measures of many individual equities. In addition, although the method was initially developed for the partition function approach, its extension to other multifractal analysis methods is straightforward.

\subsection{Structure function approach}
\label{S2:MF-SFA}

\subsubsection{Direct structure functions (MF-SF)}

Another important statistical quantity of turbulent fields is the structure function of velocity increments \cite{Kolmogorov-1962-JFM,VanAtta-Chen-1970-JFM}. Classical experiments have been carried out to measure the structure function and its nonlinear scaling behaviors \cite{Anselmet-Gagne-Hopfinger-Antonia-1984-JFM}. Such nonlinear scaling behavior is also termed as multifractality \cite{Benzi-Paladin-Parisi-Vulpiani-1984-JFM,Frisch-Parisi-1985}. The structure function approach has been also employed to investigate financial time series \cite{Muller-Dacorogna-Olsen-Pictet-Schwarz-Morgenegg-1990-JBF,Ghashghaie-Breymann-Peinke-Talkner-Dodge-1996-Nature,DiMatteo-2007-QF}. The $q$th-order structure function is also called the $q$th-order height-height correlation function in the literature \cite{Barabasi-Vicsek-1991-PRA,Barabasi-Szepfalusy-Vicsek-1991-PA}.

Consider a time series $\{X(i):i=1,\cdots, N\}$. The increments or innovations over time scale $s$ are defined as
\begin{equation}
  \Delta{X}(i,s) = X(i)-X(i-s).
  \label{Eq:MF-SF:dX}
\end{equation}
For financial assets, $X(i)$ is the logarithmic price and $\Delta{X}$ is the return over the time interval of duration $s$. The $q$th-order structure function is defined as the $q$th-order moment of the increment distribution:
\begin{equation}
  K(q,s) = \frac{\langle|\Delta{X}(i,s)|^q\rangle}{\langle|X(i)|^q\rangle} = \frac{\frac{1}{N-s}\sum_{i=s+1}^N\left|X(i)-X(i-s)\right|^q}{\frac{1}{N}\sum_i[X(i)]^q}.
  \label{Eq:MF-SF:Kqs}
\end{equation}
We stress that $q\geq0$ for time series in turbulence \cite{Anselmet-Gagne-Hopfinger-Antonia-1984-JFM}, finance \cite{DiMatteo-Aste-Dacorogna-2005-JBF,DiMatteo-2007-QF}, and other fields, because $\Delta{X}(i,s)$ can be zero such that the moments of negative orders are not defined. When $q=0$, we have $K(0,s)\equiv1$, which is independent of the scale $s$.
When $q=2$, $K(2,s)$ is proportional to the autocorrelation function $\sim \langle{X(i)X(i-s)}\rangle$ \cite{DiMatteo-Aste-Dacorogna-2005-JBF,DiMatteo-2007-QF}.

For self-similar time series, we expect to have
\begin{equation}
  K(q,s) \sim s^{\zeta(q)} \triangleq s^{qH(q)},
  \label{Eq:MF-SF:Kq:s}
\end{equation}
where $H(q)$ is the generalized Hurst exponent.
The scaling function is obtained as  \cite{Mandelbrot-Fisher-Calvet-1997}
\begin{equation}
  \tau(q) = \zeta(q)-1 = qH(q)-1.
  \label{Eq:MF-SF:tau:Hq}
\end{equation}
When $q=0$, we have
\begin{equation}
  \tau(0) = -1.
  \label{Eq:MF-SF:tau:q=0}
\end{equation}
In practice, for each fixed $q$, we calculate the $K(q,s)$ values with varying $s$ values and perform a linear least-squares regression of $\ln{K(q,s)}$ against $\ln{s}$ in a properly chosen scaling range $[s_{\min},s_{\max}]$ to obtain $qH(q)$ and $\tau(q)$. It follows that
\begin{equation}
  \tau(q) = qH(q)-1 = \frac{\sum_s\ln{s}\sum_s\ln{K(q,s)}-N\sum_s\ln{s}\ln{K(q,s)}}
      {\left(\sum_s\ln{s}\right)^2-N\sum_s(\ln{s)}^2}-1
  \label{Eq:MF-SF:tau:Hq:LSE}
\end{equation}
We can further determine the singularity function $\alpha(q)$ and the singularity spectrum $f(\alpha)$ through the Legendre transform. When $X(t)$ is monofractal, $H(q)=H$ is a constant. When $X(t)$ has a multifractal nature, $H(q)$ decreases with $q$.

According to the definition of $K(q,s)$, we have
\begin{equation}
  \frac{{\mbox{d}}\ln{K(q,s)}}{{\mbox{d}}q} = \frac{1}{K(q,s)}\frac{{\mbox{d}}K(q,s)}{{\mbox{d}}q}
   = \frac{1}{\sum_tv^q}\frac{{\mbox{d}}\sum_tv^q}{{\mbox{d}}q}
   = \frac{\sum_tv^q\ln{v}}{\sum_t{v^q}},
\end{equation}
where $\sum_t$ is the sum for $s+1\leq{t}\leq{T}$ with $v>0$. Denote $k$ the proportionality coefficient of the scaling relation (\ref{Eq:MF-SF:Kq:s}). It is obvious that $k>0$. Combining Eq.~(\ref{Eq:MF-SF:Kq:s}) and Eq.~(\ref{Eq:MF-SF:tau:Hq}), we have
\begin{equation}
  \tau'(q)k\ln{s} = \frac{{\mbox{d}}[\tau(q)+1]k\ln{s}}{{\mbox{d}}q} =\frac{{\mbox{d}}\ln{K(q,s)}}{{\mbox{d}}q}  = \frac{\sum_tv^q\ln{v}}{\sum_tv^q}.
\end{equation}
Let $g(x)=x\ln{x}$. We have $g'(x)>0$ when $x\in[1/e,+\infty)$ and $g'(x)<0$ when $x\in(0,1/e]$.

\begin{figure}[htb]
  \centering
  \includegraphics[width=0.32\linewidth]{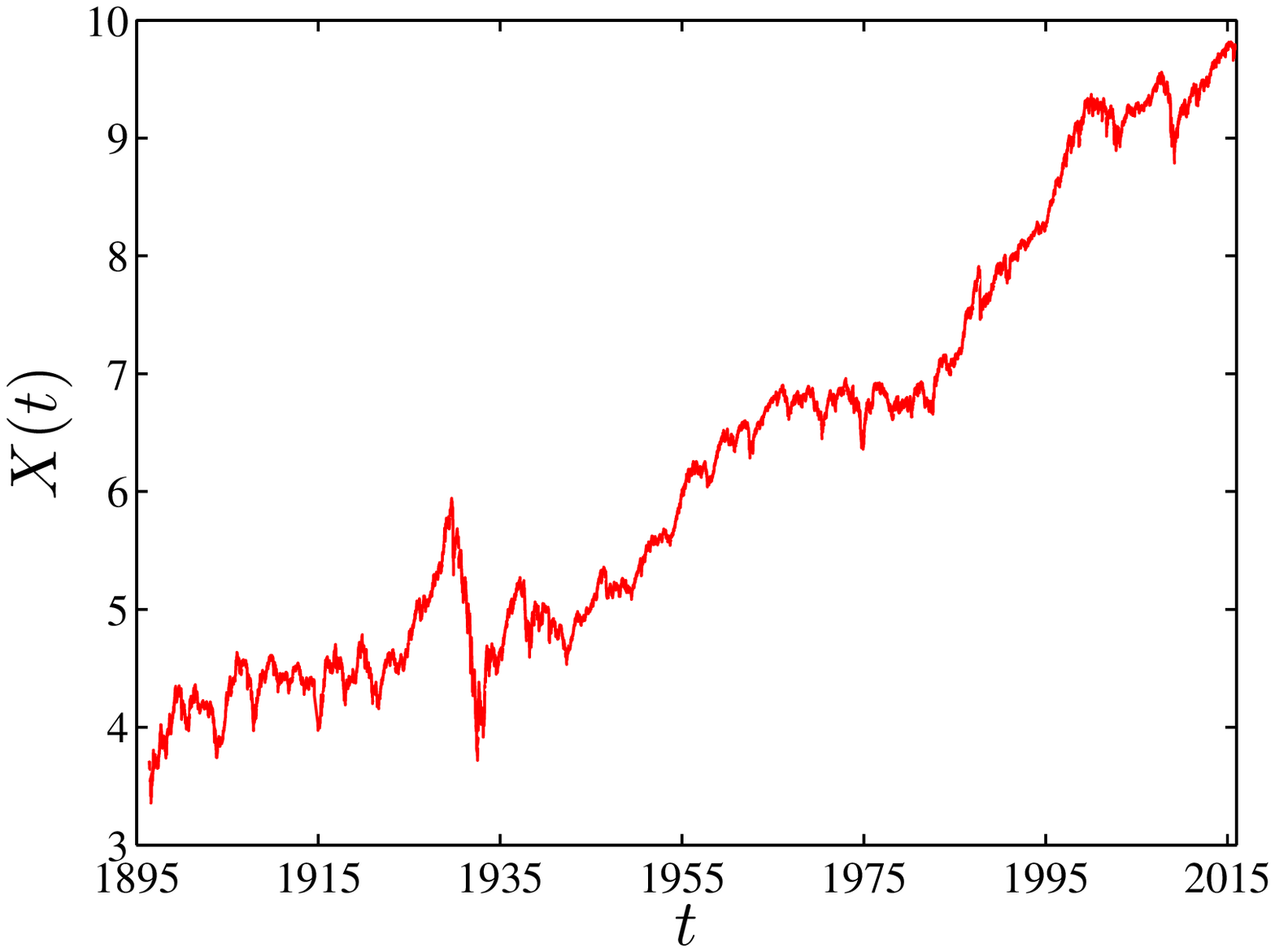}
  \includegraphics[width=0.32\linewidth]{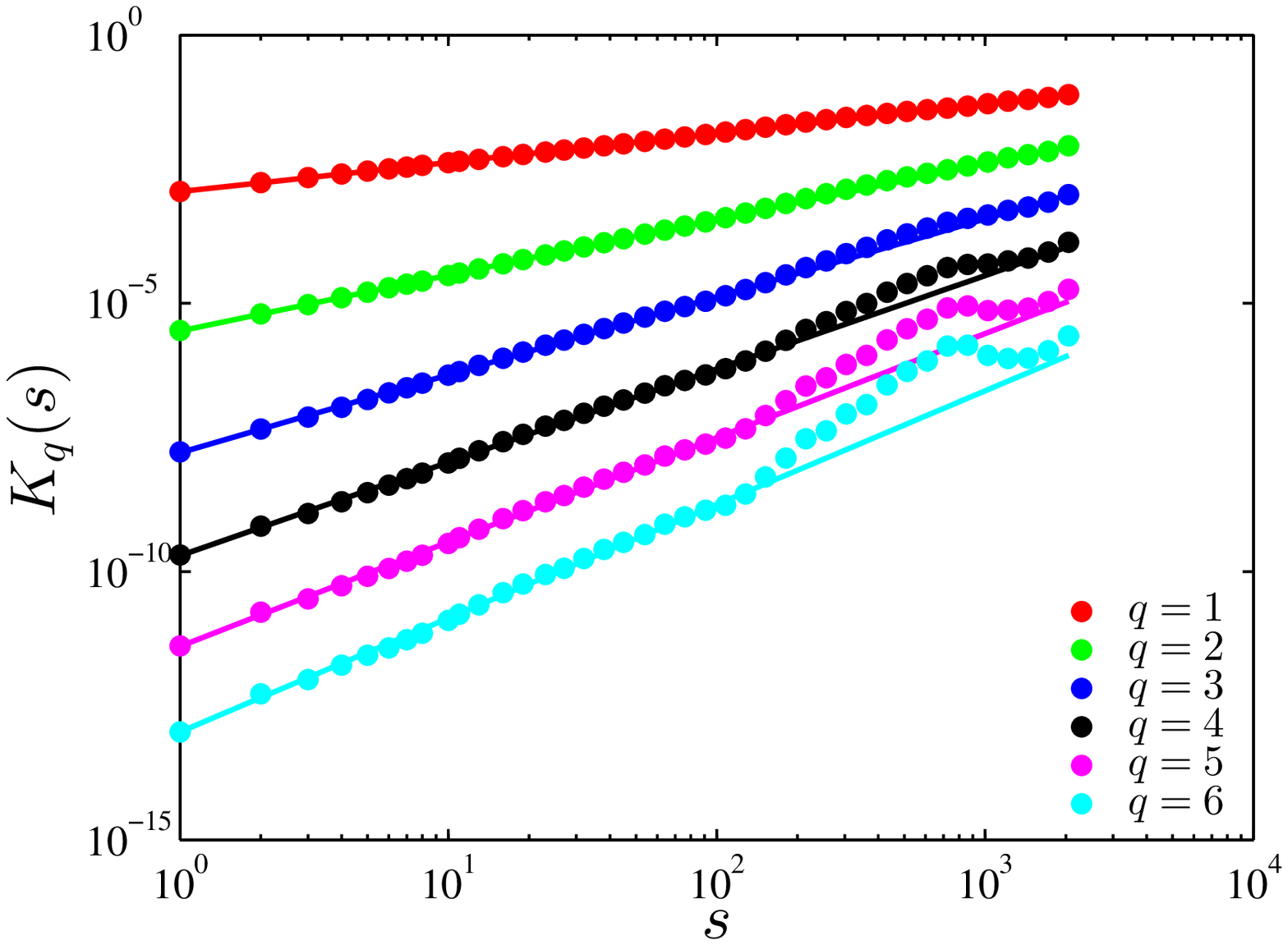}
  \includegraphics[width=0.32\linewidth]{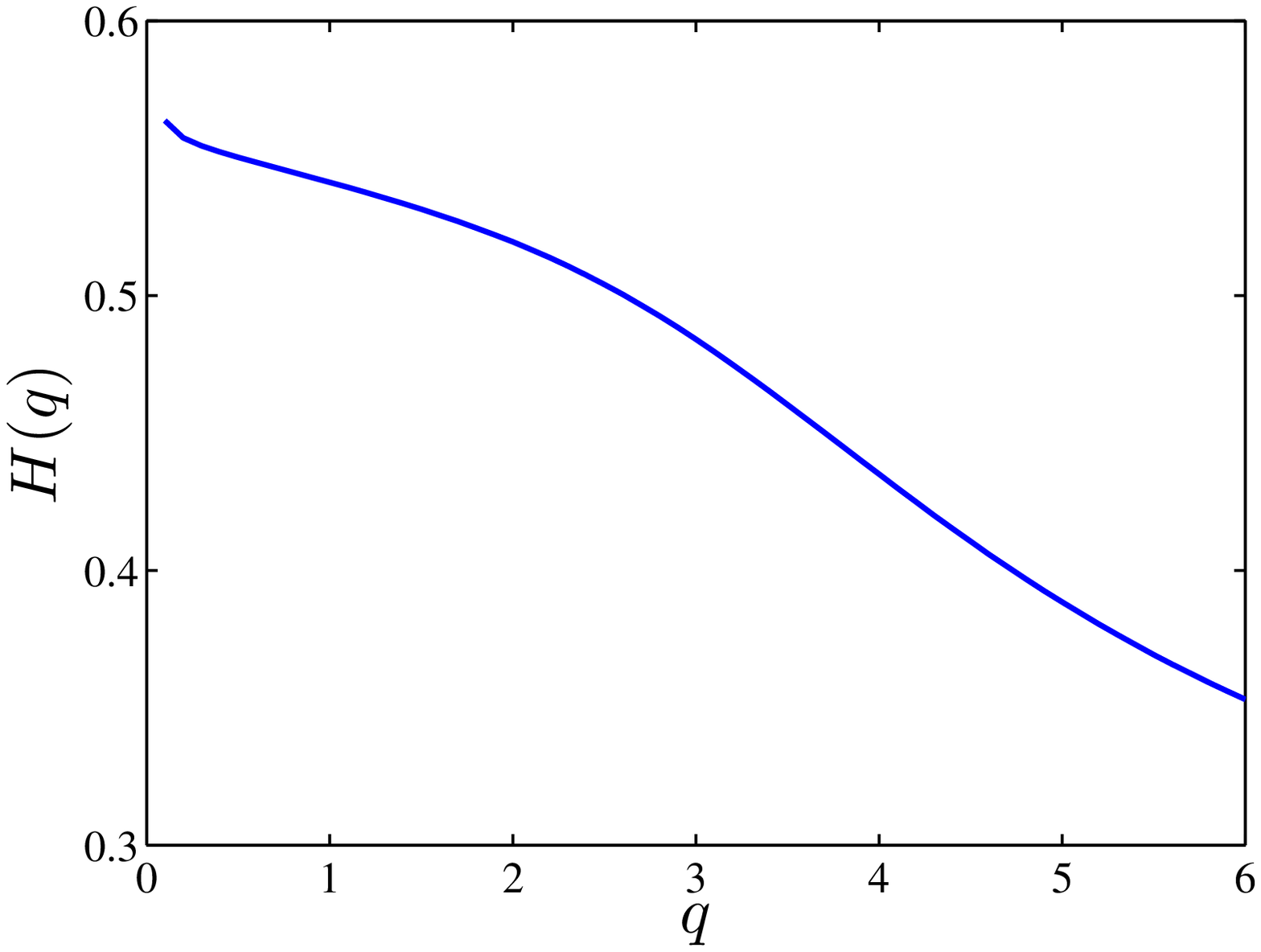}
  \includegraphics[width=0.32\linewidth]{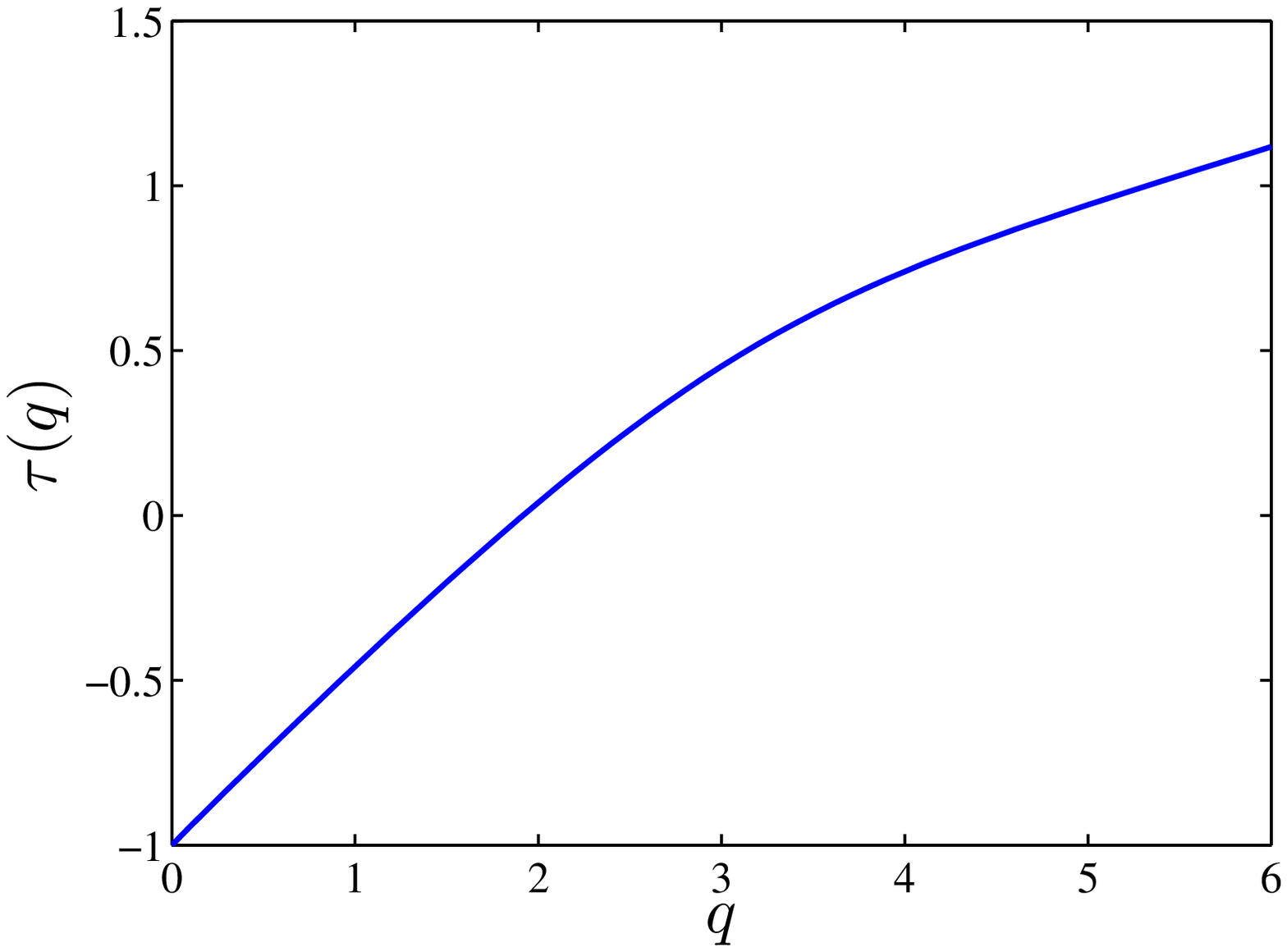}
  \includegraphics[width=0.32\linewidth]{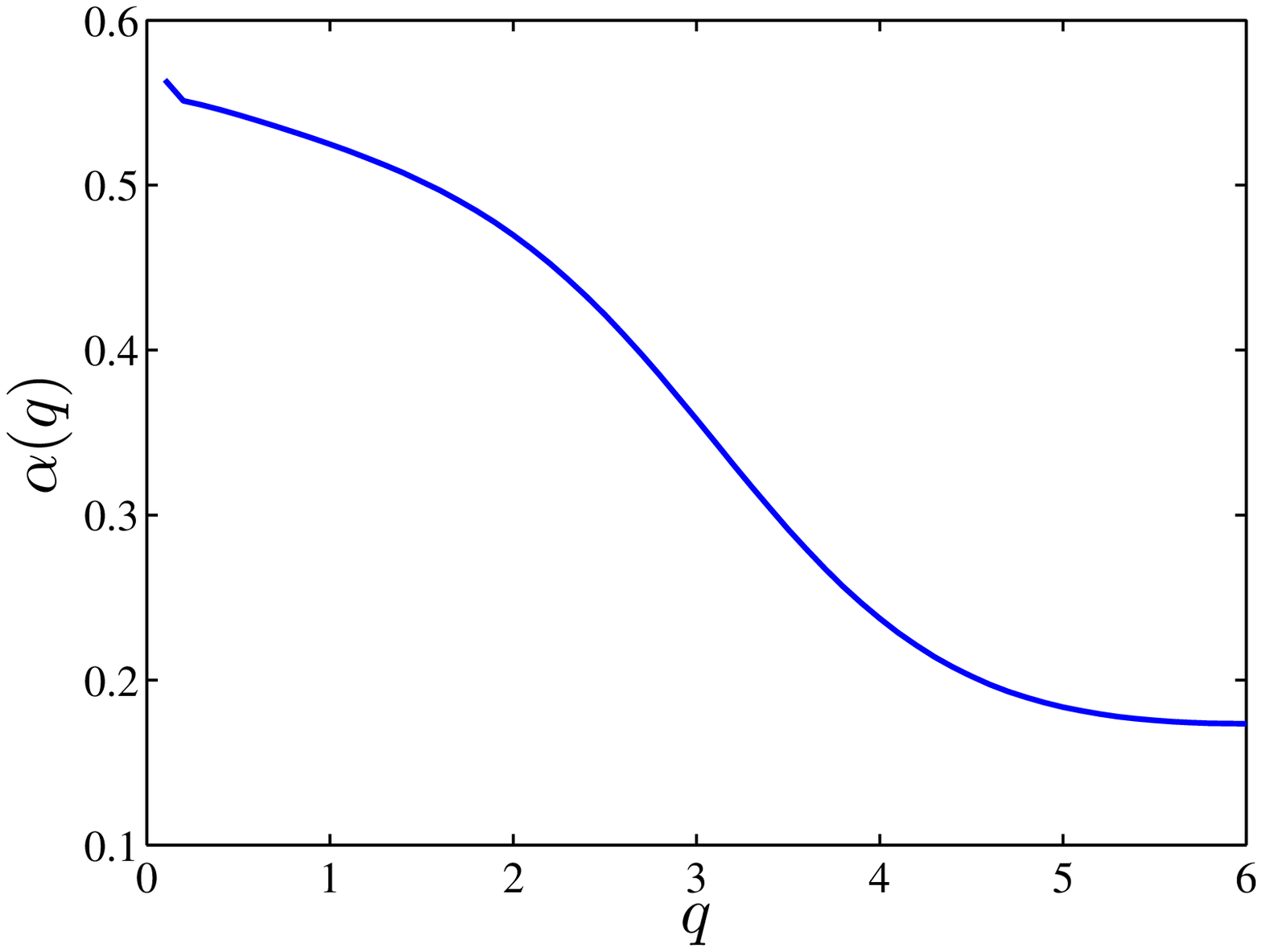}
  \includegraphics[width=0.32\linewidth]{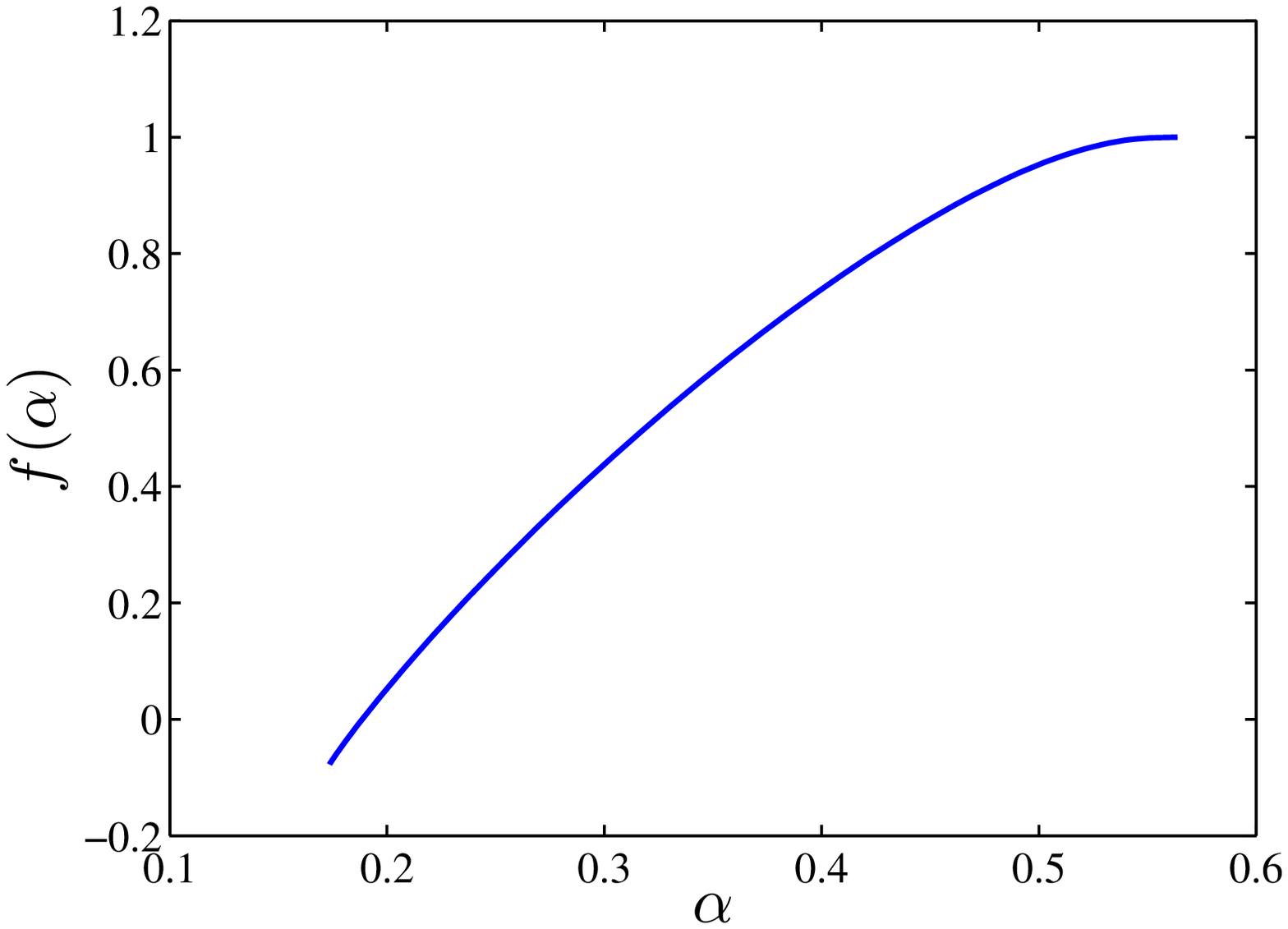}
  \caption{(color online) Multifractal analysis of the daily time series of the logarithmic price $X(t)$ of the Dow Jones Industrial Average index (DJIA, 1896-2015) based on the structure function approach. (a) The daily price time series of the DJIA index. (b) Power-law dependence of the structure function $K(q,s)$ on time lag $s$ for different $q$ values. (c) The generalized Hurst exponent $H(q)$ as a function of $q$. (d) The scaling exponent function $\tau(q)$. (e) The singularity strength function $\alpha(q)$. (f) The multifractal singularity spectrum $f(\alpha)$. }
\label{Fig:MF-SFA:DJIA}
\end{figure}

Figure \ref{Fig:MF-SFA:DJIA} presents the results of the multifractal analysis of the daily time series of the logarithmic price $X(t)$ of the Dow Jones Industrial Average index (DJIA, 1896-2015) based on the structure function approach.

In a seminal study \cite{Muller-Dacorogna-Olsen-Pictet-Schwarz-Morgenegg-1990-JBF}, M{\"u}ller et al. presented a statistical analysis of financial data
documenting scaling laws in the form of Eq.~(\ref{Eq:MF-SF:Kq:s}) for $q=1$ and $q=2$, though they did not mention ``multifractal'', ``multiscaling'' or ``structure function''. Their analysis included intra-day prices over 3 years and daily prices over 15 years of four foreign exchange spot rates (DEM/USD, JPY/USD, CHF/USD, USD/GBP) and intra-day prices over 3 years of gold (XAU/USD). Similar analyses for the first- and second-order moments have been adopted by other economists \cite{Andersen-Bollerslev-Diebold-Labys-2001-JASA}.

\subsubsection{Multifractal fluctuation analysis (MF-FA)}

The method of fluctuation analysis is a classical approach to extract the Hurst exponent of time series \cite{Peng-Buldyrev-Goldberger-Havlin-Sciortino-Simons-Stanley-1992-Nature}, which has been applied extensively in econophysics \cite{Eisler-Kertesz-2006-PRE,Eisler-Kertesz-2006-EPJB,Eisler-Kertesz-2006,Eisler-Kertesz-2007-EPL,Eisler-Kertesz-2007-PA}. The fluctuation function can be calculated as follows,
\begin{equation}
 Z(2,s) = \sigma_2^2(s) = \left\langle\left[\Delta{X(i,s)} - \langle\Delta{X(i,s)}\rangle\right]^2 \right\rangle \sim s^ {2H}.
 \label{Eq:MF-FA:Sigma:s}
\end{equation}
where $H$ is the fluctuation analysis (FA) exponent, i.e. approximatively the Hurst index. A value of the Hurst exponent $H$
larger than $0.5$ means that the time series is correlated, For $H < 0.5$, the time series is anti-correlated, and for $H = 0.5$, it is uncorrelated.

One can extend the fluctuation analysis in equation (\ref{Eq:MF-FA:Sigma:s}) to higher orders as follows \cite{Matia-Ashkenazy-Stanley-2003-EPL,Eisler-Kertesz-Yook-Barabasi-2005-EPL,Jiang-Guo-Zhou-2007-EPJB},
\begin{equation}
 Z(q,s) = \left\langle|\Delta{X(i,s)} - \langle\Delta{X(i,s)}\rangle|^q \right\rangle \sim s^{\zeta(q)}.
 \label{Eq:MF-FA:Zq:s}
\end{equation}
which enables us to understand the multifractal nature in the dynamics of markets. The relationship between the scaling exponent $\zeta(q)$ and the generalized Hurst exponent $H(q)$ can be described by Eq.~(\ref{Eq:MF-SF:tau:Hq}).
When $q = 2$, $H = H(2)$ is the Hurst exponent presented in Eq.~(\ref{Eq:MF-FA:Zq:s}). Using the Legendre transform, we can obtain the singularity strength $\alpha$ and its spectrum $f(\alpha)$.

\subsubsection{Inverse statistics and inverse structure functions}

Analogous to the inverse measure in the partition function approach \cite{Riedi-Mandelbrot-1995-AAM,Riedi-Mandelbrot-1997-AAM}, Jensen \cite{Jensen-1999-PRL} investigated the moments of exit time or first passage time. For a given increment $\Delta{X}$,
using definition (\ref{Eq:MF-SF:dX}), the exit time scale $s$ at time $i$ is defined by
\begin{equation}
  s(i, \Delta{X}) = \inf\{s^{\dag}: \Delta{X}(i,s^{\dag})\ge \Delta{X}\}~,
\end{equation}
which is the minimal time needed for the fluctuation to exceed the threshold $\Delta{X}\ge0$. The first passage time has important usage in finance.
For instance, Cho and Frees adopted it to construct an asymptotically unbiased volatility estimator \cite{Cho-Frees-1988-JF}. They argued that the natural volatility estimator focuses on how much the price changes, whereas the ``temporal'' estimator based on the first passage time focuses on how quickly the price changes \cite{Cho-Frees-1988-JF}. The inverse statistics have also some connections to the persistence probability \cite{Zheng-2002-MPLB,Ren-Zheng-2003-PLA,Ren-Zheng-Lin-Wen-Trimper-2005-PA,Constantin-DasSarma-2005-PRE}. The variable $s(i, \Delta{X})$
and derived quantities have natural interpretations in finance, such the ``time to recovery'' after a drawdown and are part of
metrics used in a standard way in evaluating performance of financial strategies and hedge-funds.
This explains that inverse statistics have been extensively applied in finance, as well as in other fields like in the study
of heart beat rate and cardiac diseases \cite{Ebadi-Shirazi-Mani-Jafari-2011-JSM}.

If $X(i)$ is a Brownian motion, the distribution of $s$ is solved analytically as a Gamma distribution \cite{Karlin-1966,Rangarajan-Ding-2000-PLA}:
\begin{equation}
  \Pr(s) = \frac{1}{\sqrt{\pi}}\frac{a}{s^{3/2}}\exp\left(-\frac{a^2}{s}\right)
  \label{Eq:FPT:PDF:BM}
\end{equation}
where $a$ is proportional to $\Delta{X}$. Since asset prices are not Brownian motions, Simonsen et al. \cite{Simonsen-Jensen-Johansen-2002-EPJB} proposed a generalized shifted Gamma distribution
\begin{equation}
  \Pr(s) = \frac{\nu}{\Gamma\left(\frac{\delta}{\nu}\right)}\frac{\beta^{2\delta}}{(s+s_0)^{\delta+1}}\exp\left[-\left(\frac{\beta^2}{s+s_0}\right)^\nu\right],
  \label{Eq:FPT:PDF:GS:Gamma}
\end{equation}
which has a power-law tail $\Pr(s)\sim s^{-(\delta+1)}$. The Gamma distribution in Eq.~(\ref{Eq:FPT:PDF:BM}) is recovered if $\nu=1$, $\delta=1/2$, $s_0=0$ and $\beta=a$. They analyzed the wavelet filtered logarithmic daily closing prices of the DJIA (26 May 1896 to 5 June 2001) and found that the distribution (\ref{Eq:FPT:PDF:GS:Gamma}) with $\delta\approx0.5$ fits the data well. This observation seems quite ubiquitous for uniformly spaced time series of indices and stocks in different markets \cite{Zhou-Yuan-2005-PA,Eisler-Kertesz-Lillo-Mantegna-2009-QF,Ebadi-Bolgorian-Jafari-2010-PA}, although there are studies reporting that $\delta$ can deviate from 0.5 \cite{Johansen-Jensen-Simonsen-2006-PA,Shayeganfar-Hoelling-Peinke-Tabar-2012-PA}. For tick-by-tick data at the transaction level, the tail exponent can be larger than 0.5, as found for five highly liquid stocks (AstraZeneca, GlaxoSmithKline, Lloyds TSB Group, Shell, and Vodafone) in the electronic market (SETS) of the London Stock Exchange (LSE) during the year 2002 \cite{Eisler-Kertesz-Lillo-Mantegna-2009-QF} and the foreign exchange rate of DEM against USD for the full year of 1998 ($\delta=1.4$) \cite{Jensen-Johansen-Petroni-Simonsen-2004-PA}.

Simonsen et al. \cite{Simonsen-Jensen-Johansen-2002-EPJB} argued that the most probable first passage time $s^*$ is the optimal investment horizon and found that $s^*$ scales with $\Delta{X}$ as a power law,
\begin{equation}
  s^* \sim (\Delta{X})^\gamma,
  \label{Eq:FPT:OIH:PL}
\end{equation}
where $\gamma\approx1.8$, deviating from $\gamma=2$ for Brownian motions. It is found that $\gamma <2$ holds for regularly spaced samples in most markets and the $\gamma$ value is larger in developed stock markets than emerging markets \cite{Zhou-Yuan-2005-PA,Karpio-ZaluskaKotur-Orlowski-2007-PA}. For tick-by-tick data that are unevenly sampled, it was reported that $\gamma>2$ \cite{Bartolozzi-Mellen-Chan-Oliver-DiMatteo-Aste-2007-PSPIE}.

For asset prices, it is natural to also consider the minimal time needed for the price to depreciate by a certain return $\Delta{X}\leq 0$.
Then, the exit time $s$ at time $i$ is defined by \cite{Jensen-Johansen-Simonsen-2003-PA}
\begin{equation}
  s(i, \Delta{X}) = \inf\{s^{\dag}: \Delta{X}(i,s^{\dag})\le \Delta{X}\}.
\end{equation}
A lot of research has shown that the optimal investment horizon is longer for gains than for losses, which is termed the gain-loss asymmetry \cite{Jensen-Johansen-Simonsen-2003-PA,Jensen-Johansen-Simonsen-2003-IJMPB,ZaluskaKotur-Karpio-Orlowski-2006-APPB,Simonsen-Ahlgren-Jensen-Donangelo-Sneppen-2007-EPJB, Grudziecki-Gnatowska-Karpio-Orlowski-ZaluskaKotur-2008-APPA,Karpio-ZaluskaKotur-Orlowski-2007-PA,Lee-Kim-Hwang-2008-JKPS,Bolgorian-2010-IJMPC,Zhang-Li-2011-PA,Sandor-Neda-2015-PA}. Several models have been proposed to reproduce the gain-loss asymmetry \cite{Donangelo-Jensen-Simonsen-Sneppen-2006-JSM,Ahlgren-Jensen-Simonsen-Donangelo-Sneppen-2007-PA,Siven-Lins-Hansen-2009-JSM}. Nevertheless, there are also assets that do not show significant gain-loss asymmetry \cite{Johansen-Jensen-Simonsen-2006-PA,Balogh-Simonsen-Nagy-Neda-2010-PRE}.

An analysis in the frequency space based on the discrete wavelet transform showed that the gain-loss asymmetry is introduced mainly by the low-frequency content of the price series and the asymmetry disappears if enough of the low-frequency content is removed \cite{Siven-Lins-Hansen-2009-JSM}. Alternatively, the gain-loss asymmetry is found to be caused by the non-Pearson-type autocorrelations in the time series \cite{Sandor-Simonsen-Nagy-Neda-2016-PRE}. Furthermore, the gain-loss asymmetry vanishes if the temporal dependence structure is destroyed by shuffling the time series \cite{Siven-Lins-2009-PRE}.

The gain-loss asymmetry can be related to the leverage effect as follows.
Recall first that the leverage effect is the fact that, after a large negative return, the
volatility increases significantly and then relaxes back, usually
roughly exponential in time \cite{Figlewski-Wang-2000,Bouchaud-Matacz-Potters-2001-PRL,Eisler-Kertesz-2004-PA}.
In contrast, after a positive return, there is no significant change
of volatility. And the causality is from (negative) returns to future volatility and
not the reverse, in the sense that a variation of volatility does not produce
any measurable change of the expected return in the future. The leverage
effect can be interpreted as due to the impact of a loss on the risk perception of the firm,
which increases as the equity over debt ratio has decreased as a result of the loss \cite{Figlewski-Wang-2000}.
It also reflects a behavioral response of investors, who after a loss frantically
reassess their risk exposure and readjust their portfolios, leading to large price moves.
The leverage effect allows one to account for both (i) the asymmetry in the average time for a price drop
of a fixed percentage versus a price gain of the same value (for instance 10
days for a loss of 5\% and 20 days for a gain of 5\% on the DJIA) and (ii)
the fact that this asymmetry in waiting times for positive vs negative objectives
is stronger for indices and weaker or absent for individual stocks \cite{Siven-Lins-2009-PRE,Siven-Lins-Hansen-2009-JSM}. ?
Consider a fixed drop target of -5\%. Such a drop is achieved by a succession of daily returns,
more negative than positive. When some negative loss occurs on one day,
the leverage effect leads to an increase of the amplitude of the next daily
return. Because we are calculating a waiting time conditional to a
cumulative drop of -5\%, the following returns will tend to be negative
and with larger amplitude due to the leverage effect. In contrast, for the waiting
time to reach a positive gain of +5\%, a majority of the daily moves will
be positive, and the leverage effect will be weaker or absent if there are
no or weak negative returns along the price path. As a consequence,
as the amplitude of the daily returns tend to be small for a sequence of
daily returns involved in a positive level (+5\%), the average waiting
time to reach a positive return (+5\%) is larger than for a negative
return goal. The asymmetry of a single stock is small because the leverage effect has a
rather short memory and the difference between the number of up and down
daily moves for reaching the target threshold is not large.
In contrast, for an index, or equivalently a portfolio of $N$ stocks,
one must account for the ubiquitous existence of
correlations between the returns of the stocks constituting
the portfolio. These correlations, which are in general positive (most stocks
move approximately together on average), produce an amplification
of the gain-loss asymmetry, because the leverage effect is stronger
by the averaging over the idiosyncratic residuals of the constituting stocks. This also leads to predict that the gain-loss asymmetry
for a portfolio is the stronger, the stronger is the average correlation
coefficients of its constituting stocks.

The moments of the exit time, called the distance structure functions \cite{Jensen-1999-PRL} or inverse structure functions \cite{Biferale-Cencini-Vergni-Vulpiani-1999-PRE,Biferale-Cencini-Lanotte-Vergni-Vulpiani-2001-PRL}, are defined by
\begin{equation}
  T_p(\Delta{X}) \equiv \langle{s^p}(\Delta{X})\rangle.
  \label{Eq:MF-SFA:ISF}
\end{equation}
Due to the duality between the structure function and the inverse structure function, one can intuitively expected that there is a power-law scaling stating that
\begin{equation}
  T_p({\Delta{X}})\sim {\Delta{X}}^{\phi(p)},
\end{equation}
where $\phi(p)$ is a nonlinear concave function \cite{Jensen-1999-PRL}.

Synthetic data for the GOY shell model of turbulence show perfect power-law dependence of the inverse structure functions on the velocity threshold \cite{Jensen-1999-PRL}. For two-dimensional turbulence, the inverse structure functions exhibit well-defined multifractal properties \cite{Biferale-Cencini-Lanotte-Vergni-Vulpiani-2001-PRL}. In contrast, for three-dimensional turbulence, the inverse structure functions of an experimental time series at high Reynolds number do not exhibit clear power law scaling \cite{Biferale-Cencini-Vergni-Vulpiani-1999-PRE}. Nevertheless, different experiments show that the inverse structure functions of three-dimensional turbulence exhibit a more general scaling behavior called
extended self-similarity (see next subsection) \cite{Beaulac-Mydlarski-2004-PF,Pearson-vandeWater-2005-PRE,Zhou-Sornette-Yuan-2006-PD}. To our knowledge, the behavior of inverse structure functions of financial asset prices has not been studied.

The inversion formula relating the direct scaling exponents $\zeta(q)$ of direct structure functions and the inverse scaling exponents $\phi(p)$ of inverse structure functions  can also be derived \cite{Hastings-2002-PRL,Schmitt-2005-PLA} and give
\begin{equation}
  \left\{
  \begin{array}{lll}
  \zeta(q) &=& -p\\
  \phi(p) &=& -q
  \end{array}
  \right.~.
  \label{Eq:ZetaTheta}
\end{equation}
These expressions are verified by the simulated velocity fluctuations from the shell model \cite{Roux-Jensen-2004-PRE}. However, this prediction (\ref{Eq:ZetaTheta}) cannot be confirmed by wind-tunnel turbulence experiments (Reynolds numbers ${\mbox{Re}}=400\sim1100$) \cite{Pearson-vandeWater-2005-PRE}, which is not surprising because the inverse structure function does not scale as a power law of the velocity increment thresholds \cite{Biferale-Cencini-Vergni-Vulpiani-1999-PRE,Pearson-vandeWater-2005-PRE}.

\subsubsection{Extended self-similarity}

To investigate the scaling properties of the inverse structure functions, one can define a set of relative exponents using the framework of extended self-similarity (ESS)
\cite{Benzi-Ciliberto-Tripiccione-Baudet-Massaioli-Succi-1993-PRE}:
\begin{equation}
  K(q,s) \sim [K_{q_0}(s)]^{\xi(q,q_0)},
  \label{Eq:MF-SF:ESS:Kq:Kq0}
\end{equation}
where $q_0$ is the order taken as a reference value. If Eq.~(\ref{Eq:MF-SF:Kq:s}) holds, we have
\begin{equation}
  \xi(q,q_0) = \frac{qH(q)}{q_0H(q_0)}.
  \label{Eq:MF-SF:ESS:xi:Hq:Hq0}
\end{equation}
When the time series is monofractal, $H(q)$ is independent of $q$, that is, $H(q)=H(q_0)$. It follows for monofractal time series that
\begin{equation}
  \xi(q,q_0) = q/q_0.
  \label{Eq:MF-SF:ESS:xi:q:q0}
\end{equation}
In the case of velocity structure functions in fluid mechanics, $q_0=3$ is a natural choice based on the exact Kolmogorov's Four-Fifth Law \cite{Kolmogorov-1941b-DAN,Zhou-Sornette-Yuan-2006-PD}. For financial time series, we suggest to choose $q_0=2$ because $H(2)$ is the Hurst index. Generally, the ESS approach provides a wider scaling range for the extraction of scaling exponents.

\begin{figure}[htb]
  \centering
  \includegraphics[width=0.32\linewidth]{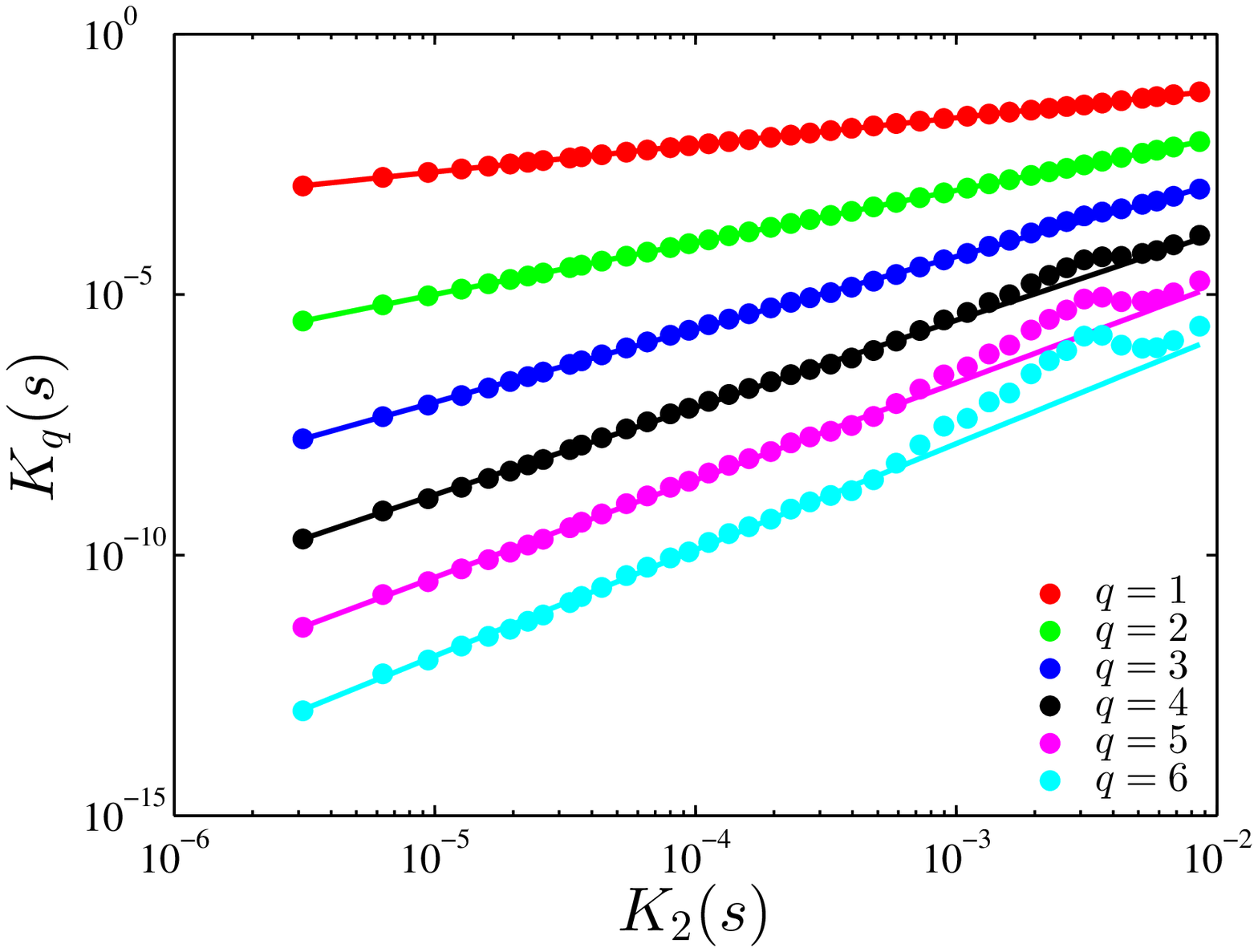}
  \includegraphics[width=0.32\linewidth]{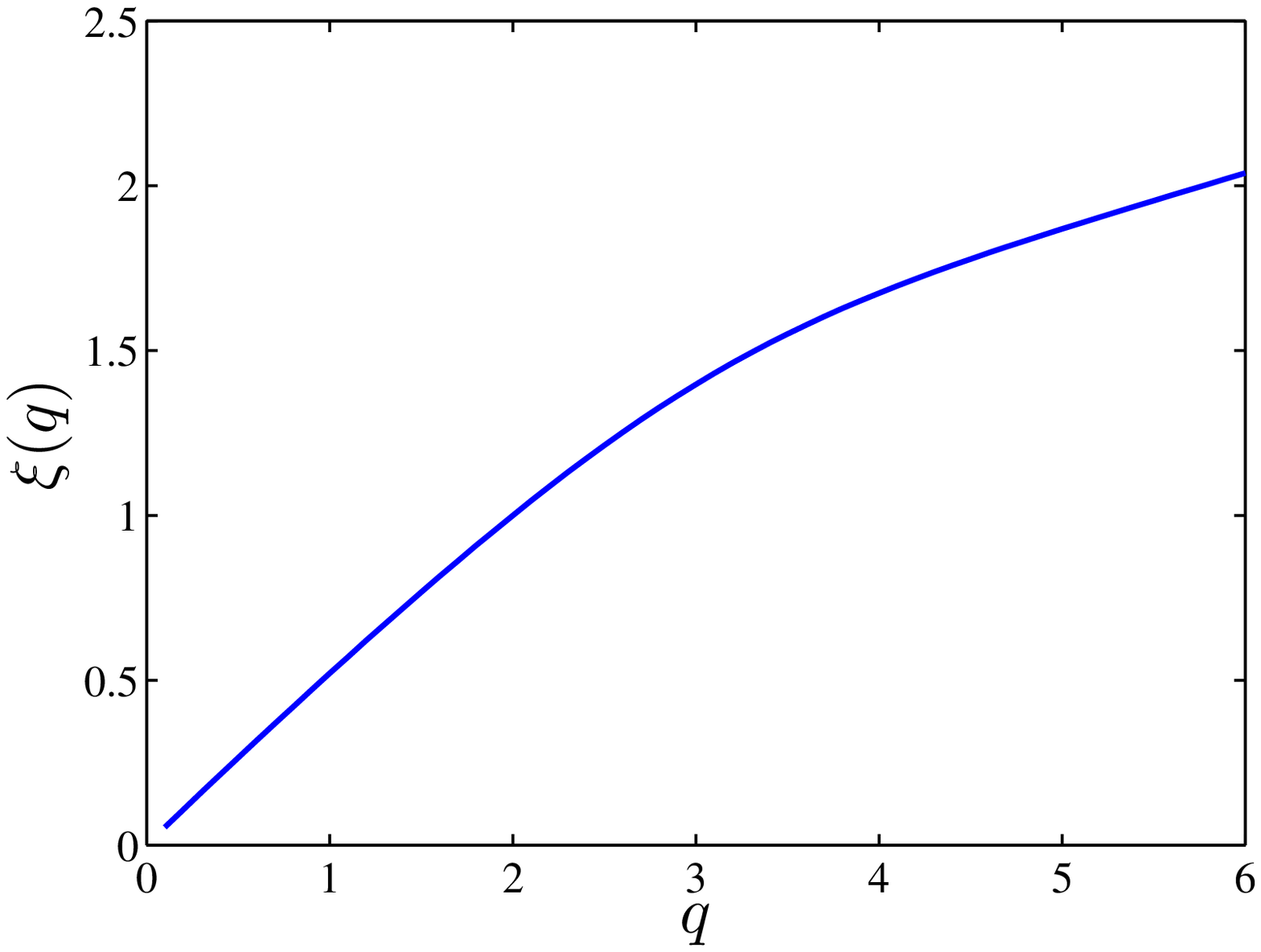}
  \includegraphics[width=0.32\linewidth]{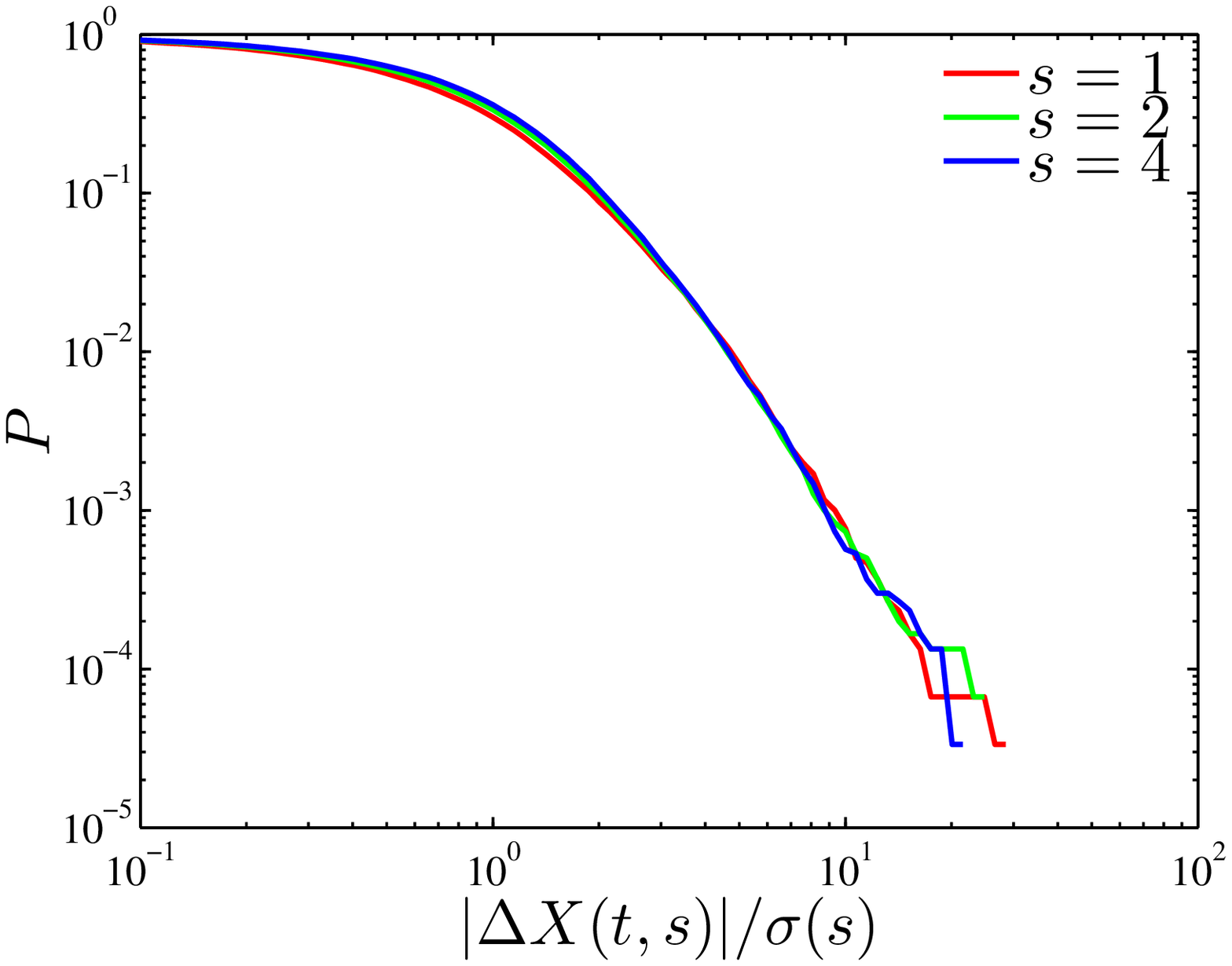}
  \caption{(color online) Extended self-similarity analysis of the daily price time series of the DJIA index (1896-2015). (a) Power-law dependence of the $q$th-order structure function $K(q,s)$ on the second-order structure function $K_2(s)$ for different $q$ values. (b) The ESS scaling exponent $\xi(q)$ as a function of $q$. (c) The distributions of normalized volatility $|\Delta{X}(t,s)|/\sigma(s)$ for $s=1$, $s=2$ and $s=4$.}
\label{Fig:MF-SF:ESS:DJIA}
\end{figure}

Fig.~\ref{Fig:MF-SF:ESS:DJIA}(a) presents the log-log plots of $K(q,s)$ against $K_2(s)$ for $q= 1, 2, \cdots, 6$ with $s \in [1,2^{11}]$. The straight lines hold over more than two orders of magnitude and over at least 3.5 orders of magnitudes for $q\leq3$, showing the existence of extended self-similarity in the structure functions. The scaling range for small $q$'s seems to be broader than for large $q$'s.
The ESS scaling exponents $\xi(q,2)$ are shown in Fig.~\ref{Fig:MF-SF:ESS:DJIA}(b).
There is a {\it{weak}} indication that $\xi(q,2)$ has a nonlinear dependence as a function of $q$, with a downward curvature making the curve depart from the linear dependence $\xi(q,2) = q/2$ observes for small $q$'s.

Whether a time series possesses monofractal or multifractal behaviors is related to the innovation distributions. For simplicity of notation, we denote $v=|\Delta{X}(t,s)|$. By definition, we have
\begin{equation}
  K(q,s) = \int_{0}^{\infty} v^q \Pr(s;v){\mbox{d}}v = \left[\sigma(s)\right]^q \int_{0}^{\infty} {\mbox{d}}x~\Phi(s;x) x^q,
  \label{Eq:MF-SF:ESS:PDF1}
\end{equation}
where $\Pr(s;v)$ is the distribution of $v$, $\sigma(s)$ is the standard deviation of $v$, $x=v/\sigma$ is the normalized volatility, and $\Phi(s;x)$ is the distribution of $x$. Eliminating $\sigma(s)$ using $q=2$, it follows that
\begin{equation}
  K(q,s) = [K_2(s)]^{q/2}~ \frac{\int_{0}^{\infty} {\mbox{d}}x~\Phi(s;x) ~x^q}{\left[\int_{0}^{\infty} {\mbox{d}}x~\Phi(s;x) ~x^2 \right]^{q/2}}.
  \label{Eq:MF-SF:ESS:PDF2}
\end{equation}
If $\Phi(s;x)$ is universal and independent of $s$, then the last term in Eq.~(\ref{Eq:MF-SF:ESS:PDF2}) is a number independent of $s$ (and thus of $K_2(s)$) and the monofractal behavior $\xi(q,2) = q/2$ follows. Fig.~\ref{Fig:MF-SF:ESS:DJIA}(c) shows the distributions of the normalized volatility $|\Delta{X}(t,s)|/\sigma(s)$ at three different scales. Although the three curves well overlap in their tails, they show clear differences in the bulk around $x=1$. This suggests that the ESS cannot be nonfractal, which is consistent with the presence of nonlinear curvature in Fig.~\ref{Fig:MF-SF:ESS:DJIA}(b).

We note that the ESS analysis is not constrained to the structure functions. Rather, it can be used for the moments of other multifractal analysis approaches. In validating and qualifying the multiscaling behaviour of the recurrence intervals of financial volatility, the relative dependence of recurrence interval moments of different orders of empirical time series has been investigated \cite{Wang-Yamasaki-Havlin-Stanley-2008-PRE}, which is in essence an ESS analysis \cite{Ren-Zhou-2008-EPL}.

\subsection{Wavelet transform approaches}
\label{S2:MF-WT}

\subsubsection{Multifractal analysis based on wavelet transform (MF-WT)}
\label{S3:MF-WT:WT}

The wavelet transform is widely used as a mathematical microscope to analyze time series \cite{Grossmann-Morlet-1984-SIAMJMA,Daubechies-Grossmann-Meyer-1986-JMP}. In particular, it can be used to analyze the singular structure of fractal and multifractal time series \cite{Holschneider-1988-JSP,Arneodo-Grasseau-Holschneider-1988-PRL,Ghez-Vaienti-1989-JSP}.

The wavelet transform of function $X(t)$ is \cite{Grossmann-Morlet-1984-SIAMJMA,Daubechies-Grossmann-Meyer-1986-JMP}
\begin{equation}\label{Eq:WT:Def}
  C_{\psi}(b,s) = \frac{1}{s}\int_{-\infty}^{+\infty}X(t)\psi\left(\frac{t-b}{s}\right){\rm{d}}t,
\end{equation}
where $b\in{\mathbb{R}}$ is the position parameter, $a\in{\mathbb{R}}^+$ is the dilation parameter, $\psi(t)$ is the analyzing ``mother'' wavelet. Gaussian wavelets are widely adopted  \cite{Mallat-1997} and are defined as derivatives of the Gaussian function:
\begin{equation}
  g_n(t) = g^{(n)}(t) = c_n\frac{{\rm{d}}^n}{{\rm{d}}t^n}{\rm{e}}^{-t^2/2},
  \label{Eq:WT:GaussianWavelets}
\end{equation}
where $c_n$ is the normalization coefficient. The second order ($n=2$) Gaussian wavelet is known as  the Mexican hat.

To perform wavelet transform on a time series $\{X(i)\}_{i=1}^N$, one usually discretizes Eq.~(\ref{Eq:WT:Def}) in which $b$ can
take any value along the sampling spacing and $s$ takes $n_s$ scales forming a geometric sequence:
\begin{equation}
  s_j=\lambda^{j-1}s_{\min}, ~~~1\leqslant{j}\leqslant{n_s},
  \label{Eq:WT:aj}
\end{equation}
where $\lambda>1$. The discretised wavelet transform of a given time series $Y(i)$ is then
\begin{equation}
   W(s,i) = \frac{1}{s} \sum_{j = 1}^N X(j) \psi \left(\frac{j-i}{s}\right),~~i=1,\cdots, N.
 \label{Eq:MFXWT:WTTransform}
\end{equation}
%
The wavelet transform is used to identify different types of singularities in the signals in the time-scale plane,
depending on the order $n$ of derivative of the mother wavelet (\ref{Eq:WT:GaussianWavelets}).
More generally, choosing a mother wavelet $\psi$ satisfying $\int x^{m+1} \psi(s) {\rm{d}} x = 0$
implies that the wavelet transform is blind to polynomial dependence up to order $m$ and only higher order
powers are detected. For instance, for $m=0$, the wavelet transform is independent of the level of the signal
and informs essentially on the local slope. For $m=1$, the wavelet transform is independent of the level of the signal
and of its local slope, and it informs on the local curvature.
We should also mention the existence of orthogonal wavelet transforms \cite{Daubechies-1988-CPAM,Daubechies-1993-SIAMjma,Mallat-1997} and biorthogonal wavelet transform \cite{Cohen-Daubechies-Feauveau-1992-CPAM}, which can be used directly to determine the Hurst index of a long-range correlated time series \cite{Muzy-Bacry-Arneodo-1994-IJBC,Arneodo-Bracy-Muzy-1995-PA,Audit-Bacry-Muzy-Arneodo-2002-IEEEtit}.

To investigate the multifractal nature of a time series, one can calculate the $q$th order moments of the wavelet coefficients \cite{Ghez-Vaienti-1992-NL,Bacry-Muzy-Arneodo-1993-JSP}
\begin{equation}
  M_q(s)=\frac{1}{N}\sum_{i=1}^N|W(s,i)|^q
  \label{Eq:WT:BO:Mq:a}
\end{equation}
and we expect that
\begin{equation}
  M_q(s) \sim s^{qH(q)}, ~~s\gg0.
  \label{Eq:WT:BO:Mq}
\end{equation}
We note that $M_q(s)$ might diverge for $q\leq0$ because some values of $|W(i,k)|$ might be close or equal to 0.

One can obtain the scaling exponents through ordinary least-squares regression. Or, we can adopt the weighted least-squares regression by putting more weight on the points with smaller variance \cite{Audit-Bacry-Muzy-Arneodo-2002-IEEEtit}.
Denoting
\begin{equation}\label{Eq:WT:BO:LqN}
  {\mathcal{L}}_{j,q}=\ln{M_q}/q,
\end{equation}
the corresponding weight $w_j$ is
\begin{equation}\label{Eq:WT:BO:wj}
  w_j \sim {\rm{Var}}({\mathcal{L}}_{j,q})^{-1}
\end{equation}
and the objective function is
\begin{equation}\label{Eq:WT:BO:Obj}
  \sum_{j}w_j{\mathrm{E}}\left[({\mathcal{L}}_{j,q}-H(q)\ln{s_j}-A)^2\right].
\end{equation}
The estimate of $H(q)$ is determined as follows:
\begin{equation}\label{Eq:WT:BO:hatH}
  \hat{H}(q)=\frac{\sum_{j}w_j\sum_{j}w_j\ln{s_j}{\mathcal{L}}_{j,q}-\sum_jw_j\ln{s_j}\sum_{j}w_j{\mathcal{L}}_{j,q}}
{\sum_{j}w_j\sum_{j}w_j\ln^2{s_j}-\left(\sum_{j}w_j\ln{s_j}\right)^2}.
\end{equation}

\subsubsection{Wavelet transform modulus maxima (WTMM)}
\label{S3:MF-WT:WTMM}

A more elegant approach, which removes the redundancy contained in continuous wavelet transforms,
is based on the wavelet transform modulus maxima (WTMM) \cite{Muzy-Bacry-Arneodo-1991-PRL,Bacry-Muzy-Arneodo-1993-JSP,Muzy-Bacry-Arneodo-1993-PRE,Muzy-Bacry-Arneodo-1994-IJBC}. At each scale $s_j$, there exists several local maxima of $|W(s_j,i)|$. These maxima form several maxima lines in the $(i,s)$ plane. The WTMM is capable of detecting all the singularities of a large class of time series \cite{Mallat-Hwang-1992-IEEEtit}. The traditional method based on partition functions may face problems in dealing with the situation of $q\leq0$. However, we will find that the multifractal analysis based on WTMM can overcome this difficulty because the wavelet coefficients on the WTMM line are local maxima that are non 0 \cite{Muzy-Bacry-Arneodo-1993-PRE}.

Assume that there are $N_j$ local maxima at scale $s_j$ and $W(i_{j,k},s_j)$  is the $k$th local maximum located at $i_{j,k}$. The $q$th order moments defined on the WTMM are
\begin{equation}
  M(q,s_j)=\frac{1}{N_j}\sum_{k=1}^{N_j}|W(i_{j,k},s_j)|^q
  \label{Eq:MF-WTMM:Mq:a}
\end{equation}
or alternatively the partition functions read
\begin{equation}\label{Eq:MF-WTMM:Zq:a}
  Z(q,s_j)=\sum_{k=1}^{N_j}|W(i_{j,k},s_j)|^q.
\end{equation}
For multifractal time series, we have
\begin{equation}
  Z(q,s)\sim s^{\tau(q)}.
  \label{Eq:MF-WTMM:tau:q}
\end{equation}
The generalized dimensions and singularity spectrum can then be calculated straightforwardly.

Analogous to the thermodynamic approximation in the direct determination of singularities and singularity spectrum in Section \ref{S3:MF-PF:DirectDetermine}, we can define the canonical measure  \cite{Muzy-Bacry-Arneodo-1991-PRL},
\begin{equation}
 \mu_k(q,s_j)=\frac{|W(i_{j,k},s_j)|^q}{Z_q(s_j)},
 \label{Eq:MF-WTMM:DC:mu}
\end{equation}
such that the singularity strength $\alpha$ is
\begin{equation}
 \alpha(q)= \lim_{s_j\to0}\frac{1}{\ln{s_j}}\sum_{i=1}^{N_j}\left[\mu_k(q, s_j)\ln{|W(i_{j,k},s_j)|}\right]
 \label{Eq:MF-WTMM:DC:alpha}
\end{equation}
and the singularity spectrum $f(\alpha)$ is
\begin{equation}
 f(\alpha) =\lim_{s_j\to0}\frac{1}{\ln{s_j}}\sum_{i=1}^{N_j}\left[\mu_k(q,s_j)\ln{\mu_k(q,s_j)}\right]
 \label{Eq:MF-WTMM:DC:f}.
\end{equation}
In applications, we use different scales $s_j$ and perform (weighted) linear regression on the log-log scales.

Using respectively the partition function approach and wavelet analysis to detect the multifractal nature of a single time series, it is found that $\tau^{\mbox{\tiny{PF}}}(q)=\tau^{\mbox{\tiny{WT}}}(q)+q$ and $\alpha^{\mbox{\tiny{PF}}}(q)=\alpha^{\mbox{\tiny{WT}}}(q)+1$ \cite{Turiel-Parga-2000-NCmp,Kestener-Arneodo-2003-PRL,Turiel-PerezVicente-Grazzini-2006-JComputP}. 

\subsubsection{Wavelet leader (MF-WL)}
\label{S3:MF-WT:WL}

A more recent multifractal formalism is based on wavelet leaders \cite{Jaffard-2004,Jaffard-2006-AFST,Wendt-Abry-Jaffard-2007-IEEEspm,Wendt-Abry-2007-IEEEtsp, Lashermes-Roux-Abry-Jaffard-2008-EPJB, Serrano-Figliola-2009-PA}. Wavelet leaders, introduced by Jaffard \cite{Jaffard-2004,Jaffard-2006-AFST}, are defined based on discrete wavelet coefficients, which decompose the time series $X(i)$ on the bases $\{ \psi_{j,k}\}_{j \in Z, k \in Z}$ composed of discrete wavelets $\psi_{j, k}$. The integers $j \in Z$ and $k \in Z$ define the scale $s = 2^j$ and location $b = k2^j$. The discrete wavelets $\{ \psi_{j,k}\}_{j \in Z, k \in Z}$ are space-shifted and scale-dilated templates of an analyzing wavelet $\psi(t)$:
\begin{equation}
  \psi_{j,k}(t) = \frac{1}{2^j} \psi \left(\frac{i-k2^j}{2^j} \right),
  \label{Eq:MF-WL:OrthoBasis}
\end{equation}
which usually uses the Daubechies wavelet of order 1 for the mother wavelet. 
The discrete wavelet coefficients are defined as
\begin{equation}
  W(j, k) = \int_i X(i) \frac{1}{2^j} \psi \left(\frac{i-k2^j}{2^j} \right).
  \label{Eq:MF-WL:DCT}
\end{equation}
One defines a dyadic interval $\lambda(j, k)$ as
\begin{equation}
  \lambda(j,k) = \left[k2^j, (k+1)2^j\right),
  \label{Eq:MF-WL:Lambda:jk}
\end{equation}
and denotes the union of interval $\lambda$ and its two adjacent neighbors as $3\lambda$,
\begin{equation}
  3\lambda(j,k) = \lambda(j, k-1) \cup \lambda(j, k) \cup \lambda(j, k+1).
  \label{Eq:MF-WL:3Lambda:jk}
\end{equation}
The wavelet leader $\ell(j, k)$ is defined by \cite{Jaffard-2004,Jaffard-2006-AFST}
\begin{equation}
  \ell(j, k) =  \sup_{\lambda' \subset 3\lambda(j, k) } |W(\lambda')|,
  \label{Eq:MF-WL:Leader}
\end{equation}
which corresponds to the largest value of the absolute wavelet coefficients $|W(j',k')|$ calculated on intervals $(k-1)2^j \le 2^{j'} k' < (k+2)2^j$ with $0 < j' \le j$. Note that, to compute the wavelet leaders, all the fine scales $2^{j'} \le 2^j$ must be considered.

The multifractal analysis based on wavelet leaders can be described as follows \cite{Jaffard-2004,Jaffard-2006-AFST,Wendt-Abry-Jaffard-2007-IEEEspm,Wendt-Abry-2007-IEEEtsp, Lashermes-Roux-Abry-Jaffard-2008-EPJB, Serrano-Figliola-2009-PA}. As usual, we can define the moments of order $q$ by
\begin{equation}
  M(q,j) = \frac{1}{n_j}\sum_{k=1}^{n_j} \ell(j,k)^q,
  \label{Eq:MF-WL:SF}
\end{equation}
where $n_j$ is the number of wavelet leaders at scale $s=2^j$. One can also expect the following scaling behavior if the underlying processes are jointly multifractal,
\begin{equation}
  M(q,j) \sim 2^{j \zeta(q)}=s^{\zeta(q)}.
  \label{Eq:MF-WL:SF:Scaling}
\end{equation}
where $\zeta(q)$ is related to $\tau(q)$ by
\begin{equation}
  \zeta(q) = \tau(q)+1 = qH(q).
\end{equation}
We can also provide a canonical approach to determine the singularity strength function $\alpha(q)$ and the multifractal spectrum $f(\alpha(q))$ directly, as for the WTMM case in Section \ref{S3:MF-WT:WTMM}.

An interesting application of wavelet leaders is the pointwise wavelet leaders entropy \cite{Rosenblatt-Serrano-Figliola-2012-IJWMIP}, which is defined over different scales ($j$) for a given point ($k$) by
\begin{equation}
  E(k) = -\sum_{j=1}^{m} \rho_{j,k}\log_2\rho_{j,k},
  \label{Eq:MF-WL:pL:Entropy}
\end{equation}
where $\rho_{j,k}\log_2\rho_{j,k}=0$ if $\rho_{j,k}=0$ and
\begin{equation}
  \rho_{j,k} =
  \left\{
  \begin{array}{lll}
  W(j,k)^2/\sum_{j=1}^{m}W(j,k)^2 &  {\mathrm{if}}~W(j,k)\neq0\\
  0                                 &  {\mathrm{if}}~W(j,k)=0
  \end{array}
  \right.
  \label{Eq:MF-WL:pL:Entropy:Rho}
\end{equation}
When applied to the DJIA index (1928-2011), the resulting wavelet leader entropy is able to identify historical large market fluctuations \cite{Rosenblatt-Serrano-Figliola-2012-IJWMIP}.

Recently, wavelet $p$-leaders are introduced to characterize the $p$-exponents of negative pointwise regularity \cite{Jaffard-Melot-Leonarduzzi-Wendt-Abry-Roux-Torres-2016a-PA}. Based on the wavelet coefficients calculated in Eq.~(\ref{Eq:MF-WL:DCT}), the wavelet $p$-leaders are defined by
\begin{equation}
  \ell^{(p)}(j, k) =  \left(\sum_{j'\leq{j},\lambda' \subset 3\lambda(j, k)} |W(j,k)|^p2^{j'-j}\right)^{1/p}.
  \label{Eq:MF-WL:p-Leader}
\end{equation}
The $p$-leader multifractal formalism relies on the sample moments of the $p$-leaders \cite{Leonarduzzi-Wendt-Abry-Jaffard-Melot-Roux-Torres-2016b-PA}:
\begin{equation}
  M^{(p)}(q,j) = \frac{1}{n_j}\sum_{k=1}^{n_j} \left[\ell^{(p)}(j,k)\right]^q,
  \label{Eq:MF-WL:pL:SF}
\end{equation}
It is found that MF-DFA characterizes local regularity via the 2-exponent, not the H{\"{o}}lder exponent, and the wavelet $p$-leader multifractal formalism can be viewed as an
wavelet-based extension of MF-DFA \cite{Leonarduzzi-Wendt-Abry-Jaffard-Melot-Roux-Torres-2016b-PA}. The $p$-leaders suffer
from finite-resolution effects, which can be resolved through an explicit and universal closed-form correction \cite{Leonarduzzi-Wendt-Abry-Jaffard-Melot-2017-IRRRtsp}.

\subsection{Detrended fluctuation approaches}
\label{S2:MF:Detrend:Methods}

\subsubsection{General framework}
\label{S3:MF:Detrend:GeneralFramework}

Consider a time series $\{X(i)|i =1,2,\cdots,N\}$, whose data points are usually the cumulative sums of a time series of innovations. Let $\widetilde{X}(t)$ be the local trend function, which is usually determined locally around the data points using some method. The detrended residuals are defined by
\begin{equation}
  \epsilon(i) = X(i)-\widetilde{X}(i).
  \label{Eq:MF:Detrend:epsilon}
\end{equation}

We divide $\{\epsilon(i)\}$ into $N_s=\mathrm{int}[N/s]$ non-overlapping segments with boxes of size $s$, where $\mathrm{int}[y]$ is the largest integer that is no larger than $y$. We denote the $v$th segment as $S_v=\{\epsilon((v-1)s+j)| j=1,2, \cdots, s\}$.
The local detrended fluctuation function $F_v(s)$ in the $v$th box is defined as the r.m.s. of the detrended residuals:
\begin{equation}
  \left[F_v(s)\right]^2 = \frac{1}{s}\sum_{i=1}^{s} \left[\epsilon((v-1)s+j)\right]^2~.
  \label{Eq:MF:Detrend:Fv:s}
\end{equation}
The $q$th-order overall detrended fluctuation is
\begin{equation}
  F_q(s) = \left\{\frac{1}{N_s}\sum_{v=1}^{N_s} {F_v^q(s)}\right\}^{\frac{1}{q}},
  \label{Eq:MF:Detrend:Fqs}
\end{equation}
where $q$ can take any real value except $q=0$.
When $q=0$, we have
\begin{equation}
  \ln[F_0(s)] = \frac{1}{N_s}\sum_{v=1}^{N_s}{\ln[F_v(s)]},
  \label{Eq:MF:Detrend:Fq0}
\end{equation}
according to L'H\^{o}spital's rule.

When the whole series $\{\epsilon(i)\}$ cannot be completely covered by $N_s$ boxes, one can use $2N_s$ boxes to cover the series from both ends of the series, where $N_s=\mathrm{int}[N/s]$ and a short part at each end of the residuals may remain uncovered \cite{Kantelhardt-Zschiegner-KoscielnyBunde-Havlin-Bunde-Stanley-2002-PA}. We denote the $v$th segment partitioned from left to right as $S_v=\{\epsilon((v-1)s+j)| j=1,2, \cdots, s\}$ for $v=1,2, \cdots, N_s$ and the $v$th segment partitioned from left to right as $S_v=\{\epsilon(N-(v-N_s)s+j)| j=1,2, \cdots, s\}$  for $v=N_s+1,N_s+2, \cdots, 2N_s$. The local detrended fluctuation function $F_v(s)$ in the $v$th box is defined as the r.m.s. of the detrended residuals,
\begin{subequations}
\begin{equation}
  \left[F_v(s)\right]^2 = \frac{1}{s}\sum_{i=1}^{s} \left[\epsilon((v-1)s+j)\right]^2~~{\mathrm{for}}~v=1,2, \cdots, N_s
  \label{Eq:MF:Detrend:Fv:s:L2R}
\end{equation}
and
\begin{equation}
  \left[F_v(s)\right]^2 = \frac{1}{s}\sum_{i=1}^{s} \left[\epsilon(N-(v-N_s)s+j)\right]^2~~{\mathrm{for}}~v=N_s+1,N_s+2, \cdots, 2N_s.
  \label{Eq:MF:Detrend:Fv:s:R2L}
\end{equation}
\end{subequations}
The $q$th-order overall detrended fluctuation is
\begin{equation}
  F_q(s) = \left\{\frac{1}{2N_s}\sum_{v=1}^{2N_s} \left[F_v(s)\right]^q\right\}^{\frac{1}{q}},
  \label{Eq:MF:Detrend:Fqs:LR}
\end{equation}
where $q$ can take any real value except for $q=0$.
When $q=0$, we have
\begin{equation}
  \ln[F_0(s)] = \frac{1}{2N_s}\sum_{v=1}^{2N_s}{\ln[F_v(s)]},
  \label{Eq:MF:Detrend:Fq0:LR}
\end{equation}
according to L'H\^{o}spital's rule.

Varying the values of the segment size $s$, we can determine the power-law relation between the function $F_q(s)$ and the size scale $s$,
\begin{equation}
  F_q(s) \sim {s}^{H(q)}.
  \label{Eq:MF:Detrend:hq}
\end{equation}
According to the standard multifractal formalism, the multifractal scaling exponent $\tau(q)$ can be used to characterize the multifractal properties, which reads
\begin{equation}
\tau(q)=qH(q)-D_f,
\label{Eq:tau:hq}
\end{equation}
where $D_f$ is the fractal dimension of the geometric support of the multifractal measure \cite{Kantelhardt-Zschiegner-KoscielnyBunde-Havlin-Bunde-Stanley-2002-PA}. For time series analysis, we have $D_f=1$. If the scaling exponent function $\tau(q)$ is a nonlinear function of $q$, the time series is regarded to have a multifractal nature.

Following the approach determining directly $f(\alpha)$ from the partition function \cite{Chhabra-Jensen-1989-PRL,Chhabra-Meneveau-Jensen-Sreenivasan-1989-PRA,Meneveau-Sreenivasan-Kailasnath-Fan-1990-PRA}, the singularity strength and its spectrum in the detrended fluctuation approaches can also be determined directly \cite{Xu-Gu-Zhou-2017-PRE}. Defining the canonical measure $\mu(q,s,v)$
\begin{equation}
  \mu(q,s,v) = \frac{F_v^q(s)}{\sum_{v=1}^{N_s} F_v^q(s)},
  \label{Eq:MF-DMA-CN:mu}
\end{equation}
we have
\begin{subequations}
\begin{equation}
  \alpha(q) = \lim_{s\to0} \frac{\sum_{v=1}^{N_s} \mu(q,s,v) \ln F_v(s)}{\ln{s}},
  \label{Eq:MF-DMA-CN:alpha}
\end{equation}
and
\begin{equation}
   f(\alpha(q)) = \lim_{s\to0}\frac{\sum_{v=1}^{N_s} \mu(q,s,v) \ln\left[\mu(q,s,v)\right]}{\ln{s}}.
  \label{Eq:MF-DMA-CN:fq}
\end{equation}
\label{Eq:MF-DMA-CN:alpha:fq}
\end{subequations}
Numerical experiments based on the $p$-model and fractional Brownian motions showed that MF-DMA and its direct determination variant have comparative performance to unveil the fractal and multifractal properties \cite{Xu-Gu-Zhou-2017-PRE}.

\subsubsection{Multifractal detrended fluctuation analysis (MF-DFA)}
\label{S3:MF-DFA}

The Detrended Fluctuation Analysis (DFA) was invented originally to investigate the long-range dependence in coding and noncoding DNA nucleotide sequences \cite{Peng-Buldyrev-Havlin-Simons-Stanley-Goldberger-1994-PRE}. The multifractal extension of DFA, MF-DFA, was developed shortly thereafter \cite{CastroESilva-Moreira-1997-PA,Weber-Talkner-2001-JGR}. Since the work of Kantelhardt et al. \cite{Kantelhardt-Zschiegner-KoscielnyBunde-Havlin-Bunde-Stanley-2002-PA}, MF-DFA has become one of the most important methods for multifractal analysis in diverse fields.

\begin{figure}[tb]
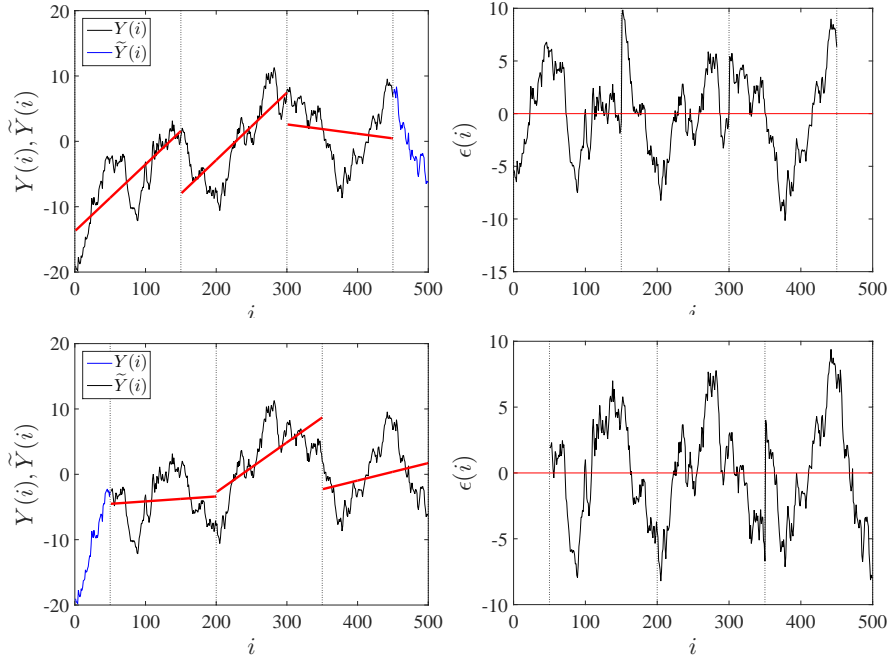

  \centering
  \includegraphics[width=0.35\linewidth]{Fig_MFDFA_Algo_a.eps}
  \includegraphics[width=0.35\linewidth]{Fig_MFDFA_Algo_b.eps}
  \includegraphics[width=0.35\linewidth]{Fig_MFDFA_Algo_c.eps}
  \includegraphics[width=0.35\linewidth]{Fig_MFDFA_Algo_d.eps}
  \caption{(color online) Detrending process in the MF-DFA method. The time series has 500 data points and the timescale is $s=150$. The first row shows the detrending process from left to right in which the data points with $i=451, 452, \cdots, 500$ should not be used in calculating $F$, and the second row illustrates the detrending process from right to left in which the data points with $i=1, 2, \cdots, 150$ should not be used.}
  \label{Fig:MF-DFA:Detrend}
\end{figure}

In the MF-DFA method, the trend function is a polynomial \cite{Kantelhardt-Zschiegner-KoscielnyBunde-Havlin-Bunde-Stanley-2002-PA}, and the linear trend is most generally adopted. For each point $i$, its trend value $\widetilde{X}(i)$ is determined together with other points in the same box $S_v$, using a polynomial fitting of order $l$ to the $s$ data points in $i\in S_v$:
\begin{equation}
  \widetilde{X}(i) = \sum_{k=0}^\ell a_k i^k, ~~~l=1,2, \cdots.
  \label{Eq:MF:Trend:Polynomial}
\end{equation}
The detrending process for $\ell=1$ is shown in Fig.~\ref{Fig:MF-DFA:Detrend}. The figure shows a case with $N_s<N/s$ in which the data points at each end of the time series should not be included to form a box. Otherwise, one obtains incorrect results with $f_{\max}=f(q=0)>1$, as will be elaborated in Section \ref{S2:MF:N/s}. In the standard procedure implementing MF-DFA, researchers remove the constant shift $\langle{\Delta{X}}\rangle$ from the increments $\Delta{X}$ and work on the cumulative profile
\begin{equation}
  x(i) = \sum_{j=1}^{i} \left[\Delta{X}(i)-\langle{\Delta{X}}\rangle\right],~~i = 1, 2, \cdots, N,
  \label{Eq:cumsum}
\end{equation}
where $\langle{\Delta{X}}\rangle$ is the sample mean of the $\Delta{X}(i)$ series. However, when the trend function is polynomial, this manipulation does not have any impact on the results.

A subtle issue of the MF-DFA method is the choice of polynomial order. Using classic mathematical models (fractional Brownian motion, L{\'e}vy process and binomial measure) and real-world time series (foreign exchange rate and literary texts), O{\'{s}}wi{\c{e}}cimka et al. found that the calculated singularity spectra could be very sensitive to the order of the detrending polynomial and this sensitivity depends on the analyzed time series \cite{Oswiecimka-Drozdz-Kwapien-Gorski-2013-APPA}. Without a deep understanding of the underlying mechanism driving the dynamics, it is hard to determine which polynomial order should be used. Anyway, for financial time series such as returns, we can check if the value of the Hurst index $H(2)$ is close to 0.5, based on the fundamental stylized fact of the absence of long memory in financial returns \cite{Cont-2001-QF}, which however might not hold at the transaction level \cite{Zhou-Gu-Jiang-Xiong-Chen-Zhang-Zhou-2017-CE}.

A possible solution is to use ``feasible trend functions'' in the multifractal feasibly detrended fluctuation analysis (MFFDFA) \cite{Rak-Zieba-2015-APPB}. One presets {\textit{a priori}} a set of {\textit{any}} trend functions, say, $\{ax^2+bx+c, a\sin(x^2)+bx+c, ax^3+bx+c\}$, and fit each segment with all the preset trend functions. For each segment, only one detrended function is chosen which has the maximum coefficient of determination $R^2$. Hence, for each time scale $s$, the trend functions for different segments are not necessary the same. Numerical experiments on fractional Brownian motions and binomial measures show that the MFFDFA method outperforms the MF-DFA method \cite{Rak-Zieba-2015-APPB}. For binomial measures, the improvement is partially due to the fact that the fluctuation functions $F_q(s)$ for MFFDFA have much smaller log-periodic oscillation amplitudes, which reduces the impact of the choice of scaling ranges \cite{Zhou-Sornette-2009b-PA}.

\subsubsection{Multifractal detrending moving average analysis (MF-DMA)}
\label{S3:MF-DMA}

Based on the moving average (MA) technique for estimating the Hurst exponent of self-affinity signals \cite{Vandewalle-Ausloos-1998-PRE}, the Detrending Moving Average (DMA) analysis considers the difference between the original signal and its moving average function \cite{Alessio-Carbone-Castelli-Frappietro-2002-EPJB,Carbone-2009-IEEE}. The DMA method can also be easily implemented to estimate the correlation properties of non-stationary series without any {\textit{a priori}} assumptions, which is widely applied to the analysis of real-world time series \cite{Carbone-Castelli-2003-SPIE,Carbone-Castelli-Stanley-2004-PA,Carbone-Castelli-Stanley-2004-PRE,Varotsos-Sarlis-Tanaka-Skordas-2005-PRE,Serletis-Rosenberg-2007-PA,Arianos-Carbone-2007-PA,Matsushita-Gleria-Figueiredo-Silva-2007-PLA,Serletis-Rosenberg-2009-CSF} and synthetic signals as well \cite{Serletis-2008-CSF}. Extensive numerical experiments unveil that the performance of DMA is comparable to DFA with slightly different priorities under different situations and both DFA and DMA remain ``The Methods of Choice'' in determining the Hurst index of time series \cite{Xu-Ivanov-Hu-Chen-Carbone-Stanley-2005-PRE,Bashan-Bartsch-Kantelhardt-Havlin-2008-PA,Shao-Gu-Jiang-Zhou-Sornette-2012-SR}.

\begin{figure}[tb]
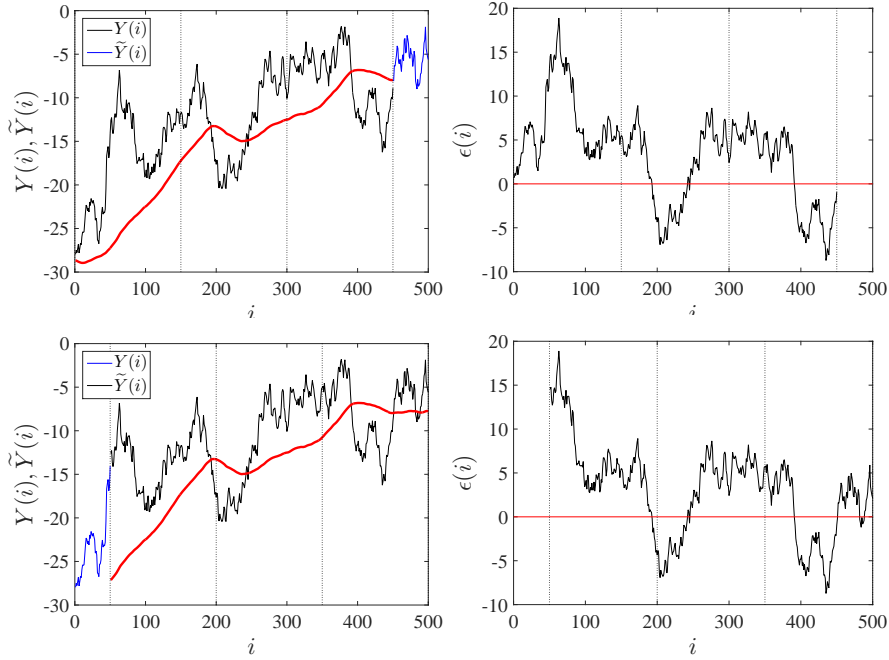

  \centering
  \includegraphics[width=0.35\linewidth]{Fig_MFDMA_Algo_a.eps}
  \includegraphics[width=0.35\linewidth]{Fig_MFDMA_Algo_b.eps}
  \includegraphics[width=0.35\linewidth]{Fig_MFDMA_Algo_c.eps}
  \includegraphics[width=0.35\linewidth]{Fig_MFDMA_Algo_d.eps}
  \caption{(color online) Detrending process in the MF-BDBA method ($\theta=0$). The time series has 500 data points and the timescale is $s=150$. The time series shown are obtained by truncating the points on both ends from a much longer time series. The first row shows the detrending process from left to right in which the data points with $i=451, 452, \cdots, 500$ should not be used in calculating $F$, and the second row from right to left in which the data points with $i=1, 2, \cdots, 150$ should not be used.}
  \label{Fig:MF-DMA:Detrend}
\end{figure}

The DMA method was generalized to MF-DMA for the investigation of the multifractal nature of time series \cite{Gu-Zhou-2010-PRE}. The main difference between the MF-DFA and MF-DMA algorithms is about the detrending procedure. In the MF-DMA approach, one calculates the moving average function $\widetilde{X}(i)$ in a moving window of size $s$ \cite{Arianos-Carbone-2007-PA},
\begin{equation}
  \widetilde{X}(i)=\frac{1}{s}\sum_{k=-\lfloor(s-1)\theta\rfloor}^{\lceil(s-1)(1-\theta)\rceil}X(i-k),
  \label{Eq:MF-DMA:trend}
\end{equation}
where $\lfloor{y}\rfloor$ is the largest integer not greater than $y$, $\lceil{y}\rceil$ is the smallest integer not smaller than $y$, and $\theta\in[0,1]$ is the position parameter. In other words, one considers $\lceil(s-1)(1-\theta)\rceil$ data points in the past and $\lfloor(s-1)\theta\rfloor$ points in the future to determine $\widetilde{X}(i)$. When $i$ is close to each of the endpoints, the size of the box might be less than $s$. One can either abandon those points or calculate the moving averages directly.
Usually, we consider three special cases, corresponding to three representative versions of MF-DMA:
\begin{itemize}
  \item MF-BDMA. The first case $\theta=0$ refers to the backward moving average \cite{Xu-Ivanov-Hu-Chen-Carbone-Stanley-2005-PRE}, in which the moving average function $\widetilde{X}(i)$ is calculated over all the past $s-1$ data points of the signal.
  \item MF-CDMA. The second case $\theta=0.5$ corresponds to the centered moving average \cite{Xu-Ivanov-Hu-Chen-Carbone-Stanley-2005-PRE}, where $\widetilde{X}(i)$ contains half past and half future information in each window. This version is also termed MF-CMA \cite{Schumann-Kantelhardt-2011-PA}.
  \item MF-FDMA. The third case $\theta=1$ is called the forward moving average, where $\widetilde{X}(i)$ considers the trend information of $s-1$ data points in the future.
\end{itemize}
We note that the size of the window used to calculate the moving averages must be identical to the segment size in calculating the detrended functions \cite{Gu-Zhou-2010-PRE,Schumann-Kantelhardt-2011-PA}. The detrending process is illustrated in Fig.~\ref{Fig:MF-DMA:Detrend}.

\subsubsection{Other detrending algorithms}
\label{S3:MF:Detrend:Others}

Although determining the intrinsic trend embedded in non-stationary time series is difficult, there are many different options \cite{Wu-Huang-Long-Peng-2007-PNAS}. Because the intrinsic trend of a time series is usually unknown, methods have different performances for different types of time series. Hence, quite a few variants of MF-DFA and MF-DMA have been proposed. Local trends can be be further classified into discontinuous ones and continuous ones. MF-DFA and MF-DMA belong respectively to these two classes (see also Fig.~\ref{Fig:MF-DFA:Detrend} and Fig.~\ref{Fig:MF-DMA:Detrend}).

Zhou and Leung proposed the multifractal temporally weighted detrended fluctuation analysis (MF-TWDFA) in moving windows \cite{Zhou-Leung-2010a-JSM}. For each $i$, they fit a linear function to the data points $(X(i-s), \cdots, X(i), \cdots, X(i+s))$ in the moving window of size $2s+1$, using weighted least-squares regression. The objective function is
\begin{equation}
  {\cal{O}}(i; s) = \sum_{j=-s}^s\left[X(i+j)-\sum_{k=0}^la_k\times(i+j)^k\right]^2w_{ij}, ~|i-j|\leq{s},
\end{equation}
where $w_{ij}$'s are weights that decrease with increasing $|i-j|$. In other words, the deviation $\left[X-a_0-a_1\times(i+j)\right]^2$ has a larger weight when $j$ is far from $i$. The weights $w_{ij}$ in MF-TWDFA are defined by \cite{Zhou-Leung-2010a-JSM}
\begin{equation}
  w_{ij} = \left[1-\left(\frac{i-j}{s}\right)^2 \right]^2.
\end{equation}
For the equally weighted case,
\begin{equation}
  w_{ij} = 1,
\end{equation}
we obtain the multifractal moving window detrended fluctuation analysis (MF-MWDFA) \cite{Zhou-Leung-2010a-JSM}. These two methods belong actually to the continuous class, since the trend changes mildly without jumps. Indeed, when $l=0$, we recover the centered MF-DMA method \cite{Zhou-Leung-2010a-JSM}.

The empirical mode decomposition (EMD) is an innovative method of analysis for nonlinear and non-stationary time series, which decomposes the time series
$\{X((v-1)s+i)\}_{i=1}^s$ into a finite number of intrinsic mode functions (IMFs) \cite{Huang-Shen-Long-Wu-Shih-Zheng-Yen-Tung-Liu-1998-PRSA}. Let $\{x(i)\}_{i=1}^s\triangleq\{X((v-1)s+i)\}_{i=1}^s$. The decomposition is a sifting process, which has six steps \cite{Wu-Huang-Long-Peng-2007-PNAS}:
(1) Identify all extrema of $x(i)$;
(2) Interpolate the local maxima to form an upper envelope $U(i)$;
(3) Interpolate the local minima to form a lower envelope $L(i)$;
(4) Calculate the mean envelope:
      \begin{equation}
         \mu(i)=[U(i)+L(i)]/2;
         \label{Eq:EMD:mu}
      \end{equation}
(5) Extract the mean from the signal
      \begin{equation}
         C_k(i) = X(i)-\mu(i);
         \label{Eq:EMD:Ck}
      \end{equation}
(6) Check whether $C_k(i)$ satisfies the IMF conditions. If the IMF conditions are fulfilled, $C_k(i)$ is an IMF, and the sifting stops. Otherwise, let $x(i)=C_k(i)$ and keep sifting.
Finally, we obtain the EMD-based local trend of the time series segment $\{X((v-1)s+i)\}_{i=1}^s$:
\begin{equation}
 r_n\triangleq\widetilde{X}((v-1)s+i)= X((v-1)s+i)-\sum_{k=1}^K C_k(i).
 \label{Eq:EMD:rn}
\end{equation}
Qian et al. used this local trend in the MF-DFA method and developed the EMD-based MF-DFA method \cite{Qian-Gu-Zhou-2011-PA}. Numerical experiments on multifractal binomial measures indicate that the EMD-based MF-DFA method has comparable performance with respect to the MF-DFA method except that the EMD-based method performs slightly better for small singularities \cite{Qian-Gu-Zhou-2011-PA}. The method has been applied to analyze stock market indices  \cite{Qian-Gu-Zhou-2011-PA,Caraiani-2012-PLoS1}, foreign exchange rates \cite{Caraiani-Haven-2015-PA}, and Certified Emission Reduction (CER) and European Union Allowances(EUA) futures prices \cite{Cao-Xu-2016-CSF}, all of which confirm the presence of multifractality in financial returns.

In order to handle time series with highly non-stationary sinusoidal and polynomial overlying trends, Horvatic et al. proposed to detrend with
polynomials of varying orders \cite{Horvatic-Stanley-Podobnik-2011-EPL}. In this method, the polynomial order $\ell$ is not fixed but increases with the scale $s$ and the detrended fluctuations at different scales should be normalized to secure an identical intercept. Although an objective criterion for the determination of the varying order is lacking, a subjective choice ($\ell=1$ for $s=4$, $\ell=2$ for $s=6$, $\ell=4$ for $s=10$, $\ell=8$ for $s=18$, and $\ell=16$ for $s>18$) seemingly works well \cite{Horvatic-Stanley-Podobnik-2011-EPL}. The underlying logic is that we need high-order polynomials to detrend properly the oscillations.

Du et al. constructed the dual-tree complex wavelet transform based multifractal detrended fluctuation analysis (DTCWT-MFDFA) for non-stationary time series \cite{Du-Tao-Gong-Gong-Liu-2016-APS}. Based on the anti-aliasing and nearly shift-invariance of the dual-tree complex wavelet transform \cite{Kingsbury-2001-ACHA,Selesnick-Baraniuk-Kingsbury-2005-IEEEspm}, the raw time series is decomposed through the pyramid algorithm to extract the scale-dependent trends and the fluctuations from the wavelet coefficients. The length of the non-overlapping segment on each time scale is computed through the Hilbert transform and each of the extracted fluctuations is partitioned into non-overlapping segments with identical size. The detrended fluctuation function for each segment can be calculated. The DTCWT-MFDFA method is found to outperform the MF-DFA method for multifractal binomial measures \cite{Du-Tao-Gong-Gong-Liu-2016-APS}.

In order to perform the MF-DFA, Manimaran et al. proposed to detrend using discrete wavelet transform \cite{Manimaran-Panigrahi-Parikh-2005-PRE}. The discrete wavelets belonging to the Daubechies family are adopted to analyze the raw time series up to level $s$. The resulting low-pass coefficients capture the polynomial trends in the data. One then reconstructs a time series using these low-pass coefficients to represent the local trends at different scales. The method has been tested for synthetical time series and applied to NASDAQ and BSE indices \cite{Manimaran-Panigrahi-Parikh-2008-PA,Manimaran-Panigrahi-Parikh-2009-PA}. Ghosh et al. used Db4 and Db6 basis sets to isolate respectively the local linear and quadratic trends at different scales from the closing prices of fifteen large-capitalization companies listed on the New York Stock Exchange (NYSE) \cite{Ghosh-Manimaran-Panigrahi-2011-PA}. Multifractality is observed in the investigated stock indices and individual stocks.

We note that other detrending methods used to preprocess the raw time series in Section \ref{S2:Preprocess} can also be used in the local detrending at different scales to construct variants of MF-DFA.

\subsection{Other methods}
\label{S2:MF:OtherMethods}

In this section, we will review several other multifractal analysis methods. It should be keep in mind that different multifractal analysis methods have been proposed in diverse fields and it is hard to mention all of them. We focus on the methods that are relevant to those discussed in the previous sections and that are utilized for financial time series analysis. Besides the ones presented below, we would like to mention the multifractal signal fluctuation analysis based on the 0-1 test \cite{Xu-Shang-Feng-2015-CSF} and the focus-based multifractal formalism \cite{Mukli-Nagy-Eke-2015-PA}.


%
%
%
%
%
%

\subsubsection{Multiplier method}
\label{S3:MF-MP}

The concept of the multiplier was introduced by Novikov to describe the intermittency and scale self-similarity in turbulent flows \cite{Novikov-1971-JAMM}. In econophysics, one can apply this concept to study volatility multipliers. For high-frequency financial data, we first construct an additive measure in the time interval $[t_1,t_2]$, which is the sum of absolute returns:
\begin{equation}
 \mu([t_1,t_2]) = \sum_{t = t_1}^{t_2}|r(t)|,
 \label{Eq:MF:VolatilityMeasure}
\end{equation}
The absolute return is actually a measure of the volatility on $[t_1,t_2]$ \cite{Bollen-Inder-2002-JEF}. The volatility time series $|r(t)|$ is divided into boxes of identical size $s$. Each {\textit{mother}} box is further partitioned into $a$ {\textit{daughter}} boxes, in which $a$ is called a base. The multiplier $m_{a,s}$ is determined by the ratio of the measure on a daughter box to that on her mother box \cite{Chhabra-Sreenivasan-1991-PRA,Chhabra-Sreenivasan-1992-PRL}.

If the multiplier is scale invariant, for any two bases $a$ and $b$, the multiplier density functions $p_a(m_a)$ and $p_b(m_b)$ are related through Mellin transform $\mathbb{M}$ as follows \cite{Chhabra-Sreenivasan-1992-PRL}:
\begin{equation}
 [\mathbb{M} \{ p_a(m_a) \}]^{1 / \ln a} = [\mathbb{M} \{ p_b(m_b) \}]^{1 / \ln{b}},
 \label{Eq:Mellin}
\end{equation}
which leads to
\begin{equation}
 \frac{\ln \int_0^1 m_a^q p_a(m_a) d m_a}{\ln a} = \frac{\ln \int_0^1 m_b^q p_b(m_b) {\mbox{d}} m_b}{\ln{b}}.
 \label{Eq:MF:multiplier:base}
\end{equation}
The mass scaling exponent $\tau(q)$ of the moment of $m_a$ can be obtained as  \cite{Chhabra-Sreenivasan-1991-PRA,Chhabra-Sreenivasan-1992-PRL}
\begin{equation}
 \tau(q) = - D_0 - \frac{\ln\langle m_a^q \rangle}{\ln a}~,
 \label{Eq:MF:multiplier:tau}
\end{equation}
where $D_0$ is the fractal dimension of the support of the measure. For time series, we have $D_0 = 1$.

The local singularity exponent $\alpha$ and its spectrum $f(\alpha)$ are related to $\tau(q)$ through the Legendre transform. It follows that
\begin{equation}
 \alpha(q) = -\frac{\langle m_a^q \ln m_a \rangle}{\langle m_a^q \rangle \ln a}
 \label{Eq:MF:multiplier:alpha}
\end{equation}
and
\begin{equation}
 f(\alpha) =  \frac{\langle m_a^q \rangle \ln \langle m_a^q \rangle
 - \langle m_a^q \ln m_a^q \rangle}{\langle m_a^q \rangle \ln
 a}~.
 \label{Eq:MF:multiplier:falpha}
\end{equation}
Eqs.~(\ref{Eq:MF:multiplier:base}-\ref{Eq:MF:multiplier:falpha}) imply that, for each $q$, each of these characteristic multifractal quantities converges to a constant in the scaling range and is independent of the base $a$.

Note that, for financial volatility, $q>-1$ \cite{Jiang-Zhou-2007-PA}, which is quite analogous to the situation in turbulence \cite{Pearson-vandeWater-2005-PRE}. Because the returns can be zero, the multiplier $m$ can be zero and the probability density $p_a(0)\neq0$ at $m_a=0$. When $m$ is smaller than a small number $\delta$, $p(m)=p(0)$ is a constant. For a given small $\delta$, $\int_{\delta}^1m^qp(m)dm = C$ is finite. However, the moment
\begin{equation}
  \int_0^1m^qp(m) {\mbox{d}}m=\int_0^{\delta}m^qp(m) {\mbox{d}}m+ C =\left.p(0)\frac{1}{q+1}m^{q+1}\right|_0^{\delta}+C
\end{equation}
diverges when $q\leq-1$. This property is verified empirically \cite{Jiang-Zhou-2007-PA}, based on the idea that the integrand $m_a^qp_a(m_a)$ diverges for a given $q$ \cite{Lvov-Podivilov-Pomyalove-Procaccia-Vandembroucq-1998-PRE,Zhou-Sornette-Yuan-2006-PD}.

The scale-invariant multiplier distribution is perceived to be more fundamental than the standard multifractal spectrum $f(\alpha)$ \cite{Chhabra-Sreenivasan-1991-PRA,Chhabra-Sreenivasan-1992-PRL}. Moreover, it needs exponentially less computational time and is more accurate than the partition function approach \cite{Chhabra-Sreenivasan-1991-PRA,Chhabra-Sreenivasan-1992-PRL}.

Using 1-min high-frequency data of the S\&P 500 index, Jiang and Zhou reported that the volatility multiplier is scale invariant and the volatility has the property of multifractality \cite{Jiang-Zhou-2007-PA}. Niu and Wang confirmed the presence of scale invariance in the multiplier distributions in the Shanghai SSEC index and in the simulated data from the model of ``voter interacting dynamic system'' \cite{Niu-Wang-2013-JAS}.

\subsubsection{Multifractal Hilbert-Huang spectral analysis}
\label{S3:MF-HHSA}

The Hilbert-Huang transform combines the EMD and the Hilbert spectral analysis \cite{Huang-Shen-Long-Wu-Shih-Zheng-Yen-Tung-Liu-1998-PRSA,Huang-Shen-2005}, in which the Hilbert transform is applied to each component $C_k(i)$ in Eq.~(\ref{Eq:EMD:Ck}). The overall trend can be expressed as
\begin{equation}
\widetilde{X}(i) = X(i)-\sum_{k=1}^nC_k(i) = X(i)-\sum_{k=1}^n\mathcal{A}_k(i)\exp\left({\mathbf{j}}\int\omega_k(\ell){\mbox{d}}\ell\right),
  \label{Eq:HHT:expression}
\end{equation}
where ${\mathbf{j}}^2=-1$.
For each mode, the Hilbert-Huang spectrum is defined as the squared amplitude
\begin{equation}
 H(\omega,i)=\mathcal{A}^2(\omega,i),
 \label{Eq:HHT:Spectrum}
\end{equation}
The Hilbert-Huang marginal spectrum of the original time series is then written as
\begin{equation}
 h(\omega)=\int H(\omega,i) \, d i,
 \label{Eq:HHT:MarginalSpectrum}
\end{equation}
which corresponds to an energy density at frequency $\omega$. This method is known as EMD-based Hilbert spectral analysis (EMD-HSA), which we term the Hilbert-Huang spectral analysis (HHSA) for simplicity.

Huang, Schmitt, Lu and Liu proposed the multifractal Hilbert-Huang spectral analysis (MF-HHSA) by generalizing the Hilbert-Huang transform from the second order to arbitrary order \cite{Huang-Schmitt-Lu-Liu-2008-EPL,Huang-Schmitt-Hermand-Gagne-Lu-Liu-2011-PRE}. The Hilbert-Huang spectrum $H(\omega,i)$ represents the original signal at the local level and can be used to define the joint probability density function $p(\omega,\mathcal{A})$ of the frequency $\omega_k$ and amplitude $\mathcal{A}_k$. The Hilbert-Huang
marginal spectrum in Eq.~(\ref{Eq:HHT:MarginalSpectrum}) can be rewritten as
\begin{equation}
 h(\omega)=\int_0^{\infty} p(\omega,\mathcal{A}) \mathcal{A}^2 {\mbox{d}} \mathcal{A},
  \label{Eq:HHT:MarginalSpectrum:2}
\end{equation}
which corresponds to the second-order moment. Huang et al. generalize Eq.~(\ref{Eq:HHT:MarginalSpectrum:2}) into arbitrary-order moments
 \begin{equation}
 \mathcal{L}_q(\omega)=\int_0^{\infty} p(\omega,\mathcal{A}) \mathcal{A}^q {\mbox{d}} \mathcal{A},
 \label{Eq:HHT:Moments}
\end{equation}
where $q \ge 0$ and $h(\omega)=\mathcal{L}_2(\omega)$. In the inertial range, we assume the following scaling relation,
 \begin{equation}
 \mathcal{L}_q(\omega) \sim \omega^{-\tau(q)},
 \label{Eq:HHT:Moments:tau:q}
\end{equation}
where $\tau(q)$ is the corresponding scaling exponent function in the amplitude-frequency space.

Li and Huang applied this MF-HHSA method to the CSI 300 index of the Chinese stock market and confirmed the presence of multifractality \cite{Li-Huang-2014-PA}.

\subsubsection{EMD-based multifractal methods}

Welter and Esquef proposed an EMD-based dominant amplitude multifractal formalism (EMD-DAMF), which is implemented as follows \cite{Welter-Esquef-2013-PRE}:
\begin{itemize}
  \item[] {\textit{Step 1:}} Perform EMD on $X(i)$ to obtain $K$ IMFs $C_k(i)$.
  \item[] {\textit{Step 2:}} Determine the amplitude $a_k(i)$ and characteristic timescale $s$ for $C_k(i)$. The characteristic timescale $s$ could be estimated in each mode by the Hilbert spectral analysis (HSA) as the reciprocal of the mean frequency, or simply by twice the mean time interval among all adjacent zero crossings.
  \item[] {\textit{Step 3:}} Determine the dominant amplitude coefficients
      \begin{equation}
         u_{k,j}:=\sup_{k'\leqslant{k}}\{\max[|a_{k'}(i\in{I}_{k,j})|]\}, ~ k=1,2,\cdots,K~~\mathrm{and}~~j=1,2,\cdots, n_k,
         \label{Eq:MF:EMD:DA:ukj}
      \end{equation}
      where $n_k$ is the number of local maxima of $a_k(i)$ and $I_{k,j}$ is a time support around the $j$th maxima of $a_k(i)$.
  \item[] {\textit{Step 4:}} Calculate the $q$th order moment for different $q$ values:
      \begin{equation}
         M_q(s_k)=\frac{1}{n_k}\sum_{i=1}^{n_k}(u_{k,j})^q,~~k=1,2,\cdots, K,
         \label{Eq:MF:EMD:DA:Sk}
      \end{equation}
      where the characteristic time scale $s_k$ can be obtained by HSA in each mode as the reciprocal of the mean (linear) frequency or simply by the mean time interval among all adjacent zero crossings.
  \item[] {\textit{Step 5:}} Estimate the scaling function
      \begin{equation}
         M_q(s_k) \sim s_k^{\tau(q)+1-q}.
         \label{Eq:MF:EMD:DA:PL}
      \end{equation}
\end{itemize}

In real applications, Welter and Esquef suggested that the number of siftings is fixed to ten (that is, $K=10$) and signal envelopes are computed through cubic spline
interpolation \cite{Welter-Esquef-2013-PRE}. They validated the method with fractional Brownian motions \cite{Mandelbrot-VanNess-1968-SIAMR}, L{\'{e}}vy processes \cite{Levy-1937}, Multifractal random wavelet cascades \cite{Benzi-Biferale-Crisanti-Paladin-Vergassola-Vulpiani-1993-PD,Arneodo-Bracy-Muzy-1998-JMP}, and multifractal multiplicative cascades from the $p$ model \cite{Meneveau-Sreenivasan-1987-PRL}. Note that the EMD-DAMF method has a small number $n_k$ of amplitude maxima at larger scale modes and a limited number of scales \cite{Welter-Esquef-2013-PRE}. The latter drawback might impact the determination of the scaling range.

Alternatively, Zhou et al. proposed the so-called EMD-DFA method \cite{Zhou-Manor-Liu-Hu-Zhang-Fang-2013-PLoS1}. The integrated time series $X(i)$ is decomposed into a set of intrinsic mode functions and a residual or trend component. All intrinsic time scales $s$ are identified by computing the number of data points between each neighboring local minima throughout each intrinsic mode function (IMF). For a given $s$, all IMFs are resampled through replacing by 0's all data points corresponding to scales not equal to $s$ and the fluctuation time series $X_s(i)$ is obtained by summing all resampled IMFs. The overall fluctuation at scale $s$ is the root mean square of $X_s(i)$:
\begin{equation*}
  F_2(s) = \left[\frac{1}{N}\sum_{i=1}^N X_s^2(i) \right]^{1/2}, 
\end{equation*}
which produces better scaling than the DFA method \cite{Zhou-Manor-Liu-Hu-Zhang-Fang-2013-PLoS1}. The EMD-DFA method can be extended for multifractal time series by working on the $q$-order moments:
\begin{equation}
  F_q(s) = \left[\frac{1}{N}\sum_{i=1}^N |X_s(i)|^q \right]^{1/q}.
\end{equation}
Note that the EMD-DFA method does not have an explicit detrending form like Eq.~(\ref{Eq:MF:Detrend:epsilon}).

\subsubsection{Multiscale multifractal analysis}
\label{S3:MF:MMA}

Gieraltowski et al. proposed the multiscale multifractal analysis (MMA) based on the MF-DFA method \cite{Gieraltowski-Zebrowski-Baranowski-2012-PRE}. After calculating all values of the overall fluctuations $F_q(s)$ by the MF-DFA method, rather than estimating $H(q)$ from Eq.~(\ref{Eq:MF:Detrend:hq}) in the scaling range, they proceeded to estimate the local generalized Hurst exponents $H(q,s)$ in moving windows around different scales. In this method, two free parameters are introduced, the moving window size and the step. Although the original MMA method is designed for MF-DFA, it can obviously be applied to other methods of multifractal analysis.

The MMA method has been applied to investigate the multifractal nature of daily returns of the Dow Jones Industrial Average (DJIA), New York Stock Exchange Index (NYSE), S\&P 500 Index, Heng Seng Index (HSI), Shanghai Stock Exchange Composite Index and Shenzhen Stock Exchange Component Index (SZCI) from 12 May 1992 to 8 May 8 2012 \cite{Lin-Ma-Shang-2015-PA,Lin-Shang-2016-PA} and nine stock market indices (CAC 40, DAX, FTSE 100, HSI, NIKKEI 225, SSEC, NASDAQ, S\&P 500) from 3 January 2000 to 1 October 2014 \cite{Jiang-Shang-Shi-2016-PA}.

More generally, this method is equivalent to calculate the local logarithmic slopes of the $F_q(s)$ functions:
\begin{equation}
  h(q,s) = \lim_{s'\to{s}}\frac{d\ln{F_q(s')}}{d\ln{s'}},
  \label{Eq:MF:Detrend:hq:Local}
\end{equation}
which is widely utilized in the study of structure functions in turbulence \cite{Biferale-Cencini-Lanotte-Vergni-Vulpiani-2001-PRL,Zhou-Sornette-Yuan-2006-PD}, as well as in DFA \cite{AlvarezRamirez-Alvarez-Solis-2010-EE}. The idea to estimate local logarithmic slopes is especially informative when the scaling range is narrow and it can serve as a method for the determination of scaling ranges. Different numerical methods can be adopted to estimate numerically the local logarithmic slopes $H(q,s)$ and the linear fitting in a moving window suggested in the MMA method is one of them. Empirical analyses have been performed on financial time series, such as the WTI crude oil market \cite{Wang-Liu-2010-EE} and stock market indices \cite{Li-Kou-Sun-2015-PA}, without referring to the name of multiscale multifractal analysis.

\subsubsection{Time-varying multifractal analysis} 
\label{S3:MF:TimeVarying}

To quantify the evolution of a multifractal spectrum, one can perform a multifractal analysis of time series in moving windows or rolling windows. We would like to stress that short window sizes will reduce the estimation accuracy of multifractal quantities and increase the uncertainty of the results. Time-varying multifractal quantities have practical implications, as discussed in Section \ref{S3:Appl:TimeVarying}.

Alternatively, Xiong and Zhang proposed the time-varying multifractal spectrum distribution (TM-MFSD) \cite{Xiong-Zhang-Liu-2012-PA}. The key idea is to perform multifractal analysis on the instantaneous autocorrelation function
\begin{equation}
  r(i, \delta) = {\mathbb{E}}[X^*(i-\Delta/2)X(i+\Delta/2)],
\end{equation}
rather than the raw time series $X(i)$, where the time-delayed conjugation $X^*(i)$ of the analyzed series $X(i)$ is selected as the window function. To obtain the time-varying multifractal spectrum distribution of a time series, one can adopt different multifractal analysis methods described in previous sections, such as WTMM \cite{Xiong-Zhang-Liu-2012-PA}, wavelet leaders \cite{Xiong-Zhang-Zhao-2014-APS,Xiong-Zhang-Zhao-Xi-2014-ND}, MF-DMA \cite{Xiong-Zhang-Zhao-2014-CSF}, and MF-DFA \cite{Xiong-Yu-Zhang-2015b-PA}. Essentially, the multifractal analysis is performed with these methods on moving windows.

\subsubsection{Phase space reconstruction}
\label{S3:MF:PhaseSpace}

A classic method in nonlinear dynamics is to embed properly the time series into a phase space and estimate the generalized dimensions $D_q$ of the data points in the phase space \cite{Grassberger-1983-PLA,Grassberger-Procaccia-1983-PD,Grassberger-Procaccia-1983-PRL,Pawelzik-Schuster-1987-PRA}, which is however less applied in econophysics. Based on the delay embedding technique \cite{Packard-Crutchfield-Farmer-Shaw-1980-PRL}, the time series $X(i)$ can be converted into a sequence of vectors:
\begin{equation}
   \overrightarrow{X}(j) = [X(j),X(j+\tau), \cdots, X(j+(d-1)\tau)],~~ j=1,\cdots,N_v,
   \label{Eq:Enbedding:Vector}
\end{equation}
where $N_v$ is the number of points (or vectors) in the phase space, $\tau$ is a suitably chosen time delay, and $d$ is the embedding dimension of the phase space. The embedding dimension $d$ and the delay $\tau$ can be determined in different ways \cite{Grassberger-Procaccia-1983-PD,Kennel-Brown-Abarbanel-1992-PRA,Buzug-Pfister-1992-PRA,Kennel-Isabelle-1992-PRA,Buzug-Pfister-1992-PD,Kim-Eykholt-Salas-1999-PD}.

The relative number of data points within a distance $R$ to point $i$ is calculated by
\begin{equation}
  c_{i}(s) = \frac{1}{N_v} \sum_{j=1,j\neq i}^{N_v} {\cal{H}}\left(s-\left|\overrightarrow{X}(i)-\overrightarrow{X}(j)\right|\right),
   \label{Eq:Enbedding:ci:R}
\end{equation}
where ${\cal{H}}$ is the Heaviside function. The generalized correlation integral $C_{q}(s)$ is \cite{Paladin-Vulpiani-1984-LNC}
\begin{equation}
   \label{e.5}
   C_{q} (s) = \left\{\frac{1}{N_{v}} \sum_{i}^{N_{v}} \left[c_i (s)\right]^{q-1}\right\}^{1/(q-1)},
\end{equation}
which is the generalization of the correlation integral \cite{Grassberger-1983-PLA,Grassberger-Procaccia-1983-PD,Grassberger-Procaccia-1983-PRL}.
The generalized dimensions are given by \cite{Pawelzik-Schuster-1987-PRA}
\begin{equation}
    \label{e.6}
    D_{q} \equiv \lim_{s \rightarrow 0} \frac{\log C_q (s)}{\log s}.
\end{equation}

Lee investigated the multifractal characteristics of the 1-min volatility of the Korea composite stock price index KOSPI from 20 March 1992 to 28 February 2007 \cite{Lee-2009-PA}. The volatility time series is first detrended with a wavelet filter and then a multifractal analysis in the phase space with $\tau=1$ and $d=1$ is performed to confirm the presence of multifractality.

\subsubsection{Asymmetric multifractal analysis}
\label{S3:MF:Asymmetric}

Inspired by the stylized fact that the correlation between two assets is usually asymmetric \cite{Bekaert-Wu-2000-RFS,Longin-Solnik-2001-JF,Ang-Chen-2002-JFE,Hong-Tu-Zhou-2007-RFS}, Alvarez-Ramirez et al. developed the asymmetric multifractal detrended fluctuation analysis (A-MF-DFA) to detect asymmetric multifractal scaling in individual time series \cite{AlvarezRamirez-Rodriguez-Echeverria-2009-PA}. In this method, the trend function is linear such that $a_i=0$ for all $i>1$ in Eq.~(\ref{Eq:MF:Trend:Polynomial}). The main idea is to calculate the upwards (downwards) fluctuation functions in which the removed linear trends have non-negative (negative) slopes. Mathematically, the $q$th-order upwards overall detrended fluctuation is
\begin{equation}
  F_q^+(s) = \left\{\frac{1}{\sum_{v=1}^{2N_s} \frac{1+k_v}{2}}\sum_{v=1}^{2N_s} \left[\frac{1+k_v}{2}F_v(s)\right]^q\right\}^{\frac{1}{q}}
           = \left\langle[F_v(s)]^q\right\rangle|_{a_1\geq0},
  \label{Eq:MF:A-MF-DFA:Fqs:LR:p}
\end{equation}
and the $q$th-order downwards overall detrended fluctuation is
\begin{equation}
  F_q^-(s) 
           = \left\langle[F_v(s)]^q\right\rangle|_{a_1<0},
  \label{Eq:MF:A-MF-DFA:Fqs:LR:n}
\end{equation}
where $k_v=1$ if the slope of the linear trend (\ref{Eq:MF:Trend:Polynomial}) is nonnegative, that is $a_1\geq0$, and $k_v=-1$ if the slope $a_1<0$. When $q=0$, we apply L'H\^{o}spital's rule as in Eq.~(\ref{Eq:MF:Detrend:Fq0:LR}). The respective generalized Hurst exponents can be determined by
\begin{equation}
  F_q^{\pm}(s) \sim {s}^{H^{\pm}(q)}.
  \label{Eq:MF:A-MF-DFA:hq}
\end{equation}
Note that the original A-MF-DFA method uses the mean absolute value of the residuals in each segment $v$ \cite{AlvarezRamirez-Rodriguez-Echeverria-2009-PA}.

Empirical studies using the A-MF-DFA method investigated and confirmed the asymmetric multifractal behavior in the daily returns of the DJIA index from 23 May 1980 to 25 August 2008 \cite{RiveraCastro-Miranda-Cajueiro-Andrade-2012-PA}, the WTI oil price from 2 January 1986 to 14 December 2010 \cite{RiveraCastro-Miranda-Cajueiro-Andrade-2012-PA}, the SSEC index from 19 December 1990 to 27 April 2012 and the SZCI index from 2 April 1991 to 27 April 2012 \cite{Cao-Cao-Xu-2013-PA}, and the DJIA, NASDAQ, NYSE, and S\&P500 indices from 1 January 1991 to 31 December 2015 \cite{Lee-Song-Park-Chang-2017-CSF}.

The A-MF-DFA method can be easily extended to design the asymmetric multifractal detrending moving average analysis (A-MF-DMA), in which moving averages serve as the trend function and the direction (or ``sign'') is determined exactly as for the A-MF-DFA method \cite{Zhang-Ni-Ni-Li-Zhou-2016-PA}. The asymmetric multiscale detrended fluctuation analysis is obtained by calculating the local exponents at different time scales, which has been applied to investigate the hourly returns of the California electricity spot prices during years 1999 and 2000 \cite{Fan-2016-PA}.

Cao et al. developed the asymmetric multifractal detrended cross-correlation analysis (MF-ADCCA) method  \cite{Cao-Cao-Xu-He-2014-PA}, as an asymmetry extension of the MF-X-DFA method (see Section \ref{S2:MF-DCCA}). The method has been applied to identify asymmetric cross-correlations between the daily returns of the SSEC index and six foreign exchange rates from 22 July 2005 to 13 January 2012 \cite{Cao-Cao-Xu-He-2014-PA}, the 5-min returns of the CSI 300 index spot and futures returns from 30 June 2011 to 7 June 2013 \cite{Cao-Han-Cui-Guo-2014-PA}, the daily returns of the gold spot prices in the Shanghai Gold Exchange (Au 99.95) and the New York Gold Exchange from 2 January 2003 to 27 April 2012 \cite{Cao-Zhao-Han-2015-IJMPB}, the EU ETS for carbon emission trading products vs. energy futures -- Brent crude oil (Brent), Richards bay coal (Coal), UK base electricity (Electricity) and UK natural gas (Gas) -- from 1 January 2008 to 31 December 2012 \cite{Xu-Cao-2016-Fractals}, the daily returns of the WTI price and eight foreign exchange rates (CAD, MXN, NOK, GBP, JPY, AUD, EUR and KRW) against the USD from 4 January 2000 to 31 December 2014 \cite{Jiang-Gu-2016a-PA}, the Shanghai SSEC and four other indices (US S\&P 500, Germany DAX, India BSESN, and Brazil BVSP) from 1 January 2002 to 26 September 2014 \cite{Cao-Han-Li-Wu-2017-PA}, and the WTI crude oil prices and eight stock merket indices (Brazil IBOVESPA, Chile IPSA, Argentina MERVAL, Mexico IPC, Colombia COLCAP, Peru SPBVL, S\&P500 and DJIA) from the 28 January 2005 to 29 December 2016 \cite{Gajardo-Kristjanpoller-2017-CSF}.

Chen and Zheng proposed the asymmetric joint multifractal analysis based on partition functions \cite{Chen-Zheng-2017-PA}, as an extension of the MF-X-PF method  (see Section \ref{S2:MF-X-PF}). This extension becomes feasible for financial volatility because one can use the associated returns to define the signs in the segments.

\subsubsection{Multifractal diffusion entropy analysis} 
\label{S3:MF:DEA}

The diffusion entropy analysis (DEA) for time series $\{X(i)\}_{i=1}^N$ is based on the Shannon entropy \cite{Scafetta-Hamilton-Grigolini-2001-Fractals,Grigolini-Palatella-Raffaelli-2001-Fractals,Scafetta-Grigolini-2002-PRE,Scafetta-Latora-Grigolini-2002-PRE}, which can be determined by taking asset prices as an example. For a given diffusion time or time scale $s$, we calculate the return series
\begin{equation*}
	\Delta{X}(i,s) = X(i)-X(i-s), ~~i=1,2,\cdots, N-s,
    \label{Eq:Return:dX(i,s)}
\end{equation*}
which denotes the total diffusion distance of $s$ jump steps. For each $s$, the values of $\{\Delta{X}(i,s)\}_{i=1}^N$ are covered by $N_s$ non-overlapping intervals of length $\epsilon(s)$. The interval length $\epsilon(s)$ is chosen as a fraction of the square root of the variance of the fluctuations $\{\Delta{X}(i,1)\}$, which is independent of the time scale $s$. We count the number $N_{j}(s)$ of the returns $\Delta{X}(i,s)$ that fall in the $j$th interval and the probability (or frequency) by means of
\begin{equation*}
  p_{j}(s) \equiv  \frac{N_j(s)}{N-s}.
  \label{Eq:MF:DEA:pj}
\end{equation*}
The Shannon entropy of the diffusion process is determined as
\begin{equation*}
  I(s) = - \sum_{j=1}^{N_s} p_j(s) \ln p_j(s).
  \label{Eq:MF:DEA:ShannonEn}
\end{equation*}
The scaling behavior is characterized by
\begin{equation*}
  I(s) \sim \delta \ln{s},
\end{equation*}
where $\delta$ is the scaling exponent.

For fractional Brownian motions, the DEA scaling exponent is equal to the Hurst index, that is, $\delta=H$ \cite{Scafetta-Grigolini-2002-PRE}, which does not necessary hold for other processes such as L{\'e}vy processes \cite{Scafetta-Grigolini-2002-PRE,Scafetta-Latora-Grigolini-2002-PRE}. For financial assets, it seems that one should work on returns to obtain approximately $\delta=H$; otherwise, $\delta$ will be greater than 0.9 \cite{Sarkar-Barat-2006-PA,Cai-Zhou-Yang-Yang-Wang-Zhou-2006-PA,Perello-Montero-Palatella-Simonsen-Masoliver-2006-JSM}. There is also an evident finite-size effect that the entropy $I(s)$ deviates from the ``theoretical'' value for small scales and converges to the theoretical value for large $s$ \cite{Scafetta-Latora-Grigolini-2002-PRE}. It is possible to overcome this drawback using the idea of integrated moments \cite{Buonocore-Aste-DiMatteo-2017-PRE} (see Section \ref{S3:MF:Properties:IntMomemnt} for details).

A natural generalization of DEA for multifractal analysis is to replace the Shannon entropy by the R{\'e}nyi entropy expressed in Eq.~(\ref{Eq:RenyiEn}), leading to the multifractal diffusion entropy analysis (MF-DEA), whose main output is a spectrum of scaling exponents $\delta(q)$ \cite{Huang-Shang-Zhao-2012-PA}. Analogous to the multiscale multifractal analysis based on the MF-DFA method \cite{Gieraltowski-Zebrowski-Baranowski-2012-PRE} (see Section \ref{S3:MF:MMA}), the multiscale multifractal diffusion entropy analysis can be designed \cite{Huang-Shang-2015-PA}. The MF-DEA method suffers some drawbacks \cite{Morozov-2013-PA}. We notice that $\delta(q)$ is not related directly to the well-known generalized dimensions $D_q$. Hence, the MF-DEA method unveils multifractality in certain mappings of the original time series, but not in original time series per se. In the DEA or MF-DEA method, the estimation of the probabilities $p_j(s)$ has a crucial impact on the estimation of the R{\'e}nyi entropy and consequently the $\delta(q)$ spectrum \cite{Jizba-Korbel-2014-PA}. In this vein, an optimal choice of $\epsilon(s)$ is required \cite{Jizba-Korbel-2014-PA}.

\subsubsection{Symbolization presentation and R{\'e}nyi dimensions}
\label{S3:MF:Symbolic}

Xu and Beck applied symbolic dynamics techniques to financial returns to determine the R{\'e}nyi dimensions or the generalized dimensions $D_q$ \cite{Xu-Beck-2017-EPL}. There are different symbolization techniques that can convert a financial time series $\{X(i)\}_{i=1}^N$ into a symbol sequence. Li and Wang proposed an approach by considering the speed of price changes, which is quantified by an angle \cite{Li-Wang-2006-CSB,Li-Wang-2007-PA},
\begin{equation}
  \theta_i = \arctan[(X(i+\ell)-X(i))/\ell],
  \label{Eq:MF:Symb:Angle}
\end{equation}
which falls within the interval $[-90^{\circ},90^{\circ}]$. To symbolize the raw time series, we partition $[-90^{\circ},90^{\circ}]$ into $2n$ subintervals, which are usually almost symmetric with respect to $\theta=0^{\circ}$. Li and Wang considered four subintervals $(-90^{\circ},-45^{\circ})$, $[-45^{\circ},0^{\circ})$, $[0^{\circ},45^{\circ})$, and $[45^{\circ},90^{\circ})$ and four corresponding symbols $D$, $d$, $r$ and $R$, representing respectively fast falls, slow falls, slow rises and fast rises of prices.

The symbolization method of Xu and Beck is similar \cite{Xu-Beck-2017-EPL}. They suggested to partition the return domain $\Delta{X_i}\in(-1,1)$ into $2n$ subintervals, mapped into $2n$ symbols. When $n=1$, the two subintervals are $(-1,0)$ and $[0,\infty)$ and two corresponding symbols $d$ and $u$, representing respectively falling down and rising of the prices. When $n=2$, the four subintervals are $(-1,-c_1)$, $[-c_1,0]$, $[0,c_2)$ and $[c_2,\infty]$, where $c_1\in(0,1)$ and $c_2\in[0,\infty]$ are parameters. Note that Xu and Beck used open intervals, which may discard some data points. In addition, a return cannot be less than $-1$. For stock markets with price limit rules, the subintervals should be adjusted accordingly.

For any manner of symbolization, if $\theta_i$ or $\Delta{X}_i$ is an element of a certain subinterval, the $i$th symbol $S_i$ is determined by the corresponding symbol. In this way, the raw time series is mapped to a symbol sequence $\{S_i\}$. The symbol sequence is divided into $N_k$ segments of equal length $k$. For any given $k$, we obtain
$\omega(k)=(2n)^k$ allowed symbol subsequences or configurations. The occurrence probability of each configuration $C_j$ of length $k$ can be determined, denoted as $p_j = p(C=C_j)$. Xu and Beck recalled the well-known  R{\'e}nyi information \cite{Renyi-1970}
\begin{equation}
I_{q} =\frac{1}{q-1} \ln \sum^{\omega(k)}_{j=1}p_j^q
\label{renin}
\end{equation}
and the R{\'e}nyi dimension
\begin{equation}
  D_q=\lim_{\varepsilon \to 0} \frac{I_q}{\ln \varepsilon}=\lim_{\varepsilon \to 0} \frac{1}{\ln \varepsilon} \frac{1}{q-1} \ln \sum^{\omega(k)}_{j=1}p_{j}^{q},
\end{equation}
where $\varepsilon=(2n)^{-k}$.

The method has been applied to the daily and 1-min price time series of several US stocks (Alcoa, Bank of America, General Electric, Intel, Johnson \& Johnson, Coca Cola and WalMart) from January 1998 to May 2013 and the generalized dimensions were calculated and compared for different values of $k$ \cite{Xu-Beck-2017-EPL}.

%
%
%


\section{Joint multifractal analysis}
\label{S1:MF-X}

\subsection{Joint multifractal analysis based on partition functions (MF-X-PF)}
\label{S2:MF-X-PF}

A complex system usually contains two or more distinct multifractal measures that are distributed simultaneously on the same geometric support. In turbulent flows, the velocity, temperature, and concentration fields are embedded in the same space as joint multifractal measures
\cite{Meneveau-Sreenivasan-Kailasnath-Fan-1990-PRA}, in which the scaling behavior of the joint moments $\langle[m_x(s,i)]^{p/2}[m_y(s,i)]^{q/2}\rangle$ of two multifractal measures $m_x$ and $m_y$ over box size $s$ is investigated. We can also call this method the multifractal cross-correlation analysis based on the partition function approach (MF-X-PF) \cite{Xie-Jiang-Gu-Xiong-Zhou-2015-NJP}. In the original framework, the orders are $p$ and $q$ rather than $p/2$ and $q/2$. When considering the degenerate case of $m_x=m_y$ and $p=q$, the joint moment becomes $\langle[m_x(s,i)]^{q}\rangle$, which recovers exactly the partition function approach presented in Section~\ref{S2:MF-PF}. The case of $p=q$ has been investigated in Ref.~\cite{Wang-Shang-Ge-2012-Fractals}. There are also mathematical treatments of the bivariate case \cite{Riedi-Scheuring-1997-Fractals,Lin-2008-PA} and the general case of many measures \cite{Olsen-2005-JMAA,Dai-2009-ATA}.

In natural science, the joint multifractal analysis has also been applied to study the joint multifractal nature between topographic indices and crop yield in agronomy \cite{Kravchenko-Bullock-Boast-2000-AJ,Zeleke-Si-2004-AJ}, nitrogen dioxide and ground-level ozone \cite{JimenezHornero-JimenezHornero-deRave-PavonDominguez-2010-EMA}, wind patterns and land surface air temperature \cite{JimenezHornero-PavonDominguez-deRave-ArizaVillaverde-2011-AR}, and temperature and nitrogen dioxide \cite{PavonDominguez-JimenezHornero-deRave-2015-SERRA}.

\subsubsection{General formalism for arbitrary \texorpdfstring{$(p,q)$}{} pairs}

For the integrated measures $m_x(s,i)$ and $m_x(s,i)$ in the $i$th box of size $s$, the local singularity strengths $\alpha_x$ and $\alpha_y$ are defined as \cite{Xie-Jiang-Gu-Xiong-Zhou-2015-NJP}
\begin{equation}
  m_x(s,i) \sim s^{\alpha_x} ~~~~~{\mbox{and}}~~~~~  m_y(s,i) \sim s^{\alpha_y}.
  \label{Eq:MF-X-PF:mxy:s:alpha}
\end{equation}
Denoting $N_s(\alpha_x,\alpha_y)$ as the number of boxes needed to cover the neighborhoods of points with singularities $\alpha_x$ and $\alpha_y$, the fractal dimension of the set is determined according to
\begin{equation}
  N_s(\alpha_x,\alpha_y) \sim s^{-f(\alpha_x,\alpha_y)},
  \label{Eq:MF-X-PF:Ns:s:f:alpha:xy}
\end{equation}
in which $f(\alpha_x,\alpha_y)$ is the joint distribution of the two singularities \cite{Meneveau-Sreenivasan-Kailasnath-Fan-1990-PRA}, or the joint multifractal spectrum.

Through the joint partition function \cite{Xie-Jiang-Gu-Xiong-Zhou-2015-NJP}
\begin{equation}
  \chi_{xy}(p,q,s)= \sum_t\left[m_x(s,i)\right]^{p/2}[m_y(s,i)]^{q/2},
  \label{Eq:MF-X-PF:pq:chi:s}
\end{equation}
we can obtain the joint mass exponent function $\tau_{xy}(p,q)$ as
\begin{equation}
  \chi_{xy}(p,q,s) \sim s^{\tau_{xy}(p,q)}.
  \label{Eq:MF-X-PF:pq:tauxy}
\end{equation}
Inserting Eq.~(\ref{Eq:MF-X-PF:mxy:s:alpha}) into the joint partition function, rewriting the sum into a double integral over $\alpha_x$ and $\alpha_y$, and then applying the steepest descent approach to estimate the integral at small $s$ values, we have
\begin{equation}
  \tau_{xy}(p,q) = p\alpha_x/2+q\alpha_y/2 - f(\alpha_x,\alpha_y),
  \label{Eq:MF-X-PF:tau:alpha:f}
\end{equation}
where
\begin{equation}
  {\partial f(\alpha_x,\alpha_y)}/{\partial \alpha_x} =  {p}/{2} ~~~~~{\mbox{and}}~~~~~  {\partial f(\alpha_x,\alpha_y)}/{\partial \alpha_y} =  {q}/{2}.
  \label{Eq:MF-X-PF:df:alphaxy:q}
\end{equation}

Taking the partial derivative of Eq.~(\ref{Eq:MF-X-PF:tau:alpha:f}) over $p$ or $q$, we have
\begin{equation}
  {\partial\tau_{xy}(p,q)}/{\partial p} = {\alpha_x}/{2}  ~~~~~{\mbox{and}}~~~~~  {\partial\tau_{xy}(p,q)}/{\partial q} = {\alpha_y}/{2}.
\end{equation}
We can then obtain the double Legendre transforms:
\begin{subequations}
\begin{equation}
  \alpha_x(p,q) = 2{\partial \tau(p,q)}/{\partial p} ~~~~~{\mbox{and}}~~~~~  \alpha_y(p,q) = 2{\partial \tau(p,q)}/{\partial q},
  \label{Eq:MF-X-PF:alphaxy:tau}
\end{equation}
and
\begin{equation}
  f(\alpha_x,\alpha_y) =  p\alpha_x(p,q)/2+q\alpha_y(p,q)/2 -\tau_{xy}(p,q).
  \label{Eq:MF-X-PF:f:alpha}
\end{equation}
\label{Eq:MF-X-PF:pq:Legendre}
\end{subequations}

From the canonical perspective, we can obtain the $f(\alpha_x,\alpha_y)$ function directly by defining two canonical measures
\begin{equation}
  \mu_{xy}(p,q,s,i) = \frac{[m_x(s,i)]^{p/2}[m_y(s,i)]^{q/2}}{\sum_t [m_x(s,i)]^{p/2}[m_y(s,i)]^{q/2}}. 
  \label{Eq:MF-X-PF:pq:muxy}
\end{equation}
The two singularity strengths $\alpha_x(p)$ and $\alpha_x(p)$ and the joint multifractal spectrum $f_{xy}(p,q)$ can be computed by linear regressions in log-log scales using the following equations:
\begin{subequations}
\begin{equation}
  {\alpha_x(p,q)} = \lim_{s\to0} \frac{\sum_i \mu_{xy}(p,q,s,i) \ln{m_x(s,i)}}{\ln{s}},
  \label{Eq:MF-X-PF:pq:alphax:mu:mx}
\end{equation}
\begin{equation}
  {\alpha_y(p,q)} = \lim_{s\to0}\frac{\sum_t \mu_{xy}(p,q,s,i) \ln{m_y(s,i)}}{\ln{s}},
  \label{Eq:MF-X-PF:pq:alphay:mu:my}
\end{equation}
\begin{equation}
  {f_{xy}(p,q)}  = \lim_{s\to0}\frac{\sum_t \mu_{xy}(p,q,s,i) \ln\left[\mu_{xy}(p,q,s,i)\right]}{\ln{s}}.
  \label{Eq:MF-X-PF:pq:fpq:mu}
\end{equation}
\label{Eq:MF-X-PF:pq:Direct}
\end{subequations}
The joint mass exponent function can be obtained by using Eq.~(\ref{Eq:MF-X-PF:tau:alpha:f}).

\subsubsection{Analytical results for binomial measures}

Xie et al. derived analytical results for binomial measures with multipliers $p_x$ and $p_y$ \cite{Xie-Jiang-Gu-Xiong-Zhou-2015-NJP}. They found that the two integrated measures $m_x(s,i)$ and $m_y(s,i)$ in boxes of size $s=2^{l}$ are related explicitly by
\begin{equation}
  m_x(s,i) =C(s) \left[m_y(s,i)\right]^{\beta} = e^{-\gamma L}s^{{\gamma}/{\ln 2}} \left[m_y(s,i)\right]^{\beta},
  \label{Eq:MF-X-PF:analytic:mxmy}
\end{equation}
where $2^L$ is the length of the time series,
\begin{equation}
  \beta=\frac{\ln p_x-\ln (1-p_x)}{\ln p_y-\ln (1-p_y)},
  \label{Eq:MF-X-PF:analytic:beta}
\end{equation}
and
\begin{equation}
  \gamma=\beta\ln (1-p_y) -\ln (1-p_x).
  \label{Eq:MF-X-PF:analytic:gamma}
\end{equation}
When $p_x+p_y=1$, we have $\beta=-1$. When $p_x=p_y$, we have $\beta=1$ and $C(s)=1$. When both $p_x$ and $p_y$ are greater than 0.5 or less than 0.5, that is, $(p_x-0.5)(p_y-0.5)>0$, we have $\beta>0$; Otherwise, when $(p_x-0.5)(p_y-0.5)<0$, we have $\beta<0$.

The joint mass exponent function is expressed as \cite{Xie-Jiang-Gu-Xiong-Zhou-2015-NJP}
\begin{equation}
  \tau_{xy}(p,q) =\frac{p\gamma}{2\ln 2}+\tau_y(Q) =\frac{p\gamma}{2\ln 2}-\frac{\ln[p_y^{Q}+(1-p_y)^{Q}]}{\ln{2}}.
  \label{Eq:MF-X-PF:analytic:tauxy:p:q}
\end{equation}
where
\begin{equation}
  Q = \beta p/2+q/2
  \label{Eq:MF-X-PF:analytic:chi:p:q:c}
\end{equation}
and
\begin{equation}
 \tau_y(Q) =- \ln[p_y^{Q}+(1-p_y)^{Q}]/\ln{2}.
 \label{Eq:MF-X-PF:analytic:tauy}
\end{equation}
It follows that
\begin{equation}
  \alpha_x = \frac{\gamma}{\ln 2}-\frac{\beta}{\ln 2}\frac{p_y^{Q}\ln p_y+(1-p_y)^{Q}\ln (1-p_y) }{p_y^{Q}+(1-p_y)^{Q}}
  \label{Eq:MF-X-PF:analytic:alphax}
\end{equation}
and
\begin{equation}
  \alpha_y = -\frac{1}{\ln 2}\frac{p_y^{Q}\ln p_y+(1-p_y)^{Q}\ln (1-p_y) }{p_y^{Q}+(1-p_y)^{Q}}.
  \label{Eq:MF-X-PF:analytic:alphay}
\end{equation}
We obtain immediately the relationship between $\alpha_x$ and $\alpha_y$,
\begin{equation}
  \alpha_x =\frac{\gamma}{\ln 2}+\beta\alpha_y,
  \label{Eq:MF-X-PF:analytic:alphaxalphay}
\end{equation}
which shows that $f_{xy}(\alpha_x,\alpha_y)$ is a curve along this line rather than a surface and the line segment (\ref{Eq:MF-X-PF:analytic:alphaxalphay}) is the projection of $f_{xy}(\alpha_x,\alpha_y)$ onto the $(\alpha_x,\alpha_y)$ plane.

Xie et al. obtained that \cite{Xie-Jiang-Gu-Xiong-Zhou-2015-NJP}
\begin{equation}
     \left\{
    \begin{aligned}
      \alpha_{y,\min} &= \lim_{Q\rightarrow \infty} \alpha_y = \min\left\{-\frac{\ln p_y}{\ln 2},-\frac{\ln(1-p_y)}{\ln 2}\right\} \\
      \alpha_{y,\max} &= \lim_{Q\rightarrow -\infty} \alpha_y = \max\left\{-\frac{\ln p_y}{\ln 2},-\frac{\ln(1-p_y)}{\ln 2}\right\}
    \end{aligned}
    \right.
    \label{Eq:MF-X-PF:properties:analytic:max:min:alphay}
\end{equation}
such that the width of the singularity spectrum of $\alpha_y$ is
\begin{equation}
  \Delta \alpha_y =\frac{\left|\ln (1-p_y) - \ln p_y\right|}{\ln 2}.
  \label{Eq:MF-X-PF:properties:analytic:deltaalphay}
\end{equation}
When $p_y=0.5$, $\Delta \alpha_y=0$. In this case, the measure is neither multifractal nor monofractal since it is uniformly distributed on the support.
Similarly, we have
\begin{equation}
     \left\{
    \begin{aligned}
      \alpha_{x,\min} &= \lim_{Q\rightarrow \infty} \alpha_x = \min\left\{-\frac{\ln p_x}{\ln 2},-\frac{\ln(1-p_x)}{\ln 2}\right\} \\
      \alpha_{x,\max} &= \lim_{Q\rightarrow -\infty} \alpha_x = \max\left\{-\frac{\ln p_x}{\ln 2},-\frac{\ln(1-p_x)}{\ln 2}\right\}
    \end{aligned}
    \right.
    \label{Eq:MF-X-PF:properties:analytic:max:min:alphax}
\end{equation}

Further, the multifractal spectrum $f_{xy}(\alpha_x,\alpha_y)$ can be expressed as  \cite{Xie-Jiang-Gu-Xiong-Zhou-2015-NJP}
\begin{equation}
  f_{xy}(\alpha_x,\alpha_y) =\frac{1}{\ln 2}\frac{Q\left(\frac{1-p_y}{p_y}\right)^{Q} \ln\left(\frac{p_y}{1-p_y}\right) +\left[1+\left(\frac{1-p_y}{p_y}\right)^{Q}\right]\ln\left[1+\left(\frac{1-p_y}{p_y}\right)^{Q}\right]} {1+\left(\frac{1-p_y}{p_y}\right)^{Q}}~.
  \label{Eq:MF-X-PF:analytic:fpq}
\end{equation}
It follows immediately that
\begin{equation}
  f_{xy}(Q=0) = 1 ~~~~~{\rm{and}}~~~~~f_{xy}(Q)=f_{xy}(-Q),
\end{equation}
where $f_{xy}(Q)\triangleq f_{xy}(\alpha_x,\alpha_y; Q)$. It implies that $f_{xy}(\alpha_x,\alpha_y)$ is symmetric with respect to the line $Q=0$. In addition, it is found that
\begin{equation}
  \lim_{Q\to\pm\infty} f_{xy}(p,q) = 0.
  \label{Eq:MF-X-PF:properties:analytic:fpq:infty}
\end{equation}
When $Q<0$, $df_{xy}(Q)/dQ>0$ so that $f_{xy}(Q)$ is a monotonically increasing function of $Q$. When $Q>0$, $df_{xy}(Q)/dQ>0$ so that $f_{xy}(Q)$ is a monotonically decreasing function of $Q$. Therefore, the maximum of $f_{xy}(Q)$ is 1 and its minimum is 0.

\subsubsection{The case \texorpdfstring{$p=q$}{}}

The multifractal cross-correlation analysis based on statistical moments (MFSMXA) proposed in Ref.~\cite{Wang-Shang-Ge-2012-Fractals} is actually a special case of MF-X-PF$(p,q)$ when $p=q$. We call it MF-X-PF$(q)$ for consistence \cite{Xie-Jiang-Gu-Xiong-Zhou-2015-NJP}. In this case, we have
\begin{equation}
  \tau_{xy}(q) = q(\alpha_x+\alpha_y)/2 - f_{xy}(\alpha_x,\alpha_y),
  \label{Eq:MF-X-PF:q:tau:ax:ay:f}
\end{equation}
where
\begin{equation}
  {\partial f_{xy}(\alpha_x,\alpha_y)}/{\partial \alpha_x} = {\partial f_{xy}(\alpha_x,\alpha_y)}/{\partial \alpha_y} = q/2.
  \label{Eq:MF-X-PF:q:df:dalphax:q}
\end{equation}
Taking the derivative of Eq.~(\ref{Eq:MF-X-PF:q:tau:ax:ay:f}) over $q$ and applying Eq.~(\ref{Eq:MF-X-PF:q:df:dalphax:q}), we have
\begin{equation}
  \frac{\mbox{d}\tau_{xy}(q)}{\mbox{d}q} = \frac{\alpha_x+\alpha_y}{2}.
  \label{Eq:MF-X-PF:q:dtau:dq:alphax:alphay}
\end{equation}
Defining that
\begin{equation}
  \alpha_{xy} \triangleq [\alpha_{x}(q) + \alpha_{y}(q)]/2,
  \label{Eq:MF-X-PF:q:alpha:alphax:alphay}
\end{equation}
we have
\begin{subequations}
\begin{equation}
  \alpha_{xy} = d\tau_{xy}(q)/dq,
  \label{Eq:MF-X-PF:q:dtau:dq:alpha}
\end{equation}
and
\begin{equation}
  f_{xy}(\alpha_{xy}(q)) \triangleq f_{xy}(\alpha_x,\alpha_y) =  q \alpha_{xy}(q) -\tau_{xy}(q),
  \label{Eq:MF-X-PF:q:f:alpha}
\end{equation}
\end{subequations}
which is the Legendre transform.

It is derived that \cite{Xie-Jiang-Gu-Xiong-Zhou-2015-NJP}
\begin{equation}
  \tau_{xy}(q)= [\tau_x(q) + \tau_y(q)]/2,
  \label{Eq:MF-X-PF:q:tauxy:taux:tauy}
\end{equation}
and
\begin{equation}
  f_{xy}(q)= [f_x(q) + f_y(q)]/2.
  \label{Eq:MF-X-PF:q:fxy:fx:fy}
\end{equation}
We define the generalized joint Hurst exponent $H_{xy}(q)$ as
\begin{equation}
  \tau_{xy}(q)= qH_{xy}(q)-1.
  \label{Eq:MF-X-PF:tauxy:Hxy}
\end{equation}
It follows that
\begin{equation}
  H_{xy}(q)= \frac{H_x(q)+H_y(q)}{2},
  \label{Eq:MF-X-PF:Hxy:Hx:Hy}
\end{equation}
where $H_x(q)$ and $H_y(q)$ are the generalized Hurst exponent of $m_x$ and $m_y$.
These relations were observed numerically using the MF-X-DFA method \cite{Zhou-2008-PRE}, the MF-X-DMA method \cite{Jiang-Zhou-2011-PRE} and the MF-X-PF method with $p=q$ \cite{Wang-Shang-Ge-2012-Fractals}.

If we can define the canonical measures
\begin{equation}
  \mu_{xy}(q,s,i) = \frac{[m_x(s,i)m_y(s,i)]^{q/2}}{\sum_t [m_x(s,i)m_y(s,i)]^{q/2}}. 
  \label{Eq:MF-X-PF:q:mu:p:q}
\end{equation}
the two singularity strengths $\alpha_x(p)$ and $\alpha_x(p)$ and the joint multifractal spectrum $f_{xy}(p,q)$ can be computed directly:
\begin{subequations}
\begin{equation}
  \alpha_{xy}(q) = \lim_{s\to0} \frac{\sum_t \mu_{xy}(q,s,i) \ln{[m_x(s,i)m_y(s,i)]^{1/2}}}{\ln{s}}
      = \frac{\alpha_{x}(q)+\alpha_{y}(q)}{2},
  \label{Eq:MF-X-PF:q:alpha:mu:mx}
\end{equation}
where Eq.~(\ref{Eq:MF-X-PF:pq:alphax:mu:mx}) and Eq.~(\ref{Eq:MF-X-PF:pq:alphay:mu:my}) are used in the second equality, and
\begin{equation}
  f_{xy}(\alpha_{xy}(q))  = \lim_{s\to0}\frac{\sum_t \mu_{xy}(q,s,i) \ln\left[\mu_{xy}(q,s,i)\right]}{\ln{s}}.
  \label{Eq:MF-X-PF:q:fpq:mu}
\end{equation}
\end{subequations}
The joint mass exponent function can be obtained by using Eq.~(\ref{Eq:MF-X-PF:q:f:alpha}).

Xiong and Shang proposed the weighted MF-X-PF$(q)$ method (W-MFSMXA) \cite{Xiong-Shang-2016-CNSNS}. Their numerical experiments on two-exponent ARFIMA processes, binomial measures and NBVP time series illustrate that W-MFSMXA slightly outperforms MF-X-PF$(q)$ for relatively short time series.


\subsection{Joint structure function analysis (MF-X-SF)}
\label{S2:MF-X-SF}

\subsubsection{General formalism for arbitrary \texorpdfstring{$(p,q)$}{} pair}

The joint multifractal analysis based on the structure function approach, termed MF-X-SF, also has its root in turbulence, in which the joint structure function of the velocity and temperature fluctuations was introduced to study the cross-correlation between the temperature and velocity dissipation fields in a heated turbulent jet \cite{Antonia-VanAtta-1975-JFM,Schmitt-Schertzer-Lovejoy-Brunet-1996-EPL}. For two self-similar time series $X(i)$ and $Y(i)$ with $i=1,2,\cdots,N$, the $(p,q)$-order joint structure function is defined as
\begin{equation}
  K_{xy}(p,q,s) = \left\langle \Delta{X(i,s)}^{p/2} \Delta{Y(i,s)}^{q/2} \right\rangle
          =\frac{1}{N-s+1}\sum_{i=s}^{N} \Delta{X(i,s)}^{p/2} \Delta{Y(i,s)}^{q/2},
  \label{Eq:MF-X-SF:Kxy:pq}
\end{equation}
where $\Delta{X(i,s)}=X(i)-X(i-s)$ and $\Delta{Y(i,s)} = Y(i)-Y(i-s)$. The special case of $p=1$ is called by some researchers as the analogous multifractal
cross-correlation analysis (AMF-XA) \cite{Wang-Liao-Li-Zou-Shi-2013-Chaos}, in which the authors were seemingly unaware of previous works \cite{Antonia-VanAtta-1975-JFM,Kristoufek-2011-EPL}. Because the innovations $\Delta{X}$ and $\Delta{Y}$ could be negative, the structure functions $K_{xy}(p,q,s)$ are usually defined on positive integers of $p$ and $q$. Therefore, it is more convenient to use absolute innovations to define the joint structure function
\begin{equation}
  K_{xy}(p,q,s) = \left\langle \left|\Delta{X(i,s)}\right|^{p/2} |\Delta{Y(i,s)}|^{q/2} \right\rangle
          =\frac{1}{N-s+1}\sum_{i=s}^{N} |\Delta{X(i,s)}|^{p/2} |\Delta{Y(i,s)}|^{q/2},
  \label{Eq:MF-X-SF:Kxy:pq:absolute}
\end{equation}
which is widely adopted in the structure function approach, as well as in the multifractal height cross-correlation analysis (MFHXA) \cite{Kristoufek-2011-EPL}. If there is a scaling relation between the joint structure function $K_{xy}(p,q,s)$ and the scale $s$, we can define the scaling exponent function $\tau_{xy}(p,q)$ as
\begin{equation}
  K_{xy}(p,q,s) \sim s^{\zeta_{xy}(p,q)} = s^{\tau_{xy}(p,q)+1}~,
  \label{Eq:MF-X-SF:Kxy:pq:s}
\end{equation}
which is an analog of the MF-X-PF$(p,q)$ approach.

We further define the singularity functions $\alpha_x(p,q)$ and $\alpha_y(p,q)$ as the partial derivatives of $\tau_{xy}(p,q)$ with respect to $p$ and $q$, respectively, as in Eq.~(\ref{Eq:MF-X-PF:alphaxy:tau}). The singularity spectrum function $f_{xy}(p,q)\triangleq f_{xy}(\alpha_x,\alpha_y)$ can be obtained as in Eq.~(\ref{Eq:MF-X-PF:f:alpha}). We note that it is hard to define a generalized Hurst index $H_{xy}(p,q)$ in a consistent way.
The joint structure function approach has been considered for the multifractal random walks \cite{Muzy-Delour-Bacry-2000-EPJB}.

\subsubsection{The case \texorpdfstring{$p=q$}{}}

A special case of the MF-X-SF$(p,q)$ approach when $p=q$ was proposed by Kristoufek, which was termed the multifractal height cross-correlation analysis \cite{Kristoufek-2011-EPL}. For consistency, we call it the MF-X-SF$(q)$ approach. In this case, the $q$th order joint structure function is defined by
\begin{equation}
  K_{xy}(q,s) = \left\langle \left|\Delta{X(i,s)}\Delta{Y(i,s)}\right|^{q/2} \right\rangle
          =\frac{1}{N-s+1}\sum_{i=s}^{N} |\Delta{X(i,s)}\Delta{Y(i,s)}|^{q/2}.
  \label{Eq:MF-X-SF:Kxy:q}
\end{equation}
If there is a scaling relation between the joint structure function $K_{xy}(p,q,s)$ and the scale $s$, we can define the scaling exponent function $\tau_{xy}(q)$ as
\begin{equation}
  K_{xy}(q,s) \sim s^{\tau_{xy}(q)+1} = s^{qH_{xy}(q)},
  \label{Eq:MF-X-SF:Kxy:q:s}
\end{equation}
where $H_{xy}(q)$ are the generalized bivariate Hurst exponents or the generalized joint Hurst exponents. Under certain conditions, we have \cite{Kristoufek-2011-EPL}
\begin{equation}
  H_{xy}(q) = \frac{H_x(q)+H_y(q)}{2}.
  \label{Eq:MF-X:Hxy:Hx:Hy}
\end{equation}

We define the joint singularity function and its spectrum through the Legendre transform:
\begin{subequations}
\begin{equation}
  \alpha_{xy} = {\mbox{d}}\tau_{xy}(q)/{\mbox{d}}q,
  \label{Eq:MF-X-SF:q:dtau:dq:alpha}
\end{equation}
and
\begin{equation}
  f_{xy}(\alpha_{xy}) =  q \alpha_{xy} -\tau_{xy}.
  \label{Eq:MF-X-SF:q:f:alpha}
\end{equation}
\end{subequations}
If Eq.~(\ref{Eq:MF-X:Hxy:Hx:Hy}) holds, we have
\begin{equation}
  \tau_{xy}(q) = \frac{\tau_x(q)+\tau_y(q)}{2}
  \label{Eq:MF-X:tayxy:taux:tauy}
\end{equation}
according to the definitions that $\tau_x(q)=qH_x(q)-1$, $\tau_y(q)=qH_y(q)-1$ and $\tau_{xy}(q)=qH_{xy}(q)-1$, and
\begin{equation}
  \alpha_{xy}(q) = \frac{\alpha_x(q)+\alpha_y(q)}{2}
  \label{Eq:MF-X:alphaxy:alphax:alphay}
\end{equation}
according to the definitions that $\alpha_x(q)={\mbox{d}}\tau_x(q)/{\mbox{d}}q$, $\alpha_y(q)={\mbox{d}}\tau_y(q)/{\mbox{d}}q$ and $\alpha_{xy}(q)={\mbox{d}}\tau_{xy}(q)/{\mbox{d}}q$. It follows from Eq.~(\ref{Eq:MF-X-SF:q:f:alpha}) that
\begin{equation}
  f_{xy}(\alpha_{xy}) = \frac{f_{x}(\alpha_x)+f_y(\alpha_y)}{2}.
  \label{Eq:MF-X:fxy:fx:fy}
\end{equation}

\subsection{Multifractal wavelet coherence analysis (MF-WCA)}

\subsubsection{Cross wavelet transformation and wavelet coherence analysis}
\label{S3:WCA}	

The cross wavelet transform (XWT) of two time series $\{X_i\}_{i=1}^N$ and $\{Y_i\}_{i=1}^N$ was introduced by Hudgins et al. to study the highly intermittent pattern in atmospheric turbulence \cite{Hudgins-Friehe-Mayer-1993-PRL}, which is defined by
\begin{equation}
  {\cal{W}}_{xy}(s,i) = {\mathcal{W}}_x(s,i){\mathcal{W}}_y(s,i).
  \label{Eq:MF-X-WT:Wxy}
\end{equation}
The cross-wavelet power is
\begin{equation}
  P_{xy}(s) = \|{\cal{W}}_{xy}(s,i)\|.
  \label{Eq:MF-X-WT:PWxy}
\end{equation}
The wavelet coherence $\rho_{xy}^2(s)$ is defined as
\begin{equation}
  \rho_{xy}^2(s) = \frac{\|{\cal{W}}_{xy}(s,i)\|}{\|{\cal{W}}_x(s,i)\|\|{\cal{W}}_y(s,i)\|}
               = \frac{\sum_{i=1}^N|W_x(s,i)W_y(s,i)|}{\left(\sum_{i=1}^NW_x^2(s,i)\right)^{1/2}\left(\sum_{i=1}^NW_y^2(s,i)\right)^{1/2}}~.
  \label{Eq:MF-X-WT:WC}
\end{equation}
The wavelet coherence is able to uncover the co-movement between two time series in the time-frequency domain, which is widely applied in finance and economics \cite{Rua-Nunes-2009-JEF,Vacha-Barunik-2012-EE,Grinsted-Moore-Jevrejeva-2004-NPG}.

\subsubsection{MF-X-WT}
\label{S3:MF-X-WT}	

Similar to the MF-X-PF method, Jiang et al. designed the multifractal cross wavelet analysis \cite{Jiang-Gao-Zhou-Stanley-2017-Fractals}. It generalizes the concept of wavelet coherence and constructs the following joint partition function:
\begin{equation}
  \chi_{xy}(p,q,s)= \sum_{i=1}^N|W_x(s,i)|^{p/2}|W_y(s,i)|^{q/2}.
  \label{Eq:MF-X-WT:Chi:p:q:s}
\end{equation}
%
%
The definition of the partition function allows us to uncover the more intricate relationship between the coherency and the scale under different scopes, which corresponds to a cross-multifractal behavior. If the underlying processes are jointly multifractal, the result is a scaling behavior:
\begin{equation}
  \chi_{xy}(p,q,s) \sim s^{{\cal{T}}_{xy}(p, q)},
  \label{Eq:MF-X-WT:Chi:Scaling}
\end{equation}
where ${\cal{T}}_{xy}(p, q)$ is the joint mass exponent function, which can be estimated by regressing $\ln \chi_{xy} (p,q,s)$ against $\ln s$ in the scaling range for a given pair $(p,q)$.

Analogous to the double Legendre transform in the joint multifractal analysis based on the partition function approach MF-X-PF$(p,q)$ \cite{Xie-Jiang-Gu-Xiong-Zhou-2015-NJP}, we define the joint singularity strength function $h_x$ and $h_y$
\begin{eqnarray}
 h_x(p,q) & = & 2\partial {\cal{T}}_{xy}(p, q) / \partial p, \label{Eq:MF-X-WT:hx}\\
 h_y(p,q) & = & 2\partial {\cal{T}}_{xy}(p, q) / \partial q, \label{Eq:MF-X-WT:hy}
\end{eqnarray}
and the multifractal spectrum $D_{xy}(h_x, h_y)$
\begin{equation}
 D_{xy}(h_x, h_y) = p h_x /2 + q h_y /2  - {\cal{T}}_{xy}. \label{Eq:MF-X-WT:Dh}
\end{equation}

When $p = 0$ and $q = 0$, the joint partition function in Eq.~(\ref{Eq:MF-X-WT:Chi:p:q:s}) is equal to the number of wavelet coefficients, or equivalently the total number of data points of the original series. In other words, we have
\begin{equation}
  \chi_{xy}(0,0,s) = N,
  \label{Eq:MF-X-WT:Chi:0:0}
\end{equation}
which results in
\begin{equation}
  {\cal{T}}_{xy}(0, 0) = 0.
  \label{Eq:MF-X-WT:Txy:0:0}
\end{equation}
In contrast, in the MF-X-PF$(p, q)$ method, $\tau_{xy} (0,0)=-1$. Therefore,
\begin{equation}
  {\cal{T}}_{xy}(p, q) \neq \tau_{xy}(p,q).
  \label{Eq:MF-X-WT:Txy:neq:tau}
\end{equation}
It follows immediately that $h_x(p,q)$, $h_y(p,q)$ and $D_{xy}(h_x, h_y)$ obtained from the MF-X-WT$(p,q)$ method are different from the joint singularity strengths $\alpha_x(p,q)$, $\alpha_y(p,q)$ and the joint multifractal spectrum $f_{xy}(\alpha_x, \alpha_y)$ obtained from the MFXPF$(p,q)$ method.

Analogous to the partition function approach \cite{Chhabra-Jensen-1989-PRL,Chhabra-Meneveau-Jensen-Sreenivasan-1989-PRA}, we can directly estimate the joint singularity strength $h_x$ and $h_y$ and the joint multifractal spectrum $D_{xy}(p,q)$ using
\begin{eqnarray}
 h_x(p,q) & = & \lim_{s \rightarrow 0} \frac{1}{\ln s} \sum_i
 \mu_{xy}(p,q,s,i) \ln |w_x(s,i)|, \label{Eq:MF-X-WT:DE:hx}\\
 h_y(p,q) & = & \lim_{s \rightarrow 0} \frac{1}{\ln s} \sum_i
 \mu_{xy}(p,q,s,i) \ln |w_y(s,i)|, \label{Eq:MF-X-WT:DE:hy} \\
 D_{xy}(p,q) &=& \lim_{s \rightarrow 0} \frac{1}{\ln s} \sum_i
 \mu_{xy}(p,q,s,i) \ln \mu_{xy}(p,q,s,i). \label{Eq:MF-X-WT:DE:Dxy}
\end{eqnarray}
where
\begin{equation*}
  \mu_{xy} (p,q,s,i) = \frac{|w_x(s,i)|^{p/2} |w_y(s,i)|^{q/2}}{\chi_{xy}(p,q,s)}. 
\end{equation*}
Thus we can directly determine the joint singularity strength functions $h_x(p,q)$ and $h_y(p,q)$ and the joint multifractal function $D_{xy}(p,q)$ from Eqs.~(\ref{Eq:MF-X-WT:DE:hx}--\ref{Eq:MF-X-WT:DE:Dxy}) through linear regressions.

Extensive numerical experiments on two multifractal binomial measures unveil connections between the theoretical joint multifractal formulas of MF-X-PF$(p,q)$ derived in Ref.~\cite{Xie-Jiang-Gu-Xiong-Zhou-2015-NJP} and the empirical joint multifractal features of MF-X-WT$(p,q)$ \cite{Jiang-Gao-Zhou-Stanley-2017-Fractals}:
\begin{eqnarray}
 {\cal{T}}_{xy}(p, q)  + p/2 + q/2-1 & = & \tau_{xy}(p, q), \label{Eq:pModel:WTTau:PFTau}\\
  h_x(p, q) + 1 & = & \alpha_x(p, q), \label{Eq:pModel:WThx:PFalphax}\\
  h_y(p, q) + 1 & = & \alpha_y(p, q), \label{Eq:pModel:WThy:PFalphay}\\
  D_{xy} (h_x, h_y) + 1 & = & f_{xy}(\alpha_x,
  \alpha_y), \label{Eq:pModel:WTDh:PFfalpha}
\end{eqnarray}
where $\tau_{xy}(p,q)$, $\alpha_x(p,q)$, $\alpha_y(p,q)$ and $f_{xy}(\alpha_x, \alpha_y)$ are given in Eq.~(\ref{Eq:MF-X-PF:analytic:tauxy:p:q}), Eq.~(\ref{Eq:MF-X-PF:analytic:alphax}), Eq.~(\ref{Eq:MF-X-PF:analytic:alphay}) and Eq.~(\ref{Eq:MF-X-PF:analytic:fpq}).

Although the MF-X-WT$(p,q)$ method is able to successfully identify the monofractal or multifractal cross correlations in time series, the results show non-negligible deviations from the theoretical results \cite{Jiang-Gao-Zhou-Stanley-2017-Fractals}. Nevertheless, empirical applications to financial returns and financial volatilities uncover evident cross-multifractality \cite{Jiang-Gao-Zhou-Stanley-2017-Fractals}.

%
%
%

\subsubsection{MF-X-WTMM}
\label{S3:MF-X-WTMM}

The implementation of a joint multifractal analysis based on WTMM is not straightforward. We can conveniently determine the two sets of modulus maxima lines ${\cal{L}}_x$ and ${\cal{L}}_y$ for $X(i)$ and $Y(i)$. However, there is no one-to-one correspondence between ${\cal{L}}_x$ and ${\cal{L}}_y$ because there are different numbers of modulus maxima for the two time series at a given scale. Lin and Sharif found a nice solution to overcome this difficulty by properly pairing up the modulus maxima lines in ${\cal{L}}_x$ and ${\cal{L}}_y$ \cite{Lin-Sharif-2007-EPJB,Lin-Sharif-2010-Chaos}. Two modulus maxima lines in ${\cal{L}}_x$ and ${\cal{L}}_y$ are paired up if they are close to each other in the time coordinate. The joint partition function can thus be defined as  \cite{Lin-Sharif-2007-EPJB}
\begin{equation}
  Z_{p,q}(s_j)= \sum_{k=1}^{N_j}|W_x(i_{j,k},s_j)|^p|W_y(i_{j,k},s_j)|^q ~,
  \label{Eq:MF-X-WTMM:Zpq}
\end{equation}
which scales as a power law with respect to the scale $s$:
\begin{equation}
  Z_{p,q}(s) \sim s^{\tau_{xy}(p,q)}
  \label{Eq:MF-X-WTMM:Zpq:s:Tau}
\end{equation}
Note that we use $p$ and $q$ instead of $p/2$ and $q/2$ following Ref.~\cite{Lin-Sharif-2007-EPJB}. In this case, the Legendre transform is expressed as \cite{Lin-Sharif-2007-EPJB}
\begin{subequations}
\begin{equation}
  {\alpha_x}=\partial\tau(p,q)/\partial p,~~~ {\alpha_y}=\partial\tau(p,q)/\partial q,
  \label{Eq:MF-X-WTMM:alpha:dtau:dq}
\end{equation}
and
\begin{equation}
  f({\alpha_x},{\alpha_y})=p{\alpha_x}(p,q) +q{\alpha_y}(p,q)-\tau(p,q),
  \label{Eq:MF-X-WTMM:f:alpha}
\end{equation}
\label{Eq:MF-X-WTMM:LegendreTrandform}
\end{subequations}
where $p=\partial{f}/\partial{\alpha_x}$ and $q=\partial{f}/\partial{\alpha_y}$.

The validity of the MF-X-WTMM method is confirmed by numerical experiments on coupled random binomial cascades \cite{Lin-Sharif-2007-EPJB}. Coupled random binomial cascades are characterized by a coupling strength parameter $g\in[0,1]$, which $g=0$ and $g=1$ stand respectively for uncoupled and fully coupled time series \cite{Meneveau-Sreenivasan-Kailasnath-Fan-1990-PRA}. Their numerical result that the $f(\alpha_x, \alpha_y)$ spectrum collapses into a one-dimensional curve in the fully coupled case is in excellent agreement with the analytical result of Eq.~(\ref{Eq:MF-X-PF:analytic:alphaxalphay}) \cite{Xie-Jiang-Gu-Xiong-Zhou-2015-NJP}.

\subsubsection{MF-X-WL}
\label{S3:MF-X-WL}

Due to the inherent drawbacks associated with the multifractal wavelet transform analysis, the MF-X-WT$(p,q)$ is not recommended for multifractal cross-correlation analysis.
In addition to the MF-X-WTMM method, a suitable alternative candidate is the wavelet leaders  \cite{Jaffard-2004,Jaffard-2006-AFST,Wendt-Abry-Jaffard-2007-IEEEspm,Wendt-Abry-2007-IEEEtsp, Lashermes-Roux-Abry-Jaffard-2008-EPJB, Serrano-Figliola-2009-PA}. Jiang et al. generalized the MF-WL method to the joint multifractal analysis based on wavelet leaders with two moment orders $p$ and $q$, termed MF-X-WL$(p, q)$ \cite{Jiang-Yang-Wang-Zhou-2017-FoP}.

One first obtains the wavelet leaders $\ell_x(j,k)$ and $\ell_y(j,k)$ of $\{X_i\}_1^N$ and $\{Y_i\}_1^N$ at different scales $s = 2^j$, and then defines the joint moment with orders $p$ and $q$ as  \cite{Jiang-Yang-Wang-Zhou-2017-FoP}
\begin{equation}
  M_{xy}(p, q, j) = \frac{1}{n_j}\sum_{k=1}^{n_j} \ell_x(j,k)^{p/2}\ell_y(j,k)^{q/2}, \label{Eq:MF-X-WL:SF}
\end{equation}
where $n_j$ is the number of wavelet leaders at scale $s=2^j$. When $X = Y$ and $p=q$, we recover the traditional multifractal formalism based on wavelet leaders presented in Section \ref{S3:MF-WT:WL}. Further scaling analyses follow the conventional way as other similar methods.

Through numerical experiments, it is found that the MF-X-WL$(p,q)$ method can successfully detect joint multifractality in binomial measures and monofractality in bivariate fractional Brownian motions \cite{Jiang-Yang-Wang-Zhou-2017-FoP}. The method has also been used to investigate the joint multifractality between the daily returns of the DJIA and NASDAQ indices, and between their daily volatilities as well. The MF-X-WL$(p,q)$ method works for both positive and negative $(p,q)$ values. However, the method seemingly cannot provide very accurate results.

\subsection{Multifractal detrended cross-correlation analysis (MF-DCCA)}
\label{S2:MF-DCCA}


Inspired by the seminal work of Podobnik and Stanley on the detrended cross-correlation analysis \cite{Podobnik-Stanley-2008-PRL}, Zhou proposed the multifractal detrended cross-correlation analysis, which is a combination of MF-DFA and DCCA \cite{Zhou-2008-PRE}. The method was initially termed MF-DXA \cite{Zhou-2008-PRE} and then MF-X-DFA so that one can distinguish it from the MF-X-DMA method which is a combination of MF-DMA and DCCA \cite{Jiang-Zhou-2011-PRE}.

Consider two time series $\{X(i)\}$ and $\{Y(i)\}$ of the same length $N$, where $i=1, 2, \cdots, N$. Each time series is covered with $N_s=\mbox{int}[N/s]$ non-overlapping boxes of size $s$. The two segments of time series within the $v$th box $[l_v+1,l_v+s]$ are denoted by $X_v(k)$ and $Y_v(k)$ with $k=1,\cdots,s$, where $l_v=(v-1)s$.
Assume that the local trend functions of $\{X_v(k)\}_{k=1}^s$ and $\{Y_v(k)\}_{k=1}^s$ are $\{\widetilde{X}_v(k)\}$ and $\{\widetilde{Y}_v(k)\}$, respectively. The cross-correlation for each box is calculated as
\begin{equation}
 F_{v}^2(s) = \frac{1}{s}\sum_{k=1}^s \left[X_v(k)-\widetilde{X}_v(k)\right]\left[Y_v(k)-\widetilde{Y}_v(k)\right]~.
\end{equation}
The $q$th order cross-correlation is calculated as
\begin{equation}
 F_{xy}^2(q,s) = \left[\frac{1}{m}\sum_{v=1}^m |F_{v}(s)|^{q}\right]^{1/q}
\end{equation}
when $q\neq0$ and
\begin{equation}
 F_{xy}^2(0,s) = \exp\left[\frac{1}{N_s}\sum_{v=1}^{N_s} \ln |F_{v}(s)|\right]~.
\end{equation}
We then expect the following scaling relation
\begin{equation}
 F_{xy}(q,s) \sim s^{H_{xy}(q)}~.
 \label{Eq:Fxy:s}
\end{equation}

There are many different methods for the determination of $\widetilde{X}_v$ and $\widetilde{Y}_v$. The local trend functions could be polynomials \cite{Peng-Buldyrev-Havlin-Simons-Stanley-Goldberger-1994-PRE,Hu-Ivanov-Chen-Carpena-Stanley-2001-PRE}, which recovers the MF-X-DFA method \cite{Zhou-2008-PRE}. The local trend functions could also be moving averages \cite{Vandewalle-Ausloos-1998-PRE,Alessio-Carbone-Castelli-Frappietro-2002-EPJB}, in which case the algorithm is called MF-X-DMA \cite{Jiang-Zhou-2011-PRE}. We can certainly use other detrending approaches as described in Section \ref{S3:MF:Detrend:Others}. When $X=Y$, MF-X-DFA reduces to MF-DFA and MF-X-DMA reduces to MF-DMA. The DCCA method based on DMA was proposed by He and Chen \cite{He-Chen-2011b-PA}, which is a special case of the MF-X-DMA method with $q\equiv2$.

%
%
%

For dependent multifractal binomial measure pairs \cite{Meneveau-Sreenivasan-1987-PRL} and dependent multifractal random walk pairs \cite{Bacry-Delour-Muzy-2001-PRE}, it is found numerically that \cite{Zhou-2008-PRE}
\begin{equation}
  H_{xy}(q)=\frac{H_x(q)+H_y(q)}{2},
  \label{Eq:MF-DCCA:hXY:hX:hY}
\end{equation}
Further, He and Chen gave an approximate derivation of the following inequality \cite{He-Chen-2011a-PA}:
\begin{equation}
  H_{xy}(q)\leqslant\frac{H_x(q)+H_y(q)}{2}.
  \label{Eq:MF-DCCA:hXY:hX:hY:2}
\end{equation}
However, both relations (\ref{Eq:MF-DCCA:hXY:hX:hY}) and (\ref{Eq:MF-DCCA:hXY:hX:hY:2}) deviate from many empirical results \cite{Zhou-2008-PRE,He-Chen-2011a-PA,He-Chen-2011-CSF,Ma-Wei-Huang-2013-PA}.

For the monofractal case with $q=2$ such as two coupled autoregressive fractional moving average (ARFIMA) signals with identical error terms, it is shown with certain approximations that \cite{Podobnik-Stanley-2008-PRL,Podobnik-Grosse-Horvatic-Ilic-Ivanov-Stanley-2009-EPJB,Gvozdanovic-Podobnik-Wang-Stanley-2012-PA}
\begin{equation}
  H_{xy}(2)=\frac{H_x(2)+H_y(2)}{2}~,
  \label{Eq:DCCA:HXY:HX:HY}
\end{equation}
where $H_x(2)$ and $H_y(2)$ are the Hurst exponents of time series $\{X_i\}$ and $\{Y_i\}$. Kristoufek derived equation (\ref{Eq:DCCA:HXY:HX:HY}) for two ARFIMA processes regardless of the correlation between error terms, as long as it remains non-zero and for an ARFIMA process and an MA process \cite{Kristoufek-2015b-PA}. Presenting the memory effects as power spectrum in the frequency domain, Kristoufek showed that the bivariate Hurst exponent cannot be larger than the average of the separate Hurst exponents \cite{Kristoufek-2015d-PA}:
\begin{equation}
  H_{xy}(2)\leqslant\frac{H_x(2)+H_y(2)}{2},
  \label{Eq:DCCA:HXY:HX:HY:2}
\end{equation}
which holds for several stochastic processes \cite{Lavancier-Philippe-Surgailis-2009-SPL,Coeurjolly-Amblard-Achard-2010-EUSIPCO,Podobnik-Jiang-Zhou-Stanley-2011-PRE,Sela-Hurvich-2012-JTSA,Amblard-Coeurjolly-Lavancier-Philippe-2013-BSMF,Kristoufek-2013-PA,Kristoufek-2015b-PA,Kristoufek-2015d-PA}.
The difference can be qualified by the DCCA coefficient differentiation \cite{Zebende-daSilva-Filho-2013-PA}.

There are also variants of MF-X-DFA. The time-delay MF-X-DFA method includes a time delay in pairing two time series such that the detrended fluctuation functions in segments read \cite{Zhao-Shang-Jin-2011-Fractals}
\begin{equation}
 F_{v}^2(\delta;s) = \frac{1}{s}\sum_{k=1}^s \left[X_v(k)-\widetilde{X}_v(k)\right]\left[Y_v(k+\delta)-\widetilde{Y}_v(k+\delta)\right].
\end{equation}
When $\delta=0$, we recover the original MF-X-DFA method. 
Shi et al. generalized the multiscale multifractal analysis \cite{Gieraltowski-Zebrowski-Baranowski-2012-PRE} to multiscale multifractal detrended cross-correlation analysis, called MSMF-DXA \cite{Shi-Shang-Wang-Lin-2014-PA}. This method is applied to the pairs of daily returns of six stock market indices (DJIA, NASDAQ, S\&P500, SSCI, SZCI and HSI) \cite{Shi-Shang-Wang-Lin-2014-PA}, return pairs (DJIA vs NYSE and DJIA vs HSI) in different time periods \cite{Lin-Shang-Zhou-2014-ND}, and pairs of opening price, closing price, highest price, lowest price and transaction volume of S\&P 500, HSI, SSCI and ASX \cite{Yang-Li-Yang-2017-PA}. Cao and Xu suggested to determine the local trends using the maximum overlap wavelet transform (MODWT) \cite{Cao-Xu-2016-PA}. Cao et al. proposed the volatility-constrained multifractal detrended cross-correlation analysis \cite{Cao-Zhang-Li-2017-PA}. Numerical investigations using mathematical models on the performance of these methods have not been conducted yet.

The MF-X-DFA method has been generalized to investigate the multifractal cross-correlation behavior among multiple time series, which leads to the coupling detrended fluctuation analysis (CDFA) \cite{Hedayatifar-Vahabi-Jafari-2011-PRE}. For multiple time series $X_j(i)$ with $j=1, 2, \cdots, n$, the local detrended fluctuation function becomes
\begin{equation}
 F_{v}^n(s) = \frac{1}{s} \sum_{k=1}^s \prod_{j=1}^n\left[X_{j,v}(k)-\widetilde{X}_{j,v}(k)\right].
\end{equation}
Other methods for joint multifractal analysis can also be extended to multiple time series in a similar way.
A similar treatment is to sum the local fluctuations of multiple time series \cite{Fan-Wu-2015-EPJB}:
\begin{equation}
 F_{v}^2(s) = \frac{1}{s} \sum_{k=1}^s \sum_{j=1}^n\left[X_{j,v}(k)-\widetilde{X}_{j,v}(k)\right]^2.
\end{equation}
This method is reported to work well for two-component ARFIMA processes and binomial measures.

The CDFA method has been applied to hourly concentrations of four air pollutants (NO$_2$, NO$_x$, total hydrocarbon content THC and O$_3$) at Tehran from 2 February 2006 to 31 July 2007 \cite{Hedayatifar-Vahabi-Jafari-2011-PRE}, weekly foreign exchange rates of three currencies (EUR/USD, GBP/USD and JPY/USD) from 14 December 1998 to 31 January 2011  \cite{Hedayatifar-Vahabi-Jafari-2011-PRE}, daily warehouse-out quantity records of steel products from three warehouses (YZ, JS and BY) from 1 January 2009 to 1 October 2013 \cite{Yao-Lin-Zheng-2017-PA}, and daily closing price data of four Asian stock markets' indices (SSEC, NIKKEI 225, KOSPI and Indian SENSEX) from 1 July 2003 to 1 March 2015 \cite{Wang-Zhu-Yang-Mul-2017-PA}.

%
%
%
%

\subsection{Multifractal cross-correlation analysis (MF-CCA)}

O{\'{s}}wi{\c{e}}cimka et al. argued that most existing methods for cross-correlation analysis often present the serious limitation of reporting spurious multifractal cross-correlations when there are none \cite{Oswiecimka-Drozdz-Forczek-Jadach-Kwapien-2014-PRE}. To remedy this drawback, they proposed the multifractal cross-correlation analysis (MF-CCA), which incorporates the sign of fluctuations $F_{v}(s)$ into the overall detrended fluctuation $F_{xy}(q,s)$:
\begin{equation}
 F_{xy}(q,s) = \left\{
 \begin{array}{lll}
  \displaystyle \left[\frac{1}{N_s}\sum_{v=1}^{N_s} {\mathrm{sign}}\left(F_{v}(s)\right)|F_{v}(s)|^{q}\right]^{1/q}, & q\neq0\\
  \displaystyle      \sum_{v=1}^{N_s} \frac{{\mathrm{sign}}\left(F_{v}(s)\right)\ln|F_{v}(s)|}{\sum_{v=1}^{N_s} {\mathrm{sign}}\left(F_{v}(s)\right)}, & q=0
 \end{array}
 \right.
\end{equation}
where $\sum_{v=1}^{N_s} {\mathrm{sign}}\left(F_{v}(s)\right)\neq{N_s}$ in many cases. When $F_{xy}(q,s)$ is negative for every $s$, one represents $-F_{xy}(q,s)$ with respect to $s$ in the log-log plot \cite{Podobnik-Stanley-2008-PRL,Oswiecimka-Drozdz-Forczek-Jadach-Kwapien-2014-PRE}. The MF-CCA method is a direct, natural multifractal generalization of the detrended cross-correlation analysis, which is recovered when $q=2$ \cite{Podobnik-Stanley-2008-PRL}. Numerical experiments on coupled ARFIMA processes and coupled Markov-switching multifractal (MSM) processes validate the suitability of MF-CCA in extracting joint monofractal or multifractal properties in two time series.

Combining the multifractal temporally weighted detrended fluctuation analysis (MF-TWDFA) \cite{Zhou-Leung-2010a-JSM} and the multifractal cross-correlation analysis (MF-CCA) \cite{Oswiecimka-Drozdz-Forczek-Jadach-Kwapien-2014-PRE}, Wei et al. designed the multifractal temporally weighted detrended cross-correlation analysis (MFTWXDFA) to quantify power-law cross-correlations in two time series and confirmed the validity of the MFTWXDFA method with numerical experiments on bivariate fractional Brownian motions, coupled ARFIMA processes, and multifractal binomial measures \cite{Wei-Yu-Zou-Anh-2017-Chaos}. The method is applied to confirm the presence of joint multifractality in three pairs of daily stock index returns (S\&P 500 and FTSE 100, S\&P 500 and NIKKEI 225, and S\&P 500 and HSI) from 4 January 2001 to 30 September 2016 \cite{Wei-Yu-Zou-Anh-2017-Chaos}.

The idea of using signed fluctuation applies to other methods, such as MF-X-DMA and MF-X-SF. Indeed, Wang et al. proposed a variant of the MF-X-SF$(q)$ method by taking into account of the information on the innovation signs in the definition of structure functions \cite{Wang-Yang-Wang-2016-PA}:
\begin{equation}
  K'_{xy}(q,s) = \left\langle \mathrm{sign}\left[\Delta{X(i,s)}\Delta{Y(i,s)}\right] \left|\Delta{X(i,s)}\Delta{Y(i,s)}\right|^{q} \right\rangle.
  \label{Eq:MF-X-SF:Kxy:q:sign}
\end{equation}
This definition should be treated with caution because $K'_{xy}(q,s)$ are not defined on all $q\in{\mathbb{R}}$ values and their values could be negative so that the scaling relation (\ref{Eq:MF-X-SF:Kxy:q:s}) does not hold. If this happens, one may argue that there is no cross-correlation scaling between the two time series \cite{Wang-Yang-Wang-2016-PA}.

\subsection{Multifractal detrended partial cross-correlation analysis (MF-DPXA)}

The observed long-range power-law cross-correlations between two time series may not be caused by their intrinsic relationship but by a common third driving force or by common external factors \cite{Kenett-Shapira-BenJacob-2009-JPS,Shapira-Kenett-BenJacob-2009-EPJB,Kenett-Tumminello-Madi-GurGershgoren-Mantegna-BenJacob-2010-PLoS1}. If the influence of the common external factors on the two time series are additive, we can use partial correlation to measure their intrinsic relationship \cite{Baba-Shibata-Sibuya-2004-ANZJS}. To extract the intrinsic long-range power-law cross-correlations between two time series affected by common driving forces, Liu developed the detrended partial cross-correlation analysis (DPXA), which combines the ideas of DCCA and partial correlation, and studied the DPXA exponents of variable cases \cite{Liu-2014}. The DPXA method has been proposed independently by Yuan et al. \cite{Yuan-Fu-Zhang-Piao-Xoplaki-Luterbacher-2015-SR}. Qian et al. generalized DPXA to analyze multifractal time series, resulting in the multifractal detrended partial cross-correlation analysis (MF-DPXA) \cite{Qian-Liu-Jiang-Podobnik-Zhou-Stanley-2015-PRE}.

%
%

Assume that two stationary time series $\{x(t):t=1,\cdots, N\}$ and $\{y(t):t=1, \cdots, N\}$, such as returns, depend on $n$ common external driving factors characterized by a sequence of time series $\{z_i(t):t=1, 2, \cdots, N\}$ with $i=1, \cdots, n$. Each time series is partitioned into $N_s=\mbox{int}[N/s]$ non-overlapping windows of size $s$. For each window, say the $v$th window $[l_v+1,l_v+s]$ where $l_v=(v-1)s$, we calibrate respectively two linear regression models for ${\mathbf{x}}_v$ and ${\mathbf{y}}_v$,
\begin{equation}
  \left\{
  \begin{array}{ccc}
     {\mathbf{x}}_v = {\mathbf{Z}}_v{\mathbf{\beta}_{x,v}} + {\mathbf{r}}_{x,v}\\
     {\mathbf{y}}_v = {\mathbf{Z}}_v{\mathbf{\beta}_{y,v}} + {\mathbf{r}}_{y,v}
  \end{array}
  \right.,
  \label{Eq:xy:z:rxy:betas}
\end{equation}
where ${\mathbf{x}}_v=[x_{l_v+1},\ldots,x_{l_v+s}]^{\mathrm{T}}$, ${\mathbf{y}}_v=[y_{l_v+1},\ldots,y_{l_v+s}]^{\mathrm{T}}$, ${\mathbf{r}}_{x,v}$ and ${\mathbf{r}}_{y,v}$ are the vectors of the residuals, and
\begin{equation}
  {\mathbf{Z}}_v
  =\left(
  \begin{array}{ccc}
     {\mathbf{z}}_{v,1}^{\rm{T}}\\
     \vdots\\
     {\mathbf{z}}_{v,p}^{\rm{T}}\\
  \end{array}
  \right)
  =\left(
  \begin{array}{ccc}
     {\mathbf{z}}_{1}(l_v+1) & \cdots & {\mathbf{z}}_{n}(l_v+1) \\
          \vdots         & \ddots &       \cdots       \\
     {\mathbf{z}}_{1}(l_v+s) & \cdots & {\mathbf{z}}_{n}(l_v+s) \\
  \end{array}
  \right)
\end{equation}
is the matrix of the $n$ external forces in the $v$th window, where $\mathbf{x}^{\rm{T}}$ is the transform of $\mathbf{x}$. We obtain the estimates $\mathbf{\hat{\beta}}_{x,v}$ and $\mathbf{\hat{\beta}}_{y,v}$ of the $n$-dimensional parameter vectors $\mathbf{\beta}_{x,v}$ and $\mathbf{\beta}_{y,v}$ and the sequence of residuals
\begin{equation}
  \left\{
  \begin{array}{ccc}
     {\mathbf{r}}_{x,v} = {\mathbf{x}}_v-{\mathbf{Z}}_v{\mathbf{\hat{\beta}}_{x,v}}\\
     {\mathbf{r}}_{y,v} = {\mathbf{y}}_v-{\mathbf{Z}}_v{\mathbf{\hat{\beta}}_{y,v}}
  \end{array}
  \right..
\end{equation}
We further obtain the residual profiles, that is,
\begin{equation}
  \left\{
  \begin{array}{ccc}
     R_{x,v}(k) = \sum_{j=1}^{k} r_x(l_v+j)\\
     R_{y,v}(k) = \sum_{j=1}^{k} r_y(l_v+j)
  \end{array}
  \right.,
\end{equation}
where $k=1,\cdots,s$.

Assuming that $\widetilde{R}_{x,v}$ and $\widetilde{R}_{y,v}$ are respectively the local trend functions of $R_{x,v}$ and $R_{x,v}$, the detrended partial cross-correlation in each window can be computed as
\begin{equation}
 F_{v}^2(s) = \frac{1}{s}\sum_{k=1}^s
 \left[R_{x,v}(k)-\widetilde{R}_{x,v}(k)
   \right]\left[R_{y,v}(k)-\widetilde{R}_{y,v}(k)\right],
\end{equation}
and the $q$th order detrended partial cross-correlation is
\begin{equation}
 F_{xy:{\mathbf{z}}}(q,s) = \left[\frac{1}{N_s-1}\sum_{v=1}^{N_s} |F_{v}^2(s)|^{q/2}\right]^{1/q}
\end{equation}
when $q\neq0$ and
\begin{equation}
 F_{xy:{\mathbf{z}}}(0,s) = \exp\left[\frac{1}{N_s}\sum_{v=1}^{N_s} \ln |F_{v}(s)|\right]
\end{equation}
when $q=0$. We then expect the scaling relation
\begin{equation}
 F_{xy:{\mathbf{z}}}(q,s) \sim s^{H_{xy:{\mathbf{z}}}(q)}.
 \label{Eq:Fxy:q:s}
\end{equation}

According to the standard multifractal formalism, the multifractal mass exponents $\tau(q)$ can be used to characterize the multifractal properties as
\begin{equation}
\tau_{xy:{\mathbf{z}}}(q)=qH_{xy:{\mathbf{z}}}(q)-D_f,
\label{Eq:MFDCCA:tau}
\end{equation}
where $D_f=1$ is the fractal dimension of the geometric support of the time series \cite{Kantelhardt-Zschiegner-KoscielnyBunde-Havlin-Bunde-Stanley-2002-PA}. If $\tau(q)$ is a nonlinear function of $q$, the partial cross-correlations of the time series are multifractal. According to the Legendre transform, we obtain the singularity strength function $\alpha(q)$ and the multifractal spectrum $f(\alpha)$ as \cite{Halsey-Jensen-Kadanoff-Procaccia-Shraiman-1986-PRA}
\begin{equation}
    \left\{
    \begin{array}{ll}
        \alpha_{xy:{\mathbf{z}}}(q)={\rm{d}}\tau_{xy:{\mathbf{z}}}(q)/{\rm{d}}q\\
        f_{xy:{\mathbf{z}}}(q)=q{\alpha_{xy:{\mathbf{z}}}}-{\tau_{xy:{\mathbf{z}}}}(q)
    \end{array}
    \right..
\label{Eq:MFDCCA:f:alpha}
\end{equation}

The MF-DPXA method is an extension to MF-DCCA methods. Different choices of the local trends result in different variants. When MF-DPXA is implemented with DFA or DMA,
we obtain MF-PX-DFA or MF-PX-DMA. Using two binomial measures contaminated by Gaussian noise with very low signal-to-noise ratio, numerical experiments show that MF-DPXA is significantly advantageous over MF-DCCA \cite{Qian-Liu-Jiang-Podobnik-Zhou-Stanley-2015-PRE}. It is also easy to extend the DPXA idea to other methods of joint multifractal analysis. In this vein, the performance of different MF-DPXA methods should be investigated and compared.



\section{Multifractal models}
\label{S1:Models}

\subsection{Multiplicative cascade models}
\label{S2:Models:Multiplicative}

\subsubsection{Deterministic multinomial models}

Multiplicative cascade models generate multinomial measures that are constructed in a recursive manner \cite{Mandelbrot-1989-PAG}. Multiplicative cascades exist in financial markets since there is evidence of a causal information cascade from large scales to fine scales \cite{Arneodo-Muzy-Sornette-1998-EPJB,Muzy-Sornette-Delour-Arneodo-2001-QF}. The multiplicative cascade process starts with an interval $[0,1]$ on with the measure is uniformly distributed. Without loss of generality, assume that the total measure on the interval is 1. At the first iteration, the interval is divided into $b$ segments whose lengths are $\{s_i:i=1,\cdots,b\}$ and the measure is redistributed on each segment such that the measure on the $i$th segment is $m_i$. At the second iteration, each segment is divided into $b$ smaller segments and the measure on it is redistributed in the same way. This procedure goes to infinity, resulting in a multiscale multinomial measure.
The mass exponent function can be solved from the following equation \cite{Halsey-Jensen-Kadanoff-Procaccia-Shraiman-1986-PRA}:
\begin{equation}
   \sum_{i=1}^b \frac{m_i^q}{s_i^{\tau(q)}} = 1.
   \label{Eq:MF-PF:Gamma}
\end{equation}
In this iterative process, the daughter segments are not necessary to cover the mother segment. Therefore, we have
\begin{equation}
 \sum_{i=1}^b s_i \leq 1 ~~~~~~\mbox{and}~~~~~~~ \sum_{i=1}^b m_i = 1.
\end{equation}

It can be proven that
\begin{equation}
  \left\{
  \begin{aligned}
    & {\mbox{d}}\tau(q)/{\mbox{d}}q>0 \\
    & {\mbox{d}}^2\tau(q)/{\mbox{d}}q^2\leqslant0
  \end{aligned}
  \right.,
\end{equation}
where the equality ``='' holds if and only if all $\ln{m_i}/\ln{s_i}$ terms are identical. In this situation, the measure is uniformly distributed on the support and is monofractal such that $\tau(q)$ is a linear function of $q$. According to Eq.~(\ref{Eq:MF-PF:tau:Dq}), we have $D_q \equiv D_0 = D_f$, indicating that $D_q$ is independent of $q$. Since $\alpha(q)=\tau'(q)$, we have $\alpha'(q)= \tau''(q)= 0$, suggesting that the singularity width $\alpha_{\max}-\alpha_{\min}=0$.

It is easy to verify that, when $q=0$ and $q=1$,
\begin{equation}
 \tau(0) = -1 ~~{\mathrm{and}}~~ \tau(1)=0
 \label{Eq:MF-PF:tau:0}
\end{equation}
are respectively solutions to Eq.~(\ref{Eq:MF-PF:Gamma}). Because $\tau(q)$ is a monotonically increasing function of $q$ for multifractals, these solutions are the unique solutions. For the $D_q$ function, we have
\begin{equation}
    {\mbox{d}}D_q/{\mbox{d}}q\leqslant0,
    \label{Eq:MF-PF:dDq:dq}
\end{equation}
where the equality holds if and only if all $\ln{m_i}/\ln{s_i}$ terms are identical. In addition, the limits $D_{+\infty}$ and $D_{\infty}$ exist when $q\to\pm\infty$, which read
\begin{subequations}
  \begin{equation}
  D_{+\infty}\triangleq\lim_{q\to+\infty}D_q=\min_i\left\{\ln{m_i}/\ln{s_i}\right\}
  \label{Eq:MF-PF:Dq:infty:p}
  \end{equation}
and
  \begin{equation}
  D_{-\infty}\triangleq\lim_{q\to-\infty}D_q=\max_i\left\{\ln{m_i}/\ln{s_i}\right\}.
  \label{Eq:MF-PF:Dq:infty:n}
  \end{equation}
  \label{Eq:MF-PF:Dq:infty}
\end{subequations}
For the $\alpha(q)$ function, we have
\begin{equation}
    \mbox{d}\alpha(q)/\mbox{d}q\leqslant0,
    \label{Eq:MF-PF:dalpha:dq}
\end{equation}
where the equality holds if and only if all $\ln{m_i}/\ln{s_i}$ terms are identical. In addition, the limits $\alpha(+\infty)$ and $\alpha(\infty)$ exist when $q\to\pm\infty$, which read \cite{Mandelbrot-1989-PAG}
\begin{subequations}
  \begin{equation}
  \alpha_{\min}=\alpha_{+\infty}\triangleq\lim_{q\to+\infty}\alpha(q)=\min_i\left\{\ln{m_i}/\ln{s_i}\right\}
  \label{Eq:MF-PF:alpha:infty:p}
  \end{equation}
and
  \begin{equation}
  \alpha_{\max}=\alpha_{-\infty}\triangleq\lim_{q\to-\infty}\alpha(q)=\max_i\left\{\ln{m_i}/\ln{s_i}\right\}.
  \label{Eq:MF-PF:alpha:infty:n}
  \end{equation}
  \label{Eq:MF-PF:alpha:infty}
\end{subequations}
Note that the $\alpha(q)$ function has a very similar shape to the $D_q$ function. For the $f(\alpha(q))$ function, the relationship $\mbox{d}f(\alpha)/\mbox{d}\alpha=q$ still holds and $f_{\max} = f(q=0) =1$. Moreover, we have
\begin{equation}
  \left\{
  \begin{aligned}
    & f(\alpha(q))\geqslant0 \\
    & {\rm{d}}^2f(\alpha)/{\rm{d}}\alpha^2\leqslant0
  \end{aligned}
  \right.,
\end{equation}
where the equality in the second formula holds if and only if all $\ln{m_i}/\ln{s_i}$ terms are identical. The proofs of these properties can be found in Ref.~\cite{Zhou-2007} and references therein. The multifractal measures discussed here are conservative. Non-conservative multifractal measures exhibit different properties \cite{Falconer-ONeil-1996-PRSA}.

When $s_i=1/b$ for all $i$'s, we have
\begin{equation}
 \tau(q) = -\frac{\ln\sum_{i=1}^b m_i^q}{\ln{b}}.
 \label{Eq:MF:Model:Deterministic:Multinomial:tau}
\end{equation}
The generalized dimension function $D_q$ and the generalized Hurst exponent function $H(q)$ can be explicitly expressed due to Eq.~(\ref{Eq:MF:Intro:tau:Dq}) and  Eq.~(\ref{Eq:MF:Intro:tau:Hq}). According to Eq.~(\ref{Eq:MF:Intro:alpha:dtau:dq}), we have
\begin{equation}
 \alpha(q) \triangleq \frac{\mbox{d}\tau(q)}{\mbox{d}q}= -\frac{\sum_{i=1}^b m_i^q\ln{m_i}}{\sum_{i=1}^b m_i^q\ln{b}}.
 \label{Eq:MF:Model:Deterministic:Multinomial:alpha}
\end{equation}
The multifractal spectrum $f(\alpha)$ can also be explicitly expressed from the parameter $q$ using Eq.~(\ref{Eq:MF:Intro:f:alpha}).

\subsubsection{The \texorpdfstring{$p$}{}-model}

The most widely used multifractal model in econophysics is the $p$-model that generates multifractal binomial measures in a recursive way \cite{Meneveau-Sreenivasan-1987-PRL}, It is also frequently used for testing the performance of different multifractal methods. The $p$-model is a special case of the deterministic multiplicative cascade model with $b=2$ and $s_1=s_2=1/2$. To construct the binomial measure, one starts with the zeroth iteration $k = 0$, where the data set $z(i)$ consists of one value, $z^{(0)}(1)= 1 $. In the $k$th iteration, the data set $\{z^{(k)}(i): i = 1, 2, \cdots, 2^k\}$ is obtained from
\begin{equation}
  \begin{array}{rcl}
    z^{(k)}(2i-1)&=& p_z z^{(k-1)}(i)\\
    z^{(k)}(2i)  &=& (1-p_z)z^{(k-1)}(i)
\end{array}
  \label{Eq:pModel}
\end{equation}
for $i = 1, 2, \cdots, 2^{k-1}$. An early construction of binomial measures can be attributed to de Wijs in 1951 \cite{deWijs-1951-GM,Mandelbrot-1983,Agterberg-2015-JSAIMM}.

When $k\to\infty$, $z^{(k)}(i)$ approaches a binomial measure. According to Eq.~(\ref{Eq:MF-PF:Gamma}), we have
\begin{equation}
 \tau(q) = -\ln[m_1^q+(1-m_1)^q]/\ln{2}
 \label{Eq:pModel:tau}
\end{equation}
The analytical expressions of $D_q$ and $H(q)$ can be obtained using Eq.~(\ref{Eq:MF:Intro:tau:Dq}) and  Eq.~(\ref{Eq:MF:Intro:tau:Hq}) respectively.
According to the Legendre transform expressed in Eq.~(\ref{Eq:MF:Intro:alpha:dtau:dq}) and  Eq.~(\ref{Eq:MF:Intro:f:alpha}), we have
\begin{equation}
 \alpha(q) = -\frac{m_1^q\ln{m_1}+(1-m_1)^q\ln(1-m_1)}{[m_1^q+(1-m_1)^q]\ln{2}}.
 \label{Eq:pModel:alpha}
\end{equation}
and
\begin{eqnarray}
   f(\alpha) &=& -q\frac{m_1^q\ln{m_1}+(1-m_1)^q\ln(1-m_1)}{[m_1^q+(1-m_1)^q]\ln{2}} + \frac{\ln[m_1^q+(1-m_1)^q]}{\ln{2}}\\
  \label{Eq:pModel:f:alpha:1}
            &=& -\frac{\alpha_{\max}-\alpha}{\alpha_{\max}-\alpha_{\min}}\log_2\left(\frac{\alpha_{\max}-\alpha}{\alpha_{\max}-\alpha_{\min}}\right)
                 -\frac{\alpha-\alpha_{\min}}{\alpha_{\max}-\alpha_{\min}}\log_2\left(\frac{\alpha-\alpha_{\min}}{\alpha_{\max}-\alpha_{\min}}\right)~.
  \label{Eq:pModel:f:alpha:2}
\end{eqnarray}
Eq.~(\ref{Eq:pModel:f:alpha:2}) was derived by Calvet et al \cite{Calvet-Fisher-Mandelbrot-1997}.

\begin{figure}[htb]
  \centering
  \includegraphics[width=0.32\linewidth]{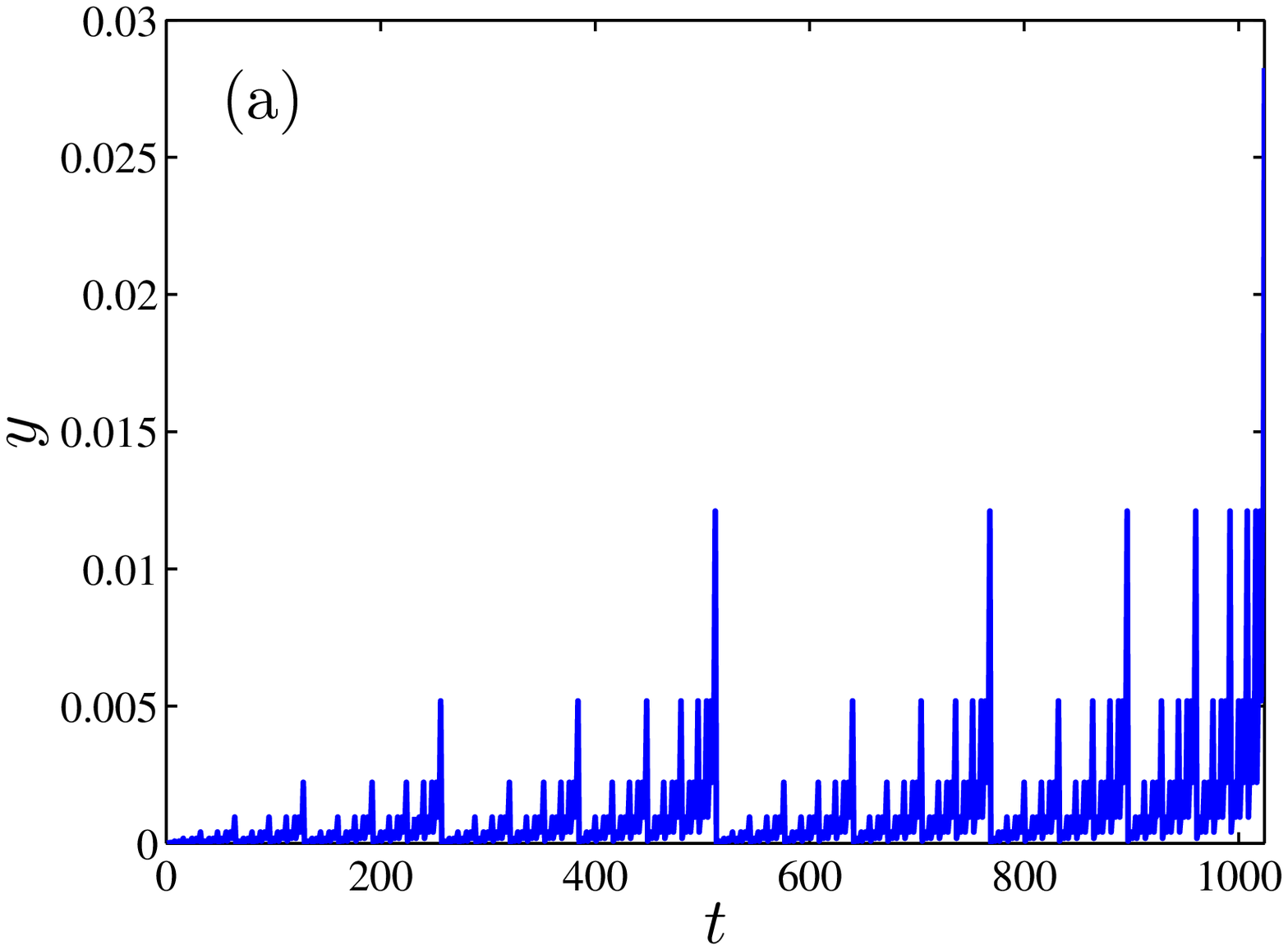}
  \includegraphics[width=0.32\linewidth]{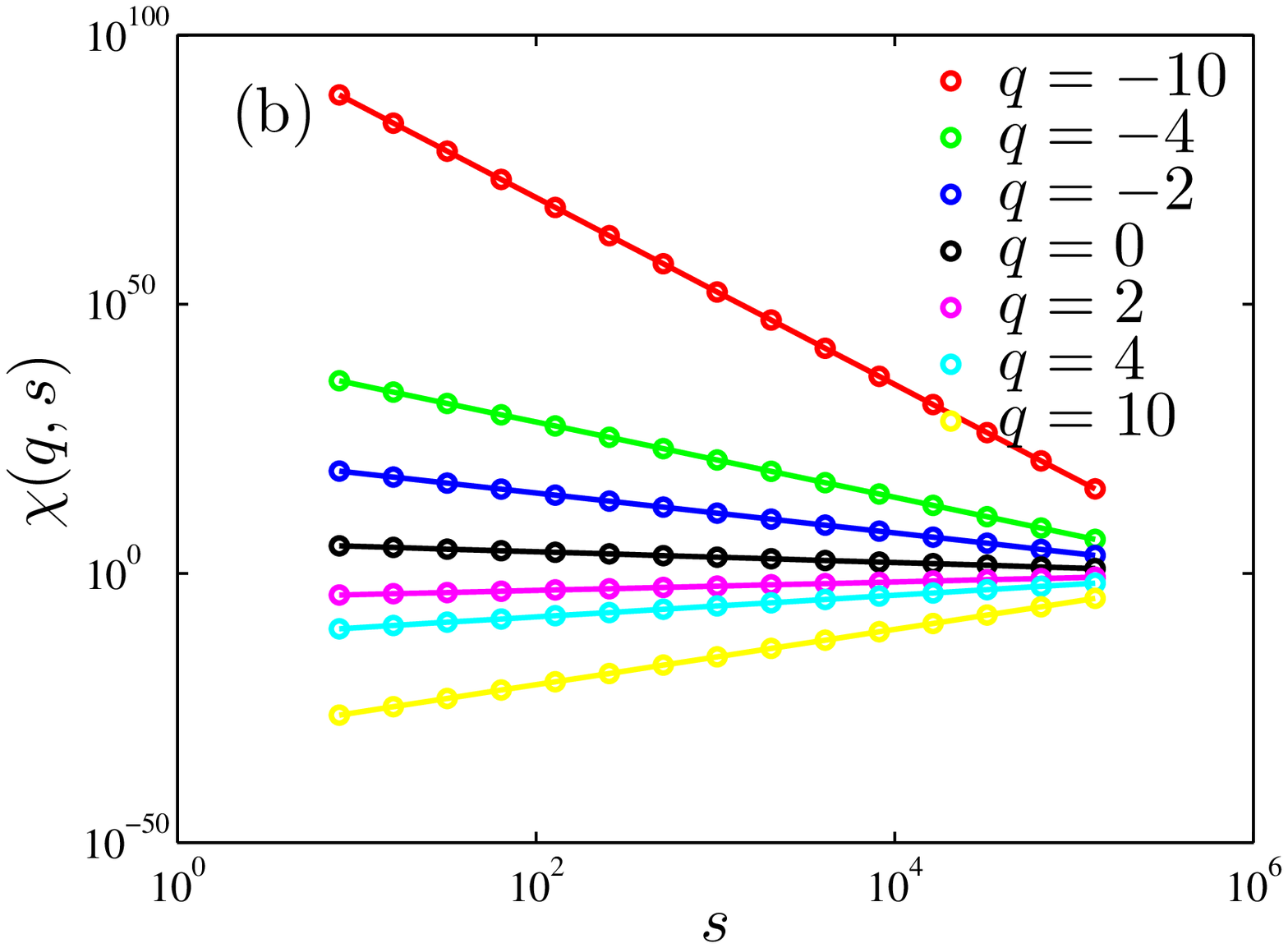}
  \includegraphics[width=0.32\linewidth]{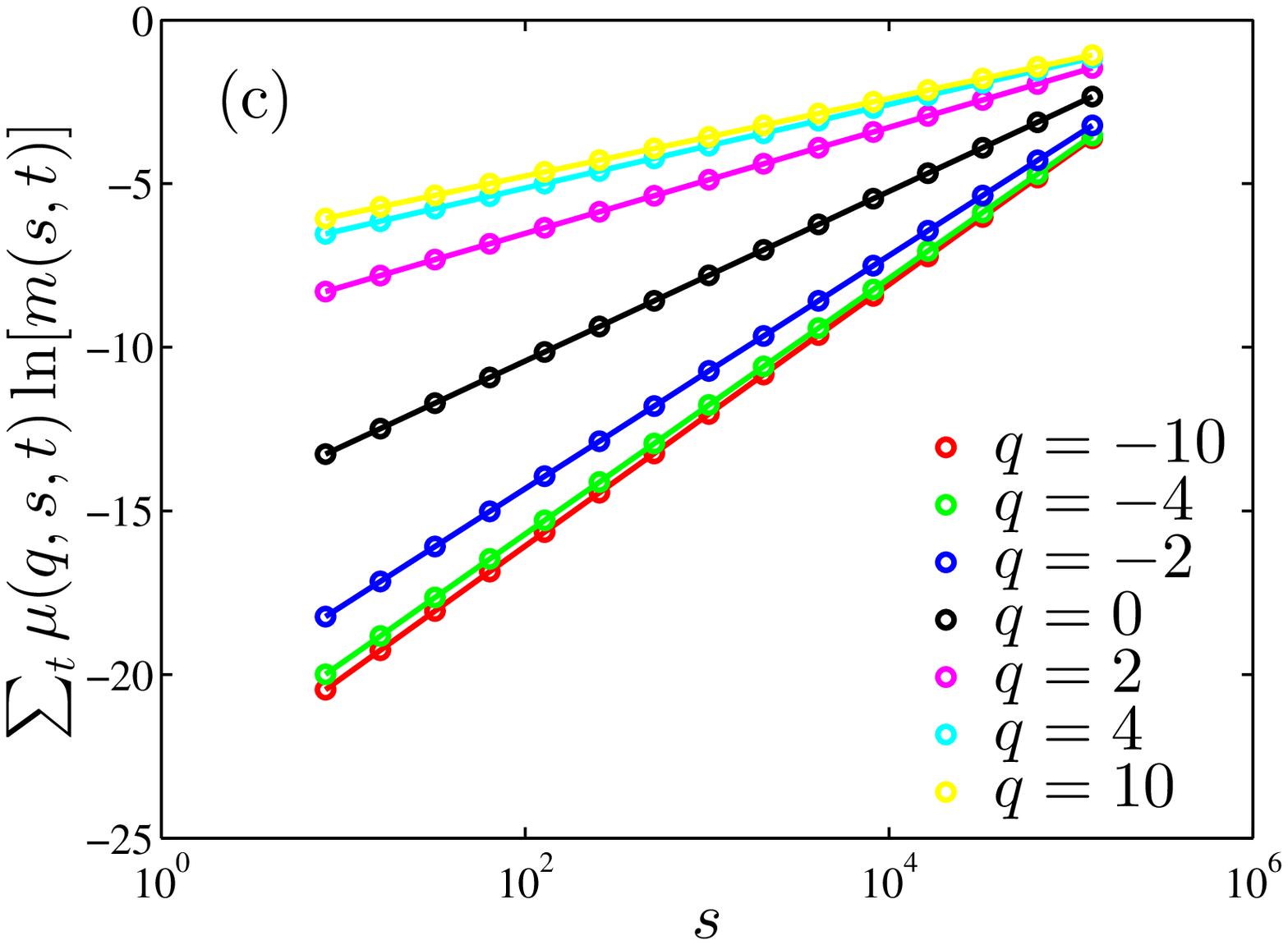}\\
  \includegraphics[width=0.32\linewidth]{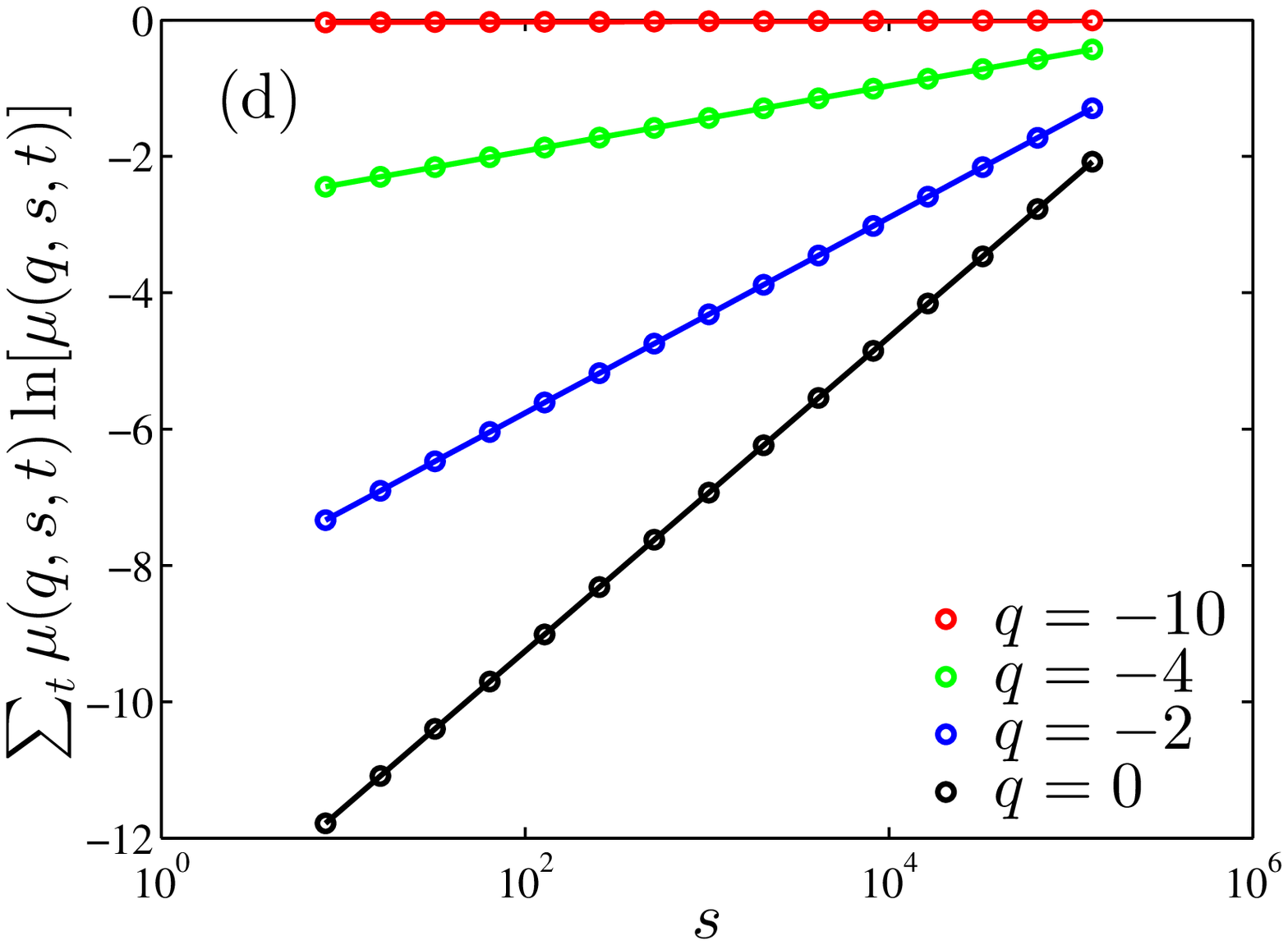}
  \includegraphics[width=0.32\linewidth]{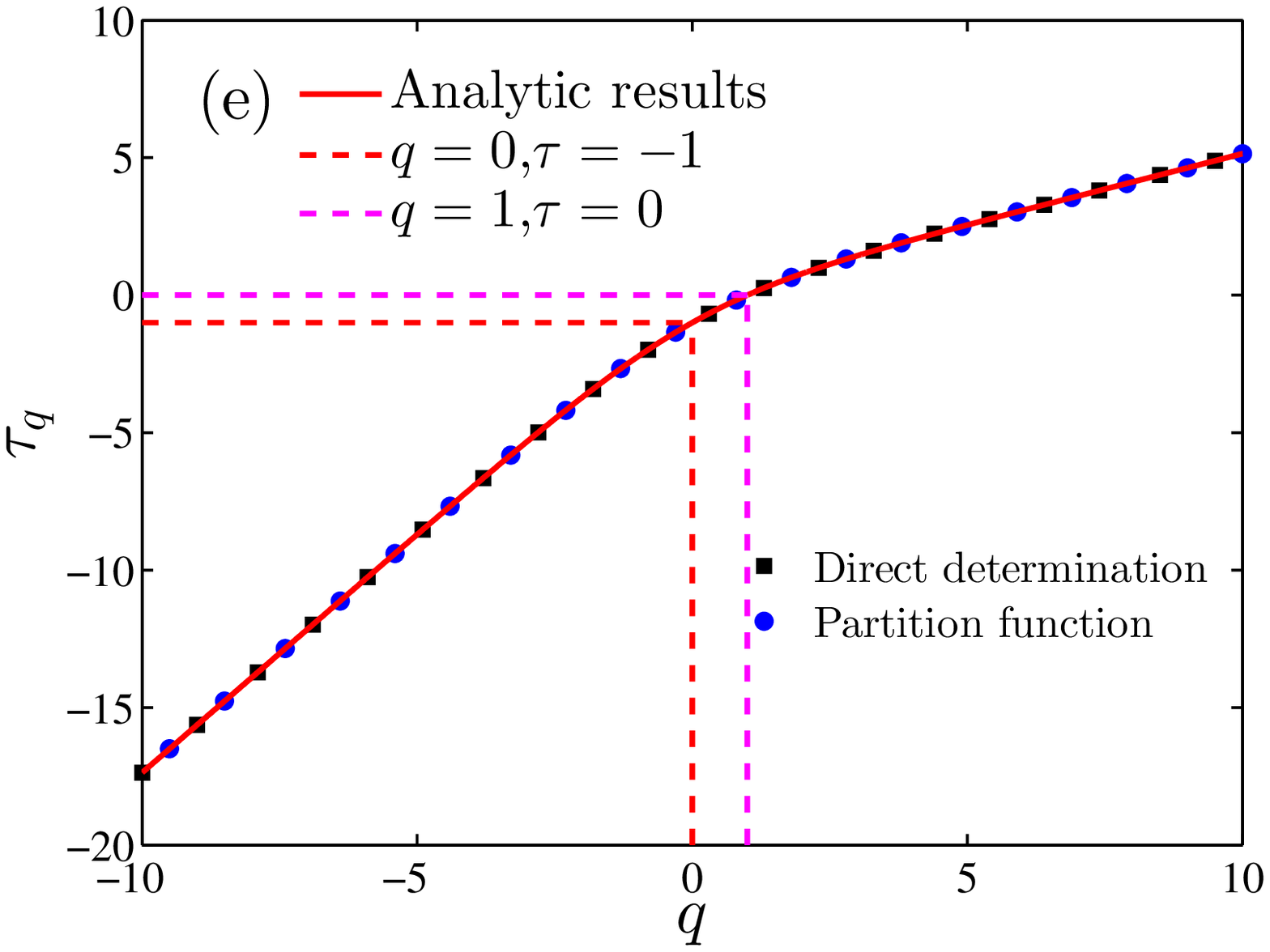}
  \includegraphics[width=0.32\linewidth]{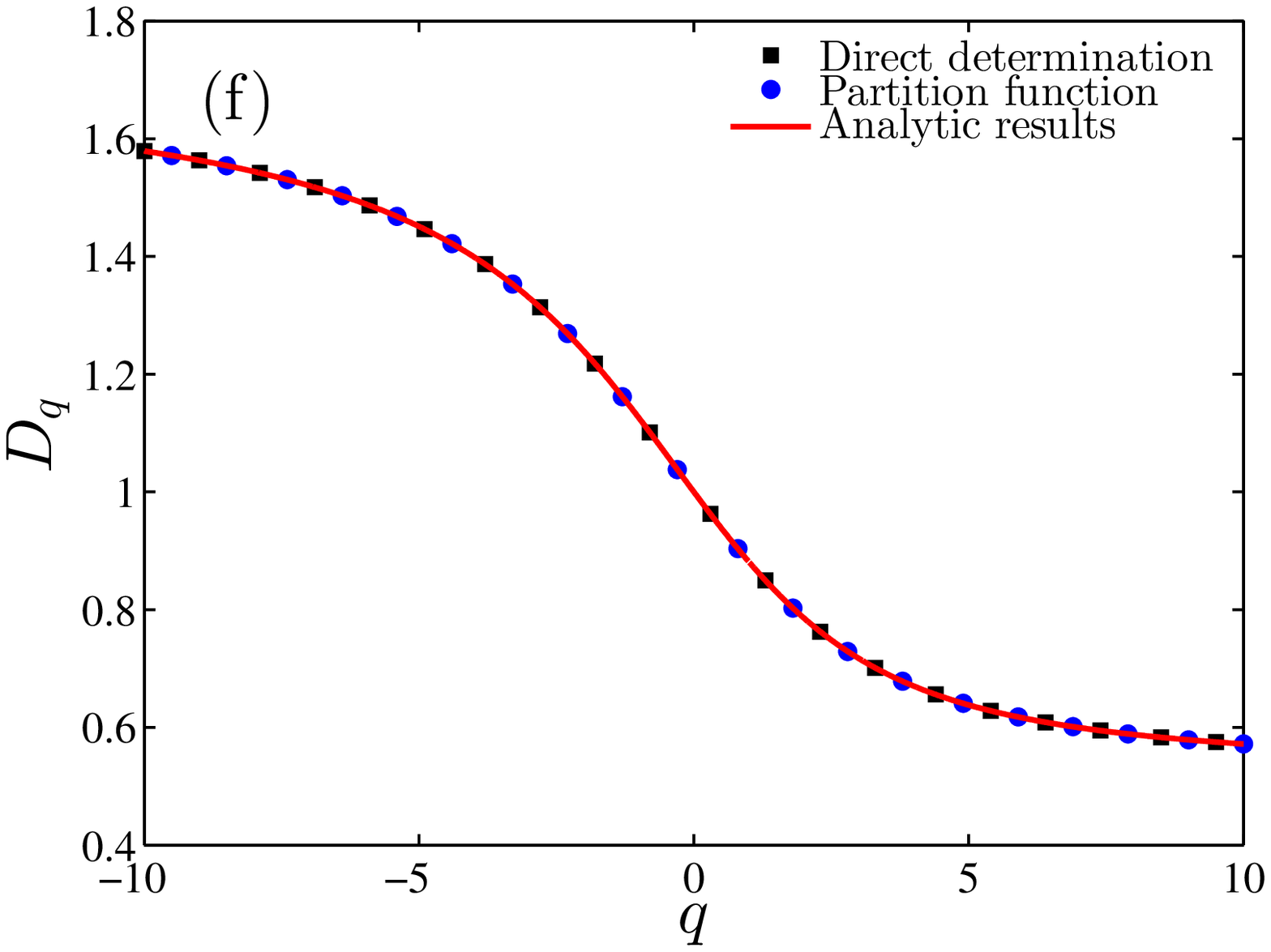}\\
  \includegraphics[width=0.32\linewidth]{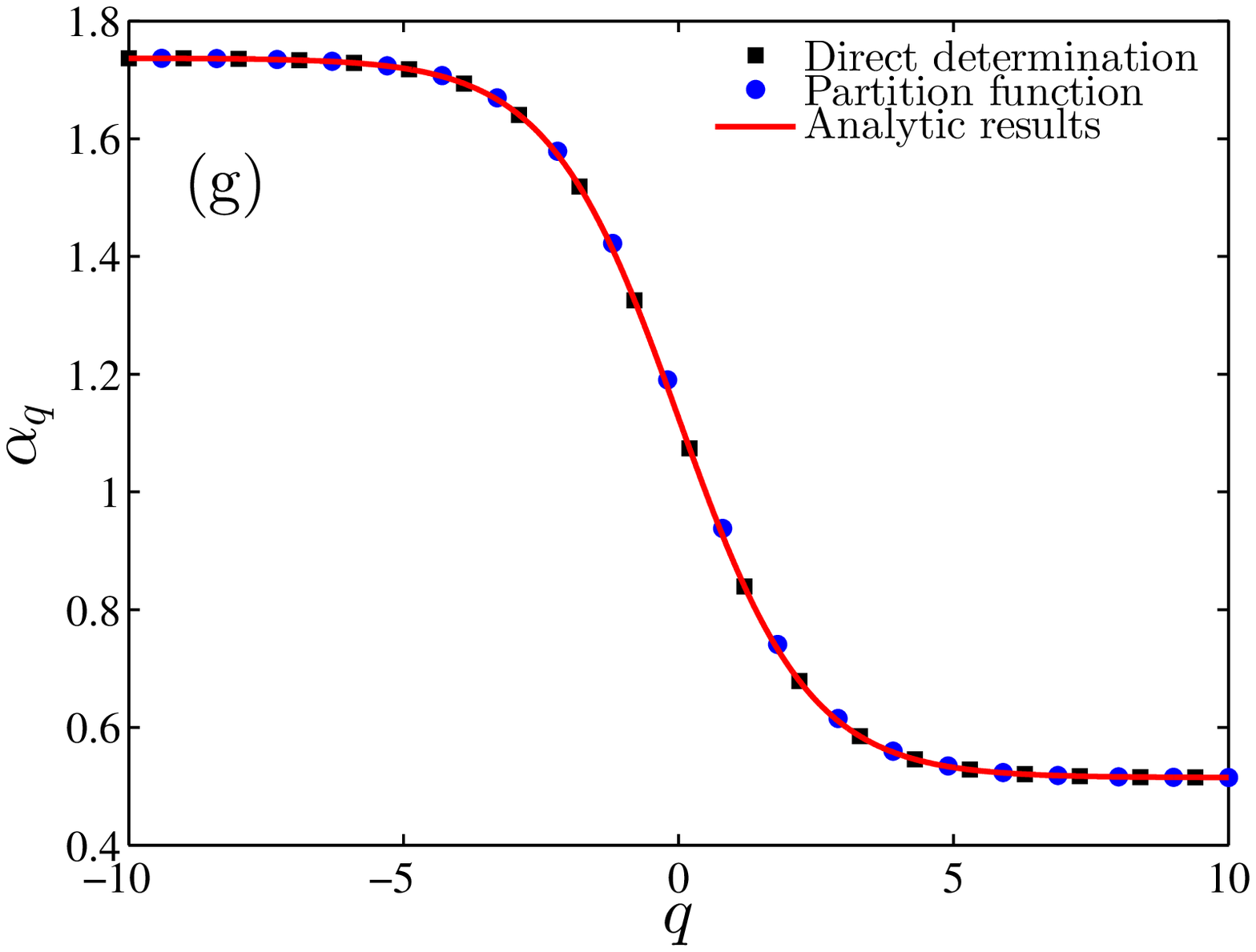}
  \includegraphics[width=0.32\linewidth]{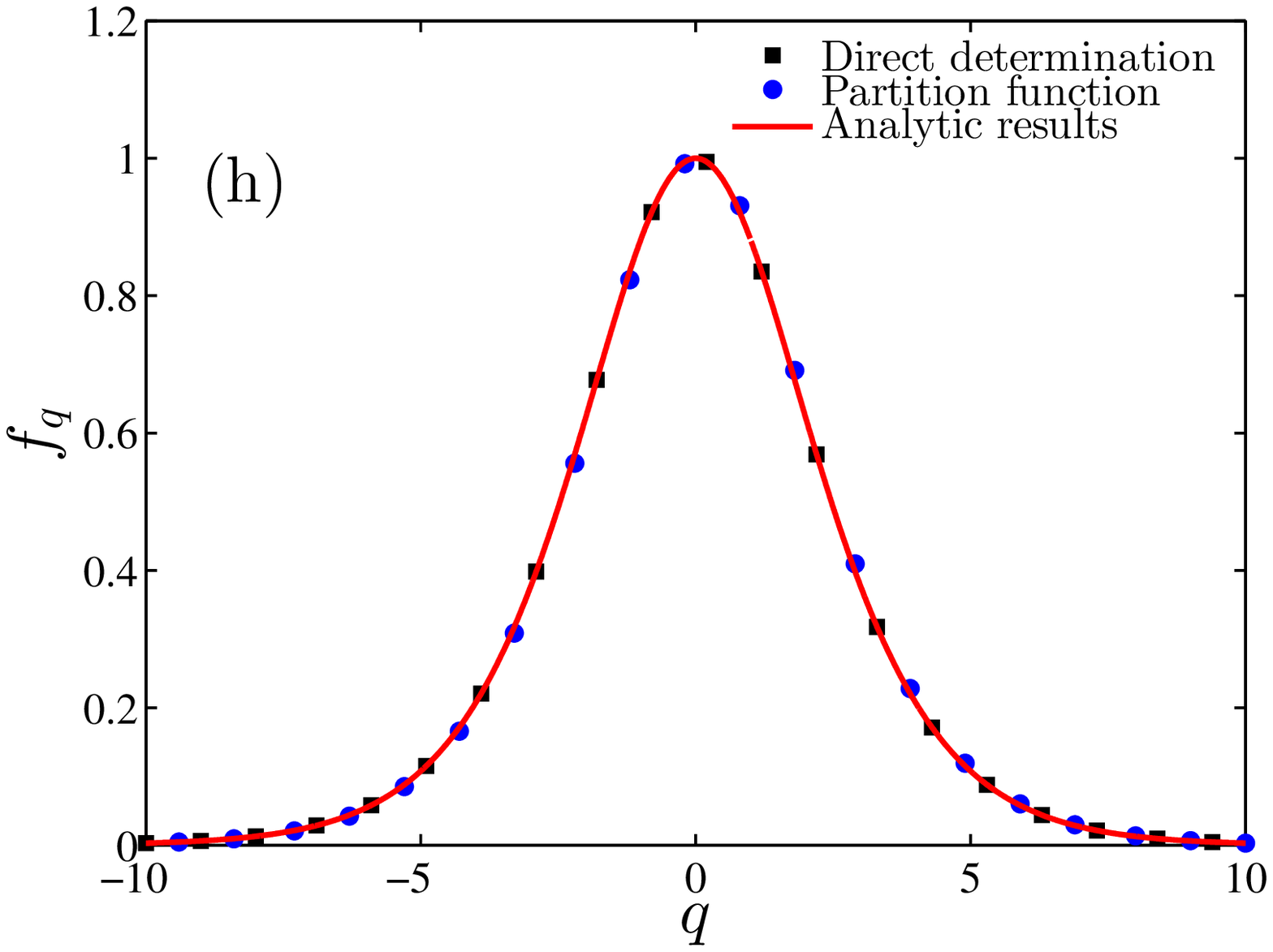}
  \includegraphics[width=0.32\linewidth]{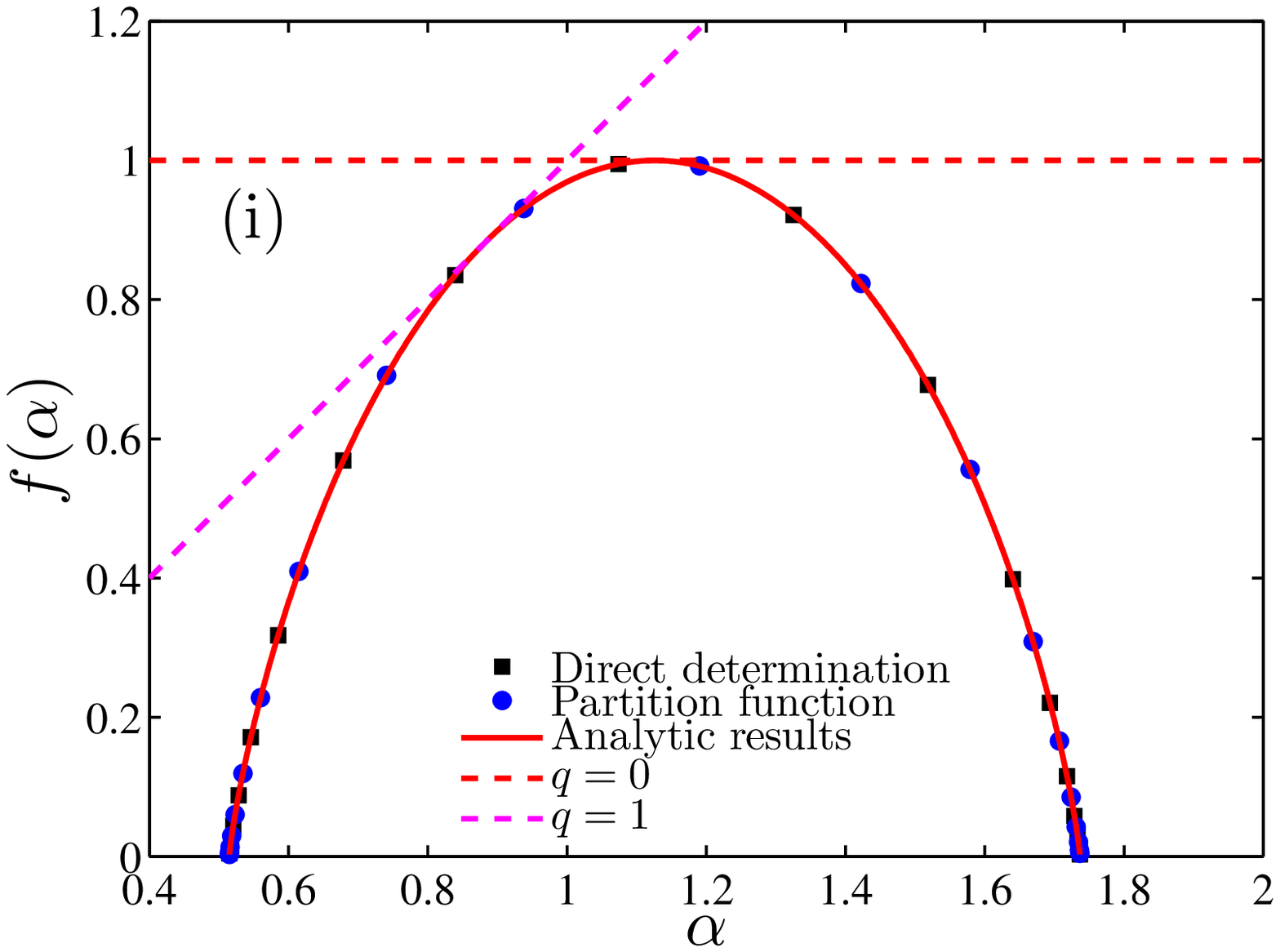}
  \caption{(color online) Multifractal analysis of a deterministic binomial measure with $m_1=0.35$. (a) First 1000 data points of the binomial measure. (b) Power-law dependence of $\chi(q,s)$ on box size $s$ for different $q$ values. (c) Linear dependence of $\sum_i \mu(q,s,i) \ln{[m(s,i)]}$ against $\ln{s}$. (d) Linear dependence of $\sum_i\mu(q,s,i)\ln[\mu(q,s,i)]$ against $\ln{s}$. (e) The mass exponent function $\tau(q)$. (f) The generalized dimensions $D_q$. (g) The singularity strength function $\alpha(q)$. (h) The $f(q)$ function. (i) The multifractal singularity spectrum $f(\alpha)$. In plots (e-i), the results presented with two types of markers are obtained from the classical partition function approach and the direct determination approach, in which we also show the corresponding analytic functions expressed in Eqs.~(\ref{Eq:pModel:tau}-\ref{Eq:pModel:f:alpha:2}).}
  \label{Fig:MF-PF:pModel}
\end{figure}

We now perform a multifractal analysis of a binomial measure with $m_1=0.3$ and length $N=2^{20}$. Figure~\ref{Fig:MF-PF:pModel}(a) presents the first 1000 data points of the measure. The partition function approach and the direct determination approach are utilized and compared. We first work on the partition function approach. Figure~\ref{Fig:MF-PF:pModel}(b) illustrates the power-law dependence of $\chi(q,s)$ on scale $s$ for different $q$ values from -10 to 10. All the curves for different $q$ values have nice power-law scaling in a broad scaling range. The slopes obtained by linear regressions of $\ln\chi(q,s)$ against $\ln{s}$ are estimates of $\tau(q)$, which are shown in Fig.~\ref{Fig:MF-PF:pModel}(e). We obtain $D_q$ using Eq.~(\ref{Eq:MF-PF:tau:Dq}), $\alpha(q)$, $f(q)$ and $f(\alpha)$ using the Legendre transform, which are presented in Fig.~\ref{Fig:MF-PF:pModel}(f-i). Based on the direct determination approach, Fig.~\ref{Fig:MF-PF:pModel}(c) plots the dependence of $\sum_i \mu(q,s,i) \ln{[m(s,i)]}$ against $s$ and Fig.~\ref{Fig:MF-PF:pModel}(d) the dependence of $\sum_i\mu(q,s,i)\ln[\mu(q,s,i)]$ against $s$ in linear-log coordinates, where $m(s,i)$ is the total measure in the $i$th box of size $s$ and $\mu(q,s,i)$ is the corresponding canonical measure. We also observe nice linearity. The slopes of the linear fits in Fig.~\ref{Fig:MF-PF:pModel}(c) and Fig.~\ref{Fig:MF-PF:pModel}(d) are estimates of $\alpha(q)$ and $f(q)$ respectively, which are also shown in Fig.~\ref{Fig:MF-PF:pModel}(g) and Fig.~\ref{Fig:MF-PF:pModel}(h). We obtain $\tau(q)$ by the Legendre transform and $D_q$ according to Eq.~(\ref{Eq:MF-PF:tau:Dq}), which are depicted in Fig.~\ref{Fig:MF-PF:pModel}(e) and Fig.~\ref{Fig:MF-PF:pModel}(f), respectively. In Fig.~\ref{Fig:MF-PF:pModel}(e), we also mark $\tau(0)=-1$ and $\tau(1)=0$, while in Fig.~\ref{Fig:MF-PF:pModel}(i), we show that $f'(\alpha)|_{q=0}=0$ and $f'(\alpha)|_{q=1}=1$. In Fig.~\ref{Fig:MF-PF:pModel}(e-i), we also show the corresponding analytical functions. It is evident that both the classical partition function approach and the direct determination approach unveil perfectly the multifractal nature of the binomial measure with very high accuracy. Note that the choice of $s$ values crucially impacts the results, because of the log-periodic oscillations in deterministic binomial measures \cite{Zhou-Sornette-2009b-PA}, which will be discussed in Section~\ref{S1:MF:Properties}.

\subsubsection{Stochastic multinomial models: Discrete multiplier distribution}

We can introduce randomness into multiplicative processes to construct stochastic multinomial models \cite{Mandelbrot-1972-LNP,Mandelbrot-1974-JFM,Mandelbrot-1989-PAG,Falconer-1994-JTP}. We focus on the uni-scale case that a segment is divided into $b$ sub-segments of identical length. At the first step, a unity segment of measure $\mu_0=1$ is divided into $b$ non-overlapping subsegments. With probability $p'_i$, measures $m'_{i,1},\cdots,m'_{i,b}$ are assigned to the subsegments. There are $k$ probabilities such that the construction matrix is expressed as
\begin{equation}
 \bf{CRM}=
    \left[
  \begin{array}{ccc|c}
   m'_{1,1} &\cdots & m'_{1,b} & p'_1 \\
   \hdots  &\hdots &\hdots  &\hdots \\
   m'_{k,1} &\cdots & m'_{k,b} & p'_k \\
  \end{array}
\right]
\label{Eq:SDMF:CRM1}
\end{equation}
At the second step, each of the $b$ sub-segments is divided into $b$ segments of length $1/b^2$. For the $i$th sub-segment, with probability $p_j'$, the measures are $m'_{i,1}m'_{j,1},\cdots,m'_{i,b}m'_{j,b}$. This process goes to infinity, resulting in stochastic multinomial measures.

Let $m_{(i-1)b+j}=m'_{ij}$, $p_{(i-1)b+j}=p'_i$ and $n=kb$, we rewrite Eq.~(\ref{Eq:SDMF:CRM1}) as
\begin{equation}\label{Eq:SDMF:CRM2}
 \bf{CRM}=
    \left(
  \begin{array}{ccccccc}
    m_1 & \cdots & m_{(i-1)b+j} & \cdots & m_n \\
    p_1 & \cdots & p_{(i-1)b+j} & \cdots & p_n \\
  \end{array}
\right)
\end{equation}
Note that $\sum_{i=1}^{n}p_i=b$.
The mass exponent function $\tau(q)$ is the solution of the following equation:
\begin{equation}\label{Eq:SDMF:Gamma}
    \Gamma(q,\tau)=\sum_{i=1}^n p_i {m_{ij}^q}{n^{\tau}} = 1~.
\end{equation}
It follows immediately that
\begin{equation}
 \tau(q) = -\frac{\ln\sum_{i=1}^n p_im_i^q}{\ln{n}}.
 \label{Eq:MF:Model:Stochastic:Multinomial:tau}
\end{equation}
The generalized dimension function $D_q$ and the generalized Hurst exponent function $H(q)$ can be explicitly expressed from Eq.~(\ref{Eq:MF:Intro:tau:Dq}) and  Eq.~(\ref{Eq:MF:Intro:tau:Hq}). According to Eq.~(\ref{Eq:MF:Intro:alpha:dtau:dq}), we have
\begin{equation}
 \alpha(q) \triangleq \frac{d\tau(q)}{dq}= -\frac{\sum_{i=1}^np_im_i^q\ln{m_i}}{\sum_{i=1}^n p_im_i^q\ln{n}}.
 \label{Eq:MF:Model:Stochastic:Multinomial:alpha}
\end{equation}
Hence, all the multifractal quantities have analytical expressions.

\begin{figure}[tb]
  \centering
  \includegraphics[width=0.32\linewidth,height=0.28\linewidth]{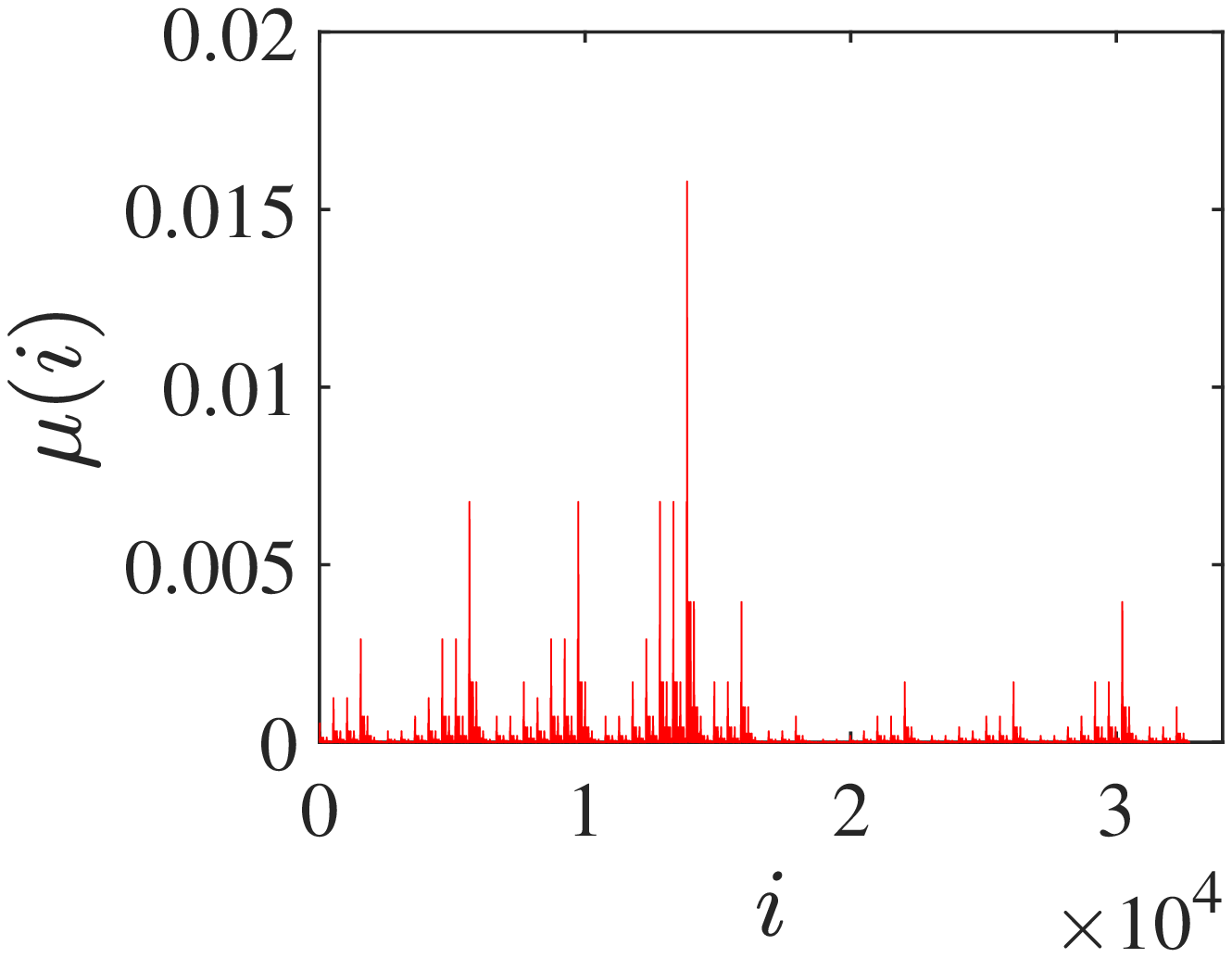}
  \includegraphics[width=0.32\linewidth,height=0.28\linewidth]{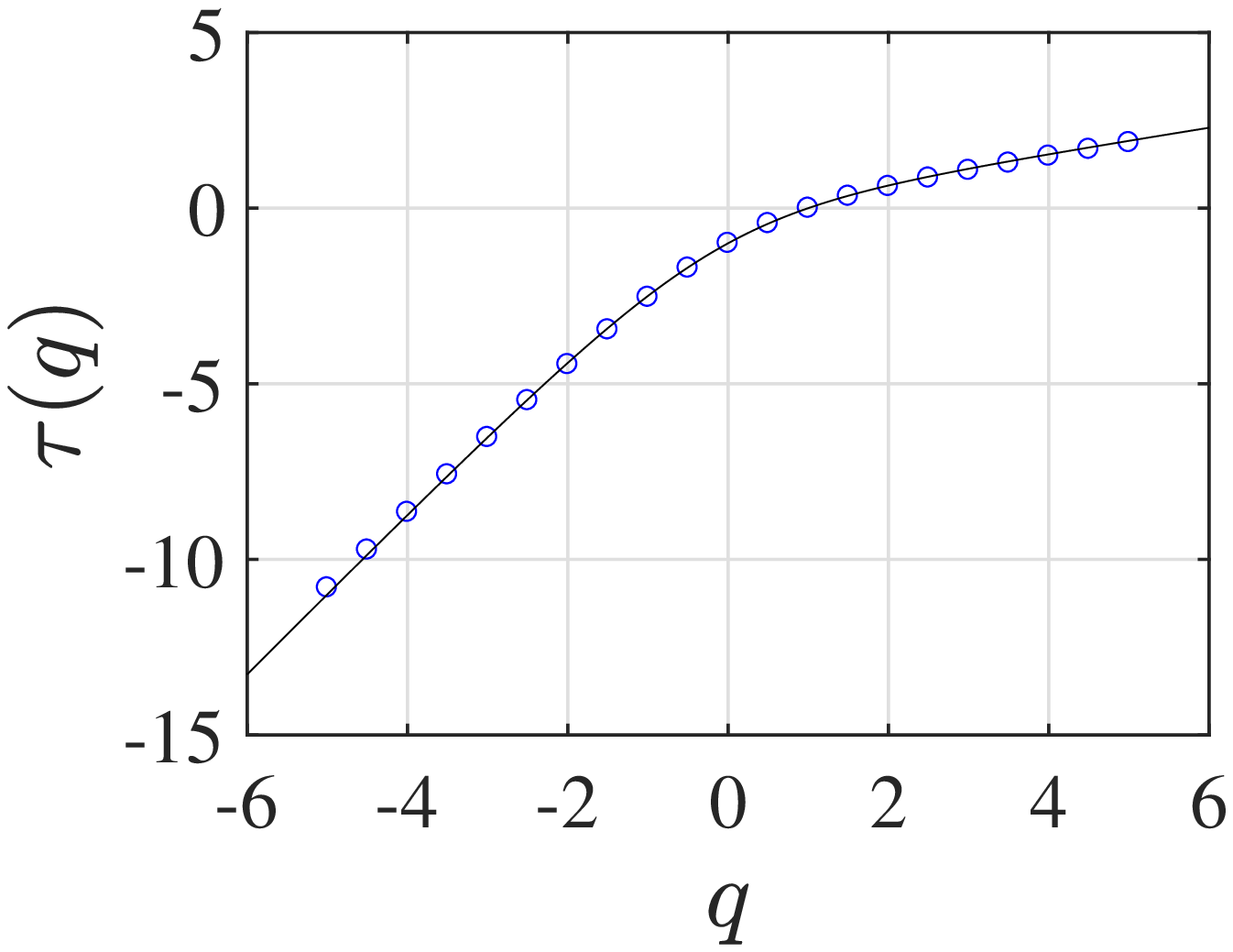}
  \includegraphics[width=0.32\linewidth,height=0.28\linewidth]{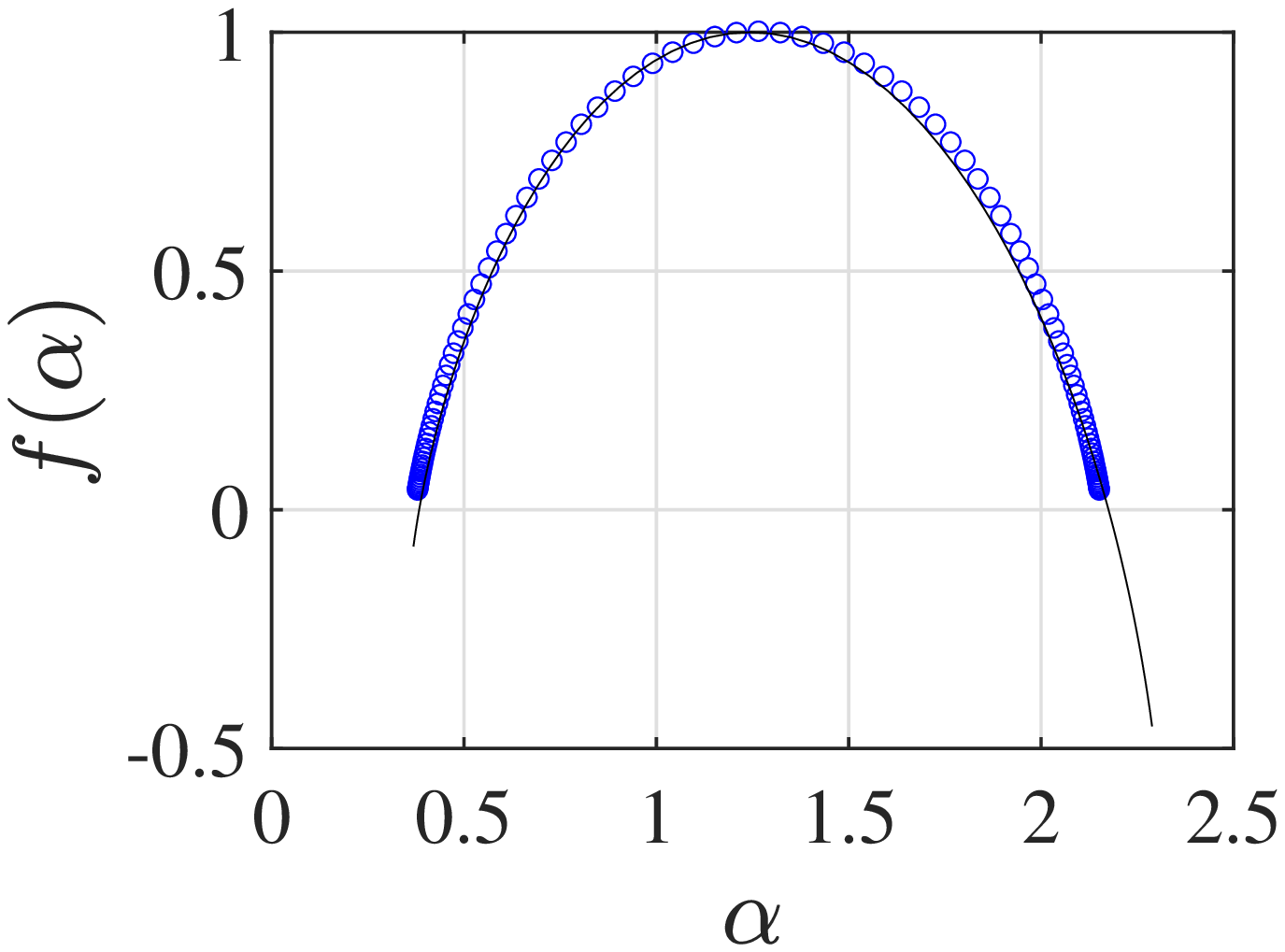}
  \caption{Multifractal properties of time series generated from a stochastic binomial model in which $m'_{1,1}=0.3$ with $p'_1=0.4$ and $m'_{2,1}=0.8$ with $p'_2=0.6$ in Eq.~(\ref{Eq:SDMF:CRM1}). (a) A realization of the stochastic binomial measure. (b) The mass exponent function $\tau(q)$. (c) The multifractal singularity spectrum $f(\alpha)$.}
  \label{Fig:MF:Model:MultiplicativeCascade}
\end{figure}

In Fig.~\ref{Fig:MF:Model:MultiplicativeCascade}, we illustrate an example of a stochastic binomial measure, together with its multifractal properties. An interesting feature is the presence of negative dimensions $f(\alpha)<0$. We will come back to discuss this issue later in Section \ref{S3:NegativeDimensions}.

\subsubsection{Stochastic multinomial models: Continuous multiplier distribution}


When the multipliers have continuous distributions, we obtain a large class of stochastic multiplicative cascade models \cite{Mandelbrot-1989-PAG}. The well-known lognormal model originated in fluid mechanics \cite{Kolmogorov-1962-JFM,Oboukhov-1962-JFM}.
For lognormal binomial measures \cite{Calvet-Fisher-Mandelbrot-1997,Calvet-Fisher-2002-RES,Muzy-Bacry-Kozhemyak-2006-PRE}, the multipliers follow the lognormal distribution in which the logarithmic multipliers have mean $\lambda$ and variance $\sigma^2$.
The mass exponent function is
\begin{equation}
  \tau(q) =-\frac{\sigma^2\ln2}{2}q^2+\lambda q -1
\end{equation}
and the singularity spectrum is
\begin{equation}
  f(\alpha) = 1-\frac{(\alpha-\lambda)^2}{2\ln2\sigma^2}.
\end{equation}
For log-Poisson binomial measures \cite{Calvet-Fisher-Mandelbrot-1997,Calvet-Fisher-2002-RES,Muzy-Bacry-Kozhemyak-2006-PRE}, the multipliers follows the Poisson distribution
\begin{equation}
  p(m) = {\rm{e}}^{-\gamma}\gamma^m/m!
\end{equation}
and the singularity spectrum is
\begin{equation}
  f(\alpha) = 1-\frac{\gamma}{\ln2}+\alpha\frac{\ln(\gamma{\rm{e}}/\alpha)}{\ln2}
\end{equation}
For log-gamma binomial measures \cite{Calvet-Fisher-Mandelbrot-1997,Calvet-Fisher-2002-RES,Muzy-Bacry-Kozhemyak-2006-PRE}, the multipliers follows the gamma distribution
\begin{equation}
  p(m) = \beta^{\gamma}m^{\gamma-1}{\rm{e}}^{-\beta{m}}/\Gamma(\gamma),
\end{equation}
where $\beta>0$ and $\gamma>0$,
and the singularity spectrum is
\begin{equation}
  f(\alpha) = 1+\gamma\frac{\ln(\alpha\beta/\gamma)}{\ln2}+\frac{\gamma-\alpha\beta}{\ln2}.
\end{equation}

Similar random cascading processes such as the $\mathcal{W}$-cascades on the dyadic trees of their orthogonal wavelet coefficients can also generate stochastic multifractal time series with explicit expressions of $\tau(q)$ and $f(\alpha)$ \cite{Arneodo-Bracy-Muzy-1998-JMP}.
For lognormal $\mathcal{W}$-cascades \cite{Arneodo-Bracy-Muzy-1998-JMP}, we have
\begin{equation}
  \tau(q) = -\frac{\sigma^2}{2\ln2}q^2-\frac{\mu}{\ln2}q-1.
  \label{Eq:MF:MathModel:W:cascades:LogNormal:tau:q}
\end{equation}
and
\begin{equation}
  f(\alpha) = 1-\frac{(\alpha+\mu/\ln2)^2\ln2}{2\sigma^2},
  \label{Eq:MF:MathModel:W:cascades:LogNormal:f:alpha}
\end{equation}
where $\mu$ and $\sigma^2$ are respectively the mean and the variance of $\ln|W|$. The minimum and maximum singularities are obtained when $f(\alpha)=0$:
\begin{equation}
  \alpha_{\min,\max} = \pm \frac{\sqrt{2}\sigma}{\sqrt{\ln2}}-\frac{\mu}{\ln2}.
  \label{Eq:MF:MathModel:W:cascades:LogNormal:alpha:min:max}
\end{equation}
For log-Poisson $\mathcal{W}$-cascades \cite{Arneodo-Bracy-Muzy-1998-JMP}, we have
\begin{equation}
  \tau(q) = \frac{\lambda(1-\delta^q)-\gamma{q}}{\ln2}-1.
  \label{Eq:MF:MathModel:W:cascades:LogPoisson:tau:q}
\end{equation}
and
\begin{equation}
  f(\alpha) = \left(\frac{\alpha}{\ln\delta}+\frac{\gamma}{\ln2\ln\delta}\right)\left[\ln\left(\frac{\alpha\ln2+\gamma}{-\lambda\ln\delta}\right)-1\right]+1-\frac{\lambda}{\ln2},
  \label{Eq:MF:MathModel:W:cascades:LogPoisson:f:alpha}
\end{equation}
where the law of $\ln|W|$ is the same as $P\ln\delta+\gamma$, and $\lambda$ is the mean and the variance of the Poisson variable $P$.
The differences of the multifractal expressions between the $\mathcal{W}$-cascades and the conventional cascade models stem from the different expressions of the multiplier distributions.

The discussions about stochastic continuous multifractals with some other multiplier distributions can be found in Ref.~\cite{Zhou-Yu-2001-PA,Zhou-Liu-Yu-2001-Fractals,Zhou-2007}, in which anomalous multifractal behavior emerges.

\subsection{Multifractal model of asset returns (MMAR)}
\label{S2:Models:MMAR}

\subsubsection{Basic MMAR model}
\label{S3::Models:MMAR:Basic}

Mandelbrot et al. proposed a seminal multifractal model of asset returns (MMAR) \cite{Mandelbrot-Fisher-Calvet-1997}, which is also termed Brownian motions in multifractal time (BMMT) \cite{Mandelbrot-2001c-QF}. In the MMAR model, the logarithmic price is assumed to be a subordinate process of a Brownian motion and a multifractal process of trading time. The MMAR model contains three assumptions:

{\textbf{Assumption 1.}} $X (t)$ is a subordinate process
\begin{equation}
  X (t) \equiv B_H[\theta(t)],
  \label{Eq:MFModels:MMAR:Compound}
\end{equation}
where $t$ is the clock time, $B(t)$ is a fractional Brownian motion with self-affine index $H$ and $\theta(t)$ is a stochastic non-decreasing trading time.

{\textbf{Assumption 2.}} The trading time (or ``subordinator'') $\theta(t)$ is the cumulative distribution function of a multifractal measure $\mu$ defined on $[0, T]$ such that it is a multifractal process with continuous, non-decreasing paths, and stationary increments.

{\textbf{Assumption 3.}} The processes $B(t)$ and $\theta(t)$ are independent.

The trading time $\theta(t)$ plays a crucial role in the MMAR model, which presumably causes the price $X(t)$ to be multifractal. It is shown that, under these three assumptions, the process $X(t)$ is multifractal with stationary increments and scaling function \cite{Mandelbrot-Fisher-Calvet-1997}.
\begin{equation}
 \tau_X(q) = \tau_\theta(Hq)~.
 \label{Eq:MFModels:MMAR:tauX:tauTheta}
\end{equation}
When $\theta(t)$ is a canonical multifractal, the increments of $X(t)$ have fat tails \cite{Mandelbrot-1972,Mandelbrot-Fisher-Calvet-1997}. When $H=1/2$, the price process displays both uncorrelated increments and long memory in volatility \cite{Mandelbrot-Fisher-Calvet-1997}. Therefore, the MMAR model with $H=1/2$ is able to reproduce the main stylized facts of financial returns \cite{Cont-2001-QF}.
A natural choice of the canonical multifractal measure $\mu$ in Assumption 2 is from the random multiplicative cascade models \cite{Mandelbrot-1989-PAG,Mandelbrot-Fisher-Calvet-1997,Calvet-Fisher-Mandelbrot-1997,Calvet-Fisher-2002-RES,Muzy-Bacry-Kozhemyak-2006-PRE}.

Recent empirical results show that the intertrade durations do have multifractal nature \cite{Oswiecimka-Kwapien-Drozdz-2005-PA,Jiang-Chen-Zhou-2009-PA,Ruan-Zhou-2011-PA,Yim-Oh-Kim-2014-PA,Heyde-2009-MMOR}, which provides strong supports for Assumption 2. Calvet et al. developed in detail the theory of multiplicative multifractals that is closely relevant to the $\theta(t)$ process \cite{Calvet-Fisher-Mandelbrot-1997}. Fisher et al. unveiled the presence of multifractal scaling laws in the DEM/USD (bid/ask quotes from 31 October 1992 to 1 September 1993 and daily data from June 1973 to December 1996) and JPY/USD exchange rates data  \cite{Fisher-Calvet-Mandelbrot-1997-SSRN}. They found that the simulated multifractal data from the MMAR generating process is consistent with the DEM/USD data, but do not appear completely consistent with the JPY/USD data \cite{Fisher-Calvet-Mandelbrot-1997-SSRN}. Calvet and Fisher confirmed the presence of multifractality and the consistency with the MMAR model in the DEM/USD data, and in the daily data from 1962 to 1998 of the value-weighted NYSE-AMEX-NASDAQ index (``CRSP Index''), Archer Daniels Midland (ADM), General Motors (GM), Lockheed-Martin, Motorola, and United Airlines (UAL) as well \cite{Calvet-Fisher-2002-RES}.

G{\"u}nay investigated the comparative performance of the lognormal MMAR model with respect to four benchmark econometric models (GARCH, EGARCH, FIGARCH and MRS-GARCH) using daily stock index returns of four emerging markets (Croatia, Greece, Poland and Turkey) from 4 January 2000 to 3 July 2014 \cite{Gunay-2016-IJFS}. He estimated the parameters of the five models and simulated each model with 1000 runs. He found that the deviation of the average mass scaling exponents of the lognormal MMAR model from those of the empirical results is the smallest.

Using the daily returns of 12 foreign exchange rates against the USD from January 1993 to February 2012 as an example, Goddard and Onali proposed a joint test for long memory and multifractality in the framework of the lognormal MMAR model \cite{Goddard-Onali-2016-PA}.

Batten et al. modified the MMAR model by representing the trading time in Assumption 2 with the series of volume ticks \cite{Batten-Kinateder-Wagner-2014-PA}, that is, the trading time $\theta(t)$  is given by the normalized volume ticks $\overline{V}_t$:
\begin{equation}
 \overline{V}_t = V_t/\sum_{t=1}^TV_t,
 \label{Eq:MFModels:MMAR:Volume:as:Theta}
\end{equation}
where $V_t\geq 0$ denotes the number of volume ticks at time $t$. The rational of this model is based on the fact that trading volumes possess multifractal nature that is universal in different markets \cite{Moyano-deSouza-Queiros-2006-PA,Eisler-Kertesz-2007-EPL,Lee-Lee-2007-PA,Mu-Chen-Kertesz-Zhou-2010-PP,Lee-2011-JKPS,Queiros-2016-CSF}. They studied the 5-min EUR/USD spot quotes and trading ticks from 5 January 2006 to 31 December 2007. They found that the model can produce admissible VaR forecasts at the 12-h forecast horizon and outperforms the historical simulation approach and the benchmark GARCH(1,1) location-scale VaR model at the daily horizon.

A caveat about the construction of some versions (including the initial one \cite{Mandelbrot-Fisher-Calvet-1997})
of the MMAR model should be stressed. Assumption 2 for the subordinator $\theta(t)$ defines it in general without
paying attention to the condition of causality. In general, a multifractal measure is a geometrical construction that exists over all times,
without a specific definition or dependence on the arrow of time. In other words, the multifractal measure pre-exists time.
The fact that the subordinator $\theta(t)$ can mimic a time flow results from the construction consisting in taking the
cumulative distribution function of this multifractal measure. However, the measure at a given instant already ``knows'' its
future values in the constructions that are for instance purely geometric without attention to causality.
While these classes of MMAR models may appear good candidates to describe financial price time series,
they should be only considered as toys, with no serious relevance to real markets. Indeed, the arguably most
important of all properties that models of financial markets should obey is that of causality. Otherwise, spurious
conclusions will derive in portfolio allocation and optimisation, in derivative pricing, investment strategies and so on.

\subsubsection{Poisson MMAR model}
\label{S3:Models:PMMAR}

In Assumption 2, when the trading time $\theta(t)$ is the c.d.f. of a Poisson multifractal measure $\mu$ defined on $[0, T]$, we obtain the Poisson MMAR model or PMM for short \cite{Calvet-Fisher-2001-JEm}. The PMM improves the MMAR in the sense that the resulting measure is grid-free since its partitioned sub-intervals do not have identical length $b^{-k}$ and it can be viewed as a stochastic volatility model with a Markov latent vector \cite{Calvet-Fisher-2001-JEm}. The PPM leads to the Markov-switching multifractal models for volatility.

\subsection{Markov-switching multifractal (MSM) models}
\label{S2:Models:MSMM}

\subsubsection{The framework of MSM models}

The Markov-switching multifractal (MSM) model designed by Calvet and Fisher is a discrete-time Markov process with multi-frequency stochastic volatility \cite{Calvet-Fisher-2001-JEm,Calvet-Fisher-2004-JFEm}. The return process is specified as
\begin{equation}
  r_i = \sigma_i \epsilon_i,
  \label{Eq:MFModels:MSM:ri:sigmai:epsiloni}
\end{equation}
where the random variables $\epsilon_i$ are i.i.d. standard Gaussians ${\mathcal{N}}(0,1)$ and $\sigma_i$ is a stochastic volatility with $\bar{k}$ volatility components $M_{1,i}, M_{2,i}, \cdots, M_{\bar{k},i}$ decaying at heterogeneous frequencies $\gamma_1, \gamma_2, \cdots, \gamma_{\bar{k}}$ such that
\begin{equation}
  \sigma_i^2 = \sigma^2 M_{1,i} M_{2,i} \cdots M_{\bar{k},i}
             = \sigma^2 \prod_{k=1}^{\bar{k}}M_{k,i}.
  \label{Eq:MFModels:MSM:ri:sigmai2:Mki}
\end{equation}
where $\sigma$ is the unconditional standard deviation of the return $r_i$ under the assumption that the multipliers $M_{1,i}$, $M_{2,i}$, $\cdots$, $M_{\bar{k},i}$ are statistically independent. The random volatility components $M_{k,i}$ are random multipliers that are persistent, non-negative and canonical such that
\begin{equation}
  {\mathbb{E}}(M_{k,i})=1
  \label{Eq:MFModels:MSM:E:Mki}
\end{equation}
for all $i$'s. It is further assumed for simplicity that the multipliers $M_{1,i}, M_{2,i}, \cdots, M_{\bar{k},i}$ at a given time $i$ are statistically independent.

In the MSM model, one stacks the volatility components in period $i$ into a $1\times\bar{k}$ vector
\begin{equation}
  {\overrightarrow{M}}_i= [M_{1,i}, M_{2,i}, \cdots, M_{\bar{k},i}]
  \label{Eq:MFModels:MSM:Mi:vector}
\end{equation}
The vector ${\overrightarrow{M}}_i$ is assumed to be a first-order Markov process and thus defines the volatility states. When the state vector ${\overrightarrow{M}}_{i-1}$ is determined in period $i-1$ , the $i$ period multiplier $M_{k,i}$ for each $k\in\{1,2,\cdots,\bar{k}\}$ is either drawn from a fixes distribution $P(M)$ with probability $\gamma_k$ or remains unchanged at its current state:
\begin{equation}
 M_{k,i} =
    \left\{
  \begin{array}{lll}
   m_{k,i}   & {\mathrm{with~probability~}}   \gamma_k, \\
   m_{k,i-1} & {\mathrm{with~probability~}} 1-\gamma_k, \\
  \end{array}
\right.
\label{Eq:MFModels:MSM:Mki:Markov}
\end{equation}
where $m_{k,i}$ is drawn from $P(M)$. The MSM model allows for a wide variety of specifications for the distribution $P(M)$ and any discrete or continuous distribution with ${\mathbb{E}}(M)=1$ defined on $M\in(0,\infty)$ could be utilized.

Introducing two parameters $\gamma_{\bar{k}}\in(0,1)$ and $b\in(1,\infty)$, the transition probabilities are specified as
\begin{equation}
  \gamma_k=1-(1-\gamma_{\bar{k}})^{b^{k-\bar{k}}},
  \label{Eq:MFModels:MSM:gammak}
\end{equation}
or equivalently
\begin{equation}
  \gamma_k = 1-(1-\gamma_1)^{b^{k-1}}.
  \label{Eq:MFModels:MSM:gammak:bis}
\end{equation}
Obviously, $\gamma_k$ is a monotonically increasing function of $k$ so that $M_{k,i}$ with small $k$ values correspond to low-frequency components and $M_{k,i}$ with large $k$ values correspond to high-frequency components. For small values of $\gamma_1$ and $k$, a Taylor polynomial approximation of Eq.~(\ref{Eq:MFModels:MSM:gammak:bis}) gives
\begin{equation}
  \gamma_k \approx \gamma_1 b^{k-1}.
  \label{Eq:MFModels:MSM:gammak:TaylorApprox}
\end{equation}
Hence, the transition probabilities of low-frequency components grow approximately at geometric rate $b$ and the growth rate of the transition probabilities of high-frequency components slows down.


The MSM model can produce the main stylized facts of financial asset returns. It contains multifrequency volatility and the price process is a multifractal jump-diffusion process \cite{Calvet-Fisher-2008-JMathE}. The changes of low-frequency multipliers induce discontinuous volatility and long memory in volatility, high-frequency multipliers produce substantial outliers and hence fat tails in returns, and the hierarchical cascade of the multiplicative process ensures the emergence of multifractality in returns. The multifractal spectrum of the price process is fully determined by the probability distribution $P(M)$ of the multipliers \cite{Calvet-Fisher-2002-RES}. 

\subsubsection{Binomial MSM model}

Calvet and Fisher proposed a parsimonious setup in which the probability distribution $P(M)$ is a binomial distribution with the random variables taking values $m_0\in[1,2]$ or $2-m_0\in[0,1]$ with equal probability, which is a binomial MSM model \cite{Calvet-Fisher-2004-JFEm}. For the binomial MSM model and other MSM models with discrete multiplier distributions, the model parameters ($\sigma$, $b$, $\gamma_{\bar{k}}$ and parameters in $P(M)$) can be estimated via the maximum likelihood estimation (MLE) method, because the close-form likelihood can be expressed and the number of frequencies $\bar{k}$ can be determined using the Akaike information criterion (AIC) or the Bayesian information criterion (BIC) by comparing the likelihoods of different MSM($\bar{k}$) models that have different numbers of free parameters \cite{Calvet-Fisher-2004-JFEm}.

Using the daily exchange rates of the Deutsche Mark (DEM), Japanese Yen (JPY), British Pound (GBP), and Canadian dollar (CAD) against the U.S. Dollar for about 30 years, Calvet and Fisher estimated binomial MSM models with four parameters and more than a thousand states. They found that the MSM model outperforms benchmark econometric models (GARCH, Markov-Switching GARCH and FIGARCH) both in-sample and out-of-sample and considerable gains in forecasting accuracy are obtained at horizons of 10 to 50 days \cite{Calvet-Fisher-2004-JFEm}. Using daily data of two stock indices (the Dow Jones Composite 65 Average Index and NIKKEI 225 Average Index) from January 1969 to October 2004, two foreign exchange rates (GBP and AUD against USD) from March 1973 to February 2004, and two U.S. treasury bond rates (TB-1y and TB-2y) from June 1976 to October 2004, it is found that the estimated model captures the temporal dependence of the data, through estimating and comparing the generalized Hurst exponents $H(q=1)$ and $H(q=2)$ for both empirical data and simulated data of the binomial MSM model \cite{Liu-DiMatteo-Lux-2007-PA}.

\subsubsection{Lognormal MSM model}

Lux studied the lognormal MSM model \cite{Lux-2008-JBES}, in which multipliers are determined by random draws from a lognormal distribution with parameters $\lambda$ and $S$:
\begin{equation}
  P(M) = \frac {1}{MS {\sqrt {2\pi}}} \exp\left(-{\frac {\left(\ln M+\lambda \right)^{2}}{2S^2}}\right).
  \label{Eq:MFModels:MSM:LN:P(M)}
\end{equation}
According to the canonical condition (\ref{Eq:MFModels:MSM:E:Mki}), we have
\begin{equation}
  S^2 = 2\lambda.
  \label{Eq:MFModels:MSM:LN:CanonicalCond}
\end{equation}
Hence, the lognormal SMS model has four parameters. Lux proposed a generalized method of moments (GMM) estimator together with linear forecasts for model calibration, which can be applied to any continuous distribution with any number of volatility components. His numerical experiments show that the GMM estimator performs well for the binomial and lognormal models.

The mass exponent function $\tau(q)$ defined through structure functions of the lognormal MSM model can be explicitly expressed as  \cite{Eisler-Kertesz-2004-PA}
\begin{equation}
  \tau(q)=\frac{q}{2}-1-\left(\frac{q^2}{4}-\frac{q}{2}\right)(\lambda-1)=-\frac{\lambda-1}{4}q^2+\frac{\lambda}{2}q-1.
  \label{Eq:MFModels:MSM:LN:tau:q}
\end{equation}
The parameter $\lambda$ can be determined by fitting Eq.~(\ref{Eq:MFModels:MSM:LN:tau:q}) to the empirically obtained $\tau(q)$ curve. Based on the DAX data, Eisler and Kert{\'e}sz showed that the MF-DFA method provides a nice estimate of $\lambda$, which is only slightly worse than the GMM estimate \cite{Eisler-Kertesz-2004-PA}.
It follows from the Legendre transform that (cf. Ref.~\cite{Oswiecimka-Kwapien-Drozdz-Gorski-Rak-2006-APPB})
\begin{equation}
  \alpha(q) = -\frac{\lambda-1}{2}q+\frac{\lambda}{2}
  \label{Eq:MFModels:MSM:LN:alpha:q}
\end{equation}
and
\begin{equation}
  f(\alpha) = 1- \frac{(\lambda-2\alpha)^2}{4(\lambda-1)}.
  \label{Eq:MFModels:MSM:LN:f:alpha}
\end{equation}
We see that $\alpha(q)$ is a linear function of $q$ without limits. When $q\geq\lambda/(\lambda-1)$, we have $\alpha\leq 0$.

Empirical analysis of five exchange rates (DEM, GBP, CND, JPY and CHF against USD) from 1 January 1979 to 31 December 2004 shows that the lognormal MSM model performs comparably to the binomial MSM model \cite{Lux-2008-JBES}. Using 1-min high frequency data of the Polish WIG20 index sampled over the period 1999-2004, O\'{s}wi\c{e}cimka et al. found that the lognormal MSM model is able to reproduce some main stylized facts including the fat-tailed distribution, the absence of correlations and multifractality in the return time series \cite{Oswiecimka-Kwapien-Drozdz-Gorski-Rak-2006-APPB}. Based on the same data sets as in Ref.~\cite{Liu-DiMatteo-Lux-2007-PA}, it is found that the lognormal MSM model performs almost identically to the binomial MSM model in reproducing the generalized Hurst exponents $H(1)$ and $H(2)$ of logarithmic prices, indicating that the parsimonious discrete specification of the binomial MSM model is flexible enough for empirical analysis \cite{Liu-DiMatteo-Lux-2008-ACS}. Similar results are obtained for absolute returns and squared returns for small $q$ values, but not for large $q$ values \cite{Liu-DiMatteo-Lux-2008-PSPIE}. The determination of the generalized Hurst exponents $H(q)$ in these works is based on the structure function approach. The situation might be improved if one uses other multifractal analyses such as MF-DFA \cite{Eisler-Kertesz-2004-PA}.

\subsubsection{Other extensions}

The time series generated from the lognormal MSM model has a vanishing leverage autocorrelation \cite{Eisler-Kertesz-2004-PA}, which is inconsistent with the well-known leverage effect in financial time series \cite{Cont-2001-QF}. Using the idea in Ref.~\cite{Pochart-Bouchaud-2002-QF}, Eisler and Kert{\'e}sz extended the MSM model by introducing leverage autocorrelations \cite{Eisler-Kertesz-2004-PA}:
\begin{equation}
  r_i = \exp\left(-\sum_{i'<i}{\mathrm{sign}}(\epsilon_{i'})K(i-i')\right)\sigma_i \epsilon_i,
  \label{Eq:MFModels:MSM:ri:sigmai:epsiloni:Leverage}
\end{equation}
where $K(i)$ is a kernel function, which is qualitatively the leverage autocorrelation function in an appropriate limit \cite{Pochart-Bouchaud-2002-QF,Eisler-Kertesz-2004-PA}. Obviously, this extension can be applied to other stochastic volatility models. Alternatively, incorporating the MSM volatility into the mean-reversion term in the square-root volatility process of the Heston model \cite{Calvet-Fearnley-Fisher-Leippold-2015-JEm}, the leverage effect emerges naturally due to the correlation between the Gaussian innovations of the volatility and the Gaussian innovations of the stock price in the Heston model \cite{Heston-1993-RFS}.

The univariate MSM model can be extended to bivariate and multivariate MSM models to investigate volatility comovement between two or many financial assets \cite{Calvet-Fisher-Thompson-2006-JEm}. For multivariate MSM models with discrete multiplier distributions, it is possible to derive the closed-form likelihoods, and parameter estimation can be conducted by maximum likelihood for state spaces of moderate size and by simulated likelihood through a particle filter in high-dimensional cases \cite{Calvet-Fisher-Thompson-2006-JEm}. The GMM estimation approach proposed in Ref.~\cite{Lux-2008-JBES} can also be designed for multivariate MSM models with discrete or continuous multiplier distributions \cite{Liu-Lux-2017-EM}. This model is homogeneous, which can be improved by allowing correlations between volatility components to form a non-homogeneous bivariate MSM model to gain better risk prediction power \cite{Liu-Lux-2015-EJF}.

Other extensions include the Copula-MSM model for characterizing spot and weekly futures price dynamics and more generally other pairs of financial assets \cite{Malo-2009-PA}, the MSM-$t$ model in which the innovations $\epsilon_i$ follow the Student distribution \cite{Lux-MoralesArias-2010-CSDA}, the MSM-skewed $t$ model in which the innovations follow a skewed Student distribution \cite{Liu-Zhang-Fu-2016-PA}, the level-MSM model for interest rates which incorporates the well-known level effect observed
in interest rates \cite{Rypdal-Lovsletten-2011-XXX}, the modified Hull-White model of interest rates in which the volatility of the short term rate is driven by an MSM model \cite{Hainaut-2013-EM}, the cyclic MSM model for insurance claims arising from climatic events in which the term $\epsilon_i$ is set to the smoothed average of claims observed during the corresponding month of the year \cite{Hainaut-Boucher-2014-EModA}, the realized volatility lognormal MSM (RV-LMSM) model in which the non-RV $\sigma_i$ in Eq.~(\ref{Eq:MFModels:MSM:ri:sigmai2:Mki}) is replaced by the realized volatility \cite{Lux-MoralesArias-Sattarhoff-2014-JFc}, and the MSM duration (MSMD) model for the analysis of inter-trade durations in financial markets \cite{Chen-Diebold-Schorfheide-2013-JEm,Zikes-Barunik-Shenai-2017-EmR,Aldrich-Heckenbach-Laughlin-2016-JEF}.

Overall, different specifications of MSM-type models have comparative performance and on average outperform benchmark econometric models in capturing the multifractal behavior of financial variables, in forecasting financial volatility, and in measuring market risks.

\subsection{Multifractal random walk (MRW)}
\label{S1:Models:MRW}

\subsubsection{The original model}
\label{S2:Models:MRW:original}

Bacry et al. introduced the multifractal random walk (MRW) to model stochastic volatility in which the volatility
can be reduced as an exponential of a long memory process \cite{Bacry-Delour-Muzy-2001-PRE}. The MRW model was inspired by the seminal idea of causal cascades in financial markets, in which a multivariate multifractal description of a portfolio of assets was also introduced and studied \cite{Arneodo-Muzy-Sornette-1998-EPJB,Muzy-Sornette-Delour-Arneodo-2001-QF}. The MRW model possesses solvable multifractal properties \cite{Bacry-Delour-Muzy-2001-PRE,Muzy-Delour-Bacry-2000-EPJB,Bacry-Delour-Muzy-2001-PA}, which belongs to the class of lognormal continuous cascade models of asset returns \cite{Bacry-Muzy-2003-CMP,Bacry-Kozhemyak-Muzy-2008-JEDC,Bacry-Kozhemyak-Muzy-2013-QF}. The MRW can be extended to the multivariate MRW \cite{Muzy-Delour-Bacry-2000-EPJB}. Bacry and Muzy further proved the equivalence between the two definitions of MRW and MMAR \cite{Bacry-Muzy-2003-CMP}.

Consider a multifractal process $X(t)$, whose discrete process are sampled at discrete times $k\Delta{t}$ is $X_{\Delta{t}}(t)=X_{\Delta{t}}(i\Delta{t})$, where $i=1,\cdots,N$.
The increment $\delta_{\Delta{t}}X_{\Delta{t}}(i\Delta{t})$ of the MRW is constructed as
\begin{equation}\label{Eq:MF:MRW:dX}
    \delta_{\Delta{t}}X_{\Delta{t}}(i\Delta{t}) =
    \varepsilon_{\Delta{t}}(i){\rm{e}}^{\omega_{\Delta{t}}(i)}
\end{equation}
and the corresponding MRW is
\begin{equation}\label{Eq:MF:MRW:X}
    X_{\Delta{t}}(t)=
    \sum_{i=1}^N\delta_{\Delta{t}}X_{\Delta{t}}(i\Delta{t}) =
    \sum_{i=1}^N\varepsilon_{\Delta{t}}(i){\rm{e}}^{\omega_{\Delta{t}}(i)}
\end{equation}
where $X(0)=0$ and $N=T/\Delta{t}$. Here, $\varepsilon_{\Delta{t}}$ and $\omega_{\Delta{t}}$ are independent, $\varepsilon_{\Delta{t}}$ is a Gaussian white noise with variance $\sigma^2{\Delta{t}}$, while $\omega_{\Delta{t}}$ is Gaussian and its covariance is
\begin{equation}\label{Eq:MF:MRW:w:cov}
    C_{\omega}(\Delta{t},s)={\rm{cov}}({\omega_{\Delta{t}}(i+s)},{\omega_{\Delta{t}}(i)})=\lambda^2\ln\rho_{\Delta{t}}[s]
\end{equation}
where
\begin{equation}\label{Eq:MF:MRW:rho}
  \rho_{\Delta{t}}[s]=
  \left\{
  \begin{array}{ll}
    \frac{T}{(s+1)\Delta{t}}, & s+1\leqslant{T}/\Delta{t} \\
    1, & {\rm{otherwise}}
  \end{array}
\right.
\end{equation}
Thus the variance is ${\rm{Var}}({\omega_{\Delta{t}}(i)})=\lambda^2\ln(T/\Delta{t})$. To ensure that the variance of $X_{\Delta{t}}(t)$ converges when  $\Delta{t}\to0$, we have
\begin{equation}\label{Eq:MF:MRW:w:mean}
    {\mathbb{E}}({\omega_{\Delta{t}}(i)})=-{\rm{Var}}({\omega_{\Delta{t}}(i)})=-\lambda^2\ln(T/\Delta{t}).
\end{equation}
It follows that the variance of $X_{\Delta{t}}(t)$ is $\sigma^2t$, which is independent of $\Delta{t}$. Because $\omega_{\Delta{t}}$ is Gaussian, this model is the lognormal MRW model and can be extended to other distributions \cite{Muzy-Bacry-2002-PRE,Bacry-Muzy-2003-CMP}.

It has been proved that the scaling exponents of the structure functions of $X_{\Delta{t}}(t)$ is \cite{Bacry-Delour-Muzy-2001-PRE}
\begin{equation}\label{Eq:MF:MRW:zeta:q:1}
    \zeta(q)=-\frac{\lambda^2}{2}q^2+\left(\frac{1}{2}+\lambda^2\right)q  
\end{equation}
when the scale $s$ is less than $N$. When $s>N$, we have
\begin{equation}\label{Eq:MF:MRW:zeta:q:2}
    \zeta(q)=q/2.
\end{equation}
Speaking differently, in the latter case, $\omega_{\Delta{t}}$ is a Gaussian white noise so that $\delta_{\Delta{t}}X_{\Delta{t}}(i\Delta{t})$ is also a Gaussian white noise. Hence, $X_{\Delta{t}}(t)$ is a Brownian motion. Lude{\~n}a provides approximate confidence intervals of $\zeta(q)$ for empirical data \cite{Ludena-2009-SPL}.
One can estimate the MRW model using the generalized method of moments \cite{Bacry-Kozhemyak-Muzy-2008-JEDC,Bacry-Kozhemyak-Muzy-2013-QF} or an approximated maximum likelihood method \cite{Lovsletten-Rypdal-2012-PRE}.

Bacry et al. \cite{Bacry-Delour-Muzy-2001-PRE} showed that this construction above has a well-defined and unique
time-continuous limit, making the MRW the single quadratic multifractal generalisation of the Fractional Brownian process with
a bona-fide continuous limit (similar to the continuous limit of a discrete random walk that defines the Wiener process).
Sornette et al. \cite{Sornette-Malevergne-Muzy-2003-Risk} presented a convenient alternative representation of the MRW
in terms of an exponential of a long memory process with a specific asymptotic $1/\sqrt{t}$ decaying kernel, which has been
generalised to arbitrary power law kernel \cite{Saichev-Sornette-2006-PRE,Saichev-Filimonov-2008-JETPL}
and to endogenous self-excited processes \cite{Filimonov-Sornette-2011-EPL}.

The MRW model is able to capture the multifractal nature of financial returns, such as the daily DJIA returns from 1900 to 2003 \cite{Bouchaud-2005-Chaos} and intraday S\&P 500 future index over the period of 1988-1999 \cite{Bacry-Kozhemyak-Muzy-2008-JEDC}. It also predicts the power-law tailed distribution of financial returns \cite{Bacry-Kozhemyak-Muzy-2008-JEDC,Saakian-Martirosyan-Hu-Struzik-2011-EPL}.

\subsubsection{Fractional MRW}

If $\varepsilon_{\Delta{t}}$ is a fractional Gaussian noise, the MRW becomes the multifractal fractional random walk (MFRW) and the scaling exponents are \cite{Bacry-Delour-Muzy-2001-PRE}
\begin{equation}\label{Eq:MF:MRW:zeta:qH}
    \zeta_{q,H}=
\left\{
  \begin{array}{lll}
    [qH-q(q-2)\lambda^2]/2, &{\rm{if}}& s\leqslant{T} \\
    qH,  &{\rm{if}}& s>T
  \end{array}
\right.
\end{equation}
Mathematical treatments for the MFRW can be found in Refs.~\cite{Ludena-2008-AnnAP,Abry-Chainais-Coutin-Pipiras-2009-IEEEtit,Perpete-2013-SD,Fauth-Tudor-2014-IEEEtit}. The application of the MFRW model is limited in econophysics since financial returns are usually uncorrelated such that $H\approx 0.5$.
Rypdal and L{\o}vsletten proposed two mean-reverting MRW models in which the anti-correlations are modeled either through a drift term as in an Ornstein-Uhlenbeck process or by using fractional Gaussian noise with $H<0.5$ for the innovations \cite{Rypdal-Lovsletten-2013-PA}. They applied the models to estimate the multifractal behavior in the 1-h electricity spot prices in Norwegian Kroner market from 4 May 1992 to 27 August 2011.

\subsubsection{Skewed MRW}

The MRW is a useful model in finance which is validated with respect to certain properties on the memory of volatility but is not validated for a fully faithful description of the stock market returns \cite{Sornette-Davis-Ide-Vixie-Pisarenko-Kamm-2007-PNAS}.

To capture the leverage effect that past price returns and future volatilities are negatively correlated, Pochart and Bouchaud proposed the skewed multifractal random walk (SMRW) \cite{Pochart-Bouchaud-2002-QF}:
\begin{equation}\label{Eq:MF:SMRW:dX}
    \delta_{\Delta{t}}X_{\Delta{t}}(i\Delta{t}) =
    \varepsilon_{\Delta{t}}(i)\exp\left[\omega_{\Delta{t}}(i)-\sum_{j<i}K(j,i)\varepsilon_{\Delta{t}}(j)\right]
\end{equation}
where the kernel $K(i,j)$ decays as a power law
\begin{equation}\label{Eq:MF:SMRW:Kij}
    K(j,i) = K_0(i-j)^{-\beta}\Delta{t}^{-\beta'}~~~~~~~~~~(j<i).
\end{equation}
They derived that
\begin{equation}\label{Eq:MF:SMRW:zeta}
    \left\{
  \begin{array}{lll}
    \zeta(2q)   = q-2q(q-1)\lambda^2 \\
    \zeta(2q+1) = q+1-2q^2\lambda^2-\beta
  \end{array}
\right.
\end{equation}
The SMRW model has been applied to understanding the volatility smiles in option pricing \cite{Pochart-Bouchaud-2002-QF}. Further mathematical treatments and empirical validation using the daily data of five stock market indices (CAC 40, DAX, FTSE 100, S\&P 500 and DJIA) from 3 December 1990 to 15 February 2010 can be found in Ref.~\cite{Bacry-Duvernet-Muzy-2012-JAP}.

%
%
%

\subsection{Exponentials of long memory processes \label{wyehtjyehw}}

\subsubsection{Alternative representation of MRW}
\label{S2:Models:MRW:Sornette}

Sornette et al. provided an alternative representation of the MRW \cite{Sornette-Malevergne-Muzy-2003-Risk},
\begin{equation}
  \delta_{\Delta{t}}X(t) =\int_{t-\Delta{t}}^t{\mbox{d}}W_1(t'){\rm{e}}^{\omega_{\Delta{t}}(t')}
  \label{Eq:MF:MRW:Sornette:dX}
\end{equation}
where $\delta_{\Delta{t}}X(t)=X(t)-X(t-\Delta{t})$ is the increment, $W_1(t)$ is a Wiener process with unit diffusion coefficient, and $\omega_{\Delta{t}}(t)$ is the log-volatility which obeys an autoregressive equation
\begin{equation}
  \omega_{\Delta{t}}(t) = \omega^0_{\Delta{t}}+\int_{-\infty}^t  \eta(t)~h_{\Delta{t}}(t-t') {\mbox{d}}t,
  \label{Eq:MF:MRW:Sornette}
\end{equation}
where $\omega^0_{\Delta{t}}$ is a constant, $\eta(t)$ denotes the process of information flow in the form of a standardized Gaussian white noise and the memory kernel $h_{\Delta{t}}(\cdot)$ is a causal function ensuring that the system is not anticipative. Thus $\omega(t)$ represents the response of the market to incoming information $\eta(t)$ up to the time $t$. At time $t$, the distribution of $\omega_{\Delta{t}}(t)$ is Gaussian with mean $\omega^0_{\Delta{t}}$ and variance $V_{\Delta{t}} = \int_0 ^\infty{h}^2_{\Delta{t}}(t'){\mbox{d}}t' = \lambda^2 \ln \left( \frac{Te^{3/2}}{\Delta t} \right)$.
The covariance of $\omega_{\Delta{t}}(t)$,
\begin{equation}
  C_{\Delta{t}}(s\Delta{t}) = \int_0 ^\infty h_{\Delta{t}}(t) h_{\Delta{t}}(t+s|\Delta{t}|) {\mbox{d}}t,
  \label{Eq:MF:MRW:Sornette:CoVar}
\end{equation}
specifies entirely the random process. When $\Delta{t}$ is sufficiently small, we have
\begin{equation}
  h_{\Delta{t}}(t') \sim h_0 \sqrt{\frac{\lambda^2 T}{t'}} ~~~~~ \mbox{for}~~ {\Delta{t}} \ll t' \ll T.
  \label{Eq:MF:MRW:Sornette:h}
\end{equation}
The long-range dependence and multifractality of the stochastic volatility are caused by the slow power-law decay of the memory kernel in Eq.~(\ref{Eq:MF:MRW:Sornette:h}).

Different from the techniques and evidence of multifractality in the domain of $f(\alpha)$ reported in this review, it can be derived from the above representation of the MRW that there is also a nice temporal signature of multifractality, in the sense that there is a full spectrum of exponent $p(A)$ describing the $1/t^{p(A)}$ relaxation of the volatility from a peak of amplitude $A$ \cite{Sornette-Malevergne-Muzy-2003-Risk}. Such multifractality has also been found in the multifractal stress activation model of earthquakes \cite{Sornette-Ouillon-2005-PRL,Ouillon-Sornette-2005-JGRse}, which has been confirmed on many catalogs \cite{Ouillon-Sornette-Ribeiro-2009-GJI,Tsai-Ouillon-Sornette-2012-BSSA}, and exogenous shocks and on endogenous multifractal responses in financial markets \cite{Sornette-Zhou-2006-PA,Zhou-Sornette-2007-EPJB}.

\subsubsection{Quasi-multifractal model}

Saichev and Sornette generalized the MRW model to a large class of continuous stochastic processes \cite{Saichev-Sornette-2006-PRE}. Similar to Eq.~(\ref{Eq:MF:MRW:Sornette:dX}), the increment process is
\begin{equation}
  \delta_{\Delta{t}}X(t) =\int_{t-\Delta{t}}^t \kappa{\rm{e}}^{\omega(t')} {\mbox{d}}t',
  \label{Eq:MF:QuasiMF:dX}
\end{equation}
where $\omega(t)$ is expressed as
\begin{equation}
  \omega(t) = \int_{-\infty}^t h(t-t'){\mbox{d}}W(t')
  \label{Eq:MF:QuasiMF:w}
\end{equation}
through a Wiener process $W(t)$ and a power-law kernel
\begin{equation}
  h(t) \sim h_0 \left(1+\frac{t}{\ell}\right)^{-0.5-\varphi}{\cal{H}}(t),
  \label{Eq:MF:QuasiMF:h}
\end{equation}
where $\varphi\geq0$ and the microscopic characteristic scale $\ell$ regularizes the singularity of the power law in the propagator $h(t)$ at $t=0$ and ${\cal{H}}(t)$ is the Heaviside function ensuring the condition of causality.
The special case $\varphi=0$ recovers the MRW model with a truncation at an integral scale.
The spectrum of the quasi-multifractal process can be computed and it is found that multifractality holds over a large range of dimensionless scales for $\varphi>0$ \cite{Saichev-Sornette-2006-PRE,Saichev-Filimonov-2007-JETP}.

Saichev and Filimonov considered the discrete version of the quasi-multifractal model and proposed a new method for the determination of the quasi-multifractal spectrum \cite{Saichev-Filimonov-2008-JETPL}.

\subsubsection{Self-excited multifractal model}

Filimonov and Sornette introduced the self-excited multifractal (SEMF) model, in which the absolute increments of the process are expressed as exponentials of a long memory of past increments \cite{Filimonov-Sornette-2011-EPL}. The SEMF model has a self-excitation mechanism combined with exponential nonlinearity such that future values of the process explicitly depends on past ones. The self-excitation captures the microscopic origin of the emergent endogenous self-organization properties of financial markets, such as the cascade in information flows, the triggering of aftershocks by previous large fluctuations and the ``reflexive'' interactions.

Without loss of generality, we denote the time increment $\Delta t \to 1$. In discrete time $t=0, 1, 2, \cdots, n$, the SEMF process reads $X_n=\sum_{i=0}^n \Delta{X}_i$, where its increments are given by the following relation
\begin{equation}
  \Delta{X}_n = \sigma \eta_n\exp\left\{-\frac{\omega_n}{\sigma}\right\},
  \label{Eq:MF:SEMF:dX}
\end{equation}
where
\begin{equation}
  \omega_n=\sum_{i=0}^{n-1}\Delta{X}_ih_{n-i-1}.
  \label{Eq:MF:SEMF:w}
\end{equation}
in which the kernel has a power-law form
\begin{equation}
  h_n=h_0 n^{-0.5-\varphi}.
  \label{Eq:MF:SEMF:h}
\end{equation}
The random variables $\eta_n$ is an i.i.d. Gaussian with zero-mean and unit variance representing a flow of external news, the parameter $\sigma$ stands for the impact amplitude of the external news and sets the dimension and scale of $\Delta{X}_n$ and $X_n$. The sum in Eq.~(\ref{Eq:MF:SEMF:w}) ensures that the dynamics is self-excited.

The discrete SEMF process can be viewed as the simplest multifractal generalization of the GARCH process \cite{Bollerslev-1986-JEm}, and in fact it is strongly related to EGARCH and other variants \cite{Nelson-1991-Em,Bollerslev-Chou-Kroner-1992-JEm}. The external flow of news $\eta_n$ can be either positive or negative, controlling the signs of the returns. The absolute returns (or volatilities) are determined by all past returns with a decaying weight as a function of their distance to the present.
The SEMF process can reproduce all the standard stylized facts found in financial time series, such as heavy tails of the return distribution, absence of autocorrelation of the returns, long-range correlation of the squared increments, multifractality and the leverage effect, which are robust to the specification of the parameters and the shape of the memory kernel \cite{Filimonov-Sornette-2011-EPL}.



The continuous-time SEMF process can be formally defined by the stochastic integro-differential equation in the Ito sense \cite{Filimonov-Sornette-2011-EPL}:
\begin{equation}
  {\mbox{d}}X(t)=\sigma\exp\left\{
  -\frac1\sigma\int_{-\infty}^t h(t-t'){\mbox{d}}X(t')
  \right\}dW(t),
  \label{Eq:MF:SEMF:dX:continous}
\end{equation}
where $dW(t)$ is the increment of the regular Wiener process and $h(t)$ is a memory kernel function. The expression (\ref{Eq:MF:SEMF:dX:continous}) is justified as the formal notation of the discrete formulation (\ref{Eq:MF:SEMF:dX}) in the limit when $\Delta{t}\to0$. When there is no memory in the kernel such that $h(t)=0$, we retrieve the standard random walk.

\subsection{Agent based models}

A promising direction is to understand the microscopic origins of apparent multifractality in financial time series through computational experiments using agent-based models (ABMs) \cite{Sornette-2014-RPP}. We provide here a parsimonious survey of such efforts without going into the details of the models.

\subsubsection{Percolation models}

Percolation models for financial markets are based on stochastic formation of opinion clusters with variant rules of cluster merging and splitting \cite{Cont-Bouchaud-2000-MeD,Eguiluz-Zimmermann-2000-PRL}. The multifractal analysis of the returns using the structure function approach illustrates that the scaling exponent function $\xi(q)$ varies with model parameters and properly determined parameters are able to produce comparable results as the DJIA index \cite{Castiglione-Stauffer-2001-PA}. Quite a few percolation models have been studied aiming at exploring the emergence of multifractal behavior in artificial financial time series, such as the stochastic cellular automata model \cite{Bartiromo-2004-PRE}, the particle-cluster aggregation model on scale-free networks \cite{Wang-Zhang-2005-IJMPC,Wang-Zhang-2005-PA}, the percolation model on Sierpinski carpet lattice \cite{Pei-Wang-2015-FNL}, and other variants of the percolation model \cite{Zeng-Wang-Xu-2017-PA,Wang-Wang-2017-IJMPC,Niu-Wang-2016-ND,Niu-Wang-2016-IJBC,Dong-Wang-2014-JSSC,Xiao-Wang-2014-IJNSNS}.

\subsubsection{Spin models}

Spin models, especially Ising models, have also been adopted to model the price formation process of financial assets \cite{Roehner-Sornette-2000-EPJB}. In a sophisticated Ising model, Sornette and Zhou incorporated global news for all agents, imitation among neighboring agents and idiosyncratic preference of individual agents with an irrational interpretation of the predictive power of global news \cite{Sornette-Zhou-2006-PA,Zhou-Sornette-2007-EPJB}. The model can generate multifractal financial time series in a self-organized way. There are also Ising models for stock markets with market makers \cite{He-Zheng-2010-Fractals,Fang-Wang-2012-IJMPC,Fang-Wang-2013-PA}, on Sierpinski carpet lattice \cite{Ko-Kim-2017-PA}, and on small-world networks \cite{Zhang-Li-Su-Zhang-2015-CPL,Zhang-Li-2015-EPJB}. These models also succeeded in generating multifractal returns.

\subsubsection{Empirical order-driven models}

Mike and Farmer introduced an empirical behavioral model for order-driven markets \cite{Mike-Farmer-2008-JEDC}, known as the Mike-Farmer model \cite{Gu-Zhou-2009-EPJB}, in which the basic rules (the long memory in order directions and a Student distribution in relative order prices) are based on the statistical regularities extracted from order flow data of real stocks. Gu and Zhou improved the Mike-Farmer model by considering further the long memory in relative order prices \cite{Gu-Zhou-2009-EPL}. The Gu-Zhou model is able to produce multifractality in the generated return series \cite{Gu-Zhou-2009-EPL} and in the recurrence intervals of returns \cite{Meng-Ren-Gu-Xiong-Zhang-Zhou-Zhang-2012-EPL}.

\subsubsection{Other models}

Thompson and Wilson found that the zero-intelligence model of Farmer et al. \cite{Farmer-Patelli-Zovko-2005-PNAS} is not able to produce multifractality in the generated returns, but the positive-intelligence model can \cite{Thompson-Wilson-2015-SSMSI}.
Crepaldi et al. showed that, although the usual minority games do not produce multifractal returns, the nonsynchronous minority games in the nonergodic phase can generate multifractal returns \cite{Crepaldi-Neto-Ferreira-Francisco-2009-CSF}. Multifractal behavior of financial time series is also observed in other types of models, such as an artificial dealer market model \cite{He-2010-CE}, a double auction model \cite{Li-Zhang-Zhang-Zhang-Xiong-2014-IS}, the stochastic voter model \cite{Deng-Wang-2014-AAA,Niu-Wang-2014-CS}, the stochastic interacting contact model \cite{Cheng-Wang-2014-AAA,Cheng-Wang-2013-EPL}, the epidemic model \cite{Lu-Wang-Niu-2015-Chaos}, and strategic models with fundamentalist agents \cite{Passos-Nascimento-Gleria-daSilva-Viswanathan-2011-EPL,Vikram-Sinha-2011-PRE}.

\section{Properties and important issues}
\label{S1:MF:Properties}

%
%
%
%
%
%
%
%
%
%
%

\subsection{Determination of multifractal quantities}
\label{S2:MF:Properties:Determination}

\subsubsection{Computational precision}

In most multifractal analyses, we need to calculate high-order moments. The problem of computational precision may arise. Consider the partition function approach as an example. When $m(s,t)\ll1$ and $q\gg1$, the corresponding values of the partition function $\chi$ will be too small such that the computer is out of memory. To overcome this problem, we can compute the logarithm of the partition function ln$\chi_{q}(s)$ rather than the partition function itself. With a simple manipulation, we obtain \cite{Jiang-Zhou-2008b-PA}:
\begin{equation}
  \ln\chi_{q}(s) = \ln\sum_{t}\left[\frac{m(s,t)}{m_{\max}}\times{m_{\max}}\right]^q
   = \ln\sum_{t}\left[\frac{m(s,t)}{m_{\max}}\right]^q +q\ln{m_{\max}},
\end{equation}
where $m_{\max}=\max \limits_{t}\{m(s,t)\}$ is the maximum of $m(s,t)$ in the boxes.

\subsubsection{What if \texorpdfstring{$N/s$}{} is not integer}
\label{S2:MF:N/s}

Multifractal analysis deals with the scaling behavior of time series at different scales. The most common way is to partition the time series of length $N$ into nonoverlapping windows of scale $s$. Hence, it is not uncommon that $N/s$ is not an integer. If one uses a window with size less than $s$, one may obtain incorrect results such as $f(\alpha)>1$, which happened sometimes in empirical studies. A usual strategy is to cover the time series from both sides and keep only the $\mathrm{int}[N/s]$ boxes of size $s$ for each covering \cite{Kantelhardt-Zschiegner-KoscielnyBunde-Havlin-Bunde-Stanley-2002-PA}.

Alternatively, we can adopt the idea of the sand box method \cite{Tel-Fulop-Vicsek-1989-PA,Vicsek-1990-PA}, in which, instead of covering the times series, one uses $M_s$ boxes of size $s$ with the centers of the boxes randomly distributed on the time series \cite{Ji-Zhou-Liu-Gong-Wang-Yu-2009-PA}. Taking the family of MF-DFA as an example, the $q$th overall fluctuation is
\begin{equation}
  F_q(s) = \left\{\frac{1}{M_s}\sum_{v=1}^{M_s} \left[F_v(s)\right]^q\right\}^{\frac{1}{q}}.
\end{equation}
This approach has an additional advantage to reduce fluctuations around the power-law scaling, especially when there are intrinsic log-periodic oscillations such as in binomial measures \cite{Zhou-Sornette-2009b-PA}.



\subsubsection{Range of \texorpdfstring{$q$}{} and convergence of moments}

A preliminary condition for multifractal analysis is the estimation accuracy of the moments. When $|q|$ is large, the estimated moment is usually unreliable since the empirical integrand $x^q\Pr(x)$ for the moment diverges, where $x$ is the measure and $\Pr(x)$ is its probability distribution in the partition function approach \cite{Lvov-Podivilov-Pomyalove-Procaccia-Vandembroucq-1998-PRE}. The divergence phenomenon has been empirically shown in turbulence data \cite{Lvov-Podivilov-Pomyalove-Procaccia-Vandembroucq-1998-PRE,Zhou-Sornette-Yuan-2006-PD} and financial data \cite{Jiang-Zhou-2007-PA}. Similarly, when $|q|$ is larger, the estimated derivatives $dF_q(s)/ds$ deviate significantly from the theoretical values at small scales and fluctuate a lot at large scales in MF-DMA \cite{Lopez-Contreras-2013-PRE}. We believe that the divergence phenomenon is ubiquitous in multifractal analysis.

The estimation accuracy of moments is worse when $|q|$ is larger and the time series length $N$ is shorter. Unfortunately, there are no theoretical results of the dependence on the length $N$ of the maximum $q$ for which the convergence of the integrand is ensured. With turbulence data, Zhou et al. estimated that $N(q=8) \approx 6 \cdot 10^{6}$
and $N(q=9) \approx 5 \cdot 10^{7}$ \cite{Zhou-Sornette-Yuan-2006-PD}. For low-frequency financial time series, such as daily returns, we suggest that the range of $q$ should be $|q|\leq6$, or even smaller.

\subsubsection{Determination of scaling range}

If a time series is fractal or multifractal, there are power-law scaling relationships between moment-like quantities and scales. In practice, taking $F_q(s) \sim {s}^{h(q)}$ in Eq.~(\ref{Eq:MF:Detrend:hq}) as an example, the scaling exponents are usually estimated using linear least-squares regressions of $\ln{F_q(s)}$ against $\ln{s}$ in a scaling range $[s_{\min},s_{\max}]$. In other words, the scaling laws hold only on the scaling range. The scaling range should be wide enough to ensure the presence of scale-invariant characteristics \cite{Malcai-Lidar-Biham-Avnir-1997-PRE,Avnir-Biham-Lidar-Malcai-1998-Science,Mandelbrot-1998-Science,Biham-Malcai-Lidar-Avnir-1998-Science}. In many cases, the change of scaling range causes dramatic variations of the estimates and sometimes very strange singularity spectrum shapes \cite{Czarnecki-Grech-2010-APPA,Yue-Xu-Chen-Xiong-Zhou-2017-Fractals}, some of which are called ``inverted multifractality'' \cite{Schumann-Kantelhardt-2011-PA}. Hence, as the scaling range plays a crucial role in the estimation of scaling exponents, many efforts have been made for its determination. Nevertheless, it remains a difficult task that has not been fully solved \cite{Shao-Gu-Jiang-Zhou-Sornette-2012-SR}. The mostly adopted method is actually eye balling.

Wang et al. argued that scaling laws hold in the scaling range but not in the two ranges outside the scaling range such that one can fit the field with three power laws by minimizing the RMS of residuals, called the three-segment method by some researchers \cite{Wang-Luo-Chen-1993-cnCJCP}. Because the assumption of two extra power laws is merely empirical, this method is not widely adopted. However, when $F_q(s)$ exhibits a crossover phenomenon as in many financial time series \cite{Fisher-Calvet-Mandelbrot-1997-SSRN,Liu-Cizeau-Meyer-Peng-Stanley-1997-PA,Janosi-Janecsko-Kondor-1999-PA,GrauCarles-2001-PA, Jiang-Ma-Cai-2007-PA,Jiang-Zhou-2008b-PA,Ma-2009-CTP}, the idea can be used to simultaneously determine the crossover scale $s_{\times}$ and the two scaling exponents $H_1$ and $H_2$ \cite{Gu-Ren-Ni-Chen-Zhou-2010-PA,Yue-Xu-Chen-Xiong-Zhou-2017-Fractals}. We rewrite the bi-power law for $q=2$ as follows
\begin{equation}
  \ln{F_2(s)}=
   \begin{cases}
   c_1+H_1\ln{s}, & s\leq s_{\times}\\
   c_2+H_2\ln{s}, & s>s_{\times}
   \end{cases}~.
   \label{Eq:MF:Crossover:TwoPowers}
\end{equation}
To estimate the three parameters $s_{\times}$, $H_1$ and $H_2$, we minimize the following function
\begin{equation}
  O(s_{\times}, H_1, H_2, c_2)= \sum\limits_{s_j\leq{s_{\times}}}\left[\ln{F_2(s_j)}-H_1\ln{s_j}-{c_1}\right]^2
    + \sum\limits_{s_j>{s_{\times}}}\left[\ln{F_2(s_j)}-H_2\ln{s_j} -{c_2}\right]^2.
  \label{Eq:DMA:ObjFun}
\end{equation}
where $c_1$ is not a free parameter due to the constraint that the two straight lines should intersect at $s_{\times}$, that is,
\begin{equation}
  {c_1}+H_1\ln{s_{\times}} = {c_2}+H_2\ln{s_{\times}}.
  \label{Eq:DMA:Fit:constraint}
\end{equation}
This method does not allow one to determine the left and right cutoffs of the two scaling ranges. The left cutoff can be determined if we combine other methods \cite{Veitch-Abry-Taqqu-2003-Fractals,Park-Park-2009-SSinica}. Alternatively, Fang et al. suggested to use a polynomial function for the second range and take the first range as the scaling range, which however fixed $s_{\min}=1$ \cite{Fang-Zhou-Zou-Zhou-Zhang-2004-JIMW}. Ge and Leung studied a method to identify crossover scales based on statistical inference \cite{Ge-Leung-2013-JGS}.

Xiong et al. argued that the optimal right cutoff $s_{\max}$ can be determined based on the intersection of the $\ln{F_{q_1}(s)}$ and $\ln{F_{q_2}(s)}$ curves, where $q_1$ and $q_2$ are respectively the minimum and maximum positive values of $q$ under investigation \cite{Xiong-Chen-Wei-Liu-Guan-2014-APS}. The candidate right cutoff is determined as follows
\begin{equation}
  \ln{F_{q_1}(s_{\max})}-\ln{F_{q_2}(s_{\max})} =\min_s\{\ln{F_{q_1}(s)}-\ln{F_{q_2}(s)}\}.
\end{equation}
In order to avoid being trapped within local optima, it is required that the variation around $s_{\max}$ should be large enough:
\begin{equation}
  \frac{\ln{F_{q_j}(s_{\max}+1)}-\ln{F_{q_j}(s_{\max})}}{\ln{F_{q_j}(s_{\max})}-\ln{F_{q_j}(s_{\max}-1)} } > 10,
\end{equation}
where $j=1$ and 2. This method works well for traffic flow time series \cite{Xiong-Chen-Wei-Liu-Guan-2014-APS}. Extensive numerical experiments are necessary to check if the method is suitable for other time series.

Berntson and Stoll proposed a {\textit{range erosion method}} based on the test of self-similarity \cite{Berntson-Stoll-1997-PRSB}. Starting from the largest range, we fit the data and performs a residual test of self-similarity, which fits second-order polynomials to the residual plots and took significant second-order polynomials as a rejection signal of self-similarity. If the self-similarity is rejected, we erode the endpoints to shorten the data and repeat the fit-and-test procedure until either too few points remained to perform a regression or the data showed self-similarity over the remaining range. This erosion is suggested to be performed in three ways: (1) removing the left endpoint, (2) removing the right endpoint, and (3) removing both endpoints. The largest remaining logarithmic range is taken as the scaling range. Interestingly, Xia et al. proposed a {\textit{range expansion method}} based on Theil's inequality coefficient \cite{Xia-Lazarou-Butler-2014-IEEEcl}. It starts from the two points in the center of the scale sequence and Theil's inequality coefficient $\delta(s_i,s_j)$ is calculated. If $\delta(s_i,s_j)$ is less than a prefixed threshold $\delta_0$, we proceed to the nest step. We include the nearest left scale $s_i-1$ to the scaling range if $\delta(s_i-1,s_j)<\delta(s_i,s_j+1)$ and $\delta(s_i-1,s_j)<\delta_0$ or the nearest right scale $s_j+1$ to the scaling range  if $\delta(s_i-1,s_j)>\delta(s_i,s_j+1)$ and $\delta(s_i,s_j+1)<\delta_0$. The iteration stops when $\delta(s_i,s_j)>\delta_0$ and the final scaling range is regarded as optimal.

Fei et al. proposed another automatic method to identify the optimal scaling range \cite{Fei-Jiang-Wang-1998-cnJXJTU}, which is implemented through the {\textit{method of exhaustion}} or {\textit{brute force}} to obtain the RMS of the residuals of the fit to the data within all potential scaling ranges $[s_i,s_j]$ with $s_j/s_i$ greater than a prefixed constant and choose the scaling range with the minimal RMS. Similarly, Gulich and Zunino proposed a criterion based on a brute-force algorithm to systematically determine optimal fitting regions in DFA and MF-DFA \cite{Gulich-Zunino-2014-PA}. Although the study of the algorithm mainly focuses on DFA and MF-DFA, it can obviously be applied to other multifractal analysis. One first determines a set of scales $\{s_i\}_{i=1}^{n}$ that are evenly spaced in the logarithmic ordinate. An empirical suggestion is to use $s_1=4$ and $s_n=[N/4]$ which embraces the scaling range of Grech and Mazur \cite{Grech-Mazur-2013-PA,Grech-Mazur-2013-PRE,Grech-Mazur-2015-APPA}. Next, one performs linear regression for each pair $[s_i,s_j]$ to obtain the R-squared statistic (or coefficient of determination) $R^2_{ij}$. To ensure that the scaling law is not spurious, the scaling range should be wide enough \cite{Malcai-Lidar-Biham-Avnir-1997-PRE}. Hence, the scaling range width $\ln{s_j}-\ln{s_i}$ should be greater than a prefixed constant, say one order of magnitude. The pair with the largest $R^2_{ij}$ is regarded as the optimal scaling range $[s_i,s_j]$. The criteria of these two methods are essentially the same. For multifractal analysis, one obtains the average $R^2_{ij}$ over different $q$ values \cite{Gulich-Zunino-2014-PA}. In addition, Leonarduzzi et al. designed a nonparametric bootstrap based procedure to achieve automated selection of scaling range, which also falls in the class of exhaustion methods \cite{Leonarduzzi-Torres-Abry-2014-SP}.

Wu suggested that a qualified fit in the estimation of fractal dimensions should satisfy three criteria \cite{Wu-2002-cnAGCS}: (1) The linearity between $\ln{F_q(s)}$ and $\ln{s}$ is significant; (2) The maximum deviation of each point does not exceed twice the standard deviation of the fit residuals; and (3) The standard deviation of the estimate is less than a preset value. After exhaustive calculations, the widest range among those for all the qualified fits (largest $\ln{s_j}-\ln{s_i}$) is used as the optimal scaling range \cite{Wu-2002-cnAGCS}. Because this method does not ask for a best goodness-of-fit, one does not need to pose a constraint on $\ln{s_j}-\ln{s_i}$. For multifractal analysis, we suggest to modify the third criterion by requiring that the relative deviation of the estimate is less than a preset percentage, say 5\%. The optimal scaling range for $q=2$ is often used for other $q$'s. However, it is not uncommon that the crossovers become evident for large $|q|$'s but not for $q=2$ when there is a crossover phenomenon or the scaling range becomes narrower for large $|q|$'s. Hence, it is better to determine the optimal scaling range based on the largest $q$ value. Alternatively, one may determine the optimal scaling range for each $q$, in which an extra constraint should be posed ensuring that the scaling ranges for different $q$'s should overlap at least partly. We can further combine the multiple linear regression (\ref{Eq:MF:Crossover:TwoPowers}) for crossover phenomena.

Grech and Mazur argued that the scaling range is related to the length $N$ of the analyzed time series, the level of long-term memory described by the Hurst exponent $H$ of the time series, and the goodness-of-fit quantified by the coefficient of determination $R^2$ or equivalently $u=1-R^2$ \cite{Grech-Mazur-2013-PA,Grech-Mazur-2013-PRE,Grech-Mazur-2015-APPA}. They provided an excellent analysis on the behavior of scaling ranges in DFA and DMA and obtained their expressions for fixed confidence levels, which is the percentage of time series matching the assumed criterion for $R^2$. In their analysis, they fixed $s_{\min}=8$. For DFA, it is found that
\begin{equation}
  s_{\max}(u,N,H) = [(a_1 H + a_2)u+ a_3]N + a_4,
\end{equation}
where $a_1=3.40$, $a_2=4.16$, $a_3=0.0097$ and $a_4=-96$ for the confidence level 97.5\% and $a_1=3.95$, $a_2=5.03$, $a_3=0.0070$ and $a_4=-106$ for confidence level 95\% \cite{Grech-Mazur-2013-PA}. For DMA, it is found that
\begin{equation}
  s_{\max}(u,N,H) = (D_0+D_1 H)N^{\eta_0-\eta_1H}u^{\xi_0-\xi_1H},
\end{equation}
where $D_0=0.558$, $D_1=0.640$, $\eta_0=1.130$, $\eta_1=0.135$, $\xi_0=1.049$ and $\xi_1= 0.657$ for the confidence level 97.5\% and $D_0=0.400$, $D_1=1.125$, $\eta_0=1.146$, $\eta_1=0.163$, $\xi_0=0.975$ and $\xi_1= 0.593$ for confidence level 95\% \cite{Grech-Mazur-2013-PRE}. In most studies, researchers adopt $s_{\max}=N/4$ recommended by Kantelhardt et al. \cite{Kantelhardt-Zschiegner-KoscielnyBunde-Havlin-Bunde-Stanley-2002-PA}, which is too large for most time series and one should use a much smaller $s_{\max}$ to ensure an more accurate estimation of $H$ \cite{Grech-Mazur-2015-APPA}.

When we are estimating the Hurst exponent of a given time series, we first choose the goodness of fit $R^2$ and the confidence level (for instance 97.5\% or 95\%). In this way, the expression of $s_{\max}(u,N,H)$ is determined, but not its value, because $H$ is unknown. We next scan different scales (denoted as $s'$) and obtain the estimate of $H$ by fitting the power law in the scaling range $[8,s']$, which is a function of $s'$, i.e., $\hat{H}(s')$. The ``optimal'' right cutoff of the scaling range is determined as follows
\begin{equation}
  s_{\max} = \min_{s'} |s_{\max}(u,N,\hat{H}(s'))-s'|,
\end{equation}
and the estimate of $H$ is
\begin{equation}
  \hat{H} = \hat{H}(s_{\max}).
\end{equation}
When we are performing MF-DFA or MF-DMA, we determine $s_{\max}$ first and utilize the same scaling range for all $q$ values. Certainly, this automatic procedure requires further validation because it is possible to have multiple $s_{\max}$ values as well as multiple $\hat{H}$ values. Moreover, the choice of $s_{\min}=8$ is rather arbitrary \cite{Shao-Gu-Jiang-Zhou-Sornette-2012-SR}.

The methods mentioned above are based on direct fitting of the data. Another class of methods are based on identifying the flat region in the local derivatives. In empirical analysis, visual inspection is frequently used \cite{Panico-Sterling-1995-JCN,Shao-Gu-Jiang-Zhou-Sornette-2012-SR,Lopez-Contreras-2013-PRE,Green-Hanan-Heffernan-2014-EPJB}.
One of the earliest method, named the self-similar ratio algorithm, was proposed by Hong and Hong in 1993 \cite{Hong-Hong-1993-cnEN}. Although the algorithm was designed to determine the scaling grange in fractal dimension estimation, it can be easily adopted for multifractal analysis. The self-similar ratio $R_i$ is defined as follows
\begin{equation}
  R_i = \frac{F_q(s_{i-1})}{F_q(s_i)}=\left(\frac{s_{i-1}}{s_i}\right)^{h(q)}.
  \label{Eq:MF:Properties:ScalingRange:SSratio}
\end{equation}
If the logarithmic scales $\{\ln{s_i}\}$ are evenly spaced, $s_{i-1}/{s_i}$ is a constant and $R_i$ is also a constant for each $q$. For each range, if $R_i=\langle{R_i}\rangle$ (or $R_i-R_{i-1}=0$) is statistically significant, the range is a qualified scaling range. The widest qualified scaling range is then the optimal scaling range. We notice that $R_i$ is in essence equivalent to the local derivative $\ln{R_i}/\ln(s_{i-1}/{s_i})$. Li et al. determined the scaling range by removing points whose local derivatives are far from the average derivative \cite{Li-Hu-Zhang-Wang-Duan-2007-CPM}. Du et al. proposed an exhaustion method on local derivatives, in which qualified scaling ranges should ensure that the correlation coefficient between the derivatives and the logarithmic scales is greater than 0.99 \cite{Du-Jia-Tang-2013-JVS}. Among all the qualified scaling ranges, the one with the minimal variance of the derivatives is selected as the optimal scaling range. Ji et al. developed an algorithm by combining the K-means algorithm and a point-slope-error algorithm \cite{Ji-Zhu-Jiang-2011-CSB}. Moreover, we can also perform tests of curvilinearity in the local derivative time series, as in Ref.~\cite{Berntson-Stoll-1997-PRSB}.

Zhou et al. proposed to consider the second-order derivatives since they should be zero or nearly zero within the scaling range \cite{Zhou-Feng-Wu-2015-APS}, which is reminiscent of testing $R_i-R_{i-1}=0$ in the self-similar ratio method \cite{Hong-Hong-1993-cnEN}. The method uses the simulated annealing genetic fuzzy C-means clustering algorithm to divide the data into three groups: positive fluctuations, zero fluctuations, and negative fluctuations around the horizontal line of zero second-order derivatives. Again, other identification methods developed for $\ln{F(s)}$ and its first-order local derivatives can also be adopted to develop new scaling range identification methods based on the second-order derivatives.

A relevant but less considered issue is the accuracy of the estimated exponent \cite{Michalski-2008-PA}, which should also be considered during the attempt to determine the scaling range.

\subsubsection{Integrated moments and factorial moments}
\label{S3:MF:Properties:IntMomemnt}

A severe problem might occur for many time series, namely that the scaling dependence of the moments on the time scales deviates from the asymptotic behavior. This happens usually together with the finite-size effect when the time series is not long enough \cite{Zhou-2012-CSF}. In addition, the discretization of a continuous process also induces deviations from the asymptotic behavior due to the effect of
auto-correlations and non-asymptotic convergence described by the Central Limit Theory \cite{Buonocore-Aste-DiMatteo-2017-PRE}. To overcome these difficulties in measuring the Generalized Hurst exponents of financial time series, Buonocore, Aste and Di Matteo introduced an integrated structure function based estimation method \cite{Buonocore-Aste-DiMatteo-2017-PRE}.

Theoretically, for monofractal or multifractal processes, the scaling of the structure functions $K(q,s)$ against the time scale $s$ expressed in Eq.~(\ref{Eq:MF-SF:Kq:s}) can be rewritten as
\begin{equation}
  \ln{K(q,s)} ={\zeta(q)}\ln{s} + A,
  \label{Eq:MF-SF:lnKq:lns}
\end{equation}
where $\ln{K(q,s)}$ is the dependent variable, $\ln{s}$ is the independent variable, and ${\zeta(q)}$ and $A$ are model parameters to be estimated. When applied to real data, Eq.~(\ref{Eq:MF-SF:lnKq:lns}) holds only asymptotically and a correction function $g(\ln{s})$ should be introduce, that is,
\begin{equation}
  \ln{K(q,s)} = g(\ln{s})+{\zeta(q)}\ln{s} + b,
  \label{Eq:MF-SF:lnKq:lns:Filter}
\end{equation}
Assuming that
\begin{equation}
  \int_0^\infty g(\ln{s})d\ln{s} = c
  \label{Eq:MF-SF:lnKq:lns:Filter:glns}
\end{equation}
and integrating Eq.~(\ref{Eq:MF-SF:lnKq:lns:Filter}), we obtain
\begin{equation}
  \int_0^\infty \ln{\hat{K}}(q,s) d\ln{s}= \frac{1}{2}{\zeta(q)}(\ln{s})^2 + b\ln{s} + c,
  \label{Eq:MF-SF:lnKq:lns:Filter:integration}
\end{equation}
where ${\hat{K}}(q,s)$ are empirically calculated from the real time series at different time scales.

The estimation proceeds as follows \cite{Buonocore-Aste-DiMatteo-2017-PRE}. For each $q$, we determine the autocorrelation length $s_{\max}$, integrate the empirical $\ln{\hat{K}}(q,s)$ for $s\in[1,s_{\max}]$, and determine $s_{\min}$ by the behavior of $c(s)$ defined in Eq.~(\ref{Eq:MF-SF:lnKq:lns:Filter:glns}). The interval $[s_{\min}, s_{\max}]$ is taken as the scaling range and we can determine the scaling exponents in the scaling range. One then checks if the correction function $g(\ln{s})$ converges to 0 and if its integral converges to a constant. With the scaling exponents $\zeta(q)$ estimated, we perform parabolic and quartic fits to Eq.~(\ref{Eq:MF:zeta:parabolic}) and Eq.~(\ref{Eq:MF:zeta:quartic}) and choose the best one to obtain a function form of $\zeta(q)$. This process is repeated for different criteria for determining $s_{\max}$ and one then chooses the one with the best goodness-of-fit.

Alternatively, Qiu et al. proposed an unbiased factorial-moment-based estimation for moments \cite{Qiu-Yang-Yin-Gu-Yang-2016-PRE}. Numerical analysis of the $p$-model (they called it the probability redistribution model) verifies that the new method can extract exactly the multifractal behavior from several hundred recordings and outperforms the basic partition function approach, the MF-DFA method and the WTMM method. They applied the method to the daily volatility of the Shanghai stock market SSEC index from 1995 to 2015 in 3y rolling windows and unveiled intriguing dynamics of the multifractal behavior. It is also found that the partition functions in the two windows 1995-1997 and 2007-2009 containing the Asian crisis and the latest Great Crash obey comparatively better power law scaling in wider scaling ranges.
%
%
Although the estimation method through integrated moments is based on the structure function approach, while the factorial-moment-based estimation is based on the partition function approach, they are both very promising, because their extensions to other multifractal analysis methods are straightforward. What is left is to check their performance.

\subsubsection{Assumed forms for \texorpdfstring{$\tau(q)$ (or $\zeta(q)$)}{} and \texorpdfstring{$f(\alpha)$}{}}

In some studies, researchers fitted the empirical curves of $\tau(q)$, $\zeta(q)$ or $f(\alpha)$. If the time series can be modelled in a multiplicative cascades, we can use the mathematical models with known multifractal properties as described in Section~\ref{S3:Models:for:Tests}. There are also other assumptions that do not relate directed to any underlying multifractal mechanisms.

%
%


In order to give a smooth presentation of the scaling exponents and a quantitative assessment of the degree of multifractality, Buonocore et al. suggested to use polynomial expressions for $\zeta(q)$ \cite{Buonocore-Aste-DiMatteo-2016-CSF,Buonocore-Aste-DiMatteo-2017-PRE}. The quadratic (or parabolic) expression is
\begin{equation}
  \zeta(q) = Bq^2+\left(\frac{1}{2}-2B\right)q
  \label{Eq:MF:zeta:parabolic}
\end{equation}
with only one parameter $B<0$, and the quartic polynomial expression is
\begin{equation}
  \zeta(q) = Dq^4+Cq^3+\frac{3C^2}{8D}q^2+\left(\frac{1}{2}-8D-4C-\frac{3C^2}{4D}\right)q
  \label{Eq:MF:zeta:quartic}
\end{equation}
with only two parameters $C$ and $D<0$. The coefficients for the terms satisfy the following constraints:
\begin{eqnarray}
    \zeta(0) = 0 & \Leftrightarrow & \tau(0) = -1,\\
    \zeta(1) = 2 & \Leftrightarrow & \tau(1) = 0,\\
    \zeta''(q)<0 & \Leftrightarrow & \tau''(q)< 0.
\end{eqnarray}
The fits to the polynomials are conducted over the range $q\geq-1$, because real return time series usually have fat tails with the tail exponent $\gamma>2$
\cite{Gopikrishnan-Meyer-Amaral-Stanley-1998-EPJB,Plerou-Stanley-2007-PRE,Mu-Zhou-2010-PRE} such that moments of returns with $q<-1$ do not exist. Similarly, a cumulant expansion of $\zeta(q)$ suggests an infinite polynomial form of $\zeta(q)$ under the multifractal framework of multiplicative cascades \cite{Muzy-Sornette-Delour-Arneodo-2001-QF} or wavelet leaders \cite{Wendt-Roux-Jaffard-Abry-2009-SP,Leonarduzzi-Wendt-Abry-Jaffard-Melot-Roux-Torres-2016b-PA}.

For the quadratic expression, based on the Legendre transform presented in Eq.~(\ref{Eq:MF:LegendreTrandform}), we have
\begin{equation*}
  \alpha(q) = 2Bq + \left(0.5-2B\right) 
  \label{Eq:MF:zeta:parabolic:alpha:q}
\end{equation*}
and
\begin{equation*}
  f(q) = Bq^2 + 1.
  \label{Eq:MF:zeta:parabolic:f:q}
\end{equation*}
It follows immediately that
\begin{equation}
  f(\alpha) = \frac{(\alpha+2B-0.5)^2}{4B} + 1.
  \label{Eq:MF:zeta:parabolic:f:alpha}
\end{equation}
We can see that $\alpha$ can be negative when $q>(0.5-2B)/2B$, $\alpha_{\min}$ and $\alpha_{\max}$ (when $q\to\pm\infty$) does not exist, the most probable singularity is $\alpha_0=0.5-2B$, and $f(\alpha)$ could be negative.

If the singularity spectrum is bounded within $[\alpha_{\min},\alpha_{\max}]$ and $f(\alpha_{\min}) = f (\alpha_{\max}) = 0$ when $q\to\pm\infty$, such as the $p$-model, the form of the $f(\alpha)$ can be assumed to be \cite{Harikrishnan-Misra-Ambika-Amritkar-2009-Chaos}
\begin{equation}
   \label{e.7}
   f(\alpha) = \left(\frac{\alpha - \alpha_{\min}}{\alpha_1 - \alpha_{\min}}\right)^{\gamma_1}
               \left(\frac{\alpha_{\max} - \alpha}{\alpha_{\max} - \alpha_1}\right)^{\gamma_2}
\end{equation}
where $\gamma_1$, $\gamma_2$, $\alpha_{\min}$, $\alpha_{\max}$ and $\alpha_1=\alpha(1)$ are a set of parameters characterising a particular $f(\alpha)$ curve. Due to the geometric properties of $f(\alpha)$, $0<\gamma_1,\gamma_2 <1$.
Combining $\frac{d}{d\alpha}f(\alpha)|_{q=1}=1$ and $\alpha_1 = f(\alpha_1)$, we have
\begin{equation}
   \frac{\gamma_1}{\alpha_1 - \alpha_{\min}} - \frac{\gamma_2}{\alpha_{\max} - \alpha_1}= \frac{1}{\alpha_1}.
\end{equation}
Hence, the $f(\alpha)$ curve has four free parameters.
There are also quadratic and quartic functional forms assumed for $f(\alpha)$ \cite{Shimizu-Thurner-Ehrenberger-2002-Fractals,MunozDiosdado-RioCorrea-2006-IEEEconf}. If we consider $\frac{d}{d\alpha}f(\alpha)|_{q=0}=0$, the linear term must vanish!



\subsection{Some important properties}
\label{S2:SomeProperties}

\subsubsection{Discrete scale invariance and log-periodicity}

Fractals and multifractal measures are scale invariant \cite{Mandelbrot-1983}. Scale invariance characterizes the relationship between the values of an observable $\mathcal{O}$ at two distinct scales $s$ and $\lambda{s}$:
\begin{equation}
 {\mathcal{O}}(s) = \kappa {\mathcal{O}}(\lambda{s}), \label{Eq:DSI:O}
\end{equation}
where $\lambda$ is the magnification factor or scaling ratio and $\kappa$ is the corresponding scaling factor of the observable. A continuous ratio $\lambda$ results in continuous scale invariance (CSI), in which Eq.~(\ref{Eq:DSI:O}) holds for arbitrary positive real values of $\lambda$. In many natural, technologic and social systems, a characteristic scale exist so that the general solution to Eq.~(\ref{Eq:DSI:O}) is not just a power simple law implied by CSI but take the form \cite{Sornette-1998}
\begin{equation}
{\mathcal{O}}(s) = s^{-D} \psi(s), \label{Eq:DSI:OSol}
\end{equation}
where $D=\ln  \kappa / \ln \lambda $ is the fractal dimension and $\psi(\lambda{s}) = \psi(s)$ is a log-periodic function
(i.e. a periodic function of $\ln s$) with period $\ln\lambda$.
In these cases, ${\mathcal{O}}(s)$ obeys the symmetry of discrete scale invarance (DSI), in which the system is scale invariant only under integer powers of specific values of the magnification factor $\lambda$ with the signature of log-periodic oscillations decorating power law scaling relationships.

Log-periodic power-law singularity (LPPLS) signals have been observed in financial time series, and has been
used for the identification of bubbles and antibubbles and the forecast of turning points \cite{Sornette-2003}.
This has been applied to the NIKKEI 225 antibubble \cite{Johansen-Sornette-1999-IJMPC}, the US housing bubble \cite{Zhou-Sornette-2006b-PA}, the Chinese stock market antibubble \cite{Zhou-Sornette-2004a-PA} and bubble \cite{Jiang-Zhou-Sornette-Woodard-Bastiaensen-Cauwels-2010-JEBO}, and the crude oil bubble of 2008 \cite{Sornette-Woordard-Zhou-2009-PA}. In addition to the ubiquitous presence of LPPLS in fractal time series \cite{Sornette-1998-PR}, Zhou and Sornette unveiled the presence of DSI in certain multifractal measures and joint multifractal measures via the methods of MF-PF and MF-X-PF \cite{Zhou-Sornette-2009b-PA}, such as the detrended fluctuation functions of multifractal binomial measures although the authors may not notice this phenomenon \cite{Kantelhardt-Zschiegner-KoscielnyBunde-Havlin-Bunde-Stanley-2002-PA,Xu-Shang-Feng-2015-CSF,Fan-Wu-2015-EPJB}. In an interesting work \cite{Xiao-Pan-Stephen-Yang-Li-Yang-2015-PA}, Xiao et al. converted the detrended cross-correlation analysis to a new form with a log-periodic oscillation modulated power law scaling and illustrated the idea using five stock market indices (DJIA, FTSE, DAX, SSEC and NIKKEI), five sectors in Shanghai stock market, and ten sectors in DJIA from 22 December 1998 to 30 April 2010. If LPPLS exists in DCCA, one can rationally expect its presence in MF-DCCA presentations.

The presence of LPPLS in the moments calculated in different multifractal analysis approaches has an important implication in the determination of the scaling exponents. This issue is closely related to the choice of the scales $s_i$ and hence the scaling range $[s_1, s_m] = [s_{\min},s_{\max}]$. A natural strategy is to ensure that the scaling range contain several complete oscillations such that $(\ln{s_{\max}}-\ln{s_{\min}})/\ln\lambda$ is an integer and each log-periodic oscillation contain the same number of scales with the same set of phases.

\subsubsection{Translation invariance}

In the detrending methods, the innovations $\Delta{X}$ is usually adjusted to have zero mean before the cumulative summation procedure. When the trend function is imposed to have a constant term, such as polynomial trends, the corresponding method (such as MF-DFA) is translation invariant, that is, $\Delta{X}$ and $\Delta{X}+c$ have the same multifractal properties, where $c$ is an arbitrary constant.

However, the situation can be very different for other detrending methods such as MF-DMA \cite{Shao-Gu-Jiang-Zhou-2015-Fractals}. The cumulative sum (or profile) of $\Delta{X}+c$ is
\begin{equation}
  U(t)= ct, 
  \label{Eq:U_p0}
\end{equation}
and the moving average at $t$ obtained from $t-s_{1}$ to $t+s_{2}$ for window size $s$ is
\begin{equation}
  \widetilde {U}(t)=c \left[t+\frac{(2\theta-1)(s-1)}{2}\right],
  \label{Eq:U_MA_p0}
\end{equation}
where $\theta$ is defined in Eq.~(\ref{Eq:MF-DMA:trend}).
Hence, the residual of the constant shift $c$ after removing the moving average is
\begin{equation}
  \epsilon _{c}(t)=c \frac{(2\theta-1)(s-1)}{2}
  \label{Eq:epsilon_p0}
\end{equation}
which is a constant for a given window size $s$. One finds that the residual $\epsilon _{c}(t)=0$ only when $\theta=0.5$. Otherwise, we will observe a crossover in the $F$ vs. $s$ scaling plots \cite{Shao-Gu-Jiang-Zhou-2015-Fractals}.

Therefore, the centered MF-DMA is translation invariant, while other MF-DMA versions are not. This analysis raises the serious issue of whether the mean should be removed from the innovations in Eq.~(\ref{Eq:cumsum}). For time series with nonzero means, this analysis explains at least partly the discrepancy of MF-DMA results for different $\theta$ values.

\subsubsection{Negative dimensions \texorpdfstring{$f(\alpha)<0$}{}}
\label{S3:NegativeDimensions}

An intriguing feature of multifractals is the possible existence of negative dimensions in the multifractal spectrum, that is, $f(\alpha)<0$ for large or small singularities $\alpha$ \cite{Mandelbrot-1989a,Mandelbrot-1990a-PA,Mandelbrot-1991-PRSA}. Negative dimensions have also been observed in the singularity spectra of financial time series, such as financial volatility \cite{Jiang-Zhou-2007-PA}, intertrade durations \cite{Jiang-Chen-Zhou-2009-PA}, WIG index returns \cite{Czarnecki-Grech-2010-APPA}, returns of the spot and futures prices of WTI crude oil \cite{Wang-Wei-Wu-2011b-PA}, foreign exchange rates \cite{Wang-Wu-Pan-2011-PA}, CSI 300 index returns \cite{Zhou-Dang-Gu-2013-PA}, MASI index returns and MADEX index returns \cite{Lahmiri-2017-PA}, and S\&P 500 index returns \cite{He-Wang-2017-PA}, and so on, for which different multifractal analysis methods have been applied.

The meaning of negative $f(\alpha)$ was firstly discussed by Cates and Witten \cite{Cates-Witten-1987-PRA}, who noted thta negative dimensions are induced by the ensemble averaging of ``supersamples''.
Negative dimensions also appear if the multiplier distribution is continuous
\cite{Chhabra-Sreenivasan-1992-PRL,Zhou-Yu-2001-PA,Zhou-Liu-Yu-2001-Fractals,Zhou-Yu-2001-PRE}, such as financial volatility multipliers \cite{Jiang-Zhou-2007-PA}. A simple illustrative analytical example shows that negative dimensions arise from either the
existence of intrinsic randomness or from a random treatment of a deterministic process \cite{Chhabra-Sreenivasan-1991-PRA}.
This coincides with Mandelbrot's insights based on his works on random multifractals \cite{Mandelbrot-1990a-PA,Mandelbrot-1991-PRSA}.

\subsubsection{Skewness of multifractal spectra}

In diffusion-limited aggregation (DLA) processes, the multifractal spectrum shows anomalies, in which there is a critical value $q_{\rm{c}}$ such that the partition functions diverge faster than power laws and the $\tau(q)$ function does not exist when $q<q_{\rm{c}}$ \cite{Lee-Stanley-1988-PRL,Blumenfeld-Aharrony-1989-PRL,Schwarzer-Lee-Bunde-Halvin-Roman-Stanley-1990-PRL}. These anomalies can be explained within the framework of left-sided multifractals based on infinite multinomial measures, in which the generating function reads \cite{Mandelbrot-1990b-PA,Mandelbrot-Evertsz-Hayakawa-1990-PRA}
\begin{equation}
  \sum_{i=1}^\infty p_i^qr_i^{-\tau}=1.
  \label{Eq:Left-sided:GenEqn}
\end{equation}
A mathematical treatment can be found in Ref.~\cite{Riedi-Mandelbrot-1995-AAM}.
When $p_i=2^{-i}$ and $r_i=i^{-\lambda}-(i+1)^{-\lambda}$ (such that $\sum_{i=1}^\infty{p_i}=\sum_{i=1}^\infty{r_i}=1$) where $\lambda>0$ is a parameter, the critical $q$ value $q_{\rm{c}}=0$ and the multifractal spectrum is one-sided:
\begin{equation}
  f(\alpha) \approx 1-\left\{
     \begin{array}{lll}
         c[\alpha_\lambda(0)-\alpha]^\kappa, & \lambda>1, &\alpha\nearrow \alpha_\lambda(0)\\
         c\exp(-c'\alpha), & \lambda=1, &\alpha\to\infty \\
         c\alpha_\kappa, & 0<\lambda<1, &\alpha\to\infty
     \end{array}
     \right.
  \label{Eq:Left-sided:falpha:case1}
\end{equation}
where $c$ and $c'$ are positive constants depending on $\lambda$ and $\kappa=\max\{2,\lambda/(\lambda-1)\}$ \cite{Mandelbrot-1990b-PA,Mandelbrot-Evertsz-Hayakawa-1990-PRA}. Hence, $f(\alpha)$ is a monotonically increasing function of $\alpha$ with $f_{\max}=1$. Numerically, we have to adopt finite approximations of the measures and the resulting multifractal spectra are left-sided and converge gradually to the exact one-sided expression. Obviously, the inverse multifractal spectra are right-sided (see Section \ref{S3:MF-PF:inverse} for inverse measures).

For empirical multifractal spectra, the skewness can be defined analogously to probability distributions. A negatively skewed spectrum has a longer or fatter left tail, which is also said to be left skewed or skewed to the left. A positively skewed spectrum has a longer or fatter right tail, which is also said to be right skewed or skewed to the right. A simple measure of spectrum skewness is the ratio of the slope of the left part of $f(\alpha)$ to the absolute slope of the right part of $f(\alpha)$ \cite{Shimizu-Thurner-Ehrenberger-2002-Fractals}. The whole spectrum is left skewed if the ratio is larger than 1. A more natural measure of skewness is \cite{MunozDiosdado-RioCorrea-2006-IEEEconf}
\begin{equation}
  skew = \frac{\alpha_{\max}-\alpha(0)}{\alpha(0)-\alpha_{\min}}
  \label{hwgqqqfvca}
\end{equation}
where $skew = 1$ for un-skewed spectra, $skew > 1$ for right-skewed spectra, and $skew < 1$ for left-skewed spectra. An un-skewed spectrum is not necessarily symmetric. This measure has also been adopted in empirical analyses \cite{Stosic-Stosic-Stosic-Stanley-2015a-PA,Mali-Mukhopadhyay-2015-PS}. In empirical applications, we suggest to impose $q_{\max}=q_{\min}$ for the estimation
of $skew$ in expression (\ref{hwgqqqfvca}). It is argued that left-skewed multifractals have more fine-structures in large fluctuations, while right-skewed multifractals have more fine-structures in small fluctuations. Skewed (or asymmetric) multifractal spectra are observed in many empirical financial systems, in which the spectra of financial returns are more likely to be left-skewed while intertrade durations develop quite systematically a right-skewed asymmetry in multifractality \cite{Drozdz-Kwapien-Oswiecimka-Rak-2010-NJP,Drozdz-Oswiecimka-2015-PRE}. It is not unexpected that there are always exceptions in social sciences \cite{Oswiecimka-Kwapien-Drozdz-2005-PA,Ruan-Zhou-2011-PA}.

\subsection{Superposed components and crossover phenomena}
\label{S2:SuperposedTrend}

Crossover phenomena in the scaling behavior of financial time series are ubiquitous \cite{Fisher-Calvet-Mandelbrot-1997-SSRN,Liu-Cizeau-Meyer-Peng-Stanley-1997-PA,Janosi-Janecsko-Kondor-1999-PA,GrauCarles-2001-PA,Jiang-Ma-Cai-2007-PA,Jiang-Zhou-2008b-PA,Ma-2009-CTP,Yue-Xu-Chen-Xiong-Zhou-2017-Fractals}. In turbulent flows, there are well-defined inertial range and viscous range \cite{Frisch-1996}. Financial markets are probably analogous to turbulence in certain sense \cite{Ghashghaie-Breymann-Peinke-Talkner-Dodge-1996-Nature,Mantegna-Stanley-1996-Nature,Arneodo-Bouchaud-Cont-Muzy-Potters-Sornette-1996-XXX}. However, financial frictions are hardly the causes of such crossover phenomena because the crossover scales are usually much larger than the ``friction scale''. Instead, there are mechanical interpretations that superposed time series can induce crossovers. The effects of trends have been studied 20 years ago \cite{Teverovsky-Taqqu-1997-JTSA}. Kantelhardt et al. performed an extensive numerical analysis on the effects of polynomial and periodic trends on the scaling behavior of DFA \cite{Kantelhardt-KoscielnyBunde-Rego-Havlin-Bunde-2001-PA}.

\subsubsection{Superposition rule}

Assume that an additive trend $u(t)$ is added to a long-range correlated time series $X(t)$ to form a new time series
\begin{equation}\label{Eq:DFA:x:hg}
   z(t) = X(t)+u(t)
\end{equation}
where $X(t)$ exhibits mono-scaling behavior without crossovers
\begin{equation}
    F_x(s) = b_0s^H.
    \label{Eq:Fx:scaling}
\end{equation}

If the detrended residuals of $X(t)$ and $u(t)$ are uncorrelated, there is a superposition law in the fluctuation behavior for DFA \cite{Hu-Ivanov-Chen-Carpena-Stanley-2001-PRE} and DMA \cite{Shao-Gu-Jiang-Zhou-2015-Fractals},
\begin{equation}
    F^2_z(s) = F^2_x(s) + F^2_u(s),
    \label{Eq:DFA:DMA:F2x:F2hF2g}
\end{equation}
which holds for MF-DFA and MF-DMA. The crossover scale $s_\times$ can be determined by the following equation:
\begin{equation}
  F_x(s_\times)=F_u(s_\times).
  \label{Eq:DFA:DMA:Crossover:Condition}
\end{equation}
In many cases, two power laws appear respectively for $s<s_\times$ and  $s>s_\times$. More generally, the appearance of crossover phenomena can be viewed as the results of the competition between multiple mechanisms, be they endogenous or exogenous, which produce power-law correlated properties \cite{Echeverria-Rodriguez-AguilarCornejo-AlvarezRamirez-2016-PA}.

\subsubsection{Polynomial trends}

In a seminal work, Hu et al. studied analytically and numerically the effects of superposed polynomial trends \cite{Hu-Ivanov-Chen-Carpena-Stanley-2001-PRE}. Consider the following polynomial trends added to the increments series:
\begin{equation}
  u(t)=\sum_{p=0}^{m} a_{p} t^{p}
  \label{Eq:polyu_def}
\end{equation}
For DFA-$\ell$, where $\ell$ is the order of the polynomial detrending function, if $\ell\geqslant{p+1}$, the superposed polynomial trend $u(t)$ is ``absorbed'' by the detrending function $\tilde{X}(t)$. Hence, $u(t)$ has no effects on the scaling behavior of $X(t)$ and there is no crossover phenomenon.

For linear superposed trend $u(t)=a_1t$, its detrended fluctuation function is
\begin{equation}\label{Eq:DFA:F2g:L}
    F_u(s) = k_0a_1s^{2},
\end{equation}
where $k_0=\sqrt{5}/60$. According to Eq.~(\ref{Eq:DFA:DMA:Crossover:Condition}), we have
\begin{equation}
  s_\times =\left(a_1 \frac{k_0}{b_0}\right)^{1/(H-2)},
  \label{Eq:DFA:sx:L}
\end{equation}
showing that the crossover scale $s_\times$ depends only on $a_1$ in a power-law form with a power-law exponent $1/(H-2)$. When $s\ll{s}_\times$, $F_x(s)$ dominates and the scaling exponent of $z(t)$ is $H$. When $s\gg{s}_\times$, $F_u(s)$ dominates and the scaling exponent of $z(t)$ is $2$.

For quadratic trends $u(t) = a_2 t^2$, the detrended fluctuation function is
\begin{equation}
    F_u(s) = k_0Na_2s^{3},
    \label{Eq:DFA:F2g:Q}
\end{equation}
where $k_0=\sqrt{15}/90$. The crossover scale is
\begin{equation}
    s_\times= \left(\frac{k_0}{b_0}Na_2\right)^{1/(H-3)}.
    \label{Eq:DFA:sx:Q}
\end{equation}
The intrinsic scaling behavior is observed for $s\ll{s}_\times$. It is found that $s_\times$ depends not only on $a_1$, but also on the length of the time series $N$. Since $H<1$, $s_\times$ is smaller for longer time series. In this case, the scaling range becomes narrower and it is more difficult to estimate $H$.

The effects of polynomial trends on DMA are different \cite{Shao-Gu-Jiang-Zhou-2015-Fractals}. For a constant trend with $p=0$, when $\theta=0.5$, $F_u(s)=0$ and there is no crossover. Recall that $\theta$ is defined in Eq.~(\ref{Eq:MF-DMA:trend}). When $\theta=0$ and $\theta=1$, we have
\begin{equation}
  F_{u} (s)= \frac{a_0(s-1)}{2}.
  \label{Eq:DMA:SuperposedTrend:p=0:Fu:theta:01}
\end{equation}
There is a crossover at
\begin{equation}
  s_{\times}=\left(\frac{2b_0}{a_0} \right)^{1/(1-H)}.
  \label{Eq:DMA:SuperposedTrend:p=0:sx:theta:01}
\end{equation}
which shows that the crossover scale $s_{\times}$ is a power-law function of $a_0$ with exponent $-1/(1-H)$.

For linear trends with $p=1$, when $1\ll s \ll N$, $F^2(s)$ behaves as
\begin{equation}
  F_{u} (s)\approx \left\{
  \begin{aligned}
  &\frac{a_1s^2}{24}, & \theta=0.5\\
  &\frac{a_1Ns}{\sqrt{12}}, & \theta=0 ~~{\mathrm{or}}~~ 1
  \end{aligned}
  \right.
  \label{Eq:DMA:SuperposedTrend:p=1:Fu}
\end{equation}
It follows that
\begin{equation}
  s_{\times}\approx \left\{
  \begin{aligned}
  &\left(\frac{24b}{a_1}\right)^{{1}/{(2-H)}}, & \theta=0.5\\
  &\left(\frac{\sqrt{12}b}{a_1N}\right)^{{1}/{(1-H)}}, & \theta=0 ~~{\mathrm{or}}~~ 1
  \end{aligned}
  \right.
  \label{Eq:DMA:SuperposedTrend:p=1:sx}
\end{equation}
When a crossover appears, the left part of the fluctuation function with $s<s_{\times}$ reflects the behavior of $X(t)$, while the right part with $s>s_{\times}$ is dominated by the polynomial trends.

\subsubsection{Periodic trends}

Montanari et al. performed numerical investigations showing that the presence of periodicity in a time series may have an influence on the estimation of the Hurst exponent $H$ and some estimators falsely detect the presence of long-range dependence \cite{Montanari-Taqqu-Teverovsky-1999-MCM}. For DFA, Hu et al. studied the effects of sinusoidal trends and found that the $F_z(s)$ function has three crossovers and exhibits a hump in the model that reflects the effects of the sinusoidal trend \cite{Hu-Ivanov-Chen-Carpena-Stanley-2001-PRE}.

For a sinusoidal trend
\begin{equation}
  u(t) = a_S \sin(2\pi{t}/T),
  \label{Eq:DFA:SuperposedTrend:sinusoidal}
\end{equation}
$F_u$ possesses two power laws with a crossover scale $s_{2\times}$.
\begin{subequations}
When $s\ll{s}_{2\times}$, we have
\begin{equation}
  F_{u1}(s) = k_1a_Ss^2/T.
  \label{Eq:DFA:SuperposedTrend:sinusoidal:Fu1}
\end{equation}
When $s\gg{s}_{2\times}$, we have
\begin{equation}
  F_{u2}(s) = k_2a_ST \sim s^0.
  \label{Eq:DFA:SuperposedTrend:sinusoidal:Fu2}
\end{equation}
\end{subequations}
It follows that
\begin{equation}
  s_{2\times} \approx T.
  \label{Eq:DFA:SuperposedTrend:sinusoidal:sx2}
\end{equation}

When $s<s_{2\times}$, the competition of $F_{u1}(s)$ and $F_x(s)$ produces a crossover at
\begin{equation}
  s_{1\times} = \left(\frac{b_0}{k_1}\frac{T}{a_S}\right)^{1/(2-H)}.
  \label{Eq:DFA:SuperposedTrend:sinusoidal:sx1}
\end{equation}
When $s>s_{2\times}$, the competition of $F_{u2}(s)$ and $F_x(s)$ introduces a third crossover at
\begin{equation}
  s_{3\times} = \left(\frac{k_2}{b_0}{T}{a_S}\right)^{1/H}.
  \label{Eq:DFA:SuperposedTrend:sinusoidal:sx3}
\end{equation}
Therefore, there are four power-law scaling regions: (1) When $s<s_{1\times}$, $F_{x}$ dominates and the scaling exponent is $H$; (2) When $s_{1\times}<s<s_{2\times}$, $F_{u1}$ dominates and the scaling exponent is $\alpha_S=2$; (3) When $s_{2\times}<s<s_{3\times}$, $F_{u2}$ dominates as a constant and the scaling exponent is $0$; and (4) When $s>s_{3\times}$, $F_{x}$ dominates and the scaling exponent is $H$.

Note that all the three crossover scales are independent of $N$. When the time series is short, the fourth scaling region might disappear. The first scaling region is usually short and is not reliable for estimating $H$. If we want to estimate $H$ based on the first and fourth regions, we need to have very long time series. These properties also applies for multifractal time series. The difficulty to have a suitable scaling range often leads to improperly selected scaling ranges, which makes fail the detection of the intrinsic multifractal spectrum \cite{Ludescher-Bogachev-Kantelhardt-Schumann-Bunde-2011-PA}.

\subsubsection{Power-law trends}

Hu et al. also investigated the effect of power-law trends \cite{Hu-Ivanov-Chen-Carpena-Stanley-2001-PRE}
\begin{equation}
  u(t) = a_P t^\lambda,
  \label{Eq:DFA:SuperposedTrend:PL}
\end{equation}
whose detrended fluctuation function reads
\begin{equation}
    F_u(s) \sim  s^{\alpha(\lambda)} N^{\gamma(\lambda)}
  \label{Eq:DFA:SuperposedTrend:PL:Fu}
\end{equation}
The function $\alpha(\lambda)$ has the following form
\begin{equation}
\alpha(\lambda)=\left\{
\begin{array}{lll}
    \ell+1,      &{\rm{for}}~~ \lambda>\ell-0.5\\
    \lambda+1.5, &{\rm{for}}~~ -1.5\leqslant\lambda\leqslant\ell-0.5\\
    0,           &{\rm{for}}~~ \lambda<-1.5
\end{array}
  \label{Eq:DFA:SuperposedTrend:PL:alpha:lambda}
\right..
\end{equation}
where $\ell$ is the order of the detrending polynomial function (\ref{Eq:MF:Trend:Polynomial}),
and the function $\gamma(\lambda)$ is
\begin{equation}
  \gamma(\lambda)=\left\{
  \begin{array}{lll}
    \lambda-\ell, &{\rm{for}}~~ \lambda>\ell-0.5\\
    -0.5,         &{\rm{for}}~~ \lambda\leqslant\ell-0.5
  \end{array}
  \right..
  \label{Eq:DFA:SuperposedTrend:PL:gamma:lambda}
\end{equation}
It follows that the crossover scale is approximately
\begin{equation}
  s_\times \approx \left[a_PN^{\gamma(\lambda)}\right]^{1/[H-\alpha(\lambda)]}.
  \label{Eq:DFA:SuperposedTrend:PL:sx}
\end{equation}
Obviously, the scaling regions depend strongly on $\ell$, as well as on $a_P$, $N$ and $\lambda$.

\subsubsection{Additive noises}

When $u(t)$ is a fractional Gaussian noise with Hurst index $H_u$, the detrended fluctuation function is
\begin{equation}
  F_u(s) = a_0 s^{H_u}~.
  \label{Eq:DFA:SuperposedTrend:fGn:Fu}
\end{equation}
Combining with Eq.~(\ref{Eq:Fx:scaling}), we obtain
\begin{equation}
  s_\times \approx \left(\frac{a_0}{b_0}\right)^{1/(H-H_u)}.
  \label{Eq:DFA:SuperposedTrend:fGn:sx}
\end{equation}
Without loss of generality, we assume that $H>H_u$. Hence, when $s<s_{\times}$, $F_u(s)$ dominates. To correctly identify the intrinsic correlation property of $X(t)$, the scaling range should be selected to the right of $s_{\times}$. Numerical studies on this case with $X(t)$ being a monofractal or a multifractal time series show the important of correctly identifying the crossover scales and the scaling range as well \cite{Ludescher-Bogachev-Kantelhardt-Schumann-Bunde-2011-PA,Gulich-Zunino-2012-PA}. When $a_0$ is sufficiently small, $u(t)$ has no effect on the scaling behavior of $X(t)$. Consider the specific case where $H-H_u=0.3$. If the magnitude of noise $u(t)$ increases 10 times, the crossover $s_\times$ will increase about 2000 times! This implies that the crossover scales are not easy to observe \cite{Ludescher-Bogachev-Kantelhardt-Schumann-Bunde-2011-PA,Gulich-Zunino-2012-PA}.

\subsubsection{Preprocessing time series with trends}
\label{S2:Preprocess}

Many high-frequency financial time series have daily cycles known as intraday patterns that occur in returns \cite{Wood-McInish-Ord-1985-JF,Harris-1986-JFE,Cornell-Schwarz-Szakmary-1995-JBF}, trading volumes
\cite{Jain-Joh-1988-JFQA,Admati-Pfleiderer-1988-RFS,Gopikrishnan-Plerou-Gabaix-Stanley-2000-PRE,Mu-Chen-Kertesz-Zhou-2010-PP}, volatility \cite{Wood-McInish-Ord-1985-JF,Andersen-Bollerslev-1997-JEF,Liu-Gopikrishnan-Cizeau-Meyer-Peng-Stanley-1999-PRE}, bid-ask spread \cite{Mcinish-Wood-1992-JF,Chan-Christie-Schultz-1995-JB,Gu-Chen-Zhou-2007-EPJB,Ni-Zhou-2009-JKPS}, intertrade duration \cite{Engle-Russell-1998-Em,Hafner-2005-QF,Anatolyev-Shakin-2007-AFE,Jiang-Chen-Zhou-2009-PA,Ruan-Zhou-2011-PA}, and so on. Hence, one can preprocess the raw time series by removing the intraday patterns or cycles and then perform multifractal analysis on the cycle-adjusted time series.

The widely used preprocessing method in econophysics is to remove the intraday patterns (or more generally seasonal patterns). Suppose that $\{Y(i): i=1, 2, \cdots, N_DK\}$ is a high-frequency time series evenly sampled on $N_D$ trading days, where $K$ is the number of points in each day. The intraday pattern is determined by averaging the values at each moment over different trading days:
\begin{equation}
  \overline{Y}(k) = \frac{1}{N_D} \sum_{n=1}^{N_D} Y((n-1)K+k),
  \label{Eq:MF:Detrend:IntradayPattern}
\end{equation}
where $k=1, 2, \cdots, K$. The adjusted data are obtained by removing the intraday pattern from the raw data \cite{Liu-Gopikrishnan-Cizeau-Meyer-Peng-Stanley-1999-PRE}:
\begin{equation}
  X((n-1)K+k) = Y((n-1)K+k)/\overline{Y}(k),
  \label{Eq:MF:Detrended:IntradayPattern}
\end{equation}
where $n=1, 2, \cdots, N_D$. For bid-ask spread \cite{Gu-Chen-Zhou-2007-EPJB}, intertrade durations \cite{Jiang-Chen-Zhou-2009-PA} and trading volume \cite{Mu-Chen-Kertesz-Zhou-2010-PP}, no much differences are observed between the scaling results of raw time series and adjusted time series. If some $\overline{Y}(k)$ is very small, we may adopt the following procedure:
\begin{equation}
  X((n-1)K+k) = Y((n-1)K+k) - \overline{Y}(k)
  \label{Eq:MF:Detrended:IntradayPattern:Minus}
\end{equation}
for all the values of $n$ and $k$.

Hu et al. proposed to adopt an adaptive detrending on the original time series and then perform MF-DFA on the detrended time series \cite{Hu-Gao-Wang-2009-JSM}. The adaptive trend is determined as follows. The raw time series is partitioned into segments of length $2n+1$, where any two successive segments overlap by $n+1$ points and the length of the last segment may be smaller than $2n+1$. For each segment $S_v$, we fit a polynomial of order $\ell$, denoted $y_v(j)$ with $j=1, 2, \cdots, 2n+1$. The adaptive trend for the overlapped region can be determined as
\begin{equation}
  y^{(a)}_v(j) = \left(1-\frac{j-1}{n}\right)y_v(n+j) + \frac{j-1}{n}y_{v+1}(j).
\end{equation}
for $j=1, 2, \cdots, n+1$. The whole adaptive trend is
\begin{equation}
  \left\{
\begin{array}{lll}
  y^{(a)}(j) &= y_1(j), & j=1, 2, \cdots, n \\
  y^{(a)}(nv+j) &= y_v^{(a)}(j), & v = 1, 2, \cdots, V ~~{\mathrm{and}}~~ j=1, 2, \cdots, n \\
  y^{(a)}(nv+j) &= y_{V+1}(j), & j=1, 2, \cdots, N-nV \\
\end{array}
  \right.
  \label{eq:xdef}
\end{equation}
where $V={\mathrm{int}}[N/n]$. The resulting time series for further multifractal analysis is
\begin{equation}
  y(i) = Y(i) - y^{(a)}(i), ~~i=1, 2, \cdots, N.
\end{equation}
The adaptive trend does not have any jumps or discontinuities. However, this preprocessing does not eliminate the jumps in the trend function $\widetilde{Y}(t)$ in MF-DFA. We stress that $n$ is irrelevant to the box size $s$ and $\ell$ is irrelevant to $l$ in Eq.~(\ref{Eq:MF:Trend:Polynomial}). Both parameters $n$ and $\ell$ are suggested to be determined by requiring that the variance of the detrended data $y(i)$ decays slowly when $\ell$ is further increased and/or $2n+1$ is further decreased \cite{Hu-Gao-Wang-2009-JSM}. Ferreira et al. used adaptive multifractal approaches (A-MF-DFA and A-MF-X-DFA) to investigate the daily indices time series of 48 stock markets from January 1995 to February 2014 \cite{Ferreira-Dionisio-Movahed-2017-PA}. In their results, the estimated Hurst indices $h_{xx}$, $h_{yy}$ and $h_{xy}$ are all larger than 1, implying that they cumulatively sum the index before removing the local trends.

There are other detrending methods designed for different types of trends, such as wavelet transform \cite{Gencay-Selcuk-Whitcher-2001b-PA}, singular-value decomposition \cite{Nagarajan-Kavasseri-2005-CSF,Nagarajan-Kavasseri-2005-PA}, chaotic singular-value decomposition \cite{Shang-Lin-Liu-2009-PA}, Fourier transform \cite{Nagarajan-Kavasseri-2005-IJBC,Chianca-Ticona-Penna-2005-PA,Zhao-Shang-Lin-Chen-2011-PA,Huang-Shang-2014-FNL}, Laplace transform \cite{Xu-Shang-Kamae-2009-CSF,Lin-Shang-2011-Fractals}, Orthogonal V-system smoothing \cite{Lin-Shang-Ma-2011-FNL}, EMD detrending \cite{Zhou-Leung-2010b-JSM,Zhao-Shang-Zhao-Wang-Tao-2012-CSF,Dong-Gao-Wang-2013-MPE}, data-driven detrending with echo state networks \cite{Maiorino-Bianchi-Livi-Rizzi-Sadeghian-2017-IS}, and so on. Essentially, any method that is designed for trend detection can be used to detrend the raw time series. A careful investigation should compare the multifractal analysis results between raw and detrended time series. It is reported that the method of removing seasonal patterns performs better than the Fourier filtering method and the adaptive detrending method in detrending a time series with regular periods or trends \cite{Zhang-Zhou-Singh-Chen-2011-JH,Zhang-Zhou-Singh-2014-HP}.

\subsection{Effects of nonlinearity and filters}

\subsubsection{Effects of nonlinearity}

Chen et al. studied the effects of non-stationarity in time series present in three ways on the outcomes of detrended fluctuation analysis \cite{Chen-Ivanov-Hu-Stanley-2002-PRE}. The first type of non-stationarity is introduced by a ``cutting'' procedure which removes segments of a given length from a time series and stitches together the remaining parts. The cutting procedure does not have any influences on the scaling behavior of persistent time series ($H>0.5$), but can warp the $F(s)$ function upwards.

The second type of non-stationarity is caused by segments with different properties, such as different standard deviations or different correlation exponents. For non-stationary persistent time series comprised of segments with two different values of the standard deviation, its scaling behavior remains the same as stationary persistent time series with constant standard deviation. For non-stationary antipersistent time series, a crossover emerges in the scaling behavior of $F(s)$. For non-stationary time series with different local correlations, the scaling behavior also changes.

The third type of non-stationarity is realized by adding to a time series a tunable concentration of random outliers or spikes with different amplitudes $a_{sp}$. The detrended fluctuation function of correlated spikes behaves as
\begin{equation}
  F_{u}(s) = k_0\sqrt{p}a_{sp}  s^\alpha,
  \label{Eq:DFA:SuperposedTrend:non-stationarity:Fu}
\end{equation}
where $k_0$ is a constant, $p$ is the concentration of spikes, and $\alpha$ is the scaling exponent to estimate. It follows that
\begin{equation}
  s_{\times} = \left(\sqrt{p}a_{sp}\frac{k_0}{b_0}\right)^{1/(H-\alpha)}.
  \label{Eq:DFA:SuperposedTrend:non-stationarity:sx}
\end{equation}
The intrinsic scaling behavior dominates for $s\gg{s_{\times}}$.

It is worth mentioning an interesting work that investigates how extreme data loss affects the scaling behavior of long-range power-law correlated and anticorrelated signals \cite{Ma-Bartsch-BernaolaGalvan-Yoneyama-Ivanov-2010-PRE}. It is found that the global scaling of persistent time series is robust to extreme data loss and the scaling exponent remains practically unchanged even for extreme data loss of up to 90\%. In contrast, the global scaling of antipersistent time series is vulnerable to extreme data loss and changes to uncorrelated behavior even with a very small fraction of extreme data loss.

\subsubsection{Effects of filters}

Chen et al. studied the effects of filters on DFA \cite{Chen-Hu-Carpena-BernaolaGalvan-Stanley-Ivanov-2005-PRE}. Investigating the effects of filters is to compare the scaling behaviors of the raw time series $X(t)$ and its transformed time series $g(t)=g(X(t))$. As expected, it is found that linear filters $g(t) = a_1X(t)+a_0$ do not change the correlation properties. In contrast, the effects of nonlinear power law $g(t) = aX^k(t)$ and logarithmic $g(t) = \ln[X(t)+\epsilon]$ filters strongly depends on $H$, $k$ and $\epsilon$. No obvious crossovers can be identified.

Wang et al. studied the effects of linear and nonlinear filters on multifractal analysis of binomial multifractal measures \cite{Wang-Shang-Dong-2013-AMC}. For linear filter, $g(t) = 2X(t)+1$, where $a$ is a constant, has the same multifractal properties as $X(t)$. When quadratic and cubic filters are applied, MF-PF and MF-DFA show that the filtered time series $g(t) = X^2(t)$ and $g(t) = X^3(t)$ exhibit multifractality, in which $\alpha_{\min}$ becomes smaller and $\alpha_{\max}$ becomes larger with the increase of the filter order. For logarithmic filters $g(t) = \ln[X(t)+\epsilon]$, with the increase of $\epsilon$, $\alpha_{\max}$ remains almost the same and $\alpha_{\min}$ decreases.

Song and Shang investigated numerically the effects of linear and nonlinear filters on MF-X-DFA \cite{Song-Shang-2011-Fractals}. They found that the linear filter $g(t) = a_1X(t) + a_0$ has no effect on the multifractal cross-correlation properties. In contrast, nonlinear filters including nonlinear polynomial filter $g(t)= aX^k(t)$, logarithmic filter $g(t) = \ln[X(t) + \epsilon]$, exponential filter $g(t) = \exp[a_1X(t) + a_0]$ and power-law filter $g(t)= [X(t) + a]^k$ have significant influence on the multifractal cross-correlation behavior.

Note that, although linear filters has no effect on DFA, MF-DFA and MF-X-DFA, they should affect the scaling behavior of DMA, MF-DMA and MF-X-DMA with $\theta\neq0.5$ due to the breaking of translation invariance in DMA.

In connection with section \ref{wyehtjyehw}, applying an exponential filter onto a long-memory process (for instance
a Hurst process) leads to multifractality. In other words, the mechanism of section \ref{wyehtjyehw} can be re-interpreted
in the context of the study of the impact of filters are creating multifractility from an initial time series that is simply
mono-fractal. This phenomenon was noticed early in Refs.~\cite{Kadanoff-1986,Halsey-Jensen-Kadanoff-Procaccia-Shraiman-1986-PRA}.

\subsection{Performance of different methods}

\subsubsection{Mathematical models as reference}
\label{S3:Models:for:Tests}

In order to compare the performance of different multifractal analysis methods, one usually applies them to mathematical models with known multifractal nature. The first class of reference models are monofractal, such as fractional Brownian motions \cite{Mandelbrot-VanNess-1968-SIAMR}, autoregressive fractionally integrated moving average (ARFIMA) processes \cite{Granger-Joyeux-1980-JTSA,Hosking-1981-Bm}. The second class of reference models are bifractal, such as ordinary L{\'e}vy processes \cite{Jaffard-1999-PTRF}, exponentially truncated L{\'e}vy processes \cite{Nakao-2000-PLA}, and power-law tailed time series \cite{Chechkin-Gonchar-2000-CSF} (see Section~\ref{S3:MF:Sources:SpuriousMF&FSE:Bifractal}). The third class of reference models have multifractal nature, including Multiplicative cascade models, MMAR models, MSM models, and MRW models (see Section~\ref{S1:Models}).

For joint multifractal analysis, one can use bivariate fractional Brownian motions (BFBMs) \cite{Lavancier-Philippe-Surgailis-2009-SPL,Coeurjolly-Amblard-Achard-2010-EUSIPCO,Amblard-Coeurjolly-Lavancier-Philippe-2013-BSMF}, two-component ARFIMA processes \cite{Podobnik-Horvatic-Ng-Stanley-Ivanov-2008-PA,Podobnik-Stanley-2008-PRL}, and two multiplicative measures \cite{Zhou-2008-PRE,Jiang-Zhou-2011-PRE}. These models can be simply extended for multivariate time series.

\subsubsection{Performance measures}

Let $H(q)$, $\tau(q)$ and $f(\alpha)$ denote the true functions representing the multifractal properties of a given system. For each mathematical model, if it is stochastic, we generate $K$ realizations. Using a certain multifractal analysis method, we obtain $H_k(q)$, $\tau_k(q)$ and $f_k(\alpha)$ for the $k$-th realization. The mean estimates are
\begin{equation}
  {\hat{H}}(q) = \frac{1}{K}\sum_{k=1}^K {\hat{H}}_k(q), ~~
  {\hat{\tau}}(q) = \frac{1}{K}\sum_{k=1}^K {\hat{\tau}}_k(q), ~~{\rm{and}}~~
  {\hat{f}}(\alpha) = \frac{1}{K}\sum_{k=1}^K {\hat{f}}_k(\alpha),
\end{equation}
where the averages are performed at fixed $q$. Their corresponding variances are
\begin{equation}
  \sigma_{\hat{H}(q)}^2 = \frac{1}{K-1}\sum_{k=1}^K \left[{\hat{H}}_k(q)-{\hat{H}}(q)\right]^2, ~~
  \sigma_{\hat{\tau}(q)}^2 = \frac{1}{K-1}\sum_{k=1}^K \left[{\hat{\tau}}_k(q)-{\hat{\tau}}(q)\right]^2, ~~{\rm{and}}~~
  \sigma_{\hat{f}(\alpha)}^2 = \frac{1}{K-1}\sum_{k=1}^K \left[{\hat{f}}_k(\alpha)-{\hat{f}}(\alpha)\right]^2,
\end{equation}
which can be used for measuring the estimation precision.

The performance of a multifractal analysis method is mainly measured by the absolute deviations of the estimates to the true values
\begin{equation}
  \sigma_{\hat{H}(q)}^2 = \frac{1}{K-1}\sum_{k=1}^K \left[{\hat{H}}_k(q)-H(q)\right]^2, ~~
  \sigma_{\hat{\tau}(q)}^2 = \frac{1}{K-1}\sum_{k=1}^K \left[{\hat{\tau}}_k(q)- \tau(q)\right]^2, ~~{\rm{and}}~~
  \sigma_{\hat{f}(\alpha)}^2 = \frac{1}{K-1}\sum_{k=1}^K \left[{\hat{f}}_k(\alpha)-f(\alpha)\right]^2,
\end{equation}
or the relative errors
\begin{equation}
  e_{\hat{H}(q)}    = \frac{\sigma_{\hat{H}(q)}}{H(q)}, ~~
  e_{\hat{\tau}(q)} = \frac{\sigma_{\hat{\tau}(q)}}{\tau(q)}, ~~{\rm{and}}~~
  e_{\hat{f}(\alpha)} = \frac{\sigma_{\hat{f}(\alpha)}}{f(\alpha)}.
\end{equation}
The absolute deviations or the relative errors allow us to identify which method performs relatively better for different $q$ values. These measures can be adopted for estimation accuracy. Certainly, the simplest one is to measure the differences between the means of estimates and the theoretical values.

To qualify the total performance of a method, we suggest to use the average errors:
\begin{equation}
  \bar{e}_{\hat{H}} = \langle{e_{\hat{H}(q)}}\rangle_q, ~~
  \bar{e}_{\hat{\tau}} = \langle{e_{\hat{\tau}(q)}}\rangle_q, ~~{\rm{and}}~~
  \bar{e}_{\hat{f}} = \langle{e_{\hat{f}(\alpha)}}\rangle_q,
\end{equation}
where $\langle{}\rangle_q$ denotes the average over all $q$ values. We argue that the averages of the absolute deviations are not suitable measures of method performance. There are certainly other performance measures. For instance, computational complexity is often used as a measure of algorithm performance. However, this is not a crucial issue nowadays. The width of the scaling range is also an important measure, which however should be used only when there is sufficient estimation accuracy.

%

\subsubsection{Length of time series (``finite size effects'')}

It can be expected that very short time series may produce ``wrong'' estimates, because the estimated multifractal properties usually deviate more for shorter time series, as confirmed with the log-Poisson binomial, log-Gamma binomial and log-normal binomial measures \cite{Lopez-Contreras-2013-PRE}. There is also an evident finite-size effect for the partition function approach \cite{Zhou-2012-CSF}.

The inpacts of time series length can be quantified at least numerically.
Weron studied numerically the confidence interval of the estimated DFA exponent of white Gaussian noise \cite{Weron-2002-PA}. He found that the confidence interval depends on the length $N=2^n$ of time series and on the scaling range. The confidence interval is wider for shorter time series. At the confidence level of 95\%, the confidence interval is
\begin{equation}
  \begin{array}{lll}
    &[0.5-25.79n^{-2.33}, 0.5+ 29.37n^{-2.46}] & {\mathrm{for}}~~ s>10,\\
    &[0.5-85.63n^{-2.93}, 0.5+ 217.0n^{-3.19}] & {\mathrm{for}}~~ s>50.
  \end{array}
\end{equation}
Similar expressions have been obtained numerically for 10 estimators \cite{Rea-Oxley-Reale-Brown-2013-MCS}.

\subsubsection{Incomplete comparison of performance}

When inventing a novel variant or novel method of multifractal analysis, we stress the importance of first
performing intensive tests of the method on {\it known} multifractal time series, as emphasised
in a slightly different context by Filimonov and Sornette \cite{Filimonov-Sornette-2012-EPJB} (see also the commentary at
\url{http://www.er.ethz.ch/media/essays/PNAS.html}). Indeed, applying a novel (and therefore
``unknown'') method on an unknown data set is likely to lead to incorrect conclusions, since
the novel phenomena that arise often result from all types of biases and distortions introduced
by the analysing method. Instead, when using a known multifractal time series, the deviation
from the theoretically known multifractal properties inform on the properties and biases introduced
by the novel analysing method.  In turn, this allows the researcher to separate the distortions
introduced by her analysis technique from the novel effects that she is hoping to discover.
Good scientific practice should require that this procedure be
performed without exception, preventing potentially many spurious claims to develop in the literature.

Following on this, there are efforts to compare different estimators for long-range dependence in time series. In several important studies, about 10 estimators including R/S analysis and DFA are compared and the (local) Whittle estimator is found to have the best performance in accuracy and precision \cite{Taqqu-Teverovsky-Willinger-1995-Fractals,Montanari-Taqqu-Teverovsky-1999-MCM,Rea-Oxley-Reale-Brown-2013-MCS}.
When the R/S analysis and DFA are compared, numerical evidence shows that DFA performs better in accuracy and precision for different lengths of time series \cite{Weron-2002-PA,Coronado-Carpena-2005-JBP}. It is also found that centered DMA and DFA are able to obtain accurate estimates \cite{Bashan-Bartsch-Kantelhardt-Havlin-2008-PA,Shao-Gu-Jiang-Zhou-Sornette-2012-SR}. However, if the form of the trend in the time series is not {\textit{a priori}} known or the long-memory scaling exponent is greater than 1 for colored noises, DFA is recommended \cite{Xu-Ivanov-Hu-Chen-Carbone-Stanley-2005-PRE,Bashan-Bartsch-Kantelhardt-Havlin-2008-PA}.

There are many variants of the classical multifractal analysis methods. However, the performance of many of them has not been validated with numerical experiments on synthetic tests and are thus open to the Filimonov-Sornette critic expressed above \cite{Filimonov-Sornette-2012-EPJB}. Most comparisons have been conducted on MF-DFA, MF-DMA, WTMM and MF-WL.

Kantelhardt et al. studied the multifractal properties of river discharge and precipitation time series using MF-DFA and WTMM and found that the results obtained are in good agreement within the error margins \cite{Kantelhardt-Rybski-Zschiegner-Braun-KoscielnyBunde-Livina-Havlin-Bunde-2003-PA}.
O{\'{s}}wi{\c{e}}cimka et al. performed multifractal analysis on financial time series using MF-DFA and WTMM and argued that MF-DFA is better for a global detection of multifractal behavior, while WTMM is the optimal tool for the local characterization of the scaling properties of signals \cite{Oswiecimka-Kwapien-Drozdz-Rak-2005-APPB}.
Figliola et al. compared the results of multifractal analysis of EEG time series from an epileptic seizure subject using MF-DFA and MF-WL and concluded that both methods are equivalent for estimating multifractal properties \cite{Figliola-Serrano-Paccosi-Rosenblatt-2010-IJBC}.
However, these comparisons are not conclusive because they are not obtained from extensive numerical experiments on time series with {\textit{a priori}} known multifractal properties. Again, the importance of performing extensive tests on synthetic time series cannot be over-emphasised.

Kantelhardt et al. found that, for deterministic multifractal binomial measures with additive linear and quadratic trends, both WTMM and MF-DFA are able to obtain nice detrending and MF-DFA have slight advantages in precision for negative $q$ values and short series \cite{Kantelhardt-Zschiegner-KoscielnyBunde-Havlin-Bunde-Stanley-2002-PA}.
O{\'{s}}wi{\c{e}}cimka et al. compared the performance of MF-DFA and WTMM using fractional Brownian motions, bifractal L{\'e}vy flights, and four sorts of multifractal binomial cascades \cite{Oswiecimka-Kwapien-Drozdz-2006-PRE}. For fractional Brownian motions, MF-DFA can correctly identify its monofractality, while WTMM often produces spurious multifractality. For bifractal L{\'e}vy flights, both methods produce spurious multifractality and MF-DFA performs relatively better. For deterministic binomial measures, MF-DFA can well capture the right part of the multifractal spectrum, but widens the left part. In contrast, WTMM can well captures the left part when a proper wavelet is adopted, but not the right part. In addition, the multifractal spectra obtained by MF-DFA with different detrending polynomials are almost the same, while those obtained by WTMM with different wavelets could be very different. Hence, MF-DFA performs relatively better than WTMM for deterministic binomial measures. MF-DFA outperforms WTMM remarkably for log-gamma and log-Poisson binomial measures. For log-normal binomial measures, the two methods perform comparably well. Murgu{\'i}a and Rosu confirmed that MF-DFA outperforms WTMM for deterministic binomial measures \cite{Murguia-Rosu-2012-PA}. Salat et al. found that MF-DFA outperforms MF-PF and WTMM for deterministic binomial measures \cite{Salat-Murcio-Arcaute-2017-PA}.

Based on log-normal multiplicative random wavelet cascades \cite{Arneodo-Bracy-Muzy-1998-JMP}, Jaffard et al. found that both WTMM and MF-WL produced the same multifractal spectrum \cite{Jaffard-Lashermes-Abry-2006}.
Turiel et al. tested the performance of MF-PF, WTMM, gradient modulus wavelet projection (GMWP) and gradient histogram (GH) and found that the GMWP method is the one attaining the best performance \cite{Turiel-PerezVicente-Grazzini-2006-JComputP}.
Serrano and Figliola compared the performance of MF-DFA and MF-WL based on Cantor sets and deterministic binomial measures \cite{Serrano-Figliola-2009-PA}. They found that MF-WL provides more accurate estimations for these two time series.

Gu and Zhou investigated the performance of MF-DFA and MF-DMA based on deterministic binomial measures \cite{Gu-Zhou-2010-PRE}. They found that the estimated multifractal scaling exponent $\tau(q)$ and the singularity spectrum $f(\alpha)$ are in good agreement with the theoretical values. In addition, the backward and forward MF-DMA methods have the best performance, MF-DFA is slightly worse especially for $q>0$, while the centered MF-DMA method has the worse performance. This conclusion is confirmed by Xi et al. \cite{Xi-Zhang-Xiong-Zhao-2015-APS}, who also found that the backward and forward MF-DMA methods are less sensitive to the length of deterministic binomial measures than MF-DFA and the centered MF-DMA is the most sensitive. Schumann and Kantelhardt found that both centered MF-DMA and MF-DFA can unveil nicely the multifractal nature of deterministic binomial measures \cite{Schumann-Kantelhardt-2011-PA}.

Huang et al. compared the performance of MF-HHSA, MF-SF, MF-DFA and MF-WL by analyzing fractional Brownian motions, stochastic multinomial measures and multifractal random walks \cite{Huang-Schmitt-Hermand-Gagne-Lu-Liu-2011-PRE}. For fractional Brownian motions, MF-SF and MF-WL provide a better estimation with narrower singularity width. For multifractal time series under investigation, it seems that the MF-HHSA and MF-DFA methods provide better singularity spectra than MF-SF and MF-WL.

Welter and Esquef compared the performance of MF-DFA, MF-DMA and EMD-DAMF in studying fractional Brownian motions, L{\'e}vy flights, multifractal random walks, and deterministic binomial measures \cite{Welter-Esquef-2013-PRE}.
For fBm processes, all the methods perform well and the estimate by EMD-DAMF is slightly more accurate than MF-DFA and MF-DMA, but with lower precision.
For L{\'e}vy processes, the EMD-DAMF outperforms the MF-DFA and MF-DMA methods, both in accuracy and precision.
For multifractal random walks, MF-DFA performs best both in accuracy and precision. MF-DMA and EMD-DAMF have similar accuracy and MF-DMA is more precise.
For deterministic binomial measures, MF-DFA and MF-DMA outperform EMD-DAMF, both in accuracy and precision, and MF-DMA is more accurate than MF-DFA with similar precision.


Comparative studies on the performance of methods for joint multifractal analysis are relatively rare. Jiang and Zhou compared MF-X-DFA and MF-X-DMA \cite{Jiang-Zhou-2011-PRE}. They found that, for bivariate fractional Brownian motions and two-component ARFIMA processes, the MF-X-DFA and centered MF-X-DMA algorithms outperform the forward and backward MF-X-DMA algorithms. For binomial measures, the forward MF-X-DMA algorithm exhibits the best performance, the centered MF-X-DMA algorithms performs worst.
Xi et al. obtained similar (but not the same) results for binomial measures \cite{Xi-Zhang-Xiong-Zhao-Yang-2017-PA}.
Cao and Shi found that, for two-component and mix-correlated ARFIMA processes, the backward and forward MF-X-DMA methods outperform centered MF-X-DMA, MF-X-DFA and MFDCCA- MODWT \cite{Cao-Shi-2018-CSF}.
Kristoufek performed an extensive Monte Carlo simulation study to compare the performance of detrended cross-correlation analysis (DCCA), height cross-correlation analysis (HXA) and detrending moving-average cross-correlation analysis (DMCA) \cite{Kristoufek-2015a-PA}. He found that DMCA strongly outperforms DCCA and HXA in the standard power-law cross-correlation case and the power-law coherency case. However, DCCA outperforms in the sense that it is less sensitive to the short-term memory.

\subsubsection{Which method should be used?}

It seems that there is no consensus on which method has the best performance for multifractal analysis. More numerical comparisons are required to gain a deeper understanding on the performance of different methods. First, more mathematical models with {\textit{a priori}} known multifractal properties should be investigated. Second, different performance criteria should be considered, such as estimation accuracy which quantifies the deviation of estimated to the true values, estimation precision which qualifies the dispersion of estimates to the mean values, size effect of time series, width of scaling ranges, and sensitivity to the choice of scaling ranges. Unfortunately, even if one is
quite convinced that a certain method performs the best for some tests of multifractal time series, we cannot extrapolate this conclusion directly to other time series.

In applications to real data, without an understanding of the underlying mechanisms that drive the dynamics of the complex system, it is hard to determine which method would be more suitable. If there is a trend that can be identified, MF-DFA with high-order detrending polynomials is recommended. In general, we suggest that different methods should be adopted such that we can compare their results qualitatively and quantitatively. For financial time series such time series of returns, we can check if the value of the Hurst index $H(2)$ is close to 0.5, based on the fundamental stylized fact of the absence of long memory in financial returns \cite{Cont-2001-QF}, which has also a strong theoretical underpinning in the concept of no-arbitrage opportunity. This can be used as a useful criterion to assess the results obtained from different methods. Indeed, there are works reporting incorrect results in the sense that the Hurst index $H(2)$ deviates significantly from 0.5.

\section{Empirical evidence of multifractal markets}
\label{S1:EmpAnal}

In the past three decades, hundreds of empirical studies have been carried out to investigate the presence of multifractality in financial markets. In the 1990's, there are only a few seminal papers, most of them applied the structure function approach \cite{Muller-Dacorogna-Olsen-Pictet-Schwarz-Morgenegg-1990-JBF,Ghashghaie-Breymann-Peinke-Talkner-Dodge-1996-Nature,Vandewalle-Ausloos-1998-EPJB,Mandelbrot-1999-SA,Ivanova-Ausloos-1999a-EPJB,Schmitt-Schertzer-Lovejoy-1999-ASMDA} or the box-counting method \cite{Vassilicos-Demos-Tata-1993}. With the development of new multifractal analysis methods, empirical studies boomed in the 21st century.

After a careful and thorough survey of the literature, we observe three trends. First, the structure function approach is widely applied in the early works and the multifractal detrended fluctuation analysis dominates later. Second, not surprisingly, the return time series of stock market indices attract most attention, while foreign exchange rates and commodities are also investigated by many researchers. Third, most financial time series studied have daily sampling frequency. The second and third trends are simply due to the availabilities of different data sets. We also find no clear evidence for the presence of universality in the multifractal properties found empirically.

There are also very few studies on economic variables other than financial variables, such as
the monthly aggregate price indices of the US consumer price index (CPI) and producer price index (PPI) from 1975 to 2011 \cite{Mulligan-2014-PA},
the daily spot rates of VLGCs on the benchmark Persian Gulf (Ras Tanura) to Japan (Chiba) route from 3 January 1992 to 24 June 2009 \cite{Engelen-Norouzzadeh-Dullaert-Rahmani-2011-EE},
the daily spot rates of ships in tanker markets from 27 January 1998 (or 1 July 2004 or 15 July 2005) to 5 August 2013 \cite{Zheng-Lan-2016-PA},
and
the daily Baltic Panamax index (BPI) and Baltic Capesize index (BCI) from 1 March 1999 to 26 February 2015 \cite{Chen-Tian-Ding-Miao-Lu-2016-PA}. 
It is not surprising that multifractal analysis is mainly carried out on financial time series because economic time series are usually recorded monthly or quarterly, which are not long enough for multifractal analysis. Therefore, there is only one study applying the MF-SF method on the monthly prices of NASDAQ for the period from 1984 to 2000 \cite{Bershadskii-2001-JPA}.

Based on these observations, we argue that it is of less significance to try to confirm again and again the presence of multifractality in financial index returns in future studies. The evidence is clear. More attention should be paid to other financial variables such as liquidity measures and to the comparison of multifractality between different assets. More importantly, one should try to unveil the financial mechanisms and implications of multifractality.

\subsection{Asset returns and prices}
\label{S2:EmpAnal:Return}

While the section is titled ``asset returns'', this is a slight misnomer as the investigated multifractal properties
refer mainly to those of the absolute values of the returns and, in general, not to the signed returns.
In this sense, since the absolute value of a return is a proxy for the volatility,  the investigated
multifractal properties are mainly those of the financial volatility of the various asset classes
enumerated below.  The subsection below dedicated specifically to volatility refers to studies that
have used the volatility estimated in different ways as their direct inputs. Indeed, there is no multifractality when the
signs are included. Either explicitly or implicitly, the return signs are removed in most multifractal analyses.

\subsubsection{Daily stock market indices}

The following researches deal with many stock market indices in single studies.
\begin{itemize}
\item Sel{\c{c}}uk investigated daily stock market indices from Argentina, Brazil, Hong Kong, Indonesia, Korea, Mexico, Philippines, Singapore, Taiwan and Turkey for 8-28 years ending at 29 December 2000 using MF-SF \cite{Selcuk-2004-PA},
\item Di Matteo et al. investigated 32 major indices of both developed and emerging markets from 1990 or 1993 to 2001 using MF-SF \cite{DiMatteo-Aste-Dacorogna-2005-JBF,DiMatteo-Aste-Dacorogna-2003-PA,DiMatteo-2007-QF},
\item Bianchi and Pianese investigated eight major stock indexes (DJIA from 1 October 1928, Bovespa from 27 April 1993, HSI from 31 December 1986, NIKKEI 225 from 4 January 1984, CAC40 from 1 March 1990, Footsie 100 from 2 April 1984, and MibTel from 19 July 1993) till October of 2004 using MF-SF \cite{Bianchi-Pianese-2007-QF},
\item Zunino et al. investigated Latin-American stock markets (Argentina, Brazil, Chile, Colombia, Mexico, Peru, Venezuela) and the US from January 1995 to February 2007 using MF-DFA \cite{Zunino-Figliola-Tabak-Perez-Garavaglia-Rosso-2009-CSF},
\item Gao et al. investigated seven indices (DJIA, S\&P500 and NASDAQ from 4 January 1986, DAX from 11 November 1990, FTSE 100 from 6 January 1986, SSEC from 4 January 2000, and HSI from 31 December 1986) till 6 September 2011 using MF-SF \cite{Gao-Cai-Wang-2012-JSM},
\item Trincado and Vindel investigated FTSE 100, CAC 40, DAX 30, IBEX 35, NASDAQ 100 NIKKEI 225, and DJIA from 1 January 1992 to 31 August 2009 using MF-SF \cite{Trincado-Vindel-2013-BJP},
\item Sensoy investigated 15 Middle East and North African (MENA) stock market indices from January 2007 to December 2012 using MF-SF \cite{Sensoy-2013-PA},
\item Lin et al. investigated DJIA from 22 January 1953 to 28 June 2012, S\&P 500 from 3 January 1950 to 13 June 2012, HSI from 31 December 1986 to 20 June 2012, FTSE from 2 April 1984 to 20 June 2012, CAC from 1 March 1990 to 3 July 2012, DAX from 26 November 1990 to 3 July 2012, SSEC from 19 December 1990 to 28 June 2012, and SZCI from 3 April 1991 to 28 June 2012 using MF-SF \cite{Lin-Chang-Li-2013-PRE},
\item Horta et al. investigated the stock market indices of Belgium, France, Greece, Japan, the Netherlands, Portugal, the UK and US from 4 January 1999 to 19 March 2013 using MF-DMA \cite{Horta-Lagoa-Martins-2014-IRFA},
\item Rizvi et al. investigated 11 developed markets (Australia, France, Germany, Hong Kong, Japan, Netherlands, Spain, Sweden, Switzerland, United Kingdom, United States) and 11 islamic markets (Bahrain, Bangladesh, Egypt, Indonesia,, Jordan, Kuwait, Malaysia, Oman, Pakistan, Saudi Arabia, Turkey) from 1 January 2001 till 31 December 2013 using MF-DFA \cite{Rizvi-Dewandaru-Bacha-Masih-2014-PA},
\item Stosic et al. investigated 13 stock market indices (AEX, BFX, BSESN, BVSP, FCHI, FTSE, GSPC, HSI, KS11, MXX, N225, SSMI, TWII) from 30 August 2006 to 1 March 2014 using MF-DFA \cite{Stosic-Stosic-Stosic-Stanley-2015b-PA},
\item Arshad et al. investigated the MSCI benchmark index of each individual market (Malaysia, Singapore, Indonesia and South Korea) from 1 January 1990 until 31 July 2013 using MF-DFA \cite{Arshad-Rizvi-2015-PA}, and
\item Ferreira et al. investigated the indices of 48 stock markets from January 1995 to February 2014 using A-MF-DFA \cite{Ferreira-Dionisio-Movahed-2017-PA}. 
\end{itemize}

Other studies focus on one or a few stock market indices over different periods. For the US market, a few indices have been studied, including
DJIA (from February 1991 to May 1997 using MF-SF \cite{Ivanova-Ausloos-1999a-EPJB}, from 1 January 1897 to 3 February 1999 using MF-SF \cite{LeBaron-2001-QF}, from 1900 to 2000 using WTMM \cite{Pavlov-Ebeling-Molgedey-Ziganshin-Anishchenko-2001-PA}, from 3 January 1928 to 18 October 2000 using MF-SF \cite{Andreadis-Serletis-2002-CSF}, from 1900 to 2005 using MF-SF \cite{Baldovin-Stella-2007-PNAS}, and from 1928 to 2012 using MF-DFA \cite{Green-Hanan-Heffernan-2014-EPJB}),
S\&P 500 (from 2 January 1980 to 31 December 1992 using MF-SF \cite{Canessa-2000-JPA}, from July 1997 to December 2007 using MF-DFA \cite{Kumar-Deo-2009-PA}, from 30 December 1927 to 3 September 2008 using MF-DFA \cite{Czarnecki-Grech-2010-APPA}, from 20 September 2004 to 10 October 2008 using MF-WT \cite{Hamrita-Abdallah-Mabrouk-2011-IJWMIP}, and from 20 December 1990 to 7 January 2016 using MF-DMA \cite{He-Wang-2017-PA}),
NYSE index (from January 1966 to December 2002 using MF-SF and MF-DFA \cite{Constantin-DasSarma-2005-PRE} and from 30 December 1927 to 3 September 2008 using MF-DFA \cite{Czarnecki-Grech-2010-APPA}),
and NASDAQ (from 30 December 1927 to 3 September 2008 using MF-DFA \cite{Czarnecki-Grech-2010-APPA}).

Quite a few indices of Chinese stock markets have been investigated, such as
SSEC (from December 20, 1990 to April 30, 2008 using MF-DFA \cite{Yuan-Zhuang-Jin-2009-PA}, from December 20, 1990 to December 30, 2010 using MF-DFA \cite{Yuan-Zhuang-Liu-2012-PA}, from 5 January 1999 to 24 February 2012 using MF-DMA \cite{Xiao-Wang-2014-IJNSNS}, from 20 December 1990 to 7 January 2016 using MF-DMA \cite{He-Wang-2017-PA}),
SZCI (from 3 April 1991 to 30 December 2010 using MF-DFA \cite{Yuan-Zhuang-Liu-2012-PA} and from 5 January 1999 to 24 February 2012 using MF-DMA \cite{Xiao-Wang-2014-IJNSNS}), %
SHSE 50 (from 2 February 1998 to 27 February 2008 using MF-DFA \cite{Bai-Zhu-2010-PA}),
SZSE 100 (from February 2, 1998 to February 27, 2008 using MF-DFA \cite{Bai-Zhu-2010-PA}),
CSI 300 (from 5 April 2006 to 9 May 2014 using MF-DFA \cite{He-Wang-Du-2014-PA}),
and CSI 800 (from 4 January 2005 to 23 March 2016\cite{Zhu-Zhang-2018-PA}).

Other investigations have been conducted on the Japanese TOPIX
(from 6 January 1986 to 23 August 1999 \cite{Katsuragi-2000-PA} using MF-SF),
the German DAX (from 1 October 1959 to 30 December 1998 using MF-SF \cite{Ausloos-Ivanova-2002-CPC} and from 29 September 1959 to 4 September 2001 using MF-SF \cite{Gorski-Drozdz-Speth-2002-PA}),
the Korean KOSPI (from April 1981 to 2003 using MF-SF \cite{Kim-Yoon-2004-PA,Yoon-Kim-Choi-2005-JKPS}),
the Iranian TEPIX (from 20 May 1995 to 18 March 2004 using MF-SF \cite{Norouzzadeh-Jafari-2005-PA}),
Taiwan's TAIEX (from January 1983 to May 2006 using MF-SF and MF-DFA \cite{Chen-Chen-Tseng-2009-IJMPB}) and OTC index (from January 1995 to May 2006 using MF-SF and MF-DFA \cite{Chen-Chen-Tseng-2009-IJMPB},
Hong Kong's HSI (from 5 January 1999 to 24 February 2012 using MF-DMA \cite{Xiao-Wang-2014-IJNSNS}, from January 1995 to May 2001 using MF-SF \cite{Bershadskii-2003-PA}) and HSCEI (from May 2 2012 to January 27 2016 using MF-DFA \cite{Ruan-Yang-Ma-2017-PA}),
India's SENSEX (from July 1997 to December 2007 using MF-DFA \cite{Kumar-Deo-2009-PA} and from 1 July 1997 to 31 December 2010 using MF-DFA \cite{Samadder-Ghosh-Basu-2013-Fractals}) and NSE index (from August 2002 to December 2007 using MF-DFA \cite{Kumar-Deo-2009-PA} and from 1 June 1990 to 31 December 2010 \cite{Samadder-Ghosh-Basu-2013-Fractals}, using MF-DFA),
the Polish WIG (from 16 April 1991 to 10 October 2008 using MF-DFA \cite{Czarnecki-Grech-2010-APPA}),
Morocco's MASI and MADEX using MF-DFA \cite{Lahmiri-2017-PA},
Turkey's ISE-100 (from 4 January 1988 to 16 July 2001 using MF-WT \cite{Balcilar-2003-EMFT}),
and the Spanish IBEX 35 from 1990 to 2000 using MF-WT \cite{Pont-Turiel-PerezVicente-2009-PA}.


Although there is overwhelming empirical evidence indicating the presence of multifractality in the daily returns of stock market indices, an exception was reported on the daily Bombay stock exchange index (BSE) data of year 2000, in which $\zeta(q)$ is found to be linear against $q$ \cite{Razdan-2002-PJP}. This is most likely due to the fact that the sample size (245 data points) is too short so that it is difficult to determine the scaling range. In addition, there are studies using prices as the variable $\Delta{X}$ in the MF-DFA analysis such that the detrending is conducted on the ``cumulative sum'' of prices \cite{Razdan-2002-PJP,Lee-Lee-2007-PA,Dutta-2010-CJP,Dutta-Ghosh-Chatterjee-2016-PA}, which either results in incorrect estimates of $\zeta(q)$ or $H(q)$ or misinterpretations of the results on reshuffling.
Indeed, the measure derived from the time series under study should be stationary in order to qualify as a genuine measure.
Prices and log-prices are not stationary and using them is well-known to lead to spurious conclusions. In regression analysis,
this is known as the spurious regression problem \cite{GrangerNewbold74,Phillips-1986-JEm}.

\subsubsection{High-frequency stock market indices}

High-frequency indices have been studied for different stock markets, such as the US market (15-min S\&P 500 index from June 24 to September 30 in 1999 using MF-SF \cite{Kleparskii-2001-ARC}, 5-min S\&P500 index from 1 February 2001 to 1 February 2002 using WTMM \cite{Budaev-2004-PA}, and 5-min DJIA from 19 September 1994 to 16 October 2002 using MF-SF \cite{Selcuk-Gencay-2006-PA}),
the French market (30s CAC 40 from 3 January 1993 until 31 December 1996 using MF-SF \cite{Brachet-Taflin-Tcheou-2000-CSF}),
the German market (1-min DAX data from 28 November 1997 to 30 December 1999  using MF-SF \cite{Gorski-Drozdz-Speth-2002-PA}),
the Russian market (30-min RTS-70 from 9 December 1996 to 22 February 2002 using MF-WT \cite{Balcilar-2003-EMFT}),
the Korean market (1-min KOSPI from 30 March 1992 to 30 November 1999  using MF-SF \cite{Lee-Lee-2005a-JKPS,Lee-Lee-2005b-JKPS,Lee-Lee-Rikvold-2006-PA,Lee-Lee-2007-JKPS,Kim-Cha-Lee-2011-CPC}),
the Polish market (1-min WIG20 from 1 April 1999 to 31 October 2005 using MF-DFA \cite{Rak-Drozdz-Kwapien-2007-PA}),
the Chinese market (1-min SSEC from 4 May 1998 to 1 June 2005 using MF-DFA \cite{Du-Ning-2008-PA}, 5-min SSEC from January 2003 to April 2008 using MF-DMA \cite{Gu-Zhou-2010-PRE}, and 5-min CSI 300 from 4 April 2005 to 19 January 2012 using MF-DMA \cite{Zhou-Dang-Gu-2013-PA}),
the Taiwan market (1-min TAIEX from February 2000 to December 2003 using MF-SF and MF-DFA \cite{Chen-Chen-Tseng-2009-IJMPB} and from 2 January 1999 to 28 August 2007 using MF-DFA \cite{Su-Wang-Huang-2009-JKPS}),
the Spanish market (13s IBEX 35 from January 2nd 2009 to December 31st 2010 using MF-DFA \cite{SuarezGarcia-GomezUllate-2014-PA}),
the 1-min Euro Stoxx 50 index from May 2008 to April 2009 using MF-DFA \cite{Green-Hanan-Heffernan-2014-EPJB},
and the daily CDS of 11 sector indices in the US stock market from 14 December 2007 to 31 December 2014 using MF-DFA \cite{Shahzad-Nor-Mensi-Kumar-2017-PA}.

\subsubsection{Stock market sector indices}

Some studies considered stock market sector indices. Sector indices, as well as stock market indices, can be viewed as portfolios of stocks. Using the MD-DFA method, Wang, Xiang and Pandey investigated the daily computer sector index and electronics sector index of China's growth enterprise market from September 2010 to August 2011 \cite{Wang-Xiang-Pandey-2012-PA}, 
Kim et al. analyzed the daily prices of 144 equity funds from 1 January 2002 to 31 December 2010 using MF-DFA \cite{Kim-Yim-Kim-Oh-2015-JKPS},
Zhuang, Wei and Ma worked on 10 sector indices of China stock market from 1 January 2005 to 3 April 2013 \cite{Zhuang-Wei-Ma-2015-PA},  
Yang, Zhu and Wang studied the daily CSI energy sub-industry index from 4 January 2005 to 15 June 2015 \cite{Yang-Zhu-Wang-2016-PA},
and Shahzad et al. researched 11 sector indices in the US stock market from 14 December 2007 to 31 December 2014 \cite{Shahzad-Nor-Mensi-Kumar-2017-PA}.

\subsubsection{Individual stocks}

The US stock market contributes the largest number of individual stocks that have been studied, containing
5-min prices of a particular stock traded on the New York Stock Exchange (NYSE) from 4 January  1993 to 30 June 1999 using MF-SF \cite{Fujiwara-Fujisaka-2001-PA},
daily prices of 2249 US stocks over 10-30 years using MF-DFA \cite{Matia-Ashkenazy-Stanley-2003-EPL},
5-min prices of Yahoo!, Microsoft and IBM  from 1 February 2001 to 1 February 2002 using WTMM \cite{Budaev-2004-PA},
daily prices of stock Intel between January 1990 and December 2002 using MF-SF and MF-DFA \cite{Constantin-DasSarma-2005-PRE},
1-min prices of stock Johnson and Johnson during the year of 2000 using MF-SF and MF-DFA \cite{Constantin-DasSarma-2005-PRE},
5-min prices of stock Intel during the year of 2000 using MF-SF and MF-DFA \cite{Constantin-DasSarma-2005-PRE},
tick-by-tick prices of 30 DJIA stocks from 1 Dec 1997 to 31 Dec 1999 using MF-DFA and WTMM \cite{Oswiecimka-Kwapien-Drozdz-Rak-2005-APPB},
daily prices of 30 DJIA stocks using MF-SF \cite{Heyde-2009-MMOR},
high-frequency prices of 30 DJIA stocks from the 1 July to 31 December 2004 using MF-DFA \cite{deSouza-Queiros-2009-CSF},
trade-by-trade returns of stock GE for calendar years 2000-2003 \cite{Thompson-Wilson-2016-MCS},
1-min to 10-min prices of 31 DJIA stocks for the years 2008-2011 using MF-DFA \cite{Rak-Drozdz-Kwapien-Oswiecimka-2015-EPL}.

Studied time series of individual stocks in other markets include
tick-by-tick data of 30 DAX companies from 1 Dec 1997 to 31 Dec 1999 using MF-DFA and WTMM \cite{Oswiecimka-Kwapien-Drozdz-Rak-2005-APPB,Oswiecimka-Kwapien-Drozdz-2005-PA},
daily prices of 202 constituent stocks of NIKKEI 225  from 4 January 2000 to 9 February 2006 using MF-DFA \cite{Cajueiro-Tabak-2009-CSF},
1-min 150 stocks in Taiwan Stock Exchange from 2 January 1999 to 28 August 2007 using MF-DFA \cite{Su-Wang-Huang-2009-JKPS},
daily prices of 172 KOSPI stocks from 1980 to 2003 using MF-DFA \cite{Oh-Kim-Eom-2010-JKPS},
trade-by-trade prices of 27 KOSPI stocks in 2008 using MF-DFA \cite{Yim-Oh-Kim-2014-PA},
daily prices of stocks in electronics and computer sectors of China's growth enterprise market from 2010 to 2011 using MF-DFA \cite{Wang-Xiang-Pandey-2012-PA},
daily prices of 13 Moroccan family business stocks using MF-DFA \cite{Lahmiri-2017-PA},
daily prices of 17 most liquid Spanish stocks from 1 January 1990 to 24 May 2001 using MF-WT \cite{Turiel-PerezVicente-2003-PA,Turiel-PerezVicente-2005-PA},
1-min prices of 4 most important Spanish stocks in 1999 using MF-WT \cite{Turiel-PerezVicente-2003-PA},
daily prices of 35 most liquid Spanish stocks from June 1996 to June 2006 using MF-WT \cite{Pont-Turiel-PerezVicente-2009-JEIC},
daily prices of 19 Spanish stocks from 1990 to 2000 using MF-WT \cite{Pont-Turiel-PerezVicente-2009-PA},
and 5-min prices of Russian Joint Stock Company and Gazprom joint stock company from 1 February 2001 to 1 February 2002 using MF-WT \cite{Budaev-2004-PA}.

\subsubsection{Equity futures}

The multifractal nature of equity futures has been confirmed on
10-sec prices of the DAX futures during 60 trading days using WTMM \cite{Pavlov-Ebeling-Molgedey-Ziganshin-Anishchenko-2001-PA},
5-min prices of the CSI 300 futures from 16 April 2010 to 12 April 2012 using MF-DFA \cite{Wang-Suo-Yu-Lei-2013-PA},
10-min prices of the CSI 300 futures IF1009 from 16 April 2010 to 17 September 2010 using MF-DFA \cite{Lu-Tian-Zhou-Li-2013-PA},
and 5-min prices of the CSI 300 futures from 30 June 2011 to 7 June 2013 using A-DFA \cite{Cao-Han-Cui-Guo-2014-PA}.

\subsubsection{Foreign exchange rates}

The global foreign exchange market is the largest financial market by transaction volumes. The multifractal nature of foreign exchange rates has been confirmed by numerous empirical studies. Some excellent works researched dozens of foreign exchange rates. For instance, Vandewalle and Ausloos \cite{Vandewalle-Ausloos-1998-IJMPC} and Di Matteo et al. \cite{DiMatteo-Aste-Dacorogna-2005-JBF,DiMatteo-2007-QF} investigated respectively 23 and 29 daily foreign exchange rates using the MF-SF method. Dro{\.{z}}d{\.{z}} et al. performed an interesting study using MF-DFA on the 1-min rates from 21:00 on 2 January 2004 to 21:00 on 30 March 2008 of two foreign exchange triangles (USD/EUR, EUR/GBP and GBP/USD) and (JPY/GBP, GBP/CHF and CHF/JPY) \cite{Drozdz-Kwapien-Oswiecimka-Rak-2010-NJP}. Pont et al. obtained the ensemble singularity spectrum of the daily returns of 9 currency exchange rates (pairs of AUD, USD, EUR, JPY and CHF except for AUD/CHF) from 1992 to 1997 using MF-WT \cite{Pont-Turiel-PerezVicente-2009-PA}.

The Japanese Yen (JPY) with respect to the USD is among the most studied. In finance, this currency pair
is particularly interesting due to its enormous transaction volume and for the carry trade that have been
implemented for decades as a major funding source. In the standard logic, one borrows in Yens at the low
interest rates that have been prevalent in Japan for decades, then sells the Yens to buy U.S. dollars, which are
invested in higher yielding instruments such as U.S. treasury bonds.
This carry trade, which contributes to pushing the Yen value lower, has been part of the strategy
of the Japanese society, which is so dependent on exports for its economic health and on the safe haven status of its currency.
Time series are recorded at different frequencies, such as daily data (
from 1 June 1973 to 1 June 1988 using MF-SF \cite{Muller-Dacorogna-Olsen-Pictet-Schwarz-Morgenegg-1990-JBF},
from 1 January 1980 to 31 December 1996 using MF-SF \cite{Vandewalle-Ausloos-1998-EPJB},
from January 1971 to June 2003 using MF-SF \cite{Kim-Choi-Yoon-2004-JKPS},
from 1 January 1990 to 2 November 2003 using MF-SF \cite{Litvin-2004b-PA},
from January 1971 to June 2003 using MF-SF \cite{Kim-Yoon-2004-PA},
from 1 January 1990 to 2 November 2003 using MF-SF \cite{Litvin-2004a-PA},
from 15 December 1998 to 9 November 2008 using WTMM \cite{Yang-Wang-Yang-Mang-2009-PA},
from 2 January 1975 to 12 November 2010 using MF-DMA \cite{Wang-Wu-Pan-2011-PA},
and from 1991 to 2005 using MF-DFA \cite{Oh-Eom-Havlin-Jung-Wang-Stanley-Kim-2012-EPJB}),
tick-by-tick data (from 1 March 1986 to 1 March 1989 using MF-SF \cite{Muller-Dacorogna-Olsen-Pictet-Schwarz-Morgenegg-1990-JBF}),
1-min data (in 2005 \cite{Kim-Lim-Chang-Kim-Lee-Park-Lee-You-Kim-2009-Fractals} and 2008 \cite{Kim-Kim-Jung-Kim-2012-JKPS} using MF-DFA), and
5-min data (from 1 February 2001 to 1 February 2002 using WTMM \cite{Budaev-2004-PA}).

For the Deutsche Mark (DEM) with respect to the USD, the MF-SF method has been applied to daily data during different periods (
from 1 June 1973 to 1 June 1988 \cite{Muller-Dacorogna-Olsen-Pictet-Schwarz-Morgenegg-1990-JBF},
from 1 January 1980 to 31 December 1996 \cite{Vandewalle-Ausloos-1998-EPJB},
from 1 January 1990 to 2 November 2003 \cite{Litvin-2004a-PA},
and from 4 June 1973 to 31 December 2001 \cite{Litvin-2004b-PA}),
tick-by-tick data (from 1 March 1986 to 1 March 1989 \cite{Muller-Dacorogna-Olsen-Pictet-Schwarz-Morgenegg-1990-JBF} and from 1 October 1992 until 30 September 1993 \cite{Baviera-Pasquini-Serva-Vergni-Vulpiani-2001-PA}),
15s data (from 1 October 1992 until 30 September 1993 \cite{Brachet-Taflin-Tcheou-2000-CSF}),
and 5-min data (from 4 January 1987 to 31 December 1998 \cite{Xu-Gencay-2003-PA}).
The MF-DFA method has also been applied to the 20s FX rates from 1 October 1992 to 30 September 1993 \cite{Nascimento-Junior-Jennings-Serva-Gleria-Viswanathan-2008-EPL}.

For the Euro exchange rates, studies have been carried out for
1-min EUR/USD in 2005 using MF-DFA \cite{Kim-Lim-Chang-Kim-Lee-Park-Lee-You-Kim-2009-Fractals},
5-min EUR/USD from 5 January 2006 to 31 December 31 2007 using MF-SF \cite{Batten-Kinateder-Wagner-2014-PA},
daily EUR/USD from 15 December 1998 to 9 November 2008 using WTMM \cite{Yang-Wang-Yang-Mang-2009-PA}
daily EUR/USD from 1 January 1990 to 2 November 2003 using MF-SF \cite{Litvin-2004a-PA},
daily EUR/USD for from January 2007 to October 2009 using MF-DFA \cite{Olemskoi-Shuda-Borisyuk-2010-EPL},
daily USD/EUR from 4 January 1999 to 12 November 2010 using MF-DMA \cite{Wang-Wu-Pan-2011-PA},
1-min EUR/USD in 2008 using MF-DFA \cite{Kim-Kim-Jung-Kim-2012-JKPS},
and
daily data of Croatian Kuna (HRK), Czech Koruna (CZK), Hungarian Forint (HUF), Polish Zlot (PLN), Romanian Leu (RON), Slovak Koruna (SKK) and Slovenian Tolar (SIT) against UER from 4 January 1999 to 31 August 2010 using MF-DFA \cite{Caraiani-Haven-2015-PA},

For the UK GBP exchange rates, multifractal analyses have been performed on
tick-by-tick USD/GBP from 1 March 1986 to 1 March 1989 using MF-SF \cite{Muller-Dacorogna-Olsen-Pictet-Schwarz-Morgenegg-1990-JBF},
daily USD/GBP from 1 June 1973 to 1 June 1988 using MF-SF \cite{Muller-Dacorogna-Olsen-Pictet-Schwarz-Morgenegg-1990-JBF},
daily GBP/USD from 1 January 1990 to 2 November 2003 using MF-SF \cite{Litvin-2004a-PA},
daily GBP/USD from 2 January 1975 to 12 November 2010 using MF-DMA \cite{Wang-Wu-Pan-2011-PA},
and
daily GBP/USD from 15 December 1998 to 9 November 2008 using WTMM \cite{Yang-Wang-Yang-Mang-2009-PA}.

For the Chinese Yuan, multifractal investigations have been conducted on
daily CNY/USD from 25 July 2005 to 12 November 2010 using MF-DMA \cite{Wang-Wu-Pan-2011-PA},
daily CNY/USD from 19 November 1997 to April 2008 and CNY/EUR from 1 January 1999 to April 2008 using MF-SF \cite{Schmitt-Ma-Angounou-2011-QF},
daily CNY/USD from 19 December 1990 to 2 August 2012 and CNY/HKD from 13 January 1994 to 2 August 2012 using MF-DFA \cite{Qin-Lu-Zhou-Qu-2015-PA},
daily prices of the onshore and offshore RMB exchange rates (CNY and CNH) from 2 May 2012 to 27 January 2016 using MF-DFA \cite{Ruan-Yang-Ma-2017-PA},
and
daily CIB-CNY Composite Index (CCI) from 21 July 2005 to 18th June 2010 using MF-DMA \cite{Wang-Yu-Suo-2012-PA}.

For the Korean Won (KRW) exchange rates, the studied time series include
daily KRW/USD from April 1981 to December 2002 using MF-SF \cite{Kim-Yoon-2004-PA,Yoon-Kim-Choi-2005-JKPS},
daily KRW/USD from 1991 to 2005 using MF-DFA \cite{Oh-Eom-Havlin-Jung-Wang-Stanley-Kim-2012-EPJB},
1-min USD-KRW in year 2008 using MF-DFA \cite{Kim-Kim-Jung-Kim-2012-JKPS},
and
the managed and independent float exchange rates using MF-DFA \cite{Stosic-Stosic-Stosic-Stanley-2015a-PA}.

For the Australian Dollar (AUD), investigations have been conducted on
1-min AUD/USD in 2005 using MF-DFA \cite{Kim-Lim-Chang-Kim-Lee-Park-Lee-You-Kim-2009-Fractals},
daily AUD/USD from 15 December 1998 to 9 November 2008 using WTMM \cite{Yang-Wang-Yang-Mang-2009-PA},
daily AUD/USD from 2 January 1975 to 12 November 2010 using MF-DMA \cite{Wang-Wu-Pan-2011-PA},
and
the managed and independent float exchange rates using MF-DFA \cite{Stosic-Stosic-Stosic-Stanley-2015a-PA}.

The multifractal nature in quite a few other currencies in different economies has also been researched, such as
daily Brazil Real (BRL/USD from 3 January 2000 to 12 November 2010 using MF-DMA \cite{Wang-Wu-Pan-2011-PA} and the managed and independent float exchange rates using MF-DFA \cite{Stosic-Stosic-Stosic-Stanley-2015a-PA}),
daily Bulgarian Lev (BGN/USD from February 1991 to May 1997 using MF-SF \cite{Ivanova-Ausloos-1999a-EPJB} and from 23 October 1995 to 11 February 2003 using MF-SF \cite{Litvin-2004b-PA}),
daily Canada Dollar (CAD/USD from 1 January 1990 to 2 November 2003 using MF-SF \cite{Litvin-2004a-PA},
from 2 January 1975 to 12 November 2010 using MF-DMA \cite{Wang-Wu-Pan-2011-PA},
and from 15 December 1998 to 9 November 2008 using WTMM \cite{Yang-Wang-Yang-Mang-2009-PA}),
daily Denmark Krone (DKK) from 2 January 1975 to 12 November 2010 using MF-DMA \cite{Wang-Wu-Pan-2011-PA},
daily French Francs (USD/FRF from 1 January 1979 to 30 November 1993 using MF-SF \cite{Schmitt-Schertzer-Lovejoy-2000-IJTAF} and CHF/FRF, DEM/FRF, GBP/FRF and JPY/FRF from 1 January 1979 to 30 November 1993 using MF-SF \cite{Schmitt-Schertzer-Lovejoy-1999-ASMDA}),
daily Hong-Kong Dollar (HKD/USD from 1991 to 2005 using MF-DFA \cite{Oh-Eom-Havlin-Jung-Wang-Stanley-Kim-2012-EPJB}),
daily Indian Rupee (INR/USD from 4 January 1999 to 12 November 2010 using MF-DMA \cite{Wang-Wu-Pan-2011-PA} and from January 1995 to December 2012 using MF-DFA \cite{Dutta-Ghosh-Chatterjee-2016-PA}),
daily Iranian Rial (IRR/USD from 24 September 1989 to 15 November 2003 using MF-DFA \cite{Norouzzadeh-Rahmani-2006-PA}),
daily Malaysian Ringgit (USD/RM, YEN/RM and SGD/RM from 2 April 1985 to 30 April 2001 using WTMM \cite{Muniandy-Lim-Murugan-2001-PA} and managed and independent float exchange rates  using MF-DFA \cite{Stosic-Stosic-Stosic-Stanley-2015a-PA}),
daily Mexican Peso (MXP/USD from 20 December 1994 to about 2001 using MF-SF \cite{AlvarezRamirez-2002-PA} and from 8 November 1993 to 12 November 2010 using MF-DMA \cite{Wang-Wu-Pan-2011-PA}),
daily New Zealand Dollar (NZD/USD from 15 December 1998 to 9 November 2008 using WTMM \cite{Yang-Wang-Yang-Mang-2009-PA} and the managed and independent float exchange rates  using MF-DFA \cite{Stosic-Stosic-Stosic-Stanley-2015a-PA}),
daily Norway Krone (NOK/USD) from 2 January 1975 to 12 November 2010 using MF-DMA \cite{Wang-Wu-Pan-2011-PA},
daily Thailand Thai Baht (THB/USD from 4 January 1999 to 12 November 2010 using MF-DMA \cite{Wang-Wu-Pan-2011-PA},
from 1991 to 2005 using MF-DFA \cite{Oh-Eom-Havlin-Jung-Wang-Stanley-Kim-2012-EPJB},
and the managed and independent float exchange rates using MF-DFA \cite{Stosic-Stosic-Stosic-Stanley-2015a-PA}),
Swiss Francs (tick-by-tick CHF/USD from 1 March 1986 to 1 March 1989 using MF-SF \cite{Muller-Dacorogna-Olsen-Pictet-Schwarz-Morgenegg-1990-JBF},
daily CHF/USD from 1 June 1973 to 1 June 1988 using MF-SF \cite{Muller-Dacorogna-Olsen-Pictet-Schwarz-Morgenegg-1990-JBF},
and daily CHF/USD from 1 January 1990 to 2 November 2003 using MF-SF \cite{Litvin-2004a-PA}),
daily Sweden Krona (managed and independent float SEK/USD exchange rates using MF-DFA \cite{Stosic-Stosic-Stosic-Stanley-2015a-PA}),
daily Taiwan New Dollar (managed and independent float TWD/USD exchange rates using MF-DFA \cite{Stosic-Stosic-Stosic-Stanley-2015a-PA}),
daily data of Romanian Leu (ROL), Slovak Koruna (SKK) and Slovenian Tolar (SIT) from 22 October 1995 to 11 February 2003 using MF-SF \cite{Litvin-2004a-PA},
and
daily Russian Ruble (RUR), Ukraine Hryvnia (UAH), Kazakhstani Tenge (KZT), Estonian Kroon (EEK), Lithuanian Litas (LTL), Latvian Lats (LVL), Croatian Kuna (HRK), Czech Koruna (CSK), Hungarian Forint (HUF), and Polish Zloty (PLZ) from 23 October 1995 to 11 February 2003 using MF-SF \cite{Litvin-2004b-PA}.

\subsubsection{Commodities}

In the commodity markets, the multifractal properties of gold prices and of returns have been studied, such as
tick-by-tick prices of gold (XAU/USD) from 1 March 1986 to 1 March 1989 using MF-SF \cite{Muller-Dacorogna-Olsen-Pictet-Schwarz-Morgenegg-1990-JBF},
daily gold prices from February 1991 to May 1997 using MF-SF  \cite{Ivanova-Ausloos-1999a-EPJB},
daily prices of gold traded in COMEX over the period of 13 July 1990 to 15 September 2009 using MF-DFA \cite{Wang-Wei-Wu-2011a-PA},
daily LME gold prices from 1968 to 2010 using MF-DFA \cite{Bolgorian-Gharli-2011-APPB}, 
daily gold prices from January 1973 to October 2011 using MF-DFA \cite{Ghosh-Dutta-Samanta-2012-APPB},
daily AU99.99 closing prices of the Shanghai Gold Exchange from 30 October 2002 and 30 November 2012 using MF-DFA \cite{Yin-Zhang-Zhang-Wei-2013-RJEF},
daily gold prices in the Indian market and daily global consumer price index from January 1985 to June 2013 using MF-DFA \cite{Mali-2014-Fractals},
daily gold consumer price index (CPI) and the gold market in China, India and Turkey from 1993 to July 2013 using MF-DFA \cite{Mali-Mukhopadhyay-2014-PA},
and
daily gold prices of China and India from 1985 to 2013 using MF-DFA \cite{Mali-Mukhopadhyay-2015-PS}.

In the oil and gas market, studies have been conducted on daily spot prices of
WTI crude oil
   (from November 1981 to about 2001 using MF-SF \cite{AlvarezRamirez-Cisneros-IbarraValdez-Soriano-2002-PA},
    from 2 January 1990 to 28 February 2001 using MF-SF \cite{Serletis-Andreadis-2004-EE},
    from 2 January 1986 to 28 May 2004 using MF-SF \cite{Balankin-Matamoros-2005-PRE},
    from 20 May 1987 to 30 September 2008 using MF-DFA \cite{Gu-Chen-Wang-2010-PA},
    from 1986 to 2008 using MF-SF \cite{Momeni-Kourakis-2010-Fractals},
    from 20 May 1987 to 14 October 2009 using MF-DFA \cite{He-Chen-2010a-PA},
    from 2 January 1991 to 31 December 2013 \cite{Liu-Ma-2014-PA}, 
    from 2 January 1986 to 28 September 2012 using MF-DFA \cite{Gu-Zhang-2016-EE},),
Brent crude oil
   (from November 1981 to about 2001 using MF-SF \cite{AlvarezRamirez-Cisneros-IbarraValdez-Soriano-2002-PA},
    from 20 May 1987 to 30 September 2008 using MF-DFA \cite{Gu-Chen-Wang-2010-PA},
    from 20 May 1987 to 14 October 2009 using MF-DFA \cite{He-Chen-2010a-PA},
Dubai crude oil from November 1981 to about 2001 using MF-SF \cite{AlvarezRamirez-Cisneros-IbarraValdez-Soriano-2002-PA},
conventional gasoline (New York Harbor) from 2 January 1991 to 31 December 2013 \cite{Liu-Ma-2014-PA},
heating oil (New York Harbor) from 2 January 1991 to 31 December 2013 \cite{Liu-Ma-2014-PA},
jet fuel (US Gulf Coast) from 2 January 1991 to 31 December 2013 \cite{Liu-Ma-2014-PA},
and
natural gas (Henry Hub, LA) from 24 January 1991 to 28 February 2001 \cite{Serletis-Andreadis-2004-EE}.
Other related studies include
 daily prices of the WTI crude oil futures traded in NYMEX from 2 January 1985 to 10 May 2011 using MF-DMA \cite{Wang-Wu-2013-CE},
 daily data of DJ Brent crude index from January 1999 to January 2015 and DJ energy index and DJ petroleum index from January 2006 to January 2015 using MF-DFA \cite{Delbianco-Tohme-Stosic-Stosic-2016-PA},
 daily crack spread of gasoline which is the difference between the spot prices of gasoline (New York Harbor) and WTI crude oil from 2 January 1986 to 26 July 2011 using MF-DMA \cite{Wang-Wu-2012-EM},
 monthly price indexes of World Bank non-fuel and World Bank energy from 1960 until 2013 using MF-DFA \cite{Delbianco-Tohme-Stosic-Stosic-2016-PA},
 and a steam coal FOB price in Qinhuangdao Port from 2006 to 2013 \cite{Zhao-Zhu-Xia-2016-Energy}.

For the electricity market, researchers have studied
 hourly spot prices in the Spain electricity exchange from 1 January 1998 to 31 May 2006 using MF-DFA \cite{Norouzzadeh-Dullaert-Rahmani-2007-PA},
 30-min electricity prices and demand data in Australia's national electricity market from December 1998 to June 2008 using MF-DFA \cite{AlvarezRamirez-EscarelaPerez-EspinosaPerez-Urrea-2009-PA},
 1-hour electricity prices and demand data from 1 January 2000 to 20 June 2008 for Alberta and from 1 May 2002 to 30 June 2008 for Ontario in Canadian electricity markets using MF-DFA \cite{AlvarezRamirez-EscarelaPerez-2010-EE},
 electricity prices of California 1999-2000 and PJM 2001-2002 using MF-DFA \cite{Wang-Liao-Li-Li-Zhou-2013-PA},
 15-min day-ahead monthly bid prices of 5 bid areas from 1 April 2012 to 31 March 2014 using MF-DFA \cite{Ghosh-Dutta-Chakraborty-2016-IJEPES},
and daily electric load records of a certain power system from 1 January 1999 to 31 December 2011 using MF-DFA \cite{Yuan-Ji-Yuan-Huang-Li-Li-2015-Fractals}.

Concerning multifractal detrended fluctuation analysis of agricultural commodities,
Matia et al. considered the daily prices of 9 commodities over 10-30 years \cite{Matia-Ashkenazy-Stanley-2003-EPL},
He and Chen studied the daily closing prices of hard winter wheat futures on the Zhengzhou Commodity Exchange from 28 December 1993 to 18 September 2009, and soy meal futures from 17 July 2000 to 18 September 2009, No. 1 soybean futures from 15 March 2002 to 18 September 2009 and corn futures from 22 September 2004 to 18 September 2009 on the Dalian Commodity Exchange \cite{He-Chen-2010b-PA},
Kim et al. researched the daily prices of leek, radish, onion, and Korean cabbage from 3 January 2001 to 23 September 2009 in the public wholesale market of Seoul Agricultural \& Marine Products Corporation \cite{Kim-Oh-Kim-2011-PA},
Wang and Hu investigated the daily prices of four agricultural commodity futures (corn, soybean, wheat and rice) on the Chicago Board of Trade (CBOT) from 1 January 2000 to 31 September 2014 \cite{Wang-Hu-2015-PA},
and
Delbianco et al. worked on the monthly data of World Bank's soybean price index and agricultural price index from 1968 until 2013 \cite{Delbianco-Tohme-Stosic-Stosic-2016-PA}.


\subsubsection{Interest rates}

Multifractal analyses on interest rates and their futures have mainly used the MF-SF method and have been conducted on daily data of
daily Korean treasury bond futures from December 2001 to June 2002 for KTB206 and from March 2002 to September 2002 for KTB209 \cite{Kim-Yoon-2004-JPSJ},
US treasury rates with different maturities from 1990 to 2001 \cite{DiMatteo-Aste-Dacorogna-2005-JBF},
daily Eurodollar interbank interest rates with different maturity dates from 1990 to 1996  \cite{DiMatteo-Aste-Dacorogna-2005-JBF},
daily Pound sterling, US dollar, Japanese Yen, Australian dollar and the EURO, LIBOR and Indonesian Rupiah for different maturities (3, 6, 9 and 12 months) from January 2000 to December 2005 using MF-SF and MF-DFA \cite{Cajueiro-Tabak-2007b-PA},
daily Eurodollar rates from 1990 to 1996 \cite{DiMatteo-2007-QF},
and US treasury bond rates from 1997 to 2001 \cite{DiMatteo-2007-QF}.

The MF-DFA method has also been applied to the interest rates of
daily Pound sterling, US dollar, Japanese Yen, Australian dollar, the EURO, LIBOR and Indonesian Rupiah for different maturities (3, 6, 9 and 12 months) from January 2000 to December 2005 \cite{Cajueiro-Tabak-2007b-PA},
daily US treasury bonds from 2 January 1970 to 30 November 2006 \cite{Lim-Kim-Lee-Kim-Lee-2007-PA},
1-min Korean Treasury Bond (KTB412) futures from 1 July 2004 to 31 December 2004 \cite{Lim-Kim-Lee-Kim-Lee-2007-PA},
daily Shanghai interbank offered rate (SHIBOR) with eight term values from 8 October 2006 to 31 December 2012 \cite{Gu-Chen-Li-2014-PA},
and daily US Effective Federal Funds from 1 January 2000 to 31 September 2014 \cite{Wang-Hu-2015-PA}.

\subsubsection{Illusionary multifractality}
\label{S2:MF:Emp:MFPF:Illustions}

Some studies applied the MF-PF method to stock market indices or asset prices such that the finest measure defined at each time $t$ is the ratio of the index price at $t$ to the sum of all index prices. Representative works are obtain from
the Hong Kong market (1-min HSI from 3 January 1994 to 28 May 1997 \cite{Sun-Chen-Wu-Yuan-2001-PA,Sun-Chen-Yuan-Wu-2001-PA} and 30 continuous trading days from 3 January 1994 \cite{Chen-Sun-Chen-Wu-Wang-2004-NJP}),
the Shanghai stock market (1-min SSEC from 19 January 1999 to 30 December 2005 \cite{Wei-Wang-2008-PA}, 5-min SSEC from January 1999 to July 2001 \cite{Wei-Huang-2005-PA}, from 4 May 1998 to 1 June 2005 \cite{Du-Ning-2008-PA}, from 4 January 2005 to 30 December 2005 \cite{Yuan-Zhuang-2008-PA}, and from 2 January 2004 to 29 June 2012 \cite{Chen-Wei-Lang-Lin-Liu-2014-PA}, and daily SSEC from 4 January 2005 to 30 December 2005 \cite{Yuan-Zhuang-2008-PA}),
the Taiwan stock market (daily TSPI from 1987 to 2002 \cite{Ho-Lee-Wang-Chuang-2004-PA} and 1-min TAIEX from 3 May 1999 to 30 November 2007 \cite{Su-Wang-2009-JKPS}),
the US market (5-min S\&P 500 from 2 January 2004 to 29 June 2012 \cite{Chen-Wei-Lang-Lin-Liu-2014-PA}),
and the oil price from 1973 to 2007 \cite{Zhao-Chang-Liu-2015-AOR}.

All these works produce illusionary multifractality simply because these indices are not singular and a characteristic is that the singularity is close to 1 such that the singularity width $\Delta\alpha$ is close to 0 \cite{Jiang-Zhou-2008a-PA,Zhou-2010-cnJMSC}. However, this illusionary multifractality seems able to capture to some extent the fluctuation characteristics of indices and has the potential power of volatility forecasting (see Section \ref{S2:Appl:VolatilityPred}).

An exception is reported for the hourly day-ahead and real-time electricity prices in the Pennsylvania-New Jersey-Maryland (PJM) electricity market from 1 January 2008 to 31 December 2012, in which the singularity width is significantly larger than 0 \cite{Liu-Chung-Wen-2014-IGTD}. This is due to the fact that the electricity prices fluctuate widely and the two investigated price time series looks like stock volatility time series (see Fig. 2 of Ref.~\cite{Liu-Chung-Wen-2014-IGTD}).

\subsection{Volatilities}

\subsubsection{Stock markets}

For the US stock market, multifractality has been reported to be present in
1-min S\&P 500 index volatility from 1 January 1982 to 31 December 1999 using MF-PF \cite{Jiang-Zhou-2009-CPL} and the multiplier method \cite{Jiang-Zhou-2007-PA},
high-frequency volatility of the 30 DJIA stocks from 1 July 2004 to 31 December 2004 using MF-DFA \cite{deSouza-Queiros-2009-CSF},
1-min volatility of the 31 DJIA stocks from 2008 to 2011 using MF-CCA \cite{Rak-Drozdz-Kwapien-Oswiecimka-2015-EPL},
daily DJIA volatility from 18 May 1995 to 18 May 2015 using MF-PF \cite{Dai-Hou-Gao-Su-Xi-Ye-2016-CSF},
and daily NASDAQ from 18 May 1995 to 18 May 2015 using MF-PF \cite{Dai-Hou-Gao-Su-Xi-Ye-2016-CSF}.

For the Chinese stock market, researchers studied
1-min volatility of 1139 Chinese stocks and 2 indexes (SSCI and SZCI) from January 2004 to June 2006 using MF-PF \cite{Jiang-Zhou-2008b-PA},
daily volatility of 36 SSE 180 stocks from 9 July 2002 to 9 July 2007 using MF-SF \cite{UrecheRangau-deRorthays-2009-JAS},
1-min SSEC volatility from 20 September 2007 to 10 May 2010 using MF-DFA \cite{Lin-Fei-Wang-2011-PA},
5-min SSEC and SZCI volatility from 4 January 2002 to 31 December 2008 using MF-DFA \cite{Chen-Wu-2011-PA},
daily volatilities of the CSI energy index from 4 January 2005 to 15 June 2015 using MF-DFA \cite{Yang-Zhu-Wang-2016-PA},
daily SSEC and SZCI volatility from 18 May 1995 to 18 May 2015 using MF-PF \cite{Dai-Hou-Gao-Su-Xi-Ye-2016-CSF},
and 5-min realized volatility of the SSEC index from 1 January 2000 to 31 May 2014 and the S\&P 500 index from 2 January 1996 to 24 June 2013\cite{Mei-Liu-Ma-Chen-2017-PA}.

Other studies considered daily DAX volatility from 1 October 1959 to 30 December 1998 using MF-PF \cite{Ausloos-Ivanova-2002-CPC},
1-min upward and downward volatility (or positive and negative returns) of the DAX index from 28 November 1997 to 30 December 1999 and from 1 May 2002 to 1 May 2004 using MF-DFA \cite{Oswiecimka-Kwapien-Drozdz-Gorski-Rak-2008-APPA},
daily volatility of the Hang Seng China Enterprises Index (HSCEI) from 2 May 2012 to 27 January 2016 using MF-DFA and MF-CCA \cite{Ruan-Yang-Ma-2017-PA},
daily volatility of 35 most liquid stocks in the Spanish stock market from June 1996 to June 2006 using MF-WT \cite{Pont-Turiel-PerezVicente-2009-JEIC},
trade-by-trade volatility of 27 highly capitalized individual companies listed on the KOSPI stock market in 2008 and 2009 using MF-DFA \cite{Yim-Oh-Kim-2014-PA},
daily volatility of CAC 40, DAX, FTSE, NASDAQ, S\&P500 and Canada TSE indices from 2 January 2007 to 31 December 2009 using MF-SF \cite{Lahmiri-2017b-Fractals},
and
daily volatility of the Casablanca market index and 111 Moroccan family business stocks from 25 March 2013 to 22 March 2016 on the Casablanca Stock Exchange (Morocco) using MF-DFA \cite{Lahmiri-2017a-Fractals}.

There are also other works dealing with cross-correlations between volatility and other financial variables (see Section \ref{S2:EmpAnal:X} below). A distinct study performed MF-DFA on the logarithmic difference of daily implied volatility for CAC 40, DAX and S\&P 500 index options from January 2002 to January 2009 \cite{Oh-2014-JKPS}.

\subsubsection{Foreign exchange rates}

On the foreign exchange markets, researchers investigated 20-sec DEM/USD volatility from 1 October 1992 to 30 September 1993 using MF-DFA \cite{Nascimento-Junior-Jennings-Serva-Gleria-Viswanathan-2008-EPL},
daily USD/JPY volatility from 10 August 2003 to 31 May 2006 using MF-WT \cite{Hamrita-Abdallah-Mabrouk-2011-IJWMIP},
daily CHY/USD volatility from 1 January 2005 to 1 November 2014 using MF-MF \cite{Dai-Shao-Gao-Sun-Su-2016-Fractals},
and
daily volatility of the onshore and offshore RMB exchange rates (CNY and CNH) from 2 May 2012 to 27 January 2016 using MF-DFA and MF-CCA \cite{Ruan-Yang-Ma-2017-PA}.

\subsubsection{Commodities}

The multifractal properties of the daily volatility time series of several commodities has been analyzed, such as
agriculture futures volatility of Hard Winter wheat from 28 December 1993 to 11 May 2009 and Strong Gluten wheat from 28 March 2003 to 11 May 2009 on the Zhengzhou Commodity Exchange, and Soy Meal from 17 July 2000 to 11 May 2009 and Soy Bean 1 futures from 15 March 2002 to 11 May 2009 on the Dalian Commodity Exchange using MF-PF \cite{Chen-He-2010-PA},
WTI crude oil from 13 July 1990 to 6 March 2009 using MF-DFA \cite{Wang-Wei-Wu-2010b-PA}, from 2 January 1990 to 9 March 2010 using MF-DFA \cite{Wang-Wei-Wu-2011b-PA} using MF-DFA and from 1 January 2005 to 1 November 2014 using MF-PF \cite{Dai-Shao-Gao-Sun-Su-2016-Fractals}, %
Brent and WTI crude oil from 20 May 1987 to 14 October 2009 using MF-PF \cite{He-Chen-2010a-PA},
and gold from 1 January 2005 to 1 November 2014 using MF-PF \cite{Dai-Shao-Gao-Sun-Su-2016-Fractals}.

\subsection{Liquidity measures}

\subsubsection{Trading volumes}
\label{S2:EmpAnal:Volume}

Trading volume is an important measure of market liquidity and the main force driving price movements \cite{Karpoff-1987-JFQA,Lillo-Farmer-Mantegna-2003-Nature,Lim-Coggins-2005-QF,Zhou-2012-QF}.
For the US stock market, Bolgorian and Raei applied MF-DFA to daily S\&P 500 trading volume from 1980 to 2009 \cite{Bolgorian-Raei-2011-PA},
Rak et al. used MF-CCA to study 1-min transaction number and trading volume of 31 DJIA stocks from 2008 to 2011 \cite{Rak-Drozdz-Kwapien-Oswiecimka-2015-EPL}, and Moyano et al. used MF-DFA to analyzed 1-min trading volume of 30 DJIA stocks from 1 July 2004 to 31 December 2004\cite{Moyano-deSouza-Queiros-2006-PA}.

For the Chinese stock market, Ureche-Rangau and de Rorthays applied MF-SF to daily trading volume of 36 SSE180 stocks from 9 July 2002 to 9 July 2007 \cite{UrecheRangau-deRorthays-2009-JAS},
Mu et al investigate 1-min trading volume of 22 Chinese stocks in 2003 using MF-DFA \cite{Mu-Chen-Kertesz-Zhou-2010-PP},
Yuan, Zhuang and Liu analyzed daily trading volume of SSEC from 20 December 1990 to 30 December 2010 and SZCI from 3 April 1991 to 30 December 2010 using MF-DFA \cite{Yuan-Zhuang-Liu-2012-PA}, 
and
Wang et al.  studied 5-min trading volume of the CSI300 index futures from 16 April 2010 to 12 April 2012 using MF-DFA \cite{Wang-Suo-Yu-Lei-2013-PA}.

On the Korean stock market,
Lee and Lee applied MF-DFA to 1-min trading volume of KOSPI from 30 March 1992 to 30 November 1999, \cite{Lee-Lee-2007-PA},
Lee used MF-PF to investigate daily trading of KOSPI from November 1997 to September 2010 \cite{Lee-2011-JKPS},
Yim, Oh and Kim performed MF-DFA on trade-by-trade trading volume of 27 highly capitalized individual companies in 2008 and 2009 \cite{Yim-Oh-Kim-2014-PA},
and
Oh adopted MF-DMA to investigate the aggregated trading volumes of buying, selling, and normalized net investor trading \cite{Oh-2017-JKPS}.

In addition, using MF-DFA, Sto{\v{s}}i{\'{c}} et al. studied daily trading volume of 13 stock market indices (AEX, BFX, BSESN, BVSP, FCHI, FTSE, GSPC, HSI, KS11, MXX, N225, SSMI, TWII) from 30 August 2006 to 1 March 2014 \cite{Stosic-Stosic-Stosic-Stanley-2015b-PA}, while Bolgorian analysed daily trading volume of individual and institutional traders on the stocks listed in the Tehran Stock Exchange from 1998 to 2009 \cite{Bolgorian-2011-PA}.

In the multifractal analysis of trading volume, the majority of studies in econophysics work on the normal volume or dollar volume. However, the values of trading volume are affected by share split. Therefore, it is recommended to investigate turnover rates instead of trading volumes \cite{Lo-Wang-2000-RFS}.

\subsubsection{Bid-ask spread}

The bid-ask spread is another important measure of market liquidity. Gu, Chen and Zhou applied MF-PF to 1-min spread of a liquid Chinese stock in 2003 and found no evidence of multifractality \cite{Gu-Chen-Zhou-2007-EPJB}. In contrast, Qiu et al. performed MF-DFA on the 1-min bid-ask spread of 4 liquid Chinese stocks (000001, 000539, 600198, 600663) from 2004 to 2006 and unveiled evidence supporting the presence of multifractality \cite{Qiu-Chen-Zhong-Wu-2012-PA}. Obviously, more research should be conducted on bid-ask spread using different multifractal analysis methods.

\subsubsection{Price gaps on order books}

Price gap is defined as the logarithmic price difference between the first two occupied price levels on the same side of a limit order book (LOB), which is a key determinant of market depth and hence market liquidity \cite{Farmer-Gillemot-Lillo-Mike-Sen-2004-QF,Zhou-2012-NJP}. Gu et al. studied 26 A-share stocks traded on the Shenzhen Stock Exchange using MF-DFA and found that the gap time series are long-range correlated and possess multifractal properties \cite{Gu-Xiong-Zhang-Chen-Zhang-Zhou-2016-CSF}. They also compared the singularity spectrum width $\Delta\alpha_{\rm{b}}$ of buy LOBs with the singularity spectrum width $\Delta\alpha_{\rm{s}}$ for sell LOBs  for all the stocks and reported a linear relationship between them:
\begin{equation}
  \Delta\alpha_{{\rm{s}}}=-0.2223+1.1452\Delta\alpha_{{\rm{b}}}.
\end{equation}
This suggests that the multifractal properties change across different stocks.

\subsection{Inter-event durations}

\subsubsection{Recurrence intervals}

Recurrence intervals are waiting times between successive events whose values exceed certain threshold. Recurrence interval analysis has been performed on financial time series of volatility \cite{Yamasaki-Muchnik-Havlin-Bunde-Stanley-2005-PNAS,Wang-Yamasaki-Havlin-Stanley-2006-PRE,Wang-Yamasaki-Havlin-Stanley-2008-PRE,Ren-Guo-Zhou-2009-PA,Jiang-Canabarro-Podobnik-Stanley-Zhou-2016-QF}, return \cite{Bogachev-Eichner-Bunde-2007-PRL,Bogachev-Bunde-2008-PRE,Ren-Zhou-2010-NJP,Jiang-Wang-Canabarro-Podobnik-Xie-Stanely-Zhou-2018-QF} and trading volume \cite{Podobnik-Horvatic-Petersen-Stanley-2009-PNAS,Ren-Zhou-2010-PRE,Li-Wang-Havlin-Stanley-2011-PRE}. Recurrence interval is often called return interval \cite{Yamasaki-Muchnik-Havlin-Bunde-Stanley-2005-PNAS}, which is however not suitable for econophysics since the term ``return'' is widely used for logarithmic price changes.

The multifractal nature of recurrence intervals of stock returns in an artificial market has been unveiled via the MF-DFA method \cite{Meng-Ren-Gu-Xiong-Zhang-Zhou-Zhang-2012-EPL}, in which the return series are generated by an empirical behavioral order-driven model \cite{Gu-Zhou-2009-EPL}. It is found that the singularity spectrum width $\Delta\alpha$ increases with the Hurst index of relative prices of the submitted orders. However, multifractal analysis on real financial time series is rare.

\subsubsection{Intertrade durations}

Intertrade duration is the waiting time between successive transactions. O{\'s}wi{\c{e}}cimka et al applied MF-DFA to the intertrade durations of 30 DAX stocks from 1 December 1997 to 31 December 1999 \cite{Oswiecimka-Kwapien-Drozdz-2005-PA},
Jiang, Chen and Zhou used MF-DFA to analyze the intertrade durations of 23 liquid Chinese stocks in 2003 \cite{Jiang-Chen-Zhou-2009-PA},
Ruan and Zhou investigated 1-min intertrade durations of a Chinese stock and its warrant from 22 August 2005 to 23 August 2006 using MF-DMA and MF-DFA \cite{Ruan-Zhou-2011-PA},
and
Yim-Oh-Kim performed MF-DFA on the intertrade durations of 27 highly capitalized individual companies listed on the KOSPI stock market in 2008 and 2009 \cite{Yim-Oh-Kim-2014-PA}.

Heyde numerically constructed the activity time (transaction time) of the S\&P 500 index and 30 DJIA consistent stocks from Ito's formula and used MF-SF to confirm the multifractal nature \cite{Heyde-2009-MMOR}.

Intertrade duration plays an important role in autoregressive conditional duration model \cite{Engle-Russell-1998-Em}, multifractal models for asset returns \cite{Mandelbrot-Fisher-Calvet-1997,Calvet-Fisher-2001-JEm}, and Markov-switching multifractal models \cite{Calvet-Fisher-2002-RES}. These empirical results confirm the presence of multifractality in the intertrade durations, providing solid backup for the main assumption in MMAR models and MSM models.

\subsubsection{Other microstructure variables}

Ni et al. performed MF-DFA on inter-cancelation durations calculated from the limit order book data of 22 liquid Chinese stocks in 2003 \cite{Ni-Jiang-Gu-Ren-Chen-Zhou-2010-PA}, while Yue et al. applied MF-DMA to the time series of the order aggressiveness of 43 Chinese stocks (see \cite{Yue-Xu-Chen-Xiong-Zhou-2017-Fractals} for the definition of ``aggressiveness''). These studies confirmed the presence of multifractality in certain microstructure variables.

\subsection{Two time series: MF-X}
\label{S2:EmpAnal:X}

In the last decade, some joint multifractal analyses have been performed on distinct pairs of financial variables. Most of these studies adopt the MF-X-DFA method \cite{Zhou-2008-PRE}. Without further justification, the method used in the empirical analyses reviewed below is MF-X-DFA.

\subsubsection{Return-return pairs}
\label{S3:EmpAnal:X:Price:Pair}

The multifractal nature in the cross-correlations in pairs of asset returns has been widely studied. For stock markets, researchers have investigated
the daily returns of A-share and B-share market indices in Shanghai and Shenzhen from 2 January 1996 to 1 April 2010 \cite{Wang-Wei-Wu-2010a-PA},
the daily returns of SSEC and SZCI indices from 4 April 1991 to 13 April 2009 \cite{Zhao-Shang-Jin-2011-Fractals}, 
the daily returns of SSEC and SZCI indices from 4 January 2002 to 31 December 2008 \cite{Zhao-Shang-Shi-2014-PA}, 
the daily returns of SSEC and surrounding markets (HSI, NIKKEI 225 and KOSPI) from 1 January 1997 to 31 December 2011 \cite{Ma-Wei-Huang-2013-PA},
the six daily return pairs of Amman AMX, Egyptian EGX, Casablanca MASI and Tunis TUNINDEX indices in 2837 trading days from 1 January 2000 \cite{ElAlaoui-Benbachir-2013-PA},
the daily returns of IBM and MSFT from 1987 to 2012 \cite{Mehraban-Shirazi-Zamani-Jafari-2013-EPL},
the 1-min returns of two pairs of German (CBK-DBK and DBK-EON) using MF-X-DFA and MF-CCA \cite{Oswiecimka-Drozdz-Forczek-Jadach-Kwapien-2014-PRE},
the daily returns of DJIA and DAX from 12 January 1990 to 12 October 2013 using MF-X-DFA and MF-CCA \cite{Oswiecimka-Drozdz-Forczek-Jadach-Kwapien-2014-PRE},
the daily CSI 300 spot and futures returns from 16 April 2010 to 17 February 2012 using MF-X-DMA \cite{Wang-Xie-2013-ND},
the daily returns between market return and 6 portfolio excess returns constructed from all stocks in NYSE, AMEX and NASDAQ based on the companies' sizes and book-to-market (BM) ratios from 2 January 1990 to 29 January 2016 \cite{Chen-Wang-2017-PA}, 
the daily instantaneous and average instantaneous cross-correlations between the US and the Chinese stock markets \cite{Qiu-Chen-Zhong-Lei-2011-PA},
%
%
the daily returns of the CDS and 11 sector indices in the US stock market from 14 December 2007 to 31 December 2014 \cite{Shahzad-Nor-Mensi-Kumar-2017-PA}.

For oil and emission markets, scholars have worked on
the daily returns of WTI crude oil spot and futures prices from 2 January 1990 to 9 March 2010 \cite{Wang-Wei-Wu-2011b-PA}, 
the daily WTI crude oil and Brent crude oil from 3 January 1995 to 13 August 2013 \cite{Pal-Rao-Manimaran-2014-PA},
four pairs (CER-WTI, CER-Brent, EUA-WTI and EUA-Brent) between daily carbon prices (Certification Emission Reduction CER from 14 March 2008 to 15 December 2012 and EU emission Allowance EUA from 22 April 2005 to 15 December 2012) and daily spot prices of WTI and Brent crude oil \cite{Zhuang-Wei-Zhang-2014-PA},
three pairs among daily time series of oil (Brent Crude Oil Futures), gas (New York Mercantile Exchange Henry Hub Natural Gas Futures), and CO$_2$ (European Climate Exchange EU Allowance EUA Futures) from 22 April 2005 to 30 April 2013 \cite{Wang-Xie-Chen-Han-2014-MPE},
daily returns between two crude oil prices (Brent and WTI) and the Baltic Dry Index BDI from 19 October 1988 to 3 February 2015 \cite{Ruan-Wang-Lu-Qin-2016-PA},
and
daily returns of WTI oil and Baltic Exchange Dirty Tanker Index (BDTI) from 4 January 2000 to 4 November 2015 \cite{Chen-Miao-Tian-Ding-Li-2017-PA}.

The relationship between daily returns of crude oils and stock markets has also attracted much interest, such as
WTI and three major US stock market indices (DJIA, NASDAQ and S\&P 500) from 2 January 2002 to 29 June 2012 \cite{Wang-Xie-2012-APPB},
WTI and BRIC stock market indices (Brazil IBOVESPA, Russian RTS, Indian SENSEX and Chinese SSEC) from 1 January 2002 to 30 September 2012 \cite{Ma-Wei-Huang-Zhao-2013-PA},
WTI and ten CSI 100 sector indices of China's stock market from 1 January 2005 to 3 April 2013 \cite{Zhuang-Wei-Ma-2015-PA},
WTI and ten CSI 300 sector indices from 4 January 2005 to 2 November 2015 \cite{Yang-Zhu-Wang-Wang-2016-PA},
and
WTI and six GCC stock market indices (Saudi Arabian Tadawul All-Share Index, Qatar Doha Securities Market, Dubai General Index, Abu Dhabi General Index, Kuwait Stock Exchange Index, Dubai General Index, and Oman MSM30 Index) as well as S\&P 500 index from 5 January 2004 to 31 August 2013 \cite{Ma-Zhang-Chen-Wei-2014-PA}.
A related energy-stock pair is the electricity prices in the Indian Energy Exchange and stock index SENSEX from 1 April 2009 to 31 March 2014, in which prices are used as the input $\Delta{X}$ \cite{Ghosh-Dutta-Chakraborty-2015-PA}.

The joint multifractal nature between crude oil prices and other financial assets has also been studied and confirmed, such as
two pairs of daily returns of the crude oil (WTI and Brent) and USD/INR from 3 January 1995 to 13 August 2013 \cite{Pal-Rao-Manimaran-2014-PA},
daily returns of WTI and five exchange rates (AUD, CAD, MXN, RUB and ZAR) from 1 January 1996 to 31 December 2014 \cite{Li-Lu-Zhou-2016-PA}, 
daily returns of WTI crude oil and US agricultural commodities (corn, soybean, oat and wheat) from 3 January 1994 to 31 December 2012 \cite{Liu-2014-PA},
two pairs of daily returns of the crude oil (WTI and Brent) and gold from 3 January 1995 to 13 August 2013 \cite{Pal-Rao-Manimaran-2014-PA},
daily returns of WTI versus the 3-month Treasury bill rate and the US dollar index from 4 February 1994 to 29 February 2016 \cite{Sun-Lu-Yue-Li-2017-PA},

Other pairs are
daily data of SSEC and CNY/USD from 1 August 2005 to 20 October 2011 \cite{Cao-Xu-Cao-2012-PA},
daily returns of Hang Seng HSCEI index and RMB exchange rates (CNY and CNH) from 2 May 2012 to 27 January 2016 (MF-X-DFA and MF-CCA) \cite{Ruan-Yang-Ma-2017-PA},
daily data of market fear gauge VIX and two Japanese Yen exchange rates (USD/JPY and AUD/JPY) and from 5 January 1998 to 18 April 2016 \cite{Lu-Sun-Ge-2017-PA}, 
daily prices of India SENSEX and USD/INR from January 1995 to December 2012 \cite{Dutta-Ghosh-Chatterjee-2016-PA}),
daily returns of same futures in the Chinese and US commodity futures markets (wheat from 1 November 1999 to 13 December 2010, soybean from 4 January 1999 to 13 December 2010, soy meal from 17 July 2000 to 13 December 2010, and corn from 22 September 2004 to 13 December 2010) in the US and China \cite{Li-Lu-2012-PA},
daily returns of gold and USD/INR from 3 January 1995 to 13 August 2013 \cite{Pal-Rao-Manimaran-2014-PA},
daily returns of spot and futures markets of nonferrous metals (aluminum, copper, nickel and zinc) in the London Mercantile Exchange \cite{Liu-Wang-2014-PA},
daily returns between London and Shanghai gold prices from 2 January 2003 to 27 April 2012 (MF-X-DFA and MF-X-DMA) \cite{Cao-Han-Chen-Yang-2014-MPLB},
daily returns of RMB nominal effective exchange rate index (NEER) and four Euronext Rogers international commodity indices (whole, energy, metals, agriculture) from 21 July 2005 to 15 March 2016 \cite{Lu-Li-Zhou-Qian-2017-PA},
5-min returns of Chinese treasury futures contracts and treasury ETF from 6 September 2013 to 26 September 2014 \cite{Zhou-Chen-2016-PA},
daily returns of the 3-month Treasury bill rate and the US dollar index from 4 February 1994 to 29 February 2016 \cite{Sun-Lu-Yue-Li-2017-PA},
daily CNY/USD and CNH/USD from 1 May 2012 to 29 February 2016 \cite{Xie-Zhou-Wang-Yan-2017-FNL},
daily data of US gold price and Bombay Stock Exchange SENSEX from January 1990 to May 2013 \cite{Dutta-Ghosh-Samanta-2014-PA},
daily returns of the US interest rate (Effective Federal Funds Rate) and four CBOT agricultural commodity futures (corn, soybean, wheat and rice) from 1 January 2000 to 31 September 2014 \cite{Wang-Hu-2015-PA},
and
daily returns of Argentina stock market Merval Index and the international prices of three commodities (soybeans, corn and wheat) from 1 August 2000 to 20 April 2012 \cite{Figliola-Catalano-2016-JAS}.

\subsubsection{Volatility-volatility pairs}
\label{S3:EmpAnal:X:Volatility-Volatility}

The joint multifractal nature between two volatility time series has also been investigated. For stock market indices, some studies have been done on
the pair of daily DJIA and NASDAQ from July 1993 to November 2003 \cite{Zhou-2008-PRE},
the pair of daily A-share and B-share market indices in Shanghai and Shenzhen from 2 January 1996 to 1 April 2010 \cite{Wang-Wei-Wu-2010a-PA},
the pair of daily SSEC and SZCI from 4 April 1991 to 13 April 2009 \cite{Zhao-Shang-Jin-2011-Fractals} and from 4 January 2002 to 31 December 2008 \cite{Zhao-Shang-Shi-2014-PA}, 
the six pairs of daily SSEC, SZCI, DJIA and NASDAQ from 18 May 1995 to 18 May 2015 using MF-X-PF \cite{Dai-Hou-Gao-Su-Xi-Ye-2016-CSF},
and the pair of daily DJIA and NASDAQ from 2 January 1997 to 29 December 2014 using MF-X-PF \cite{Xiong-Shang-2016-CNSNS}.

Other studies consider
the daily volatility pair of the corresponding agricultural futures in the Chinese and US commodity futures markets (wheat from 28 December 1993 to 3 June 2010, soy meal from 17 July 2000 to 3 June 2010, soybean from 15 March 2002 to 3 June 2010, and corn from 22 September 2004 to 3 June 2010) \cite{He-Chen-2011-CSF} 
and (wheat from 1 November 1999 to 13 December 2010, soybean from 4 January 1999 to 13 December 2010, soy meal from 17 July 2000 to 13 December 2010, and corn from 22 September 2004 to 13 December 2010) in the US and China \cite{Li-Lu-2012-PA},
the pair of positive series and negative series of Leu-EUR and Leu-USD from June 1999 to November 2012 \cite{Scarlat-Cristescu-Cristescu-2013-UPBSBAMP}, 
the three pairs of daily volatility of crude oil, gold and CHY/USD exchange rate from 1 January 2005 to 1 November 2014 using MF-X-PF \cite{Dai-Shao-Gao-Sun-Su-2016-Fractals},
the daily WTI crude oil and US agricultural commodities (corn, soybean, oat, wheat) from 3 January 1994 to 31 December 2012 \cite{Liu-2014-PA},
the daily US interest rate (Effective Federal Funds Rate) and four CBOT agricultural commodities (corn, soybean, wheat and rice) from 1 January 2000 to 31 September 2014 \cite{Wang-Hu-2015-PA},
the daily Hang Seng China Enterprises Index (HSCEI) versus the daily onshore and offshore RMB exchange rates (CNY and CNH) from 2 May 2012 to 27 January 2016 (MF-X-DFA and MF-CCA) \cite{Ruan-Yang-Ma-2017-PA},
and
the daily spot and futures prices of nonferrous metals (aluminum, copper, nickel and zinc) traded on the London Mercantile Exchange \cite{Liu-Wang-2014-PA}.

\subsubsection{Price-volume pairs}
\label{S3:EmpAnal:X:price-volume}

The joint multifractal analysis of asset price and trading volume investigates the temporal features of the price-volume relationship. In finance, the price-volume relationship considers either return and volume or volatility and volume. Analogously, Wang et al. studied the pair of 5-min return and volume time series of the CSI 300 index futures from 16 April 2010 to 12 April 2012 \cite{Wang-Suo-Yu-Lei-2013-PA}, while Lin researched the pair of daily volatility and volume of NASDAQ and S\&P 500 using MF-X-PF \cite{Lin-2008-PA}. We note that the trade-by-trade price returns and waiting times of E.ON (EOA) and Deutsche Bank (DBK) from 28 November 1997 to 31 December 1999 have also been studied using MF-X-DFA and MF-CCA \cite{Oswiecimka-Drozdz-Forczek-Jadach-Kwapien-2014-PRE}.

In the econophysics community, however, the overwhelmingly largest number of investigations have been conducted on pairs of return and volume variation, stimulated by the seminal work by Podobnik et al. on the S\&P 500 index \cite{Podobnik-Horvatic-Petersen-Stanley-2009-PNAS}. Subsequent studies cover
hourly returns of day-ahead spot price of electricity and log-difference of trading volume in the Czech Republic from 1 January 2009 and 30 November 2012 using MF-CCA \cite{Fan-Li-2015-PA},
daily Shanghai SSEC index from 20 December 1990 to 30 December 2010 and Shenzhen SZCI index from 3 April 1991 to 30 December 2010 \cite{Yuan-Zhuang-Liu-2012-PA},
daily SSEC from 20 December 1990 to 30 December 2010 and SZCI from 3 April 1991 to 30 December 2010 \cite{Yuan-Zhuang-Liu-2012-PA},
daily data of four agricultural commodity futures traded on China's Zhengzhou Commodity Exchange (hard winter wheat from 28 December 1993 to 12 March 2010) and  Dalian Commodity Exchange (soy meal from 17 July 2000 to 12 March 2010, No. 1 soybeans from 15 March 2002 to 12 March 2010, corn from 22 September 2004 to 12 March 2010) and four agricultural commodity futures traded on the Chicago Board of Trade (wheat from 1 July 1978 to 12 March 2010, soy meal from 1 July 1978 to 12 March 2010, soybeans from 1 July 1978 to 16 March 2010, and corn from 1 July 1978 to 12 March 2010) \cite{He-Chen-2011a-PA}, 
daily data of the Moroccan stock market index MASI from 1 January 2000 to 20 January 2017 \cite{ElAlaoui-2017-PA},
daily data of 13 stock market indices (AEX, BFX, BSESN, BVSP, FCHI, FTSE, GSPC, HSI, KS11, MXX, N225, SSMI, TWII) from 30 August 2006 to 1 March 2014 \cite{Stosic-Stosic-Stosic-Stanley-2015b-PA}, 
daily data of the 50 constituent stocks of the Nifty index of National Stock Exchange (Mumbai) from 01 January 2002 to 31 December 2014 \cite{Hasan-Salim-2017-PA},
and
daily data of Au99.99 on the Shanghai Gold Exchange (SGE) from 30 October 2002 to 30 January 2015 and the gold futures on the Shanghai Futures Exchange (SHFE) from 9 January 2008 to 30 January 2015 \cite{Ruan-Jiang-Ma-2016-PA}. 

There are also studies on price and volume that are added up before the detrending procedure, such as
daily data of copper futures and aluminum futures traded on the Shanghai Futures Exchange (SHFE) from 15 January 2004 to 4 July 2011 and copper and aluminum futures on the London Metal Exchange (LME) \cite{Guo-Huang-Cheng-2012-Kybernetes}, 
daily copper futures series of Shanghai Futures Exchange (SHFE) from 15 September 1993 to 4 July 2011 and aluminum from 5 October 1998 to 4 July 2011 \cite{Cheng-Huang-Guo-Zhu-2013-TNMSC},
daily electricity prices and loads in the California electricity market 1999-2000 using adaptive MF-X-DFA and MF-X-SF \cite{Wang-Liao-Li-Zou-Shi-2013-Chaos,Wang-Yang-Wang-2016-PA}, 
and
1-hour and daily prices and loads data of California power market (1999-2000) and PJM power market (2001-2002) \cite{Wang-Liao-Zhou-Shi-2013-ND}.

\section{Sources of apparent multifractality}
\label{S1:MF:Sources}

Overwhelming empirical evidence shows that various financial time series of assets and their derivatives exhibit multifractality. However, the underlying physical mechanisms are not clear. This raises an important and subtle issue on the sources of the observed multifractality. It is usually argued that the sources of
multifractality in financial time series are the fat tails and/or the long-range temporal correlations \cite{Kantelhardt-Zschiegner-KoscielnyBunde-Havlin-Bunde-Stanley-2002-PA}. However, possessing linear correlations or long memory in time series is not sufficient for the emergence of multifractality and a nonlinear process is required to have intrinsic multifractality \cite{Saichev-Sornette-2006-PRE}. Further investigations consider the impacts of linear and nonlinear correlations on the observed multifractality \cite{Zhou-2009-EPL,Zhou-2012-CSF}.

An even more critical question is to ask whether the empirical extracted multifractality is intrinsic or apparent. Indeed, there are many studies showing that empirical multifractal analysis of time series generated from monofractal financial models and mathematical models can produce spurious multifractality \cite{Bouchaud-Potters-Meyer-2000-EPJB,vonHardenberg-Thieberger-Provenzale-2000-PLA}. Moreover, the estimated generalized Hurst exponents may deviate significantly from the expected values, especially for short time series \cite{Norouzzadeh-Jafari-2005-PA,Wang-Xiang-Pandey-2012-PA}. Therefore, statistical tests are absolutely necessary in empirical multifractal analysis.

It is necessary to propose a standardized terminology here for a consistent picture. Throughout this review, the empirical multifractal properties ($\tau(q)$ or $f(\alpha)$) extracted from time series using multifractal analysis methods is called {\emph{apparent multifractality}}, although it is used to represent spurious multifractality by some researchers. Then, we refer to the apparent multifractality obtained from monofractal time series as {\emph{spurious multifractality}}. Time series generated from nonlinear, multiplicative processes possess {\emph{intrinsic}} or {\emph{true multifractality}}, whose nonlinear component in the apparent multifractal is the {\emph{effective multifractality}} \cite{Zhou-2012-CSF}.

\subsection{Spurious multifractality and finite-size effect}
\label{S2:MF:Sources:SpuriousMF&FSE}

\subsubsection{Bifractality of fat-tailed time series}
\label{S3:MF:Sources:SpuriousMF&FSE:Bifractal}

Mantegna and Stanley defined the truncated L{\'{e}}vy flight (TLF)  \cite{Mantegna-Stanley-1994-PRL}, and Kopoken introduced the exponentially truncated L{\'{e}}vey flight (ETLF)  \cite{Kopoken-1995-PRE}. Nakao studied the asymptotical scaling behavior of the structure functions of uncorrelated time series whose amplitudes follow the exponentially truncated L{\'{e}}vey distribution and derived analytically the mass scaling exponents \cite{Nakao-2000-PLA}:
\begin{equation}
  \tau(q)=\left\{
    \begin{array}{lll}
       q/\gamma-1 & {\rm{for}}~~q\leq\gamma,\\
       0          & {\rm{for}}~~q>   \gamma,
    \end{array}
  \right.
  \label{Eq:MF-SF:LevyFlight:tau}
\end{equation}
where $0<\gamma\leq 2$ is the stable index, and the singularity strength and singularity spectrum are given by
\begin{equation}
  \alpha(q)=\left\{
    \begin{array}{lll}
       1/\gamma  & {\rm{for}}~~q\leq\gamma,\\
       0         & {\rm{for}}~~q>\gamma,
    \end{array}
  \right.~~~
  f(\alpha)=\left\{
    \begin{array}{lll}
       1  & {\rm{for}}~~q\leq\gamma,\\
       0  & {\rm{for}}~~q>\gamma,
    \end{array}
  \right.
\end{equation}
The bifractality also holds for ordinary L{\'e}vy processes \cite{Jaffard-1999-PTRF} and time series for which the innovations have a power-law distribution with tail index $0<\gamma<2$ \cite{Chechkin-Gonchar-2000-CSF}.

%

In addition, consider the class of stochastic process $X(t)$ whose increments form a strictly stationary sequence having heavy-tailed marginal distribution with index $\gamma$, and which satisfies the strong mixing property with an exponentially decaying rate and zero mean.
For $\gamma > 1$, the scaling exponents of the structure functions can be expressed as  \cite{Heyde-2009-MMOR,Grahovac-Leonenko-2014-CSF,Grahovac-Leonenko-Taqqu-2015-JSP,Grahovac-Jia-Leonenko-Taufer-2015-Stat} 
\begin{equation}
  \zeta(q) = \left\{
  \begin{array}{lll}
    \frac{q}{\gamma}, & {\mathrm{if}}~~0<q\leq\gamma~~{\mathrm{and}}~~\gamma\leq 2,\\
           1, & {\mathrm{if}}~~q>\gamma     ~~{\mathrm{and}}~~\gamma\leq 2,\\
    \frac{q}{2}, & {\mathrm{if}}~~0<q\leq\gamma~~{\mathrm{and}}~~\gamma> 2,\\
    \frac{q}{2}+\frac{2(\gamma-q)^2(2\gamma+4q-3\gamma{q})}{\gamma^3(2-q)^2}, & {\mathrm{if}}~~q>\gamma~~{\mathrm{and}}~~\gamma> 2.
  \end{array}
  \right.
  \label{Eq:MF-SF:PL:tau}
\end{equation}

For fractional L{\'{e}}vy flights with tail index $\gamma$ and Hurst exponent $H$, the scaling exponents of the structure functions are \cite{Heyde-Sly-2008-PA}
\begin{equation}
  \zeta(q) = \left\{
  \begin{array}{lll}
    Hq, & {\mathrm{if}}~~0<q\leq\gamma,\\
    1+q(H-1/\gamma), & {\mathrm{if}}~~\gamma<q<\frac{1}{1/\gamma-H},\\
    0, & {\mathrm{if}}~~\frac{1}{1/\gamma-H}<q,
  \end{array}
  \right.
  \label{Eq:MF-SF:Levy:tau}
\end{equation}
which depends on both the fat-tailedness and the linear correlations.

\subsubsection{Spurious multifractality}
\label{S3:MF:Sources:SpuriousMF}

Bouchaud et al. built an asymptotically monofractal model and found spurious multifractality through the structure function approach \cite{Bouchaud-Potters-Meyer-2000-EPJB}.
LeBron proposed a simple three-factor stochastic volatility (TFSV) model, which is asymptotically monofractal, and observed numerically spurious multifractality  \cite{LeBaron-2001-QF}. Synthetical time series generated from the model after calibrating with daily DJIA returns and the DJIA data share very similar structure functions. Bianchi calibrated multifractional Brownian motions with the daily S\&P 500 index from 20 October 1982 to 27 May 2004 and the daily NIKKEI 225 index from 04 January 1984 to 27 May 2004 and found that the raw time series and their multifractional Brownian motion surrogates have very similar multiscaling behavior in the partition functions \cite{Bianchi-2005-AEL}. Similar results were obtained for other six major stock indices \cite{Bianchi-Pianese-2007-QF}. These results show that we cannot distinguish monofractality and multifractality in the time series generated from stochastic volatility models and multifractional Brownian motion models, and further tests for nonlinearity are called for.

Spurious multifractality has also been observed in the numerical multifractal analysis of time series generated from monofractal or bifractal mathematical models. For instance, for L{\'e}vy flights, the estimated $\tau(q)$ function is close to the bifractal expression (\ref{Eq:MF-SF:LevyFlight:tau}), but is more or less smoothed around $q=\gamma$, which results in a spectrum of singularities. There are well-conducted examples. For uncorrelated time series with a non-Gaussian exponential amplitude distribution, spurious multifractality is numerically observed \cite{vonHardenberg-Thieberger-Provenzale-2000-PLA}.
The linear absolute increments series of Devil's staircases show weak spurious multifractality under the local singularity method and partition function approach but exhibit nice monofractality under the punctual singularity method \cite{Pont-Turiel-PerezVicente-2006-PRE}.
The linear absolute increments series of fractional Brownian motions show strong spurious multifractality when using the local singularity method and prudent singularity method 
and relatively weak spurious multifractality when using the partition function approach \cite{Pont-Turiel-PerezVicente-2006-PRE}.

Furthermore, numerical experiments show that the singularity width of the spurious multifractality depends on the power law
tail exponent $\gamma$ and on the Hurst exponent $H$.
For uncorrelated time series with $q$-Gaussian distributions that are bifractal, the singularity width $\Delta\alpha$ increases with decreasing tail exponent and the width values are quite large in the L{\'{e}}vy regime under MF-DFA and WTMM \cite{Drozdz-Kwapien-Oswiecimka-Rak-2009-EPL}.
The fact that $\Delta\alpha$ decreases with increasing $N$ has also been observed in the MF-DFA analysis of monofractal time series \cite{Czarnecki-Grech-2010-APPA}.
For Fourier transform surrogates of stochastic binomial measures, the singularity width $\Delta\alpha$ obtained from MF-DFA and centered MF-DMA increases with the Hurst index $H$ \cite{Schumann-Kantelhardt-2011-PA}.

\subsubsection{Finite-size effect}
\label{S3:MF:Sources:FSE}

Zhou generated random numbers of different sizes ($N$) from the empirical distribution of DJIA volatilities, and different linear correlations with different Hurst exponents ($H$) were added by manipulating the generated time series using the IAAFT algorithm \cite{Zhou-2012-CSF}. The surrogate time series thus do not have multifractal nature by construction. Numerical experiments unveil that the singularity width depends on $N$ and $H$ as follows
\begin{equation}
 \Delta\alpha(H,N) \approx N^{-2(1-H)} {\mathrm{e}}^{10(1-H)}.
 \label{Eq:FSE}
\end{equation}
This means that $\Delta\alpha(H,N)\to 0$ for long time series and the decay is faster for small $H$. Therefore, the linear correlation component $\Delta\alpha_{\rm{LM}}$ is the outcome of the finite-size effect and the linear correlation component of the apparent multifractal is
\begin{equation}
 \Delta\alpha_{\rm{LM}}=\Delta\alpha_{\rm{FSE}}.
 \label{Eq:dA:LM:FSE}
\end{equation}

Similarly, for autocorrelated time series with $H>0.5$ generated based on the Fourier filtering method \cite{Makse-Havlin-Schwartz-Stanley-1996-PRE}, Grech and Pamu{\l}a determined numerically the dependence of $\Delta{h}$ with respect to the Hurst exponent $H$ and the sample size $N$ \cite{Grech-Pamula-2012-APPA}:
\begin{equation}
 \Delta{h}(H,N) \approx 1.206N^{-0.175}(1-H)+0.453N^{-0.124}(2H-1),
 \label{Eq:FSE:Dh}
\end{equation}
or, almost equivalently \cite{Grech-Pamula-2013-PA},
\begin{equation}
 \Delta\alpha(H,N) \approx 1.372N^{-0.126}(1-H)+0.572N^{-0.089}(2H-1),
 \label{Eq:FSE:Dalfa}
\end{equation}
both of which also suggest that the singularity width decreases with increasing sample size and decreasing Hurst exponent. There are corrections to these formulae when $q$ is not sufficiently large \cite{Pamula-Grech-2014-EPL}.

Overall, time series display apparent multiscaling behaviour as a consequence of a slow transition phenomenon on finite time scales.

\subsection{Empirical investigation of different sources}

\subsubsection{Impacts of linear and nonlinear correlations (Shuffle)}
\label{S3:MF:Sources:Shuffling}

To investigate and distinguish different sources of apparent multifractality, the simplest and most adopted method is to shuffle or reshuffle the raw time series and compare the singularity widths of the raw time series ($\Delta\alpha$) and the shuffled time series ($\Delta\alpha_{\rm{shuf}}$). Since the shuffling procedure destroys any linear or nonlinear long-range correlations while preserving the distribution of the values of the time series, $\Delta\alpha_{\rm{shuf}}$ qualifies the contribution of the broad probability distribution and probably the finite-size effect as well \cite{Kantelhardt-Zschiegner-KoscielnyBunde-Havlin-Bunde-Stanley-2002-PA}
\begin{equation}
  \Delta\alpha_{\rm{PDF}} = \Delta\alpha_{\rm{shuf}}~.
\end{equation}
The impact of correlations on the apparent multifractality is qualified by \cite{Kantelhardt-Zschiegner-KoscielnyBunde-Havlin-Bunde-Stanley-2002-PA}
\begin{equation}
  H_{\rm{corr}}(q) = H(q)- H_{\rm{shuf}}(q),
\end{equation}
which can be abbreviated into a single value \cite{Norouzzadeh-Rahmani-Norouzzadeh-2007-IJMPC,Zunino-Figliola-Tabak-Perez-Garavaglia-Rosso-2009-CSF}
\begin{equation}
  \Delta{H}_{\rm{corr}} = \Delta{H}- \Delta{H}_{\rm{shuf}},
\end{equation}
where $\Delta{H}=H_{\max}-H_{\min}$.
Equivalently, we have
\begin{equation}
  \Delta\alpha_{\rm{corr}} = \Delta\alpha-\Delta\alpha_{\rm{shuf}}.
\end{equation}
where $\Delta\alpha=\alpha_{\max}-\alpha_{\min}$.
A comparison of the values of $\Delta\alpha_{\rm{shuf}}$ and $\Delta\alpha_{\rm{corr}}$ can determine which source dominates.

Most empirical analyses reported a significant decrease of the apparent multifractality after shuffling, indicating the nontrivial impact of long-range correlations. Empirical examples contain
the daily returns of 29 commodities and 2445 stocks \cite{Matia-Ashkenazy-Stanley-2003-EPL},
the trade-by-trade returns and the intertrade durations of 30 DJIA stocks and 30 DAX stocks from 1 December 1997 to 31 December 1999 \cite{Oswiecimka-Kwapien-Drozdz-Rak-2005-APPB},
the Iranian rial-US dollar exchange rates from 24 September 1989 to 15 November 2003 ($\Delta\alpha = 3.54$ and $\Delta\alpha_{\rm{shuf}} = 0.60$) \cite{Norouzzadeh-Rahmani-2006-PA},
the 1-min WIG20 index returns from 4 January 1999 to 31 October 2005 ($\Delta\alpha \approx 0.14$ and $\Delta\alpha_{\rm{shuf}} \approx 0.04$) \cite{Rak-Drozdz-Kwapien-2007-PA}, %
the 1-min index returns of S\&P 500, DAX and WIG 20 from May 2004 to May 2006 \cite{Drozdz-Forczek-Kwapien-Oswicimka-Rak-2007-PA}, 
the 20s returns of the DEM/USD exchange rates from 1 October 1992 to 30 September 1993 ($\Delta\alpha \approx 0.69$ and $\Delta\alpha_{\rm{shuf}} \approx 0.34$) \cite{Nascimento-Junior-Jennings-Serva-Gleria-Viswanathan-2008-EPL},
the 20s volatilities of the DEM/USD exchange rates from 1 October 1992 to 30 September 1993 ($\Delta\alpha \approx 2.5 $ and $\Delta\alpha_{\rm{shuf}} \approx 0.37$) \cite{Nascimento-Junior-Jennings-Serva-Gleria-Viswanathan-2008-EPL}, 
the daily DJIA returns from 26 May 1896 to 27 April 2007 ($\Delta\alpha = 0.22$ and $\Delta\alpha_{\rm{shuf}} = 0.18$) \cite{Zhou-2009-EPL},
the daily returns of the Indian BSE index from July 1997 to December 2007 ($\Delta\alpha = 0.442$ and $\Delta\alpha_{\rm{shuf}} = 0.399$), the Indian NSE index from August 2002 to December 2007 ($\Delta\alpha = 0.558$ and $\Delta\alpha_{\rm{shuf}} = 0.473$), and the S\&P 500 index from July 1997 to December 2007 ($\Delta\alpha = 0.548$ and $\Delta\alpha_{\rm{shuf}} = 0.328$) \cite{Kumar-Deo-2009-PA},
the daily returns of the SENSEX index from January 2003 to December 2009 \cite{Dutta-2010-CJP},
the daily volatility of spot WTI and Brent oil prices from 20 May 1987 to 14 October 2009 ($\Delta\alpha = 0.439$ and $\Delta\alpha_{\rm{shuf}} = 0.142$ for WTI and $\Delta\alpha = 0.371$ and $\Delta\alpha_{\rm{shuf}} = 0.159$ for Brent) \cite{He-Chen-2010a-PA},
the daily returns of four foreign exchange rates (JPY/USD, HKD/USD, KRW/USD and THD/USD) from 1991 to 2005 \cite{Oh-Eom-Havlin-Jung-Wang-Stanley-Kim-2012-EPJB},
the daily returns of the Czech PX index from April 1994 to December 2010 ($\Delta\alpha = 0.69$ and $\Delta\alpha_{\rm{shuf}} = 0.44$) \cite{Caraiani-2012-PLoS1}, 
the daily gold price returns from 1973 to 2011 \cite{Ghosh-Dutta-Samanta-2012-APPB},
the daily returns of the Athens Stock Exchange index, the Irish Stock Exchange index and the Portuguese Stock Exchange index \cite{Siokis-2014-PA},
the daily index returns for the US and seven Asian markets from 1 July 2002 to 31 December 2012 \cite{Hasan-Salim-2015-PA},
the daily returns of the CSI energy sub-industry index (ESI) from 4 January 2005 to 15 June 2015 ($\Delta\alpha = 0.505$ and $\Delta\alpha_{\rm{shuf}} = 0.427$) \cite{Yang-Zhu-Wang-2016-PA}, 
and so on.


There are also examples for which the reshuffled time series has a broader singularity width than the raw time series. For instance, it is reported that $\Delta\alpha = 0.773$ and $\Delta\alpha_{\rm{shuf}} = 1.029$ for the daily volatilities of the CSI energy sub-industry index (ESI) from 4 January 2005 to 15 June 2015 \cite{Yang-Zhu-Wang-2016-PA}, and $\Delta\alpha = 0.44$ and $\Delta\alpha_{\rm{shuf}} = 0.50$ for the daily returns of the Polish WIG index from June 1993 and December 2010 \cite{Caraiani-2012-PLoS1}.
%
The observation that $\Delta\alpha_{\rm{shuf}}$ is approximately equal to or even greater than $\Delta\alpha$ should be interpreted with caution. The basic explanation is that the apparent multifractality detected in the time series under investigation is independent of any correlation structures in the data and is thus fully determined by the usually non-Gaussian distribution. However, due to the ubiquitous presence of finite-size effects in multifractal analysis, even multifractal time series generated from Markov-Switching Multifractal models may lead to the same result \cite{Barunik-Aste-DiMatteo-Liu-2012-PA}. Hence, the observation $\Delta\alpha_{\rm{shuf}} \geq \Delta\alpha$ per se does not deny the presence of intrinsic multifractality. One should either perform statistical tests based on the IAAFT algorithm or choose a proper scaling range. Indeed, such an anomaly disappears when the scaling range contains larger time scales \cite{Buonocore-Aste-DiMatteo-2016-CSF}. The shuffling test does show that $H(1)=H(2)$ is rejected for time series generated from the multifractal model of asset returns \cite{Lee-Chang-2015-CSF}.

The adopted multifractal analysis method could also have impacts on the results. Caraiani found that, for the daily returns of the Hungarian BUX index from June 1993 to December 2010, $\Delta\alpha = 0.54$ and $\Delta\alpha_{\rm{shuf}} = 0.34$ for the MF-DFA method and $\Delta\alpha = 0.49$ and $\Delta\alpha_{\rm{shuf}} = 0.65$ for the EMD-MF-DFA method \cite{Caraiani-2012-PLoS1}. Although the results of the shuffling test could be different, there is a general result showing that $\alpha_{\rm{shuf}}(q=0)\approx 0.5$ for shuffled time series.

\subsubsection{Impact of nonlinear correlations (Surrogate)}


A time series has only linear correlations if all the correlation structures of the time series is contained in the autocorrelation function, or equivalently, in the Fourier power spectrum. In this case, its dynamics can be captured by AR or ARMA  models. If the correlation structures of a time series cannot be captured by the power spectrum, it contains nonlinearity. There are several methods to generate linear surrogates without nonlinearity.
The most commonly used methods for generating surrogates include the Fourier transform algorithm \cite{Theiler-Eubank-Longtin-Galdrikian-Farmer-1992-PD}, the amplitude adjusted Fourier transform (AAFT) algorithm \cite{Theiler-Eubank-Longtin-Galdrikian-Farmer-1992-PD}, and the iterative amplitude adjusted Fourier transform (IAAFT) algorithm \cite{Schreiber-Schmitz-1996-PRL}.

The Fourier transform surrogate time series are constructed to have the same linear correlations as the raw data, while any nonlinear correlations are eliminated and the amplitude distribution becomes Gaussian \cite{Provenzale-Villone-Babiano-Vio-1993-IJBC}. One performs discrete Fourier transform of the raw time series, randomizes the phases of the amplitude through multiplying the discrete Fourier transform of the data by random phases, and performs an inverse Fourier transform to create phase-randomized surrogates \cite{Theiler-Eubank-Longtin-Galdrikian-Farmer-1992-PD}.
The phase randomization approach, adopted in the multifractal analysis of financial time series, was introduced for the multifractal analysis of heartbeat intervals \cite{Ivanov-Amaral-Goldberger-Havlin-Rosenblum-Stanley-Struzik-2001-Chaos}. Because the phase randomization procedure weakens non-Gaussianity in time series, the difference $ H_{\rm{PDF}}(q) = H(q)- H_{\rm{FT}}(q)$ between $H(q)$ and $H_{\rm{FT}}(q)$ is usually argued to reflect the impact of broad distributions \cite{Movahed-Jafari-Ghasemi-Rahvar-Tabar-2006-JSM,Norouzzadeh-Dullaert-Rahmani-2007-PA}. This can also be represented as
 \cite{Norouzzadeh-Rahmani-Norouzzadeh-2007-IJMPC,Zunino-Figliola-Tabak-Perez-Garavaglia-Rosso-2009-CSF}
\begin{equation}
  \Delta{H}_{\rm{PDF}} = \Delta{H}- \Delta{H}_{\rm{FT}},
\end{equation}
or equivalently
\begin{equation}
  \Delta\alpha_{\rm{PDF}} = \Delta\alpha- \Delta\alpha_{\rm{FT}}.
\end{equation}
However, by construction, both the impacts of PDF and nonlinear correlations are removed from the Fourier transform surrogates. The difference $\Delta\alpha- \Delta\alpha_{\rm{FT}}$ contains both PDF and nonlinearity effects.

Numerous studies have implemented the shuffling test and the Fourier transform surrogate test to understand the sources of the empirically estimated apparent multifractality in financial time series. The majority of them found that
\begin{equation}
  \Delta\alpha > \Delta\alpha_{\rm{shuf}} > \Delta\alpha_{\rm{FT}}.
\end{equation}
An almost complete list includes
the 5-min stock returns for the 100 largest US companies traded on NYSE or NASDAQ from 1 December 1997 to 31 December 1999 \cite{Kwapien-Oswiecimka-Drozdz-2005-PA},
the daily returns of the Iranian rial-US dollar exchange rates from 24 September 1989 to 15 November 2003 \cite{Norouzzadeh-Rahmani-2006-PA}, 
the daily TEPIX returns from 20 May 1995 to 19 January 2005 \cite{Norouzzadeh-Rahmani-Norouzzadeh-2007-IJMPC},
1-min returns of the TAIEX index and 150 stocks of the TSEC Taiwan 50 and TSEC Taiwan Mid-Cap 100 indices from January 2001 to August 2007 \cite{Su-Wang-Huang-2009-JKPS},
the daily return and volatility time series of market indices for Argentina, Brazil, Chile, Colombia, Mexico, Peru, Venezuela and the US from January 1995 to February 2007 \cite{Zunino-Figliola-Tabak-Perez-Garavaglia-Rosso-2009-CSF},
the daily returns of 172 individual stocks included in the Korean KOSPI index from 1980 to 2003 \cite{Oh-Kim-Eom-2010-JKPS}, 
the daily averaging prices of four Korean agricultural commodities (leek, radish, onion, and Korean cabbage) from 3 January 2001 to 23 September 2009 \cite{Kim-Oh-Kim-2011-PA},
the daily logarithmic variations of trading volume of the S\&P 500 index from 1980 to 2009 \cite{Bolgorian-2011-PA},
the daily returns of spot and futures prices of WTI crude oil from 2 January 1990 to 9 March 2010 \cite{Wang-Wei-Wu-2011b-PA},
the daily exchange rate returns of USD/AUD, USD/EUR and CNY/USD \cite{Wang-Wu-Pan-2011-PA},
the daily and weekly spot rates of VLGCs on the benchmark Persian Gulf (Ras Tanura) to Japan (Chiba) route from 3 January 1992 to 24 June 2009 \cite{Engelen-Norouzzadeh-Dullaert-Rahmani-2011-EE},
the daily returns of four commodity futures (hard winter wheat from 28 December 1993 to 18 September 2009, soy meal from 17 July 2000 to 18 September 2009, No. 1 soybean from 15 March 2002 to 18 September 2009, and corn from 22 September 2004 to 18 September 2009) in China (it is more complicated for the USA commodity futures) \cite{He-Chen-2010b-PA},
the 1-min realized volatility of SSEC from 4 January 2000 to 31 December 2007 \cite{Jia-Cui-Li-2012-PA},
the daily returns of CIB-CNY Composite Index (CCI) compiled by the China Industrial Bank Research from 21 July 2005 to 30 June 2008 \cite{Wang-Yu-Suo-2012-PA},
the daily returns of 144 Korean equity and balanced funds from 1 January 2002 to 31 December 2010 \cite{Kim-Yim-Kim-Oh-2015-JKPS},
the daily returns of the CSI energy sub-industry index from 4 January 2005 to 15 June 2015 \cite{Yang-Zhu-Wang-2016-PA},
and
the daily volume changes of the Moroccan MASI index from 1 January 2000 to 20 January 2017 \cite{ElAlaoui-2017-PA}.

There are also examples showing that
\begin{equation}
  \Delta\alpha > \Delta\alpha_{\rm{FT}} > \Delta\alpha_{\rm{shuf}},
\end{equation}
such as
the 5 min returns of the SSEC and SZCI from 4 January 2002 to 31 December 2008 \cite{Chen-Wu-2011-PA},
the daily returns of the world gold prices from 1968 to 2010 \cite{Bolgorian-Gharli-2011-APPB},
the daily logarithmic variations of individual and institutional traders' trading volume in Tehran Stock Exchange from 1998 to 2009 \cite{Bolgorian-2011-PA},
the daily exchange rates of CAD/USD, JPY/USD, USD/GBP, DKK/USD, NOK/USD, MXN/USD, BRL/USD, INR/USD and THB/USD \cite{Wang-Wu-Pan-2011-PA},
the daily returns of CIB-CNY Composite Index from 30 June 2008 to 18 June 2010 \cite{Wang-Yu-Suo-2012-PA},
the 10-minute returns of the CSI 300 futures IF1009 from 16 April 2010 to 17 September 2010 \cite{Lu-Tian-Zhou-Li-2013-PA},
the daily spot rates of ships in tanker markets from 27 January 1998 (or 1 July 2004 or 15 July 2005) to 5 August 2013 \cite{Zheng-Lan-2016-PA},
the daily returns of WTI oil and Baltic Exchange Dirty Tanker Index from 4 January 2000 to 4 November 2015 \cite{Chen-Miao-Tian-Ding-Li-2017-PA},
and
the daily returns of the Moroccan MASI index from 1 January 2000 to 20 January 2017 \cite{ElAlaoui-2017-PA}.


%
%
%
%

The impacts of different sources of apparent multifractality could change for different sample periods. 
Zhou et al. investigated the 5 min high-frequency returns of China's CSI 300 index from 4 April 2005 to 19 January 2012 and found that $\Delta\alpha > \Delta\alpha_{\rm{FT}} > \Delta\alpha_{\rm{shuf}}$ \cite{Zhou-Dang-Gu-2013-PA}. They further found that $\Delta\alpha > \Delta\alpha_{\rm{FT}} > \Delta\alpha_{\rm{shuf}}$ for the period from 8 April 2005 to 16 April 2010 and $\Delta\alpha > \Delta\alpha_{\rm{shuf}} > \Delta\alpha_{\rm{FT}}$ for the period from 19 April 2010 to 19 January 2012, where 16 April 2010 is the date when the CSI 300 stock index futures were first listed on the China Financial Future Exchange. Even more complicated situations are observed for the daily gold price returns in the Indian market and the daily global consumer price index returns \cite{Mali-2014-Fractals,Mali-Mukhopadhyay-2014-PA}.


In contrast, the AAFT algorithm aimed at generating surrogate time series having the same Fourier spectrum and distribution as the raw data \cite{Theiler-Eubank-Longtin-Galdrikian-Farmer-1992-PD}. The values of the raw data are first replaced by random numbers taken from the normal distribution based on the rank-ordering technique. The rank-ordered time series has the same rank ordering as the raw series such that their $i$th elements have the same rank in both time series. The Fourier transform surrogate of the rank-ordered time series is obtained and transformed back to have the same amplitude distribution as the raw time series through rank ordering.
However, the AAFT algorithm can produce the same power spectrum only for time series with infinite length in the limit $N\to\infty$ and does not usually result in the same sample power spectra \cite{Schreiber-Schmitz-1996-PRL}. To overcome this shortcoming, the IAAFT algorithm is introduced, which improves the AAFT algorithm \cite{Schreiber-Schmitz-1996-PRL}. In the IAAFT algorithm, the values of the raw time series $\{x(i)| i=1,2,\cdots,N\}$ are sorted, resulting in a new sequence $\{y_N\}$. We obtain the squared amplitudes $\{Y_k^2\}$ of the Fourier transform of $\{y_N\}$. The initial sequence $\{y_N^{(0)}\}$ of the iteration is a random shuffle of $\{y_N\}$. In the $j$th iteration, the squared amplitudes $\{Y_k^{2,(j)}\}$ of the Fourier transform of $\{y_N^{(j)}\}$ are obtained and replaced by $\{Y_k^2\}$, which are transformed back, and then the resulting series are replaced by $\{x_N\}$ with rank ordering. The iteration stops until a given accuracy is reached. For both AAFT and IAAFT, the surrogates still contains Fourier phase correlations \cite{Rath-Gliozzi-Papadakis-Brinkmann-2012-PRL}.
Unfortunately, the IAAFT surrogate test has been applied only in very few studies \cite{Zhou-2009-EPL,Zhou-2012-CSF,Kumar-Deo-2015-PJP,Chen-Tian-Ding-Miao-Lu-2016-PA}. We note that there are also wavelet-based methods to generate multifractal surrogates preserving the multifractal nature of the raw time series \cite{Palus-2008-PRL,Keylock-2017-PRE}.




Similar studies have been conducted on the multifractal cross correlations of pairs of time series, such as
the daily returns of copper futures and aluminum futures traded on the Shanghai Futures Exchange (SHFE) from 15 January 2004 to 4 July 2011 and copper and aluminum futures on the London Metal Exchange (LME) \cite{Guo-Huang-Cheng-2012-Kybernetes},
the daily returns of the BDI-Brent pair and the BDI-WTI pair \cite{Ruan-Jiang-Ma-2016-PA},
the daily exchange rates CNY/USD and CNH/USD from 1 May 2012 to 29 February 2016 \cite{Xie-Zhou-Wang-Yan-2017-FNL},
the daily returns and volume changes of the Moroccan MASI index from 1 January 2000 to 20 January 2017 \cite{ElAlaoui-2017-PA},
and
the daily returns of WTI oil and Baltic Exchange Dirty Tanker index from 4 January 2000 to 4 November 2015 \cite{Chen-Miao-Tian-Ding-Li-2017-PA}, in which the singularity widths reduce for the shuffled and surrogate data.
There are also cases in which an even larger width is observed after shuffling or surrogate, such as
the daily data of market fear gauge VIX and two Japanese Yen exchange rates (USD/JPY and AUD/JPY) from 5 January 1998 to 18 April 2016 \cite{Lu-Sun-Ge-2017-PA}, 
the daily returns of RMB nominal effective exchange rate index (NEER) and four Euronext Rogers international commodity indices (whole, energy, metals, agriculture) from 21 July 2005 to 15 March 2016 \cite{Lu-Li-Zhou-Qian-2017-PA},
and the pairs between the steel warehouse-out quantity of three warehouses (YZ, JS and BY) in China \cite{Yao-Lin-Zheng-2017-PA}.
There anomalous results should be interpreted with caution and further studies are needed.

\subsubsection{Impact of broad distributions}

For log-normal $\mathcal{W}$-cascades \cite{Arneodo-Bracy-Muzy-1998-JMP}, according to Eq.~(\ref{Eq:MF:MathModel:W:cascades:LogNormal:alpha:min:max}), the singularity width is expressed as
\begin{equation}
  \Delta\alpha = {2\sqrt{2}\sigma}/{\sqrt{\ln2}},
  \label{Eq:MF:MathModel:W:cascades:LogNormal:Delta:alpha}
\end{equation}
where $\sigma$ is the standard deviation of the $\ln(w)$ random variable. which is normally distributed. It follows immediately that $\Delta\alpha$ increases with $\sigma$. Oh et al. performed extensive numerical analysis to validate this relation between singularity width and extreme values \cite{Oh-Eom-Havlin-Jung-Wang-Stanley-Kim-2012-EPJB}. Specifically, they generated multifractal time series from log-normal $\mathcal{W}$-cascades with different means $\mu$ and different standard deviations $\sigma$. The tail exponents $\gamma$ of the multifractal data were obtained using the maximum likelihood estimator of Clauset et al. \cite{Clauset-Shalizi-Newman-2009-SIAMR}. The tail exponent is smaller if the tail is fatter and there are more extreme values. It is found that $\Delta\alpha$ is independent of $\mu$ and increases when $\gamma$ decreases.

To investigate the impact of the distribution in empirical studies, Zhou adopted the rank-ordered remapping technique to generate surrogate time series by replacing the raw data with random numbers drawn from a prescribed distribution while keeping the original orders \cite{Zhou-2009-EPL}. The algorithm is described as follows. For a given distribution, we generate a sequence of random numbers $\{z_0(i)| i=1,2,\cdots,N\}$ \cite{Press-Teukolsky-Vetterling-Flannery-1996}, which are rearranged such that the rearranged series $\{z(i)| i=1,2,\cdots,N\}$ has the same rank ordering as the raw series $\{x(i)| i=1,2,\cdots,N\}$. In other words, $z(i)$ should be the $n$th largest number in sequence $\{z(i)| i=1,2,\cdots,N\}$ if and only if $x(i)$ is the $n$th largest number in the $\{x(t)| i=1,2,\cdots,N\}$ sequence \cite{Bogachev-Eichner-Bunde-2007-PRL,Zhou-2009-EPL}. The series $\{z(i)\}$ is rescaled to have the same standard deviation
$\sigma$ of the returns $\{x(i)\}$:
\begin{equation}
 z(i) \to z(i)\times\sigma/\sigma_z+\mu,
 \label{Eq:zi:rescale}
\end{equation}
where $\sigma_z$ is the standard deviation of $\{z(i)\}$ and $\mu$ is the sample mean of $\{x(i)\}$.

Zhou investigated two families of distributions with fat tails in the MF-DFA analysis of daily DJIA returns \cite{Zhou-2009-EPL}. The first one is a family of ``double'' Weibull distributions
\begin{subequations}
\begin{equation}
 p(x) = \beta x^{\beta-1}e^{-|x-\mu|^\beta},
 \label{Eq:pdf:Weibull}
\end{equation}
where the shape parameter $\beta$ describes the heaviness of the tails and we require that $\beta<1$. The second one is a family of Student's t distributions
\begin{equation}
 p(x) = \frac{\Gamma\left(\frac{\gamma+1}{2}\right)}{\sqrt{\gamma\pi}\Gamma(\frac{\gamma}{2})}
        \left[1+\frac{(x-\mu)^2}{\gamma}\right]^{-(\gamma+1)/2},
 \label{Eq:pdf:Student}
\end{equation}
which have power-law tails with tail exponent $\gamma$.
\end{subequations}
Numerical simulations showed that $\Delta\alpha$ decreases with increasing $\beta$ for the Weibull family and increasing $\gamma$ for the Student family \cite{Zhou-2009-EPL}. Speaking differently, time series with heavier tails (or smaller $\beta$ or $\gamma$) exhibit stronger multifractality. Qualitatively similar results are obtained for the MF-PF method \cite{Zhou-2012-CSF}.

He and Wang investigated the impact of the distribution on the multifractality of SSEC and S\&P 500 returns from 20 December 1990 through 7 January 2016 using MF-DFA \cite{He-Wang-2017-PA}. The singularity width of the apparent multifractality is $\Delta\alpha = 0.593$ for SSEC and $\Delta\alpha = 0.597$ for S\&P 500. When Gaussian surrogates are used, the singularity width decreases significantly to $\Delta\alpha_{\rm{surr}} = 0.294$ for SSEC and $\Delta\alpha_{\rm{surr}} = 0.241$ for S\&P 500. In contrast, shuffling the raw time series has less impact and $\Delta\alpha_{\rm{shuf}} = 0.645$ for SSEC and $\Delta\alpha_{\rm{shuf}} = 0.400$ for S\&P 500.

\subsubsection{Impact of extreme values}
\label{S3:MF:Sources:Extrema}

The multifractal behavior of time series is influenced by large fluctuations, especially extreme events such as market crashes, because the small singularities characterize large fluctuations unveiled by large $q$ values. Different types of investigation have been designed and applied.


Studies show that removing a single market crash has a large influence on the singularity width.
Using the structure function approach, LeBaron compared the multifractal properties of the daily DJIA index time series sampled from 1 January 1897 to 3 February 1999 and the same sample with the crash on 19 October 1987 removed \cite{LeBaron-2001-QF}. He found that the removal of the Black Monday crash had significant influence mainly on multifractal properties for positive $q$'s. The minimum singularity $\alpha_{\min}$ became smaller and the singularity width $\Delta\alpha$ narrower. Based on value-weighted NYSE-AMEX-NASDAQ index (``CRSP Index'') and five individual stocks from July 1962 to December 1998, Calvet and Fisher showed that high-order structure functions of the whole samples vary considerably, with striking linearity after simply removing the crash on 19 October 1987 from the samples \cite{Calvet-Fisher-2002-RES}.

One can also compare the multifractal behavior of crisis periods and calm periods.
Using MF-DFA, Siokis compared the multifractal behavior of daily DJIA returns around two crashes (from August 1928 to January 1931 and August 1986 to December 1988 respectively) and a ``non-crisis'' calm period in 1964. The two crashes happened on 28 October 1929 with one-day loss of 14.5\% and on 19 October 1987 with a one-day loss of 25.6\%  \cite{Siokis-2013-PA}. It is found that, with $q\in[-5,5]$, $\Delta\alpha = 1.04$ for the 1929 crash, $\Delta\alpha=0.97$ for 1987, and $\Delta\alpha=0.19$ for the calm period.
Redelico and Proto investigated the NASDAQ 100 index and reported that the singularity spectrum widened from the calm period before 1 April 1997 to the bubble period before 03/27/2000 \cite{Redelico-Proto-2012-PA}.

The multifractal behavior of the whole sample and of the time series around crises also differs.
Caraiani studied the multifractal behavior of the daily return time series of three emerging European stock market indices (Czech PX index from April 1994 to December 2010, Hungarian BUX index from June 1993 to December 2010, and the Polish WIG index between June 1993 and December 2010) \cite{Caraiani-2012-PLoS1}. The results obtained from MF-DFA and EMD-based MF-DFA show that the singularity width $\Delta\alpha$ is larger during the crisis period (2008-2009) than during the whole sample, with one exception. Although the crisis does influence the overall shape of the multifractal spectrum, no significant influence of the crisis can be found on the multifractality of the series according to the $\chi^2$-test.

There are also significant differences in the multifractal behavior before and after crashes.
Yalamova investigated the multifractal behavior of seven stock market indices (DJIA, S\&P500, AUS, TSX, NIKKEI, NASDAQ and FTSE) before and after the 1987 crash \cite{Yalamova-2009-Fractals}. The most probable singularity exponent $\alpha_0$ decreases for DJIA, S\&P500, TSX, NASDAQ and FTSE, increases for AUS, and remains stable for NIKKEI.
The intermittency $\lambda$ estimated from the log-normal MMAR model increases, and $\alpha_{\min}$ decreases (most pronounced for AUS and NIKKEI).
Siokis compared the multifractal behavior of daily DJIA returns 300 trading days before and after the two crashes (1929, 1987) and found that the singularity width increases dramatically after the crashes \cite{Siokis-2013-PA}.
Oh et al. investigated how the 1997 Asian crisis affected the multifractality of four Asian foreign exchange rates (JPY/USD, HKD/USD, KRW/USD and THB/USD) from 1991 to 2005 \cite{Oh-Eom-Havlin-Jung-Wang-Stanley-Kim-2012-EPJB}. They found that, for the two sub-periods from 1991 to 1996 (before the crisis) and from 1998 to 2005 (after the crisis), the singularity width became larger after the crisis.
Yim, Oh and Kim performed multifractal analysis of the return time series of 27 highly capitalized TAQ stock data sets in the KOSPI market and obtained the singularity widths in three periods, that is, 2 January 2008 to 16 September 2008 (before crisis), 17 September 2008 to 26 November 2008 (during crisis) and 27 November 2008 to December 2009 (after crisis). They found counter-intuitively that the singularity width in the crisis period is smaller than those before and after crisis periods \cite{Yim-Oh-Kim-2014-PA}.
The multifractal strength of the daily returns of sector indices increased after the great financial crisis in 2008, as shown with nine Dow Jones sector ETF indices from 12 June 2000 to 23 July 2015 \cite{Tiwari-Albulescu-Yoon-2017-PA} and ten DJIM sector indices of Islamic stock markets from 9 November 1998 to 5 March 2015 \cite{Mensi-Tiwari-Yoon-2017-PA}.


Replacing extreme values with non-extreme values can also be used to illustrate the impacts of extreme values on the degree of multifractality. Oh et al. created new time series versions by replacing values above a certain threshold $M$ in units of the standard deviation $\sigma$ of the time series by linear interpolation \cite{Oh-Eom-Havlin-Jung-Wang-Stanley-Kim-2012-EPJB}. Taking the daily return time series of four foreign exchange markets JPY/USD, HKD/USD, KRW/USD and THD/USD from 1991 to 2005 as examples, they showed that $\Delta\alpha$ increases with $M$.
Zhou adopted a slightly different replacement strategy by using returns sampled randomly from the return series with $|r(t)| <M\sigma$ to analyse daily DJIA returns from 26 May 1896 to 27 April 2007 and obtained the same qualitative conclusion \cite{Zhou-2009-EPL}, which was confirmed by Siokis \cite{Siokis-2013-PA}. Alternatively, Green et al. proposed to replace any returns with $|r(t)|\leq{M}\sigma$ by $\mbox{sign}(r)M\sigma$ for the daily returns of the DJIA from 1928 to 2012 and 1-min returns of the Dow Jones Euro Stoxx 50 from May 2008 to April 2009 and found the similar effect \cite{Green-Hanan-Heffernan-2014-EPJB}. Similar results are observed for the daily index returns for the US and seven Asian markets from 1 July 2002 to 31 December 2012 \cite{Hasan-Salim-2015-PA} and for the S\&P 500, NASDAQ, FTSE 100 and NIKKEI 225 during highly anxious time periods around three major crises (the Black Monday, Dot-Com and the latest Great Recession) \cite{Siokis-2017-Fractals}.

He and Wang investigated the effects of extreme events on the apparent multifractality of the daily return time series from 20 December 1990 through 7 January 2016 for China's SSEC index and the S\&P 500 index \cite{He-Wang-2017-PA}. They replaced the 2.5\% largest (positive) returns and 2.5\% smallest (negative) returns with returns randomly from the remaining 95\% returns. They found that the removal of the 5\% extreme returns mainly impacts the behavior for positive $q$'s, which reduces the minimum singularity $\alpha_{\min}$ and narrows the singularity width $\Delta\alpha$. Zhou et al. observed similar narrowing effect after replacing the 5\% extreme returns of the 5-min CSI 300 index for the periods from 8 April 2005 to 16 April 2010 and from 4 April 2005 to 19 January 2012, but a widening effect with $\Delta{\alpha}$ from 0.29 to 0.33 for the period from 19 April 2010 to 19 January 2012 \cite{Zhou-Dang-Gu-2013-PA}.

When a high-frequency intraday price or index time series is analyzed with the partition function approach, the overnight jumps may influence the apparent multifractality. Lee et al. considered the 1-min time series of Korean KOSPI from 30 March 1992 through 30 November 1999 and found that the removal of overnight jumps changed the $H(q)$ curve \cite{Lee-Lee-Rikvold-2006-PA}.

Indeed, the local H{\"o}lder exponents localized at the extreme events are also outliers when compared with the local H{\"o}lder exponents of other non-extreme events \cite{Struzik-Siebes-2002-PA}. Therefore, removing the extreme events is equivalent to removing singularity outliers and thus narrows the width of the singularity spectrum. Identifying local H{\"o}lder exponent outliers also provides a possible method for the detection of outliers in time series \cite{Struzik-Siebes-2002-PA}.

%

\subsection{Components of apparent multifractality}

The sources of apparent multifractality in empirical time series are mainly due to the linear and nonlinear correlations and the fat-tailedness of probability distribution \cite{Moyano-deSouza-Queiros-2006-PA,Zhou-2009-EPL,Zhou-2012-CSF}. It is important to further discriminate, decompose and quantify the degree of influence of different sources. However, it is very difficult to discriminate and decompose the quantitative influence of different sources because they are nonlinearly entangled. A feasible strategy is to assume that the impacts of these sources are independent.

Using the MF-DFA method, Moyano et al. quantified the multifractal components with the generalized Hurst exponents \cite{Moyano-deSouza-Queiros-2006-PA}. The correlation component $\Delta{h}_{\rm{corr}}$ is obtained through shuffling the raw time series to eliminate linear and nonlinear correlations such that \cite{Kantelhardt-Zschiegner-KoscielnyBunde-Havlin-Bunde-Stanley-2002-PA,Kwapien-Oswiecimka-Drozdz-2005-PA,Moyano-deSouza-Queiros-2006-PA}
\begin{equation}
 \Delta{H}_{\rm{corr}} = \Delta{H}-\Delta{H}_{\rm{shuf}}.
\end{equation}
The influence of a non-Gaussian PDF is determined by comparing the multifractal behavior of shuffled time series and their Fourier transform surrogates. The shuffled time series differ from their Fourier transform surrogates only in the PDFs, as the Fourier transform surrogates have Gaussian distributions. Hence, the PDF component of multifractality is (see also \cite{deSouza-Queiros-2009-CSF})
\begin{equation}
 \Delta{H}_{\rm{PDF}}=\Delta{H}_{\rm{shuf}}-\Delta{H}_{\rm{shuf-FT}}.
\end{equation}
They further suggested that the nonlinearity component is the multifractality embedded in Fourier transform surrogates of shuffled time series:
\begin{equation}
  \Delta{H}_{\rm{nlin}} = \Delta{H}_{\rm{shuf-FT}} =\Delta{H} -\Delta{H}_{\rm{corr}} -\Delta{H}_{\rm{PDF}}.
\end{equation}
Moyano et al. studied the 1-min trading volume time series of the 30 DJIA stocks from the 1st of July until the 31st December of 2004 and found that $\Delta{H}=0.675$, $\Delta{H}_{\rm{corr}}= 0.027$, $\Delta{H}_{\rm{PDF}}=0.445$, and $\Delta{H}_{\rm{nlin}}=0.203$ \cite{Moyano-deSouza-Queiros-2006-PA}.

For the partition function method of multifractal analysis, Zhou assumed that the apparent multifractality can be decomposed into three components caused by the nonlinear correlation ($\Delta\alpha_{\rm{NL}}$), the linear correlation ($\Delta\alpha_{\rm{LM}}$) and the fat-tailed PDF ($\Delta\alpha_{\rm{PDF}}$) \cite{Zhou-2012-CSF}:
\begin{equation}
 \Delta\alpha =
 \Delta\alpha_{\rm{NL}}+\Delta\alpha_{\rm{LM}}+\Delta\alpha_{\rm{PDF}}.
 \label{Eq:Decomposition}
\end{equation}
The intrinsic multifractal nature is characterized by the effective multifractality $\Delta\alpha_{\rm{eff}}$ composing of the nonlinearity component $\Delta\alpha_{\rm{NL}}$ and the PDF component $\Delta\alpha_{\rm{PDF}}$:
\begin{equation}
 \Delta\alpha_{\rm{eff}} = \Delta\alpha-\Delta\alpha_{\rm{LM}}=\Delta\alpha_{\rm{NL}}+\Delta\alpha_{\rm{PDF}},
 \label{Eq:dA:eff}
\end{equation}
where the linear correlation component $\Delta\alpha_{\rm{LM}}$ can be determined by IAAFT surrogate time series of the raw time series. Note that the concepts of efficient multifractality $\Delta\alpha_{\rm{eff}}$ and linear correlation component $\Delta\alpha_{\rm{LM}}$ are essentially equivalent to observed multifractality and multifractal bias studied by Grech and Pamu{\l}a \cite{Grech-Pamula-2013-APPA}.

It is found that the shuffled time series have vanishing singularity width, namely, $\Delta\alpha_{\mathrm{shuf}} \approx 0$, which means that the fat-tailedness of PDF alone cannot produce any apparent multifractality \cite{Zhou-2012-CSF}. To determine $\Delta\alpha_{\rm{NL}}$ and  $\Delta\alpha_{\rm{PDF}}$, Zhou assumed that the PDF component of Gaussian surrogates is negligible, that is, $\Delta\alpha_{\rm{norm,PDF}}=0$, where Gaussian surrogates are normally distributed and have the same rank ordering as the raw data \cite{Zhou-2012-CSF}. Hence, the nonlinearity component does not change when the PDF is replaced by a Gaussian distribution:
\begin{equation}
 \Delta\alpha_{\rm{NL}} = \Delta\alpha_{\rm{norm,NL}}.
 \label{Eq:Original:dA:NL}
\end{equation}
According to Eq.~(\ref{Eq:dA:eff}), the effective singularity width of the Gaussian surrogates is
\begin{equation}
 \Delta\alpha_{\rm{norm,eff}}=\Delta\alpha_{\rm{norm}}-\Delta\alpha_{\rm{norm,LM}}=\Delta\alpha_{\rm{norm}}-\Delta\alpha_{\rm{norm,FSE}}=\Delta\alpha_{\rm{norm,NL}}.
\end{equation}
It follows immediately that
\begin{equation}
 \Delta\alpha_{\rm{PDF}} = \Delta\alpha_{\rm{eff}}-\Delta\alpha_{\rm{norm,eff}}.
 \label{Eq:Original:dA:PDF}
\end{equation}

Zhou determined the multifractal components of the daily DJIA volatility time series \cite{Zhou-2012-CSF}. Chen et al. used the framework of Eq.~(\ref{Eq:Decomposition}) to determine the multifractal components of the daily return time series of Baltic Panamax Index and Baltic Capesize Index from 1 March 1999 to 26 February 2015 \cite{Chen-Tian-Ding-Miao-Lu-2016-PA}, in which they proposed an alternative, self-consistent calculation method based on $\Delta\alpha_{\rm{PDF}}=\Delta\alpha_{\rm{shuf}}$ obtained from the shuffled time series and $\Delta\alpha_{\rm{PDF}}+\Delta\alpha_{\rm{LM}}$ is the apparent multifractality of the IAAFT surrogates. Chen et al. also obtained the three components of the multifractal cross correlations between the daily returns of WTI oil and Baltic Exchange Dirty Tanker Index from 4 January 2000 to 4 November 2015 \cite{Chen-Miao-Tian-Ding-Li-2017-PA}.
Cao and Xu studied the sources of multifractality in the daily returns of carbon emission rights in futures markets of Certified Emission Reduction and European Union Allowances from 14 March 2008 to 31 December 2012 \cite{Cao-Xu-2016-CSF}. For surrogates
generated with shuffling and rank ordering with Gaussian noise, they found that $\Delta\alpha>\Delta\alpha_{\rm{shuf}}>\Delta\alpha_{\rm{surr,G}}$, suggesting that the fat-tailed PDF has a stronger impact than the correlation component.

As a general remark, we stress that the assumption of independent components in the above decomposition methods is only a first-order assumption. In addition, the decomposition methods may not apply to other multifractal analysis methods \cite{Zhou-2012-CSF}.


\subsection{Statistical tests}

\subsubsection{Test based on parameterized multifractal models}

Lux proposed a parametric reshuffle test \cite{Lux-2004-IJMPC}. Assume that the dynamics of the financial time series under investigation follow certain multifractal cascades with the mass exponent function $\tau(q; \lambda)$ and the multifractal spectrum $f(\alpha; \lambda)$ having explicit expressions, where $\lambda$ is the parameter of the multifractal model and there exist a critical value $\lambda_c$ such that the time series becomes monofractal when $\lambda=\lambda_c$. Suitable candidate multifractal models include the $p$-model \cite{Meneveau-Sreenivasan-1987-PRL}, the log-normal MMAR model \cite{Calvet-Fisher-Mandelbrot-1997}, and so on. Because the reshuffled time series do not contain any true multifractality, the corresponding value of $\lambda$ is $\lambda_c$. The null hypothesis is thus
\begin{equation}
  {\bf{H_0}}: \lambda=\lambda_c
\end{equation}
and the null model is the reshuffled time series with any temporal correlations eliminated.  The test can be conducted based on the mass exponent function $\tau(q; \lambda)$ or the multifractal spectrum $f(\alpha; \lambda)$.

For the raw time series and a number of reshuffled time series, the empirical expressions of $\tau(q)$ (or $f(\alpha)$) are obtained and fitted to $\tau(q; \lambda)$ (or $f(\alpha; \lambda)$) to obtain the estimate of $\lambda$ and the distribution of $\lambda_c$. One-sided or two-sided test can be designed following the standard procedure. The $p$-value of the one-sided test is the probability that $\lambda>\lambda_c$. If the $p$-value falls below the prefixed significance level, the null hypothesis is rejected (meaning that the deviation of the measured value of $\lambda$ from $\lambda_c$ is unlikely to be just a statistical fluctuation of the non-multifractal null model).

Lux performed empirical analyses on the daily time series of the New York Stock Exchange Composite Index (01/1966-12/1998) and the German DAX (10/1959-12/1998), the DEM/USD exchange rates (01/1974-12/1998), and the gold prices of the London Precious Metal Exchange (01/1978-12/1998) \cite{Lux-2004-IJMPC}. Although the estimated $\lambda$ values from $\tau(q; \lambda)$ and $f(\alpha; \lambda)$ differ to some extent, the corresponding $p$-values are very close to each other. Surprisingly, the results show that the null model cannot be rejected. Lux argued that the overall conclusion is not necessarily a demonstration of the absence of multifractality, but could also be an implication of the poor power of the adopted multifractal analysis \cite{Lux-2004-IJMPC}. One should note that the $p$ model and the MSM log-nrmal model have parabolic shapes and are symmetric, while empirical multifractal spectra are often skewed \cite{Drozdz-Oswiecimka-2015-PRE}. Hence, a third possibility is that the dynamics of the time series cannot be properly described by the mathematical model adopted in the test. In other words, the null hypothesis is not rejected because the multifractal model is misspecified.

\subsubsection{Tests based on \texorpdfstring{$f(\alpha)$}{}}

Multifractality of a time series is quantified by measures of multifractal strength with the multifractal strength defines such that it vanishes for monofractal time series. Hence, the natural idea is to use a multifractal strength as the test statistic and check if it is significantly different from zero. The null hypothesis is that the time series is monofractal and the alternative hypothesis is that
the raw time series possesses multifractal properties.

Intuitively, the test consists in asking if the empirical multifractal spectrum is undistinguishable from a spectrum shrinking to the
neighborhood of $(\alpha,f)=(0,1)$.
The most common measure of multifractal strength is thus the singularity width $\Delta\alpha$. The associated null hypothesis is \begin{equation}
   {\bf{H_0}}: \Delta\alpha \leqslant \Delta\alpha_{\rm{null}},
\end{equation}
where $\Delta\alpha_{\rm{null}}$ is the singularity width of monofractal time series generated from the null model.
The $p$-value is the associated probability of false alarm for multifractality, which is determined as \cite{Zhou-2010-cnJMSC,Jiang-Zhou-2008a-PA,Jiang-Zhou-2008b-PA}
\begin{equation}
  p = \Pr(\Delta\alpha \leqslant \Delta\alpha_{\rm{null}}) = \frac{1}{n}\sum_{i=1}^n {\bf{\rm{I}}}(\Delta \alpha \leqslant \Delta\alpha_{\rm{null}}),
  \label{Eq:StatTest:dA:p}
\end{equation}
where $n$ is the number of realizations generated from the null model and ${\bf{\rm{I}}}$ is the indicator function. Similarly, defining $F = [f(\alpha_{\min}) +
f(\alpha_{\max})]/2$, an analogous null hypothesis is \cite{Zhou-2010-cnJMSC,Jiang-Zhou-2008a-PA,Jiang-Zhou-2008b-PA}
\begin{equation}
  {\bf{H_0}}: F \leqslant F_{\rm{null}},
\end{equation}
where the $p$-value is
\begin{equation}
  p = \Pr(F \leqslant F_{\rm{null}})=\frac {1}{n} \sum_{i=1}^n {\bf{\rm{I}}}(F \leqslant F_{\rm{null}}).
  \label{Eq:StatTest:F:p}
\end{equation}
We can also use other measures of multifractal strength (see Section \ref{S2:Appl:MFmeasure}) such as $\Delta{D}=D_{\max}-D_{\min}$, $\Delta{H}=H_{\max}-H_{\min}$ or $\Delta{H}(1,2)$ as the statistic and perform the same statistical test. Because all these measures are positive, we adopt the one-side test.

Because the analytical expressions of the asymptotic distributions of the multifractal strength measures are unknown, numerical determination of the distributions is needed. To proceed with the statistical tests, we need to determine the null model. An early null model is Gaussian white noises for the test of long-range correlations \cite{Weron-2002-PA,Couillard-Davison-2005-PA}, which is however not suitable for multifractality tests. The widely used null model is the shuffled time series \cite{Zhou-2010-cnJMSC,Jiang-Zhou-2008a-PA,Jiang-Zhou-2008b-PA}. A somewhat similar idea is to use bootstrap and block bootstrap \cite{Efron-1979-AS,Efron-Gong-1983-AmStat,Efron-1987-JASA}, which is conducted on the residuals of the AR model \cite{GrauCarles-2006-PA}. These methods apply well for uncorrelated time series such as financial returns \cite{Jiang-Xie-Zhou-2014-PA,Sukpitak-Hengpunya-2016-PA}. However, for long-range correlated time series, these methods destroy both linear and nonlinear correlations and their null hypothesis implies that there is not any temporal correlations in the tested time series.

The proper null model for testing the presence of multifractality should generate surrogate time series that have the same distribution and the same linear correlations as the raw time series, but do not possess any nonlinear correlations. Indeed von Hardenberg et al. have already discussed this issue and suggested to use the AAFT algorithm or the further refined IAAFT algorithm \cite{vonHardenberg-Thieberger-Provenzale-2000-PLA}. Although they did not provide any empirical examples, we argue that the IAAFT null model is the one that should be adopted.

\subsubsection{Tests based on \texorpdfstring{$\tau(q)$}{}}

The multifractal strength based on the $\tau(q)$ curve can also be defined. The main idea is to test if $\tau(q)$ is a linear function of $q$. If the Hurst exponent of the raw time series is known, a more specific statistic is \cite{SuarezGarcia-GomezUllate-2014-PA}:
\begin{equation}
  d = \left\langle|\tau(q)-qH+1|\right\rangle.
\end{equation}
Inspired by the test of self-similarity \cite{Berntson-Stoll-1997-PRSB}, we can assume a parabolic form for $\tau(q)$:
\begin{equation}
  \tau(q)= A+Bq+Cq^2,
\end{equation}
and test if the second-order coefficient $C$ is significantly different from 0. Since $\tau(q)$ of multifractals are concave, that is, $C<0$, the null hypothesis reads
\begin{equation}
  {\bf{H_0}}: -C \leqslant -C_{\rm{null}}.
\end{equation}
Other candidate statistics can also be adopted, such as the area enclosed by $\tau(q)$ and the diagonal line through $(q_{\min},\tau(q_{\min}))$ and $(q_{\min},\tau(q_{\min}))$, the average distance between $\tau(q)$ and the diagonal line, the maximum distance between $\tau(q)$ and the diagonal line, and so on. Again, the IAAFT surrogate algorithm is used as the null model.

Alternatively, the bootstrapped wavelet leader test has been thoroughly studied, which uses jointly wavelet leaders, log-cumulants and bootstrap procedures \cite{Wendt-Abry-2007-IEEEtsp,Wendt-Abry-Jaffard-2007-IEEEspm}. Under the leader-based multifractal formalism, the function $\zeta(q)$ is expanded as a polynomial of $q$
\begin{equation}
  \zeta(q)= \sum_{p=1}^{\infty}c_p\frac{q^p}{p!},
\end{equation}
where $c_p$ are the log-cumulants. At each scale, $n$ bootstrap samples are generated from the original sample of wavelet leaders, drawn blockwise and with replacement. This approach is used together with the MF-WL method for the multifractal analysis.

%
%
%
%
%
%
%


\section{Applications}
\label{S1:Applications}

%

In addition to the multifractal analysis methods presented in Section \ref{S1:EmpAnal} to investigate the presence of multifractality in financial time series, we now survey miscellaneous applications of multifractality. Multifractal strengths can be used as measures to quantify market inefficiency and market risks. The rational is that a perfectly efficient market should be characterised by the absence of any
pattern, so that relative price changes (returns) are pure white noise. In contrast, the presence of multifractality
reveals remarkable features and thus the strength of multifractality can be interpreted as a measure of the deviation
from perfect efficiency.

We will show below that
the most convincing empirical result is that multifractality-based methods usually perform better in forecasting volatility and Value-at-Risk than conventional econometric models.

\subsection{Measures of multifractality strength}
\label{S2:Appl:MFmeasure}

\subsubsection{Measures based on multifractal spectrum}

The most widely adopted measure of multifractal strength is the difference between the maximum singularity $\alpha_{\max}$ and the minimum singularity $\alpha_{\min}$,
\begin{equation}
  \Delta\alpha = \alpha_{\max}-\alpha_{\min}=\alpha(-\infty)-\alpha(\infty),
\end{equation}
which is the width of the singularity spectrum or the singularity width for short. In empirical analysis, the value of $\Delta\alpha$ depends on $q$ and it is not realistic to hope to obtain $\alpha_{\min}$ and $\alpha_{\max}$ with $q\to\pm\infty$ due to the finite size of time series. Two methods can then be adopted. The first one is to calculate $\Delta\alpha$ for a given interval $[q_{\min},q_{\max}]$,  
\begin{equation}
  \Delta\alpha=\alpha_{\max}-\alpha_{\min}=\alpha(q_{\min})-\alpha(q_{\max}),
\end{equation}
which is used in the majority of the literature. Usually, a symmetric interval of $q$ with $q_{\min}+q_{\max}=0$ is adopted.
 This method is reasonable because one usually compares the multifractal strengths for different time series on the same interval of $q$, be they from real systems or surrogates. The second method assumes an explicit expression of the singularity spectrum $f(\alpha)$, fits the empirical data to it, and determine $\alpha_{\min}$ or $\alpha_{\max}$ as solutions of
\begin{equation}
  f(\alpha)=0.
\end{equation}
This method does not work when the assumed expressions of $f(\alpha)$ for different time series do not match the empirical multifractal spectra. Moreover, it suffers the shortcoming that the equation $f(\alpha)=0$ may have no solution or only one solution such as for multinomial measures \cite{Halsey-Jensen-Kadanoff-Procaccia-Shraiman-1986-PRA}.

An alternative measure is to quantify how $f(\alpha)$ deviates from $f=1$ \cite{Zhou-2010-cnJMSC,Jiang-Zhou-2008a-PA,Jiang-Zhou-2008b-PA}, by estimating
\begin{equation}
  F = \left[f(\alpha_{\min})+f(\alpha_{\max})\right]/2.
\end{equation}
This measure is not recommended because it performs worse than $\Delta\alpha$. Different binomial measures could have different $\Delta\alpha$ values but with the same value of $F=1$ \cite{Halsey-Jensen-Kadanoff-Procaccia-Shraiman-1986-PRA}.

\subsubsection{Measures based on generalized Hurst exponents}

A similar measure of multifractal strength based on the generalized Hurst exponents is also utilized in some studies \cite{Zunino-Tabak-Figliola-Perez-Garavaglia-Rosso-2008-PA,Yuan-Zhuang-Jin-2009-PA,Yuan-Zhuang-Liu-2012-PA}, which reads
\begin{equation}
  \Delta{H}=H_{\max}-H_{\min}=H(q_{\min})-H(q_{\max}).
  \label{Eq:dH}
\end{equation}
This measure is qualified since $H(-\infty)=\alpha(-\infty)$ and $H(\infty)=\alpha(\infty)$ for many multifractal models. Gu et al. argued that this measure is not appropriate for some financial time series when there are $q\in[q_{\min},q_{\max}]$ such that $ H(q)>H(q_{\min})$ or $H(q)<H(q_{\max})$ \cite{Gu-Shao-Wang-2013-PA}. This anomalous phenomenon corresponds to the situation that $H(q)$ is not a strictly decreasing function of $q$, which has been observed for some financial time series \cite{Czarnecki-Grech-2010-APPA,Gu-Shao-Wang-2013-PA} or synthetic time series contaminated by trends  \cite{Ludescher-Bogachev-Kantelhardt-Schumann-Bunde-2011-PA,Gulich-Zunino-2012-PA}, and often leads to twisted multifractal spectrum \cite{Czarnecki-Grech-2010-APPA,Ludescher-Bogachev-Kantelhardt-Schumann-Bunde-2011-PA}.
Hence, Gu and Zhang proposed to calculate $\Delta{H}$ as follows \cite{Gu-Zhang-2016-EE}
\begin{equation}
  \Delta{H}=\max_q H(q)-\min_q H(q).
\end{equation}
In empirical analysis, however, such results should be treated with caution. It is possible that the anomalous phenomenon stems from a improper determination of the multifractal properties, such as a bad choice of the scaling range or an unsuitable preprocessing of the data.

Wang et al. defined the following measure by taking monofractals as the reference \cite{Wang-Liu-Gu-Cao-Wang-2010-PA}
\begin{equation}
  \overline{\Delta{H}(q,2)} = \left\langle|H(q)-0.5|\right\rangle_q
  \label{Eq:MFStrength:Hq:0.5:ave}
\end{equation}
which is the average of $|H(q)-0.5|$ over all $q$ values used in the multifractal analysis.
Grech proposed another measure considering the average distance between $H(q)$ and $H(2)$ for the respective multifractal and monofractal cases \cite{Grech-2016-CSF}
\begin{equation}
  \overline{\Delta{H}(q,2)}=\frac{1}{q}\int_{-q}^q|H(q)-H(2)|{\rm{d}}q,
  \label{Eq:MFStrength:Hq:H2:ave}
\end{equation}
which can be modified when multifractal biases are taken into account \cite{Grech-2016-CSF}. These two measures are essentially similar. However, using $H(2)$ instead of $0.5$ allows the measure to investigate long-range correlated time series such as volatility in a more reasonable way.

Morales et al. used the following measure \cite{Morales-DiMatteo-Gramatica-Aste-2012-PA,Morales-DiMatteo-Aste-2013-PA}
\begin{equation}
  \Delta{H}(1,2) = H(1)-H(2),
\end{equation}
which is very useful when the moments for negative $q$'s are not defined. Some researchers use $H(1)$ to study market efficiency, while some others use
\begin{equation}
  \delta{H}(q) = H(q)-0.5
\end{equation}
to measure multifratal strength and hence market inefficiency, where $q=1$ and $q=2$ are often chosen \cite{Sensoy-2013b-CSF}.
Wang et al. defined the following measure by taking monofractals as the reference \cite{Wang-Liu-Gu-Cao-Wang-2010-PA}
\begin{equation}
  D=\frac{1}{2}\left(|H(q_{\min})-0.5|+|H(q_{\max})-0.5|\right)
  \label{Eq:MFStrength:D10}
\end{equation}
which is a generalized form suggested in Ref.~\cite{Wang-Liu-Gu-2009-IRFA} with $q_{\max}=-q_{\min}=10$. Rizvi et al. used $q_{\max}=-q_{\min}=4$ \cite{Rizvi-Dewandaru-Bacha-Masih-2014-PA,Rizvi-Arshad-2017-PA}.
When $H(2)\approx0.5$, these quantities reduce to $\frac{1}{2}\Delta{H}$ with $-q_{\min}=q_{\max}=10$ or 4 in Eq.~(\ref{Eq:dH}). Note that $D$ is a simplified form of $\overline{\Delta{H}(q,2)}$ in Eq.~(\ref{Eq:MFStrength:Hq:0.5:ave}).

\subsubsection{Intermittency}

The intermittency of a time series can also be used to quantify the strength of multifractality \cite{Marshak-Davis-Cahalan-Wiscombe-1994-PRE,Schertzer-Lovejoy-Schmitt-Chigirinskaya-Marsan-1997-Fractals,Bickel-1999-PLA}. The expression of intermittency can be derived through the partition function approach \cite{Ausloos-Ivanova-2002-CPC}. Let us define a measure $\mu_{i,1}$ as
\begin{equation}
  \mu_{i,1} = \frac{|\Delta{X}(i,1)|}{\sum_{j=1}^N|\Delta{X}(j,1)|},
\end{equation}
which is the normalized volatility at the finest scale if $X(i)$ is the logarithmic price. The average measure in interval $[i,i+s-1]$ of length $s$ is
\begin{equation}
  \nu_{i,s} = \frac{1}{s}\sum_{j=i}^{i+s-1}\mu_{j,1} =\frac{1}{s}\mu_{i,s} ,~~~ i = 1, 2, \cdots, N-s+1,
\end{equation}
where $\mu_{i,s}$ is the total measure in the $i$-th interval as defined in Eq.~(\ref{Eq:disctrete:measure}).
The scaling properties are obtained through
\begin{equation}
  \langle\nu_{i,s}^q\rangle \sim s^{-\kappa(q)},
\end{equation}
which defines a scaling exponent function $\kappa(q)$. Combining with the scaling relation (\ref{Eq:MF-PF:tau:chi:s}), we have
\begin{equation}
  \langle\nu_{i,s}^q\rangle\sim \frac{1}{N}
  \sum \nu_{i,s}^q
  = \frac{1}{N} \sum_i\left[\frac{1}{s}\mu_{j,s}\right]^q
  \sim s^{1-q}\sum_i \mu_{j,s}^q
  \sim s^{1-q} s^{\tau(q)}
  \sim s^{1-q} s^{\tau(q)}
  = s^{(q-1)(D_q-1)},
\end{equation}
which leads to the following equality
\begin{equation}
  \kappa(q)=q-1-\tau(q).
\end{equation}
The intermittency is then defined as \cite{Marshak-Davis-Cahalan-Wiscombe-1994-PRE,Schertzer-Lovejoy-Schmitt-Chigirinskaya-Marsan-1997-Fractals,Bickel-1999-PLA,Ivanova-Ausloos-1999a-EPJB,Ausloos-Ivanova-2002-CPC}
\begin{equation}
  C_1\triangleq\frac{{\rm{d}}\kappa(q)}{{\rm{d}}q}\bigg|_{q=1}
  =\left.-\frac{{\rm{d}}H(q)}{{\rm{d}}q}\right|_{q=1}+H(1)+1 = 1-D_1.
\end{equation}
The intermittency is also known as the sparseness or the information codimension because $D_1$ is the information dimension. There are applications to foreign exchange rates \cite{Vandewalle-Ausloos-1998-IJMPC,Ivanova-Ausloos-1999a-EPJB,Richards-2000-PA,Ausloos-Ivanova-2002-CPC}, gold prices \cite{Ivanova-Ausloos-1999a-EPJB}, and stock market indices \cite{Ivanova-Ausloos-1999a-EPJB,Bickel-Lai-2001-CSDA}.



\subsubsection{Relationship between different measures}

The multifractal strength measures discussed above are used to qualify market inefficiency. However, there is no consensus on which measure performs best in qualifying market inefficiency. As we will see below, different market inefficiency indexes result in different ranks of markets. Fortunately, some common features exist. For instance, developed markets are more efficient than emerging markets on average,
over the set of different measures.

The relationship between different measures is an important problem, which is of theoretical and practical significance. An empirical analysis of 34 stock market indices from 1 January 1995 to 20 September 2010 shows that there is not evident correlation between $\delta{H}(2)=H(2)-0.5$ and $\Delta{H}$ \cite{Gu-Shao-Wang-2013-PA}. In contrast, a further analysis of the daily closing WTI crude oil spot prices from 2 January 1986 to 28 September 2012 unveils a weak correlation between $\delta{H}(2)$ and $\Delta{H}$ with a Pearson correlation coefficient of 0.373 when the window size is 1 year and a strong correlation with a Pearson correlation coefficient of 0.807 when the window size is 4 years
\cite{Gu-Zhang-2016-EE}. The latter results highlight the importance of choosing proper estimation methods since smaller windows lead to noisier estimates.

\subsection{Measuring market inefficiency}
\label{S2:Appl:Inefficiency}

The study of market efficiency has a long history. The efficient markets hypothesis has attracted numerous theoretical and empirical interest  \cite{Fama-1970-JF,Fama-1991-JF}. Traditional procedures in the weak-form EMH literature check linear serial correlations with the variance ratio tests, unit root, nonlinear serial dependence, long memory mainly quantified by the Hurst exponent, martingale difference sequence tests, time reversibility test, and so on \cite{Lim-Brooks-2011-JES}. A market may also have different efficiency at different periods and more generally time-varying efficiency \cite{Lim-Brooks-2011-JES}, which is reminiscent of Lo's ``adaptive markets hypothesis'' \cite{Lo-2004-JPM}
and Sornette's ``Emerging Market Intelligence Hypothesis'' \cite{Sornette-2014-RPP}.

\subsubsection{Market inefficiency index based on fractality measures}

There are may measures used to qualify the degree of long memory in time series, such as the Hurst exponent $H$, the fractal dimensions $D_0$ and the first order autocorrelation $\rho(1)$ estimated from different methods. Denoting $M_i$ as the $i$th measure estimated from the $i$th method, which is defined on the interval $[M_{i,\min},M_{i,\max}]$, and $M_i^*$ the corresponding theoretical value for uncorrelated time series, the distance $M_i-M_i^*$ is a measure of market inefficiency. Kristoufek and Vosvrda proposed a combined measure for market inefficiency
\cite{Kristoufek-Vosvrda-2013-PA}
\begin{equation}
  IE = \sqrt{\sum_{i=1}^m\left(\frac{M_i-M_i^*}{M_{i,\max}-M_{i,\min}}\right)^2}~.
\end{equation}
They call this combined measure the ``efficiency index'' with notation $EI$. However, intuitively, we recommend the name of ``inefficiency index'' together with the notation $IE$. In this way, the market with larger $IE$ is more inefficient.

The market inefficiency index has been used to analyze the efficiency of 41 stock market indices for the period from 2000 to August of 2011 \cite{Kristoufek-Vosvrda-2013-PA,Kristoufek-Vosvrda-2014-EPJB}, 38 stock market indices between January 2000 and August 2011 with different methods \cite{Kristoufek-Vosvrda-2014-EPJB}, 25 commodities in the period between 1 January 2000 to 22 July 2013 \cite{Kristoufek-Vosvrda-2014-EE}, and 142 worldwide currencies from 1 January 2011 to 30 November 2014 \cite{Kristoufek-Vosvrda-2016-PA}. It correctly identifies that developed markets are more efficient while emerging markets are inefficient.

%

There is also cumulating empirical evidence suggesting that multifractal analysis can be used to quantify the degree of inefficiency of markets in the sense that more developed stock markets have weaker multifractality \cite{Zunino-Tabak-Figliola-Perez-Garavaglia-Rosso-2008-PA,Zunino-Figliola-Tabak-Perez-Garavaglia-Rosso-2009-CSF} and an emerging market evolves with time to be more efficient with a narrowing singularity width \cite{Wang-Liu-Gu-2009-IRFA}.  Therefore, the measures of multifractal strength can be used as qualifiers of market inefficiency. It is also common in the literature that the multifractal strength is used as the measure of market efficiency. Similarly, we suggest to treat the multifractal strength as a measure of the market inefficiency rather than of the market efficiency \cite{Zunino-Tabak-Figliola-Perez-Garavaglia-Rosso-2008-PA,Zunino-Figliola-Tabak-Perez-Garavaglia-Rosso-2009-CSF}.

When there are two scaling ranges delimited by a crossover scale, we focus on the behavior at large scales since the scaling behavior at small scales is often dominated by the finite-size effects discussed in Section \ref{S3:MF:Sources:FSE}. In addition, we compare the multifractal strength of the apparent multifractality because the true multifractality is not always available in the literature. For inefficiency indexes defined from fractal analysis, it is better to perform statistical tests to ensure that the estimated measure is significantly different from that of the uncorrelated counterpart \cite{Jiang-Xie-Zhou-2014-PA,Ma-Li-Li-2016-PA}.

\subsubsection{Ranking market efficiency}
\label{S3:Appl:Markets}

Di Matteo et al. calculated the generalized Hurst exponents $H(1)$ and $H(2)$ with the structure function approach and used these exponents to differentiate stock markets in their development stage \cite{DiMatteo-Aste-Dacorogna-2005-JBF}. They ranked 29 foreign exchange rates, 32 Stock market indices, 12 Treasury bond and bill rates with different maturity dates and 16 Eurodollar rates with the maturity dates ranging from 3 months to 4 years. A strong relation between the Hurst exponents and the development stage of the market is observed such that less developed markets have in general higher Hurst exponents $H(1)$ and $H(2)$. Sensoy and Tabak ranked 27 stock markets in the European Union using $H(1)$ calculated from the daily data from 2 January 1999 to 25 February 2013 \cite{Sensoy-Tabak-2015-PA} and 17 stock markets from the daily data of from 02 January 2003 to 31 January 2013 \cite{Sensoy-Tabak-2016-IRFA}.

Todea and Plesoianu used the inefficiency measure $\delta{H}(1)=|H(1)-0.5|$ as the dependent argument in a regression model to investigate the influence of foreign portfolio investment on informational efficiency in Central and Eastern European stock markets \cite{Todea-Plesoianu-2013-EM}. Sensoy obtained the time-varying inefficiency measure $\delta{H}(1)$ of 19 members of the Federation of Euro-Asian Stock Exchanges (FEAS) using daily closing prices from January 2007 to December 2012 and used the average $IE$ to rank these markets \cite{Sensoy-2013b-CSF}. Sensoy obtained the time-varying inefficiency measure $\delta{H}(1)$ of the 19 FEAS member stock markets using daily closing prices from January 2007 to December 2012 and used the average $IE$ to rank these markets \cite{Sensoy-2013b-CSF}.

Zunino at al. performed multifractal analysis on the daily return time series of 32 stock market indices from 2 January 1995 to 23 July 2007 \cite{Zunino-Tabak-Figliola-Perez-Garavaglia-Rosso-2008-PA}. They found that on average the group of developed markets have lager values of $\Delta{H}$ and $\Delta\alpha$ than the group of emerging markets.
Zunino at al. ranked 8 Latin-American market (Argentina, Brazil, Chile, Colombia, Mexico, Peru, Venezuela and the US) by $\Delta{H}$ of the market index returns from January 1995 to February 2007 and identified the US market as the least inefficient \cite{Zunino-Figliola-Tabak-Perez-Garavaglia-Rosso-2009-CSF}.
Wang and Wu studied the daily returns of four crude oil futures contracts with the maturities of 1 to 4 months traded on the NYMEX from 2 January 1985 to 10 May 2011 and found that $\Delta{H}$ and $\Delta\alpha$ decrease for longer maturity \cite{Wang-Wu-2013-CE}.
Rizvi et al. used $D$ with  $q_{\max}=-q_{\min}=4$ to rank 11 developed markets and 11 Islamic markets from 1 January 2001 till 31 December 2013 and found that developed markets have on average relatively higher efficiency than Islamic markets \cite{Rizvi-Dewandaru-Bacha-Masih-2014-PA}.

There are also studies devoted to rank the efficiency of industrial sectors. Zhuang et al. compared the multifractal behavior of 10 sector indices of the China stock market from 1 January 2005 to 3 April 2013 and found that the industrial sectors of Information Technology, Health Care and Telecommunication Services are more efficient, while those of Energy, Materials and Consumer Discretionary are less efficient \cite{Zhuang-Wei-Ma-2015-PA}.
Tiwari et al. analyzed the daily closing spot price data for nine Dow Jones sector ETF indices from 12 June 2000 to 23 July 2015 and found that the utilities sector is the more efficient, while the financial sector and the telecommunications sector are the most inefficient \cite{Tiwari-Albulescu-Yoon-2017-PA}.
Mensi et al. studied the daily returns of 10 DJIM sector indices of Islamic stock markets from 9 November 1998 to 5 March 2015 and found that the consumer goods sector is the most efficient while the financial sector is the most inefficient \cite{Mensi-Tiwari-Yoon-2017-PA}.

\subsubsection{Changing efficiency at different periods}
\label{S3:Appl:DifferentPeriods}

As discussed in Section \ref{S3:MF:Sources:Extrema}, extreme events have a strong influence on the apparent multifractality of time series, especially for market crashes. In most cases, the market became more inefficient with a wider singularity width after a crash. We survey below the efficiency of markets in different periods. Special attention is paid to political or external events.

The degree of market efficiency changes with different market states. O{\'s}wi{\c{e}}cimka et al. observed different multifractal behavior in the positive and negative returns of 1-min DAX index in two market states, bull (28 November 1997 to 30 December 1999) and bear (1 May 2002 to 1 May 2004), and found that the bear market was more inefficient than the bull market \cite{Oswiecimka-Kwapien-Drozdz-Gorski-Rak-2008-APPA}. Arshad and Rizvi compared four East Asian stock market (Malaysia, Singapore, Indonesia and South Korea) from 1 January 1990 until 31 July 2013 and found that South Korea was the most efficient and the four markets were more efficient in booms than in recessions \cite{Arshad-Rizvi-2015-PA}. Rizvi and Arshad divided the NIKKEI index from 1990 to 2013 into 14 successive booms and recessions and found that the market inefficiency index $D$ fluctuated from time to time \cite{Rizvi-Arshad-2017-PA}.


The Chinese currency RMB experienced two major reforms. Before July of 2005, China's RMB was pegged to the USD with very small fluctuations. On 21 July 2005, the People's Bank of China (PBC) launched a reform to take a managed float regime. On 19 June 2010, the PBC resumed its foreign exchange reform to enhance the RMB exchange rate flexibility after halting the RMB appreciation during the global economic recession. Qin et al. investigated the evolution of RMB exchange efficiency during the three periods using daily RMB/USD rates from 19 December 1990 to 2 August 2012 and RMB/HKD rates from 13 January 1994 to 2 August 2012 \cite{Qin-Lu-Zhou-Qu-2015-PA}. They found that the values of $\Delta{H}$ in the three periods are 4.918, 0.415 and 0.248 for the RMB/USD and 2.952, 1.232 and 0.569 for the RMB/HKD. It is evident that the RMB markets became more efficient.
Sto{\v{s}}i{\'{c}} investigated the change of market efficiency of eight foreign exchange rates (Australia, Brazil, Malaysia, New Zealand, South Korea, Sweden, Taiwan, and Thailand), which transited from managed exchanged rates to independent float exchange rates \cite{Stosic-Stosic-Stosic-Stanley-2015a-PA}. They found that the transition from managed to independent float regime caused a decrease of the singularity width $\Delta\alpha$, indicating an increase in market efficiency, except for Malaysia ($\Delta\alpha$ increased) and South Korea (comparable $\Delta\alpha$).

Some researcher investigated the effect of the price-limit rule on market efficiency. Wang, Liu and Gu investigated the impact of the price-limit reform taking effect on 16 December 1996 on the efficiency of the Shenzhen stock market \cite{Wang-Liu-Gu-2009-IRFA}. The stock prices could fluctuate freely before the reform. Based on the daily closing price returns of the SZCI index from 3 April 1991 to 15 December 2008, they found that the market inefficiency index $D$, as defined in Eq.~(\ref{Eq:MFStrength:D10}), decreased from 0.89 to 0.36 in the long term, indicating an improved market efficiency after the price-limit reform. Similar results were observed for the Shanghai stock market \cite{Wang-Liu-Gu-Cao-Wang-2010-PA}. However, the improvement of market efficiency is more likely attributed to the very high inefficiency in the initial years after the establishment of the markets, so that the price-limit reform played a less significant role \cite{Wang-Liu-Gu-Cao-Wang-2010-PA}.

Other events also influence market efficiency.
Zhou et al. studied the 5-min high-frequency returns of CSI 300 from 4 April 2005 to 19 January 2012 and found that the introduction of the CSI 300 futures (CSI300IF) on 16 April 2010 improved the efficiency of the Chinese stock market, in which $\Delta{H}$ decreased from 0.406 to 0.180 and $\Delta\alpha$ decreased from 0.559 to 0.291 \cite{Zhou-Dang-Gu-2013-PA}. He et al. worked on the daily returns of CSI 300 from 5 April 2006 to 9 May 2014 and found that the width of the multifractal spectrum became narrower from $\Delta\alpha=0.395$ to $\Delta\alpha=0.158$ after the introduction of CSI300IF implying that the market became more efficient \cite{He-Wang-Du-2014-PA}.
Gu, Chen and Wang studied the daily returns of the WTI and Brent crude oil prices from 20 May 1987 to 30 September 2008 and divided the time series into three periods separated by the two Gulf Wars that broke out on 24 February 1991 and 30 March 2003 \cite{Gu-Chen-Wang-2010-PA}. They found that the two markets became more efficient along the three periods, that is, $\Delta{H}_{\rm{I}} > \Delta{H}_{\rm{II}} \gtrapprox \Delta{H}_{\rm{III}}$.
Wang et al. investigated the daily closing data of gold prices traded on the COMEX from 13 July 1990 to 15 September 2009 and found that the singularity width $\Delta\alpha$ decreased from 0.558 in the early period before July 1999 to 0.411 in the late period \cite{Wang-Wei-Wu-2011a-PA}.

Recently, Ruan et al. conducted an interesting study by considering the effect of the Shanghai-Hong Kong Stock Connect (implemented on 17 November 2014) on the efficiency of both markets using the daily closing prices of SSEC and HSI from 1 January 2007 to 31 December 2016 \cite{Ruan-Zhang-Lv-Lu-2018-PA}. They found that, after the Connect, the multifractality degree quantified by $\Delta{H}$ and $\Delta\alpha$ decreased for the SSEC index and increased for the HSI index, indicating that the Shanghai market became more efficient while the Hong Kong market became less efficient. Their quantitative results further suggest that the Hong Kong market became less efficient than the Shanghai market after the implementation of the Connect, which is counter-intuitive. It suggests that the apparent multifractality may be a biased measure of market inefficiency.


\subsubsection{Time-varying efficiency}
\label{S3:Appl:TimeVarying}

Wang et al. studied the evolution of the inefficiency $D$ of the daily returns of gold traded on the COMEX from 13 July 1990 to 15 September 2009 \cite{Wang-Wei-Wu-2011a-PA}. The inefficiency became smaller along time. However, external shocks increased market inefficiency.
Ghosh et al. calculated the singularity width $\Delta\alpha$ in successive five-year windows of the daily gold price returns from January 1973 to October 2011 and observed an overall decreasing trend \cite{Ghosh-Dutta-Samanta-2012-APPB}, which signals the improvement of the gold market efficiency.

Wang et al. studied the daily return and volatility time series of the Shanghai SSEC index from 19 December 1990 to 15 December 2008 and determined the values of inefficiency indexes $\overline{\Delta{H}(q,2)}$ and $D$ in moving windows of size 1008 trading days (about four years) \cite{Wang-Liu-Gu-Cao-Wang-2010-PA}. The two market inefficiency indexes evolved very similarly and exhibited an overall decreasing trend. The inefficiency indexes were very high at the early stage of the Shanghai stock market.

Sensoy and Tabak studied the time-varying efficiency of 27 European stock markets in the European Union using $H(1)$ calculated from the daily data from 2 January 1999 to 25 February 2013 \cite{Sensoy-Tabak-2015-PA}. They found that the 2008 global financial crisis caused more inefficiency on almost all stock markets while the Eurozone sovereign debt crisis decreased only the efficiency of the markets in France, Spain and Greece. In addition, joining EU for the late members did not have a uniform effect on their markets' efficiency. Sensoy and Tabak also studied 17 global stock markets from 2 January 2003 to 31 January 2013 and confirmed the impact of the 2008 crisis on market efficiency \cite{Sensoy-Tabak-2016-IRFA}.

Wang and Wu calculated the evolving singularity width $\Delta\alpha$ of the return and volatility time series of gasoline crack spread, which is the difference series between daily spot price data of WTI crude oil and gasoline (New York Harbor) from 2 January 1986 to 26 July 2011, and found that the second Gulf War that broke out on 20 March 2003 caused a sharp increase in market inefficiency while the first Gulf War did not \cite{Wang-Wu-2012-EM}.

There are also studies of the time-varying multifractal strength of cross correlations, such as pairs of WTI crude oil and BRIC stock markets \cite{Ma-Wei-Huang-Zhao-2013-PA} and pairs of US interest rates and agricultural commodity markets \cite{Wang-Hu-2015-PA}. However, there is no simple interpretation of the joint multifractality concerning the efficiency of markets.

\subsection{Multifractal volatility}
\label{S2:Appl:VolatilityPred}

\subsubsection{Definitions of multifractal volatility}
\label{S3:Appl:MFV:Definition}

Let $r(t)$ be the daily return and $RV(t)$ be the realized variance constructed by taking the sum of squared intraday returns. The scaled realized variance is a recommended proxy for volatility \cite{Hansen-Lunde-2006-JEm}:
\begin{equation}
  \sigma(t) = \left[\frac{1}{n}\sum_{j=1}^n\frac{r^2(j)}{RV^2(j)}\right]RV(t),
  \label{Eq:scaledRV}
\end{equation}
where the sum is carried over the $n$ days of the sample.
Inspired by this measure, Wei and Wang introduced the multifractal volatility ($MFV$) \cite{Wei-Wang-2008-PA}
\begin{equation}
  \sigma(t) = \left[\frac{\sum_{j=1}^nr^2(j)}{\sum_{\ell=1}^n\Delta\alpha(\ell)}\right]\Delta\alpha(t),
\end{equation}
where $\Delta\alpha(t)$ is the singularity width obtained from the intraday data on the $t$-th day. To have a full analog to the definition (\ref{Eq:scaledRV}), the multifractality could be
\begin{equation}
  \sigma(t) = \left[\frac{1}{n}\sum_{j=1}^n\frac{r^2(j)}{\Delta\alpha(j)}\right]\Delta\alpha(t).
\end{equation}
Alternatively, Chen and Wu used realized volatilities $RV(t)$ to replace squared returns $r(t)$ in the definition of multifractal volatility \cite{Chen-Wu-2011-PA}
\begin{equation}
  \sigma(t) = \left[\frac{\sum_{j=1}^nRV^2(j)}{\sum_{\ell=1}^n\Delta\alpha(\ell)}\right]\Delta\alpha(t).
\end{equation}
The logic behind these definitions of multifractal volatility is that $\Delta\alpha(t)$ can be viewed as a measure of asset risk or market risk.

The singularity width $\Delta\alpha$ is estimated empirically by analyzing asset price time series using the partition function approach \cite{Wei-Wang-2008-PA} or MF-DFA \cite{Chen-Wu-2011-PA}. Although the former method provides illusionary multifractality, the multifractal volatility is found to be strongly correlated with the realized volatility and the multifractal volatility models still show promising predictive power \cite{Wei-Wang-2008-PA}.

\subsubsection{Volatility forecasting}
\label{S3:Appl:MFV:Forcasting}

The multifractal volatility defined in Section \ref{S3:Appl:MFV:Definition} can be incorporated into existing econometric volatility models for volatility forecasting. Consider the ARFIMA$(1, d, 1)$ model as an example, which is expressed as
\begin{equation}
  (1-\phi{L})(1-L)^d[RV(t)-\mu]=(1-\theta{L})\epsilon(t),
  \label{Eq:ARFIMA}
\end{equation}
where $\mu$ is the mean of the realized volatility, $L$ is the lag operator, coefficients $d$, $\phi$ and $\theta$ are fixed and unknown, and $\epsilon(t)$ is Gaussian white noise with zero mean and variance $\sigma^2_{\epsilon}$.
Based on the ARFIMA$(1, d, 1)$ model, Wei and Wang proposed the multifractal volatility model as  \cite{Wei-Wang-2008-PA}
\begin{equation}
  (1-\phi{L})(1-L)^d[MFV(t)-\mu]=(1-\theta{L})\epsilon(t),
\end{equation}
which shares the same form as Eq.~(\ref{Eq:ARFIMA}). Extensions to other financial econometric volatility models are similar, but less studied.
Using high-frequency stock market index data, it was shown that the multifractal volatility model shows better performances in one-day ahead volatility forecasting than the realized volatility model, stochastic volatility model and GARCH-type models \cite{Wei-Wang-2008-PA,Chen-Wu-2011-PA,Wei-Wang-Huang-2011-PA,Chen-Wei-Lang-Lin-Liu-2014-PA,Liu-Ma-Long-2015-PA,Liu-Ye-Ma-Liu-2017-PA}. In contrast, the heterogeneous autoregressive model of realized volatility (HAR-log(RV)) performs better than the multifractal volatility model in predicting future volatility \cite{Ma-Wei-Huang-Chen-2014-PA}.
The multifractal volatility model can also be integrated into copula models to construct the so-called copula-multifractal volatility model, where the multifractal volatility model is used to describe the marginal distributions of returns and the dynamic copulas model the time-varying dependence structure \cite{Wei-Wang-Huang-2011-PA,Chen-Wei-Lang-Lin-Liu-2014-PA}. Empirical analysis on high-frequency financial time series shows that in general the copula-multifractal volatility model has better hedging effectiveness than the GARCH-type copula models \cite{Wei-Wang-Huang-2011-PA,Chen-Wei-Lang-Lin-Liu-2014-PA}.
The model can also be used to describe the contagion effect between financial markets \cite{Chen-Wei-Lang-Lin-Liu-2014-PA,Chen-Wei-Zhang-Yu-2014-PA}.

Another important direction concerning multifractal volatility forecasting is based on the multifractal model of asset returns (MMAR) and in particular the Markov-switching multifractal (MSM) model (see an excellent composition by Calvet and Fisher \cite{Calvet-Fisher-2008}, and also Section \ref{S2:Models:MMAR} and Section \ref{S2:Models:MSMM}). In their seminal work, Calvet and Fisher formulated
a powerful framework of volatility forecasting \cite{Calvet-Fisher-2001-JEm}. Lux and Morales Arias performed extensive numerical experiments to show that the MSM model outperforms GARCH, FIGARCH, the stochastic volatility model (SV) and the long memory stochastic volatility model (LMSV) in volatility forecasting \cite{Lux-MoralesArias-2013-QF}.
Lux and Kaizoji compared the performance of GARCH, FIGARCH, ARFIMA and lognormal MSM models in volatility forecasting using daily data of two samples of stocks in the Tokyo Market from 1975 to 2001 \cite{Lux-Kaizoji-2007-JEDC}. Based on the relative mean squared error (MSE) and mean absolute error (MAE), they found that the long memory models ARFIMA and FIGARCH outperform the short memory GARCH and ARMA models, while the log-normal MSM model performs the best. The performance differences of these models are more significant for longer forecasting horizon. The conclusion is the same for sub-periods (1986-1990, 1991-1995, and 1996-2001), as well as for volume forecasting.
Ben Nasr et al. investigated the daily data of the Global Dow Jones Islamic Market World Index from 1 January 1996 to 2 September 2013 and found that the MSM model outperforms long memory GARCH-type models (FIGARCH and FITVGARCH), which in turn performs better than the short-memory GARCH model \cite{BenNasr-Lux-Ajmi-Gupta-2016-IREF}.
Lux and Morales-Arias studied the daily data of 25 all-share stock indices, 11 10-year government bond indices and 12 real estate security indices at the country level from January 1990 to January 2008 \cite{Lux-MoralesArias-2010-CSDA}. They found that the Binomial MSM and lognormal MSM with either normal or Student innovations have comparable performance in volatility forecasting and perform better than GARCH and FIGARCH. They also found that the combined predictors perform even better than single models and the FIGARCH-MSM model seems to perform the best.
Chuang et al. found the MSM model performed better than implied, GARCH and historical volatilities in realized volatility forecasting of the daily data of the S\&P 100 index option and equity option from 3 January 2000 to 31 October 2009 \cite{Chuang-Huang-Lin-2013-NAEF}.
Wang et al. considered the daily spot data of WTI and Brent crude oil from 4 January 1993 to 9 September 2013 and found that MSM models provide better volatility forecasts than GARCH-class models and the historical volatility model \cite{Wang-Wu-Yang-2016-IJF}. Lux et al. confirmed this conclusion on daily data of WTI oil prices from 2 January 1875 to 31 December 1895 and from 2 January 1985 to 24 March 2014 \cite{Lux-Segnon-Gupta-2016-EE}.
Segnon et al. confirmed the better performance of the MSM model over GARCH and FIGARCH on the daily carbon dioxide emission allowance prices from 16 January 2009 to 20 January 2015 \cite{Segnon-Lux-Gupta-2017-RSER}.
Liu et al. compared Binomial MSM models with normal and Student-$t$ innovations and GARCH-type models (GARCHN, GARCH-$t$ and GARCH-Skewed $t$) using daily closing prices of SSEC from 15 July 1991 to 8 June 2015 and found that the Binomial MSM model with Skewed-$t$ innovations dominates \cite{Liu-Zhang-Fu-2016-PA}.

There are also other models that are superior to the MSM models.
Wang et al. proposed a support vector machine (SVM) based MSM approach to forecast volatility of the SSEC index from 2 January 1991 and ending on 31 December 2010 and the SZCI from 2 May 1991 and ending on 31 December 2010 and found that the SVM-MSM model improves the MSM model and both models are superior to the GARCH(1,1) model \cite{Wang-Huang-Wang-2013-NCA}.
Lux et al. found that the RV-ARFIMA model performs the best over the RV-MSM model and other volatility models, especially in tranquil periods, while the RV-MSM model performs much better during turmoil periods \cite{Lux-MoralesArias-Sattarhoff-2014-JFc}.
Cordis and Kirby developed a flexible class of discrete stochastic autoregressive volatility (DSARV) models \cite{Cordis-Kirby-2014-JBF}. They performed empirical analysis of the daily data of five individual US stocks (Alcoa, General Electric, Coca Cola, JP Morgan Chase and Exxon Mobil) and the S\&P 500 index from 1 February 1976 to 31 January 2001 and found that the simple first- and second-order DSARV models outperform MSM models in forecasting volatility.
Using daily WTI and Brent crude oil price data from 6 January 1992 to 31 December 2014, Charles and Darn{\'e} found that asymmetric models estimated on jump filtered returns outperform the GARCH, GAS, GJR-GARCH, EGARCH and MSM models on raw returns \cite{Charles-Darne-2017-EE}.

The Markov-switching multifractal duration (MSMD) model \cite{Chen-Diebold-Schorfheide-2013-JEm} has also been applied to volatility forecasting. Aldrich et al. studied the transaction-level data of the Chicago Mercantile Exchange near-month E-mini S\&P 500 Futures contract from 1 January 2014 to 31 December 2014 and found that the compound MSMD model outperforms the compound exponential model, the compound ACD model and the ACD-GARCH(1,1) model in out-of-sample forecasting of realized volatility \cite{Aldrich-Heckenbach-Laughlin-2016-JEF}. {\v{Z}}ike{\v{s}} et al. studied the transaction-level data of three major foreign exchange futures contracts (CHF, EUR and JPY) traded on the Chicago Mercantile Exchange from 9 November 2009 to 29 January 2010 and found that the MSMD models exhibit higher volatility forecasting ability than the long-memory stochastic duration model and the short-memory ACD model \cite{Zikes-Barunik-Shenai-2017-EmR}.

MRW models also have good performance in volatility forecasting. Bacry et al. compared the performance of the MRW model and normal and $t$-Student GARCH(1,1) models in volatility forecasting using the daily prices of four FX rates (CAD, JPY, CHF and GBP) against the USD \cite{Bacry-Kozhemyak-Muzy-2008-JEDC}. For each time series, they calibrated the models using in-sample data from 1 July 1977 to 28 December 1989 and performed out-of-sample prediction from 29 December 1989 to 20 March 2006 at four different prediction scales (1, 5, 20 and 50 days). They found that the MRW model provides smallest mean absolute prediction errors for all time series and at all prediction scales. The advantage of the MRW model over the GARCH models in volatility forecasting has been confirmed for the daily returns of 29 French CAC40 constituent stocks from 1990 to 2005 \cite{Bacry-Kozhemyak-Muzy-2013-QF}. Duchon et al. obtained precise prediction formulas for volatility forecasting and option pricing by avoiding the problem of the estimation of the so-called ``integral length'' (a key parameter of the MRW model)  \cite{Duchon-Robert-Vargas-2010-MF}, as highlighted by Saichev and Sornette \cite{Saichev-Sornette-2006-PRE}.

\subsubsection{Value-at-Risk forecasting}
\label{S3:Appl:MFV:VAR}

Wei et al. proposed a method to measure daily Value-at-Risk (VaR) through the combination of a multifractal volatility based econometric model and extreme value theory \cite{Wei-Chen-Lin-2013-PA}. Based on high-frequency SSEC data, VaR backtesting techniques show that the proposed VaR measures outperform many linear and nonlinear GARCH-type models at high-risk levels \cite{Wei-Chen-Lin-2013-PA}.

Lee et al. used the MMAR model for VaR forecasting \cite{Lee-Song-Chang-2016-PA}. Empirical analysis of the daily data of KOSPI, S\&P500, and USD/KRW foreign exchange from 1990 to 2012 shows that MMAR generates more stable and accurate VaR forecasting than Gaussian VaR, $t$-distribution VaR, GARCH VaR, and historical VaR  \cite{Lee-Song-Chang-2016-PA}.
Batten et al. studied the 5-min returns of EUR/USD spot quotes and trading ticks from 5 January 2006 to 31 December 2007 and found that a modified MMAR model performs better than the historical simulation approach and the GARCH(1,1) location-scale VaR model \cite{Batten-Kinateder-Wagner-2014-PA}.

Lux et al. worked on the daily data of WTI oil prices from 2 January 1875 to 31 December 1895 and from 2 January 1985 to 24 March 2014 and found that MSM gives better VaR forecasting than GARCH models \cite{Lux-Segnon-Gupta-2016-EE}.
Ben Nasr et al. studied the daily data of the Global Dow Jones Islamic Market World Index from 1 January 1996 to 2 September 2013 and found that the MSM model performs better than long memory GARCH-type models (FIGARCH and FITVGARCH), which in turn outperform the short-memory GARCH model \cite{BenNasr-Lux-Ajmi-Gupta-2016-IREF}.

Malo investigated the Nord Pool electricity spot prices from March 1998 to January 2006 and found that the Copula-MSM model has better VaR prediction ability than  GARCH models \cite{Malo-2009-PA}.
Liu and Lux considered the daily data of two stock market indices (the Dow Jones composite 65 average index and the NIKKEI 225 average index) from 6 January 1970 to 30 December 2008, two foreign exchange rates (USD/BPG and DEM/GBP) from 1 March 1973 to 31 December 2008, and a bond portfolio of U.S. 1- and 2-year treasury constant maturity bond rates from 1 June 1976 to 31 December 2008 \cite{Liu-Lux-2015-EJF}. They found that the bivariate MSM model with heterogeneous volatility correlations provides better forecasting results than the homogeneous benchmark and the bivariate DCC-GARCH model \cite{Liu-Lux-2015-EJF}.
Herrera et al. proposed a MSM Peaks-Over-Threshold (MSM-POT) model and compared the performance of the MSM-POT model with self-exciting models and a GARCH-EVT approach on the extreme returns of six daily commodity futures prices (Brent and WTI crude oil, cocoa, cotton, copper, and gold) \cite{Herrera-Rodriguez-Pino-2017-EE}. They found that the MSM-POT model provides the most accurate VaR forecasts.

Bacry et al. compared the performances of MRW and normal and $t$-Student GARCH(1,1) models in 1-day conditional VaR forecasting using the daily prices of four foreign exchange rates (CAD, JPY, CHF and GBP) against the USD \cite{Bacry-Kozhemyak-Muzy-2008-JEDC}. Compared to GARCH-type models, the MRW model provides comparative or better results at different VaR levels from 0.5\% to 10\%.
The advantage of the MRW model over GARCH-type models in VaR forecasting has been confirmed for the daily returns of  five stock market indices (CAC40, FUTSEE, DAX, DJIA and NIKKEI) from 1990 and 2005 \cite{Bacry-Kozhemyak-Muzy-2013-QF}. 

\subsection{Complex networks}
\label{S2:MF:Appl:Networks}

\subsubsection{Multiscale cross-correlation coefficients}
\label{S3:MF:Appl:DCCA:Coef}

The superposition law $F^2_{x+y}(s) = F^2_{x}(s) + F^2_{x}(s)$ presented in Eq.~(\ref{Eq:DFA:DMA:F2x:F2hF2g}) is derived based on the assumption that the two time series are independent. Inspired by the superposition law, Yang proposed a new, maybe the first, multiscale correlation coefficient \cite{Yang-2005-MPLB}
\begin{equation}
  \rho_{xy}(s) =\frac{F_{x+y}^2(q,s)-F^2_{x}(s) - F^2_{x}(s)}{F^2_{x}(s) + F^2_{x}(s)}.
\end{equation}
The correlation $\rho_{xy}(s)=1$ at all scales if $x=y$.

Zebende defined the multiscale detrended cross-correlation coefficient (known as DCCA coefficient for short) as  \cite{Zebende-2011-PA}
\begin{equation}
  \rho_{\rm{DCCA}}(s) =\frac{F_{xy}^2(q)}{F_{xx}(q)F_{yy}(q)}
  \label{Eq:MF:DCCA:Coefficient}
\end{equation}
where $F_{xy}(s)$ and $F_{xx}(s)$ are respectively the overall fluctuation functions from the DFA-based DCCA method and the DFA method. The coefficient $\rho_{\rm{DCCA}}(s)$ can be easily extended to define other multiscale cross-correlation coefficient based on different cross-correlation analysis, such as the DCCA with general local trends besides the polynomial trends \cite{Podobnik-Jiang-Zhou-Stanley-2011-PRE}, the detrending moving-average cross-correlation ($\rho_{\rm{DMCA}}$) \cite{Kristoufek-2014b-PA}, the cross-correlation analysis ($\rho_{\rm{CCA}}$) with signs \cite{Kwapien-Oswiecimka-Drozdz-2015-PRE}, the joint structure function approach ($\rho_{\rm{HXA}}$) \cite{Kristoufek-2017-CNSNS}. There are also wavelet-based cross-correlation coefficients or wavelet coherence as discussed in Section \ref{S3:MF-X-WT}. To assess the significance level of the empirically estimated $\rho(s)$ values, one should perform statistical tests with proper null models \cite{Podobnik-Jiang-Zhou-Stanley-2011-PRE}.

The properties of the DCCA coefficient and its advantage over conventional Pearson coefficient have been extensively studied through numerical experiments with mathematical models \cite{Zebende-2011-PA,Podobnik-Jiang-Zhou-Stanley-2011-PRE,Kwapien-Oswiecimka-Drozdz-2015-PRE,Kristoufek-2014b-PA,Kristoufek-2017-CNSNS,Balocchi-Varanini-Macerata-2013-EPL,Kristoufek-2014a-PA,Zhao-Shang-Huang-2017-Fractals}. It has been proved for $\rho_{\rm{DCCA}}$ \cite{Podobnik-Jiang-Zhou-Stanley-2011-PRE}, $\rho_{\rm{DMCA}}$ \cite{Podobnik-Jiang-Zhou-Stanley-2011-PRE,Kristoufek-2014b-PA}, and $\rho_{\rm{CCA}}$ \cite{Kwapien-Oswiecimka-Drozdz-2015-PRE} that
\begin{equation}
  -1 \leq \rho(s) \leq 1,
\end{equation}
where $\rho(s)=-1$ if $X=-Y$, $\rho(s)=1$ if $X=Y$, and $\rho(s)=0$ if $X$ and $Y$ are uncorrelated. It is reported that the CCA coefficient $\rho_{\rm{CCA}}$ performs better than the DCCA coefficient $\rho_{\rm{DCCA}}$ as $\rho_{\rm{CCA}}$ can capture better the intrinsic cross correlation between time series \cite{Kwapien-Oswiecimka-Drozdz-2015-PRE} and the DMCA coefficient $\rho_{\rm{DMCA}}$ outperforms $\rho_{\rm{DCCA}}$ and $\rho_{\rm{HXA}}$ since the variance of $\rho_{\rm{DMCA}}$ decays the fastest \cite{Kristoufek-2017-CNSNS}.

Kwapie{\'{n}} et al. extended the multiscale DCCA coefficient to the $q$-dependent detrended cross-correlation coefficient (or multiscale $q$DCCA coefficient) based on MF-DFA and MF-CCA \cite{Kwapien-Oswiecimka-Drozdz-2015-PRE}:
\begin{equation}
  \rho_{xy}(s,q) =\frac{F_{xy}^q(q,s)}{\sqrt{F_{xx}^q(q,s)F_{yy}^q(q,s)}}.
  \label{Eq:MF:qDCCA:Coefficient}
\end{equation}
When $q=2$, Eq.~(\ref{Eq:MF:qDCCA:Coefficient}) reduces to Eq.~(\ref{Eq:MF:DCCA:Coefficient}). They proved that
\begin{equation}
  -1 \leq \rho_{xy}(s,q) \leq 1,
\end{equation}
when $q>0$, which does not hold when $q<0$. The performance of the $q$DCCA coefficient has been investigated with cross-correlated, partially cross-correlated and uncorrelated time series \cite{Kwapien-Oswiecimka-Drozdz-2015-PRE}.

Similarly, the DPXA or DPCCA coefficients are defined as  \cite{Yuan-Fu-Zhang-Piao-Xoplaki-Luterbacher-2015-SR,Qian-Liu-Jiang-Podobnik-Zhou-Stanley-2015-PRE}
\begin{equation}
  \rho_{\rm{DPXA}}(s) =\frac{F_{xy:z}^2(q)}{F_{xx:z}(q)F_{yy:z}(q)},
  \label{Eq:MF:DPXA:Coefficient}
\end{equation}
which considers common external driving forces on time series $X$ and $Y$. Considering the USD index as a common external factor of the prices of crude oil and gold, it is found that the intrinsic multiscale cross-correlation coefficients between gold and oil are lower than their crude counterparts $\rho_{{\rm{g}},{\rm{o}}:{\rm{d}}}(s) <\rho_{{\rm{g}},{\rm{o}}}(s)$ \cite{Qian-Liu-Jiang-Podobnik-Zhou-Stanley-2015-PRE}. Lin et al. studied the net cross-correlations and influence among five major world gold markets (London, New York, Shanghai,
Tokyo and Mumbai) at different time scales and identified the London gold market as the most influential \cite{Lin-Wang-Xie-Stanley-2018-PA}. There are other similar measures like partial wavelet coherence \cite{Mihanovic-Orlic-Pasaric-2009-JMS,Ng-Chan-2012-JAOT}.

\subsubsection{Multiscale DCCA networks}

In contrast to the Pearson correlation coefficients, the scale-dependent DCCA coefficients are less sensitive to contaminated noises and the amplitude ratio between slow and fast components \cite{Piao-Fu-2016-SR}. Hence, the DCCA coefficients are used to construct multiscale DCCA matrices for a bundle of financial time series, where the element for scale $s$ is $\rho_{ij}(s)$ \cite{Wang-Xie-Chen-Yang-Yang-2013-PA,Wang-Xie-Chen-Chen-2013-Entropy,Wang-Xie-Chen-2017-JEIC}. Here, the term ``DCCA'' refers to the general meaning that is not constrained to detrended cross-correlation analysis \cite{Podobnik-Stanley-2008-PRL} and different types of DCCA coefficients presented in Section~\ref{S3:MF:Appl:DCCA:Coef} can be adopted. Such a construction results in multiscale DCCA matrices.

The multiscale DCCA correlation or distance matrices can be studied under the framework of random matrix theory, mainly to identify the information contents in eigenvalues deviating from their distribution predicted from the null hypothesis and their corresponding eigenvectors \cite{Laloux-Cizean-Bouchaud-Potters-1999-PRL,Plerou-Gopikrishnan-Rosenow-Amaral-Stanley-1999-PRL,Plerou-Gopikrishnan-Rosenow-Amaral-Guhr-Stanley-2002-PRE}.
Wang et al. compared the results of DCCA matrices and Pearson correlation matrices using the daily closing prices of 462 constituent stocks of S\&P 500 index from 3 January 2005 to 31 August 2012 \cite{Wang-Xie-Chen-Yang-Yang-2013-PA}. They found that the eigenvalue distributions of random DCCA matrices are different from those of random Pearson correlation matrices. The minimum and maximum eigenvalues and the ensemble eigenvalue distributions can be empirically obtained for a given sample of financial time series. Similar to the case of the correlation matrix \cite{Plerou-Gopikrishnan-Gabaix-Stanley-2002-PRE}, the largest eigenvalue of the DCCA matrix reflects the movement mode of the whole market.
Conlon et al. investigated the dynamics and the Epps effect \cite{Epps-1975-AER,Toth-Kertesz-2009-QF,Saichev-Sornette-2014-IJMPC} with
time-varying multiscale MODWT cross-correlation matrices of daily returns of 49 Euro Stoxx 50 stock from May 1999 to August 2007 and the 1-min returns of these stocks from May 2008 to April 2009 \cite{Conlon-Ruskin-Crane-2009-ACS}. They found that the optimal portfolio depends not only on scale but also on the moving window size.
Sun and Liu used the DCCA matrix to study the optimal portfolio strategy, in which the matrix element is the average DCCA coefficient over different scales \cite{Sun-Liu-2016-PA}.

The clustering features and hierarchical structures of multiscale DCCA distance matrices are often investigated through their minimum spanning trees (MSTs) \cite{Mantegna-1999-EPJB} or planar maximally filtered graphs (PMFGs) \cite{Tumminello-Aste-DiMatteo-Mantegna-2005-PNAS}.
Wang et al. investigated the daily FX rates of 44 major currencies in the period of 2007-2012 and found that the exchange rates exhibit geographic features in the MSTs. In addition, the normalized tree length, the average shortest path length and the maximum degree decrease with $s$ \cite{Wang-Xie-Chen-Chen-2013-Entropy}.
Wang et al. studied the multiscale DCCA MSTs and PMFGs of the daily closing prices of 457 constituent stocks of the S\&P 500 Index during the period from 3 January 2005 to 31 December 2012, where the DCCA coefficients are obtained through the maximal overlap discrete wavelet transform (MODWT) \cite{Wang-Xie-Chen-2017-JEIC}. They found that the network structure fluctuates at different timescales and there is an evident sector clustering effect for stocks.

Multifractal extension of DCCA coefficients results in $q$-dependent DCCA coefficients $\rho_{ij}(q,s)$ or MF-DCCA coefficients \cite{Kwapien-Oswiecimka-Drozdz-2015-PRE}, which can be used to form  MF-DCCA correlation or distance matrices \cite{Kwapien-Oswiecimka-Forczek-Drozdz-2017-PRE}. Empirical analysis of the 1-min returns of the 100 largest stocks traded on the New York Stock Exchange over the period of 1998-1999 shows that the structures of the $q$-MST graphs vary across different $q$ and $s$ values, such that the $q$-MSTs complement the MF-DCCA coefficients $\rho_{ij}(q,s)$  in disentangling ``hidden'' correlations that cannot be uncovered from the DCCA MST graphs \cite{Kwapien-Oswiecimka-Forczek-Drozdz-2017-PRE}. In addition, more diverse cross-correlations among return time series lead to more variable tree topology for different $q$'s.

The basic DCCA and MF-DCCA matrices are fully connected, which can be manipulated to construct other networks. Unweighed networks can be deduced by posing thresholds on the DCCA coefficients. Two nodes are connected if and only if the absolute DCCA coefficient exceeds the threshold. Alternatively, a more natural threshold is determined by the critical value that the DCCA coefficient is significantly different from zero. Moreover, we can also consider the signs of the DCCA coefficients, which should have influences on the clustering of nodes.

\subsubsection{Multifractality-based networks}
\label{S3:MF:Appl:Networks:MF}

Due to the bivariate nature of the generalized Hurst exponent function $H_{ij}(q)$, of the joint mass scaling exponent function $\tau_{ij}(q)$ and of the joint multifractal spectrum $f_{ij}(\alpha)$, different networks can be constructed based on these functions. The main concern lies in their performance compared with other methods.

Catalano and Figliola constructed the matrix whose element is the bivariate Hurst exponent $H_{ij}$ between the return time series of two assets $i$ and $j$ \cite{Catalano-Figliola-2015-QQ}. An unweighted, undirected network is then introduced whose element $M_{ij}=1$ if $|H_{ij}-0.5|>u$ and 0 otherwise. They chose the threshold $u$ to be $u=0.04$ and studied the daily price data of 6 commodity indices (industrial metal, precious metal, grain, energy, softs, and livestock) constructed from 24 commodities by Dow Jones, which cover the period from 2 January 1991 to 23 April 2012. They found that the two networks in the last seven years are more densely connected.
While this idea of network construction is suggestive, one should remember that, for financial returns, the $H_{ij}$ exponents are usually close to 0.5. Therefore, athreshold of $u=0.04$ does not guarantee a statistically significant deviation from 0.5.

Lin et al. defined the average Hurst surface distance as  \cite{Lin-Shang-Zhou-2014-ND}
\begin{equation}
  d_{ij} = \frac{\langle[H_{ii}(q,s)-H_{jj}(q,s)]^2\rangle^{\frac{1}{2}}}{\langle{H_{ij}(q,s)}\rangle}~,
\end{equation}
where the average is taken over $q$ and $s$ \cite{Lin-Shang-Zhou-2014-ND}. The distance matrix is able to cluster stock market indices in different geographic regions, but not for individual stocks \cite{Lin-Shang-Zhou-2014-ND}. Natural extensions of the average Hurst surface distance matrix is to vary the order $q$ and the scale $s$.

Xu and Beck introduced the R{\'e}nyi difference matrix $R_{ij}$ \cite{Xu-Beck-2017-EPL} (see also Sec.~\ref{S3:MF:Symbolic}),
\begin{equation}
  D_{ij} = \frac{1}{q_{\max} -q_{\min}} \int_{q_{\min}}^{q_{\max}} |K_i(q)-K_j(q)|^\kappa {\rm{d}}q
\end{equation}
where $q_{\max}$ and $q_{\min}$ are the maximum and minimum values of $q$ under consideration, $\kappa$ is a parameter, and $K(q)$ is the R\'enyi entropy
\begin{equation}
  K(q)=\lim_{k\to\infty}\frac{-I_q^{(k)}}{k}
  = \lim_{k\to\infty}\frac{1}{1-q}\frac{1}{k}\ln\sum^{\omega(k)}_{j=1}p_j^q.
\end{equation}
Although this method is initially introduced from the symbol sequences mapped from financial time series, it is generally suitable for other multifractal analysis methods. The R{\'e}nyi difference matrix has the potential to identify stocks with unusual behaviors \cite{Xu-Beck-2017-EPL}.

%

Similarly, we can define the multifractality difference matrix based on $H(q)$, $\tau(q)$ and $f(\alpha)$. Taking $H(q)$ as an example, we have
\begin{equation}
  D_{ij} = \frac{1}{q_{\max} -q_{\min}} \int_{q_{\min}}^{q_{\max}} |H_i(q)-H_j(q)| {\rm{d}}q,
\end{equation}
which represents the difference of multifractality in time series $i$ and $j$. This could be an interesting topic for future research.

\subsubsection{Time series of network variables}
\label{S3:MF:Appl:Networks:TS}

The network nodes are characterized by many local variables, such as degree and cluster coefficient. If the growth process of the network (or the time stamps of nodes) is known, we can obtain the time series of local variables. These time series can be investigated with multifractal analysis and the results might contain useful information about the networks.

One example concerns the networks mapped from time series such that the nodes have time stamps corresponding to the raw time series. Indeed, there are many methods that map a time series to a complex network, in which a data point or a segment of data points converts to a node and an edge is formed due to certain relationship between nodes \cite{Li-Wang-2006-CSB,Zhang-Small-2006-PRL,Xu-Zhang-Small-2008-PNAS,Lacasa-Luque-Ballesteros-Luque-Nuno-2008-PNAS,Marwan-Donges-Zou-Donner-Kurths-2009-PLA}. The statistical properties of the time series of node degree or clustering coefficient can be investigated. Using the nearest neighbour approach \cite{Xu-Zhang-Small-2008-PNAS}, Caraiani constructed three complex networks mapped from the daily returns of the Czech PX from April 1994 to December 2010, the Hungarian BUX from June 1993 to December 2010, and Polish WIG from June 1993 to December 2010 and found that the time series of node clustering coefficient possess multifractality \cite{Caraiani-2012-PA}.

Another example is to label time stamps for the nodes following the movements of a random walker abiding to specific rules \cite{Bonaventura-Nicosia-Latora-2014-PRE,Nicosia-deDomenico-Latora-2014-EPL,Weng-Zhao-Small-Huang-2014-PRE}. Multifractal cross-correlation analysis (MF-CCA) of the time series of degree, clustering coefficient, and closeness centrality can distinguish between Erd{\"{o}}s-R{\'e}nyi, Barab{\'a}si-Albert, and Watts-Strogatz networks \cite{Oswiecimka-Livi-Drozdz-2016-PRE}. Also, the degree of right-sided asymmetry of multifractal spectra of the time series of these local node variables is linked with the degree of small-worldness present in networks \cite{Oswiecimka-Livi-Drozdz-2018-CNSNS}. This method can be applied to financial networks.

\subsubsection{Eigenportfolios}
\label{S3:MF:Appl:Networks:Eigenportfolios}

Denote $\lambda_k$ and ${\mathbf{u}}_k=[u_{k1}, \cdots, u_{kN}]^{\rm{T}}$ as the $k$th largest eigenvalue of the correlation matrix and it associated eigenvector. For each eigenvector, we can construct its eigenportfolio, whose returns are calculated as
\begin{equation}
 R_k(t) = \frac{{\mathbf{u}}_k^{\rm{T}}{\mathbf{r}}(t)}{\sum_{i=1}^{N} u_{ki}}
\label{Eq:Partial:Factor}
\end{equation}
where ${\mathbf{r}}(t)=[r_1(t),\cdots,r_i(t),\cdots, r_N(t)]^{\rm{T}}$, and ${\mathbf{u}}_k^{\rm{T}}{\mathbf{r}}$ denotes the projection of the return time series on the eigenvector ${\mathbf{u}}_{k}$. The use of the normalization term is to ensure that the constructed return $R_k$ is the return of a portfolio. It is of interest to investigate the statistical properties of the eigenportfolios. An analysis of the 5-min return time series of 100 stocks in the American market (NYSE and NASDAQ) shows that the eigenportfolio return series for different eigenvectors have different multifractal spectra and the one for the largest eigenvalue $\lambda_1$ is to the right of the average spectrum of all other eigenportfolios \cite{Kwapien-Drozdz-Oswiecimka-2005-APPB,Kwapien-Drozdz-Oswiecimka-2006-PA,Drozdz-Kwapien-Oswiecimka-2007-APPB}.

\subsubsection{Correlation matrix structure and multifractality}
\label{S3:MF:Appl:Networks:withMF}

Correlation matrices are constructed from a bundle of time series and the structure of correlation matrices is expected to be related to the properties of the time series. In an interesting work, Morales et al. unveiled the relationship between correlation matrix structure and multifractal strength \cite{Morales-DiMatteo-Aste-2014-SR}. They constructed a hierarchical tree using clustering algorithms. For each node (or time series), its hierarchical order $n$ can be determined as the number of nodes on the unique path from the root node to the node under consideration. Based on the daily returns of the 342 most capitalised stocks continuously traded on the NYSE  from 2 January 1997 to 31 December 2012, a significant correlation between $n$ and $\Delta{H(1,2)}$ is observed up to certain level of hierarchy.

\subsection{Other applications}
\label{S2:Appl:Others}

\subsubsection{Calibration of agent-based models}
\label{S3:Appl:ModelCalibration}

Agent-based modelling provides a computational view for the understanding of complex human behavior in socioeconomic systems \cite{Sornette-2014-RPP}, which might be ``a better way to help guide financial policies'' \cite{Farmer-Foley-2009-Nature}. Agent-based models are constructed with a set of microscopic rules. These microscopic rules could be incomplete or misspecified. Hence, it is of crucial important to perform model calibrations by comparing the macroscopic properties between the financial time series generated by computational models and by real markets. Besides the main stylized facts, and the multifractal nature of real markets is also considered in model calibration by some researchers.

The Cont-Bouchaud model is a percolation model built on a $d$-dimensional lattice $L^d$, in which the probability $a$ that a cluster of traders want to buy is a key model parameter \cite{Cont-Bouchaud-2000-MeD}. The multifractal analysis of the returns using the structure function approach illustrates that the scaling exponent function $\xi(q)$ varies with model parameters and properly determined parameters are able to produce results comparable to the DJIA index \cite{Castiglione-Stauffer-2001-PA}. An incomplete list of other studies include the multifractal analysis of the return recurrence intervals \cite{Meng-Ren-Gu-Xiong-Zhang-Zhou-Zhang-2012-EPL} in the modified Mike-Farmer model \cite{Mike-Farmer-2008-JEDC,Gu-Zhou-2009-EPL}, the multiscaling behavior of cumulative absolute returns in a double auction model \cite{Li-Zhang-Zhang-Zhang-Xiong-2014-IS}, and multifractality in the time series in a continuum percolation model \cite{Xiao-Wang-2014-IJNSNS}.

\subsubsection{Detection of outliers}
\label{S3:Appl:OutlierDetection}

There are efforts to use multifractal models for the detection of outliers in time series. Based on the bivariate MSM model of Calvet, Fisher and Thompson \cite{Calvet-Fisher-Thompson-2006-JEm}, Idier derived indicators for the detection of crises and extreme volatilities and applied the method to pairs of stock market indices (NYSE, FTSE, DAX and CAC) \cite{Idier-2011-EJF}.

There are also studies attempting to provide early warning signals of outliers using multifractal strength.
Cristescu et al. explored the possibility of using intermittency to predict the occurrence of financial crises \cite{Cristescu-Stan-Scarlat-Minea-Cristescu-2012-PA}.
da Fonseca et al. defined the area variation rate based on the analogous specific heat $-\tau''(q)$ and suggested
that it has potential power in identifying historical crashes \cite{daFonseca-Ferreira-Muruganandam-Cerdeira-2013-PA}.
Morales et al. calculated the dynamical generalized Hurst exponents through the weighted structure function approach and found hints that  the multifractal strength $\Delta{H(1,2)}=H(1)-H(2)$ can be a tool to monitor unstable periods in financial time series \cite{Morales-DiMatteo-Gramatica-Aste-2012-PA,Morales-DiMatteo-Aste-2013-PA}.
Fern{\'{a}}ndez-Mart{\'{i}}nez et al. suggested to use $H(1)$ as an indicator for the identification of herding behavior and bubble onset \cite{FernandezMartinez-SanchezGranero-Torrecillas-Mckelvey-2017-Fractals}

%


\subsubsection{Trading strategy}
\label{S3:Appl:PricePred}

There are quite a few applications of the partition function approach to asset prices or indices directly, showing illusionary multifractality (see Sec.~\ref{S2:MF:Emp:MFPF:Illustions}). A characteristic feature of the illusionary multifractality is that the singularity spectrum shrinks to the vicinity of $(\alpha,f)=(1,1)$ such that the singularity width $\Delta\alpha$ is close to 0. However, there seems to be information content in illusionary multifractality. Maillet and Michel found that the singularity width positively correlates to the market shock index quantifying the probability of exceeding the current volatility values, in which the Spearman correlation coefficient is equal to 0.87 \cite{Maillet-Michel-2003-QF}.
Some researchers report that the width of the estimated multifractal spectrum is correlated to the price fluctuation in the future and thus can be used to predict price fluctuations \cite{Sun-Chen-Wu-Yuan-2001-PA,Wei-Huang-2005-PA,Su-Wang-2009-JKPS}. Analogous to $\Delta\alpha$, some researchers also used another quantity \cite{Sun-Chen-Wu-Yuan-2001-PA,Sun-Chen-Yuan-Wu-2001-PA,Wei-Huang-2005-PA,Su-Wang-2009-JKPS,Guo-Huang-Cheng-2012-Kybernetes}
\begin{equation}
  \Delta{f} = f(\alpha_{\max})-f(\alpha_{\min}),
\end{equation}
which is reported to be able to uncover some information about price fluctuations.

Alternatively, Dewandaru et al. used the inefficiency index $D$, together with momentum factors to construct factor models for the forecasting of future returns \cite{Dewandaru-Masih-Bacha-Masih-2015-PA}. The analysis was conducted on daily returns in moving windows of three years of a sample of Islamic stocks listed in the Dow Jones Islamic market from 1996 to 2011. They found that the strategy based on multifractal strength produced monthly excess returns of 6.12\%, outperforming other momentum strategies with higher Sharpe ratio.

%

\section{Perspectives}
\label{S1:Perspectives}

Multifractal analysis for univariate and multivariate financial time series has attracted a strong interest in the econophysics community, caused mainly by the invention of new methods and the easier accessibility to huge amount of financial data. There are still many new multifractal analysis methods being proposed, most of which are variants of existing classical methods. In this direction, we argue that one must assess the performance of the new methods with extensive numerical experiments on mathematical models and compares their performance with other competing methods. It is not easy to invent new methods with better performance. However, such methods are still valuable in deepening our understanding. A pivotal task in multifractal analysis of empirical time series is to choose a method. However, there is no consensus on which method is the best. Although it is possible to rank the performance of different methods for a specific mathematical model, the results often changes for different mathematical models. Summarising the existing literature, overall, we can recommend methods based on DFA, DMA and wavelet leaders.

Another pivotal task in empirical multifractal analysis is to determine the suitable scaling range, over which the scaling exponents are estimated. Usually, a small change of the scaling range will result in significant variations in the estimated exponents. It is not uncommon that the mass scaling exponents and the multifractal spectrum extracted from empirical time series are problematic, especially when there are linear or nonlinear trends in the time series. Consequently, a wrong estimate of the degree and nature of multifractal is obtained, which has harmful effects for our understanding the behavior of financial markets. The choice of the scaling range is more important and difficult for short time series, which is common for low-frequency financial time series. Although many subjective (by eye-balling) and objective criteria have been proposed, further studies on the determination of the scaling range are required. The main reason is that the existing criteria are specific to different degrees to the cases under study and thus lack universality.

Theoretically and practically, it is of great significance to dissect the sources of apparent multifractality for empirical time series. Macroscopically, one usually considers the linear correlations, the nonlinear correlations and the non-Gaussian distribution of the time series. Although it is not clear if the intrinsic multifractality is introduced by the nonlinearity only, it is clear that a time series with only linear correlations and fat-tailed distribution cannot produce any multifractality. In empirical studies, measures of multifractality strength are calculated based on the apparent multifractality. We argue that further scrutiny is necessary on this issue. Otherwise, for example, one may obtain an unreliable ranking of the efficiency of markets. On the other hand, researchers have also proposed different classes of agent-based models (ABMs) for financial markets to generate time series that possesses multifractality. Indeed, agent-based modelling provides us a nice viewpoint to understand the sources of multifractality in financial time series from the microscopic perspective, which is unfortunately less explored. Studies can be conducted to identify microscopic ABM rules that cause the emergence of multifractality in macroscopic time series. Furthermore, it helps uncover the relationship between human behavior and multifractality in financial markets.

So far, there is a large consensus on the presence of multifractality in most financial time series, such as returns, volatilities, trading volumes, recurrence intervals and inter-trade durations. Hence, it is not more the priority to continue confirming and adding
novel evidence on the presence of multifractality in these financial quantities. Rather, one should work on other financial quantities. Furthermore, in the study of multifractal cross correlations, common driving forces can be considered, in order to extract the intrinsic multifractal cross-correlations between time series. Other applications of multifractality, which are more demanding, illustrate the usefulness of multifractal analyses in financial markets, such as asset pricing and risk management.

Another very promising direction is to relate multifractal analysis to complex financial networks. One the one hand, we can construct equity networks based on the multifractal nature of financial time series. Some efforts have already made. There are however much work to do. On the other hand, we can study the multifractal nature of complex financial networks and it implications as well. We expect to witness this
research direction to flourish in the coming years.

\section*{Acknowledgments}

\addcontentsline{toc}{section}{Acknowledgments}

We thank Rui-Qi Han, Jun-Chao Ma, Wen-Bo Tang, Yu-Lei Wan, Yan-Hong Yang, Peng Yue, and Xiao Zhang, especially Xing-Lu Gao and Peng Wang for assistance. This work was partially supported by the National Natural Science Foundation of China [grant numbers 71532009, 11375064] and the Fundamental Research Funds for the Central Universities [grant number 222201718006].

\addcontentsline{toc}{section}{References}

\bibliography{Bibliography}

\end{document}